\documentclass[fleqn,10pt]{wlscirep}
\usepackage[utf8]{inputenc}
\usepackage[T1]{fontenc}
\usepackage{subcaption}
\title{Can We `Feel' the Temperature of Knowledge? Modelling Scientific Popularity Dynamics via Thermodynamics}

\author[1,+]{Luoyi Fu}
\author[1,+]{Dongrui Lu}
\author[1]{Qi Li}
\author[1,*]{Xinbing Wang}
\author[2,*]{Chenghu Zhou}

\affil[1]{Electronic Engineering, Shanghai JiaoTong University, China}
\affil[2]{State Key Laboratory of resources and environmental information system, Institute of Geographical Sciences and Natural Resources Research, Chinese Academy of Sciences, China}

\affil[]{yiluofu@sjtu.edu.cn}
\affil[]{cissyldr@sjtu.edu.cn}
\affil[]{liqilcn@sjtu.edu.cn}

\affil[*]{Corresponding author. E-mail: xwang8@sjtu.edu.cn}
\affil[*]{Corresponding author. E-mail: zhouch@lreis.ac.cn}
\affil[+]{these authors contributed equally to this work}

\begin{abstract}

Just like everything in the nature, scientific topics flourish and perish. While existing literature well captures article’s life-cycle via citation patterns, little is known about how scientific popularity and impact evolves for a specific topic. It would be most intuitive if we could ‘feel’ topic’s activity just as we perceive the weather by temperature. Here, we conceive knowledge temperature to quantify topic overall popularity and impact through citation network dynamics. Knowledge temperature includes 2 parts. One part depicts lasting impact by assessing knowledge accumulation with an analogy between topic evolution and isobaric expansion. The other part gauges temporal changes in knowledge structure, an embodiment of short-term popularity, through the rate of entropy change with internal energy, 2 thermodynamic variables approximated via node degree and edge number. Our analysis of representative topics with size ranging from 1000 to over 30000 articles reveals that the key to flourishing is topics’ ability in accumulating useful information for future knowledge generation. Topics particularly experience temperature surges when their knowledge structure is altered by influential articles. The spike is especially obvious when there appears a single non-trivial novel research focus or merging in topic structure. Overall, knowledge temperature manifests topics’ distinct evolutionary cycles.

\end{abstract}
\begin{document}

\flushbottom
\maketitle
%
%
\thispagestyle{empty}


\section*{}

\noindent Scientific impact assessment helps shape scientific development from aspects including investment \cite{citation-and-grants}$^{,}$\cite{interdisciplinary-funding}, promotion policy \cite{citation-and-policy}$^{,}$\cite{diffusion-credit} and individual career \cite{faculty-hiring}$^{,}$\cite{career-choice}. Thanks to its significance and widespread applications, measuring scientific impact has always been one of the most discussed topics in communities of all disciplines. Citation-based analysis always occupies a predominant role for impact assessment because of the quantitative characteristics of citations and more importantly, the positive correlation between citation and scientific influence \cite{early_citation_accumu_impact}$^{,}$\cite{perceived_impact_publication}.
For an article, citation dynamics reveals its temporal evolution of impact \cite{long-term-sci-impact}$^{,}$\cite{atypical-cite}$^{,}$\cite{aging-cite-network} and popularity \cite{popularity-poisson}. For a researcher, the evolution of individual citation statistics portraits his or her activity \cite{hot-streak}, scholar impact dynamics \cite{h-index}$^{,}$\cite{g-index}$^{,}$\cite{citation_cmp} and research interest pattern\cite{quantify-interest-evo}.
For a scientific topic, however, individual or article citation dynamics modeling fails to characterize its life-cycle because this one-dimensional indicator is not capable of exploiting the interplay among academic entities. This raises a fundamental question: how to depict the rise and fall of a scientific topic by leveraging its citation information?\\

\begin{figure}[htbp]
\centering
\begin{subfigure}{0.45\textwidth}
\centering
\includegraphics[width=0.9\linewidth]{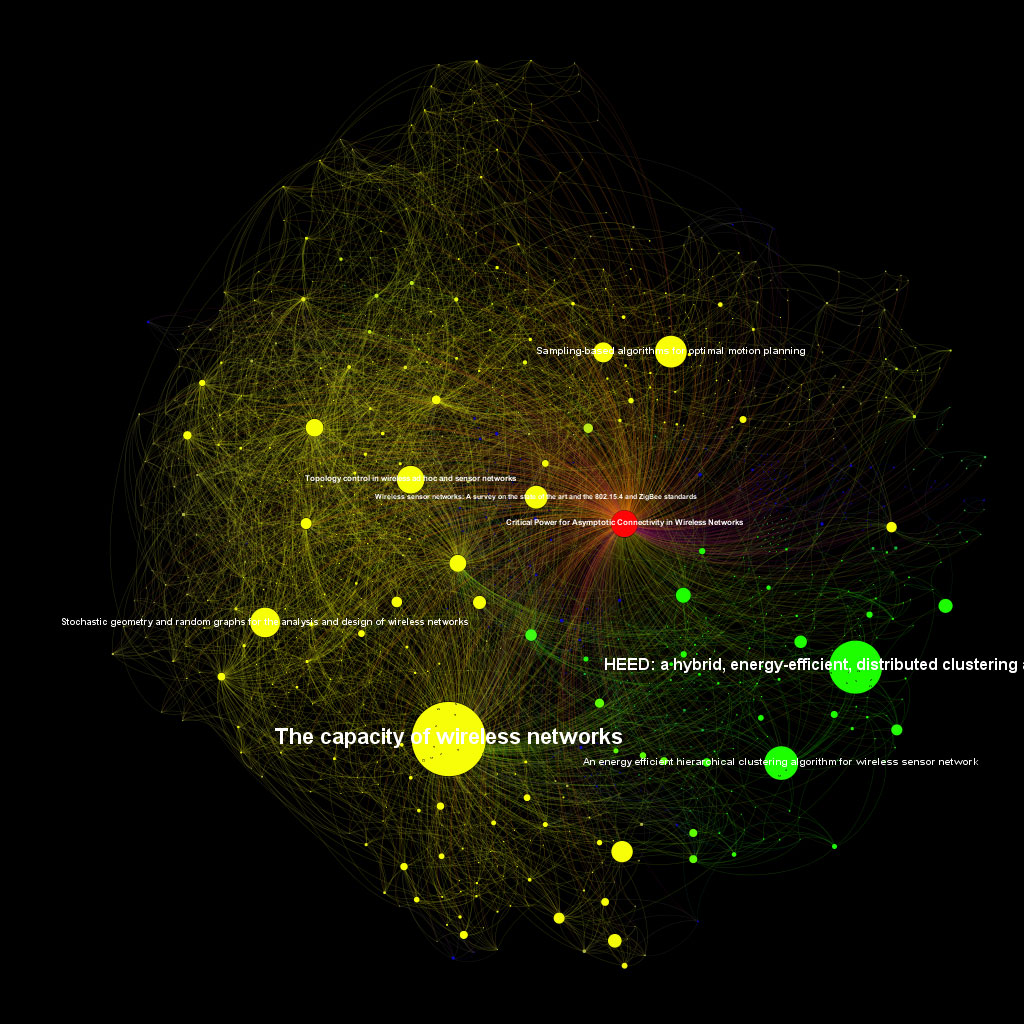}
\caption{}
\end{subfigure}
\begin{subfigure}{0.45\textwidth}
\centering
\includegraphics[width=0.9\linewidth]{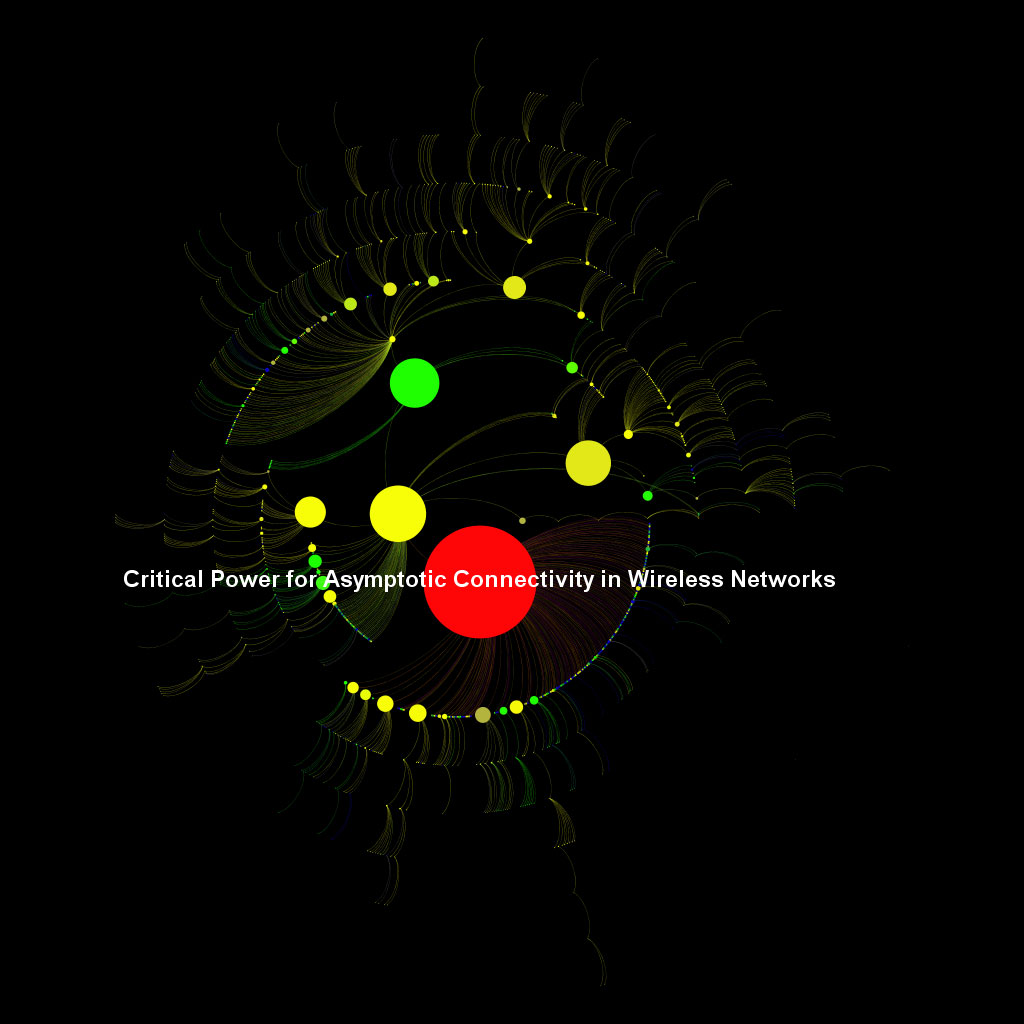}
\caption{}
\end{subfigure}
\begin{subfigure}{0.45\textwidth}
\centering
\includegraphics[width=0.9\linewidth]{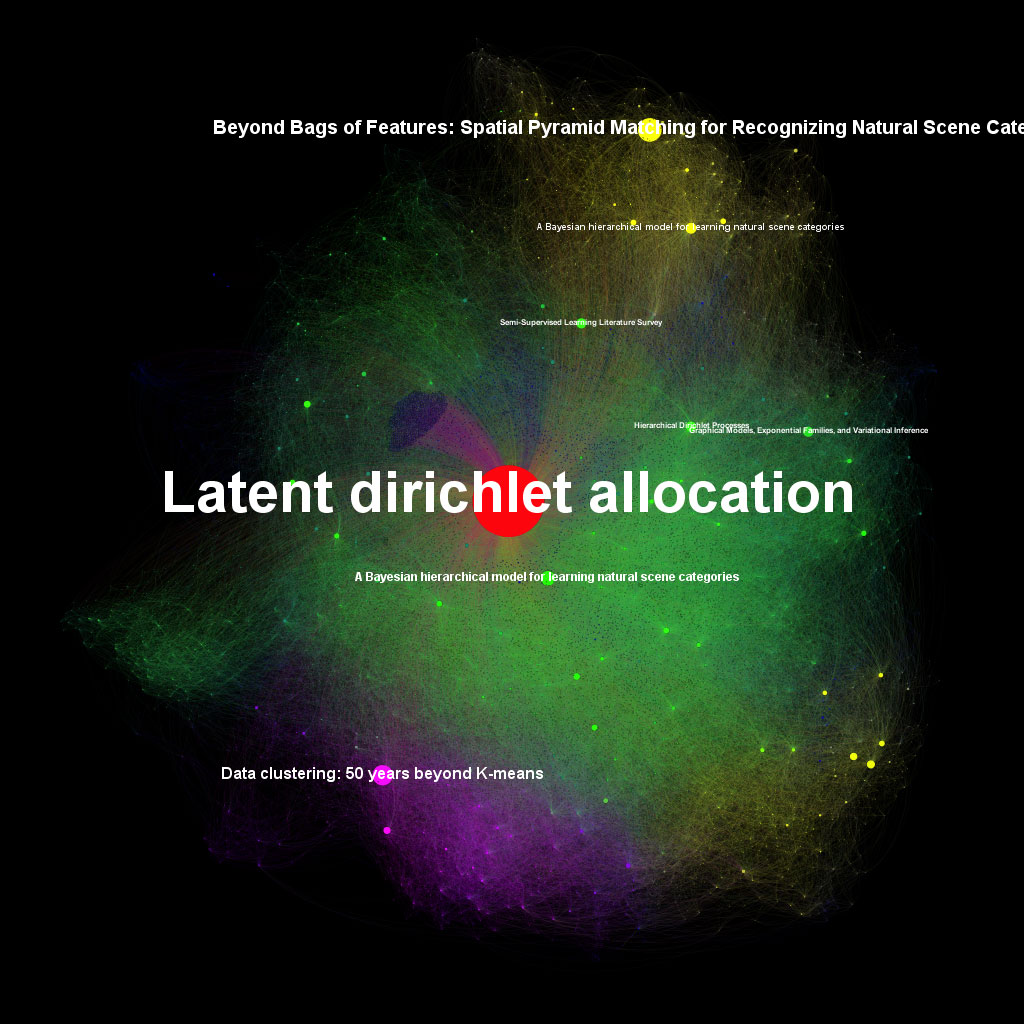}
\caption{}
\end{subfigure}
\begin{subfigure}{0.45\textwidth}
\centering
\includegraphics[width=0.9\linewidth]{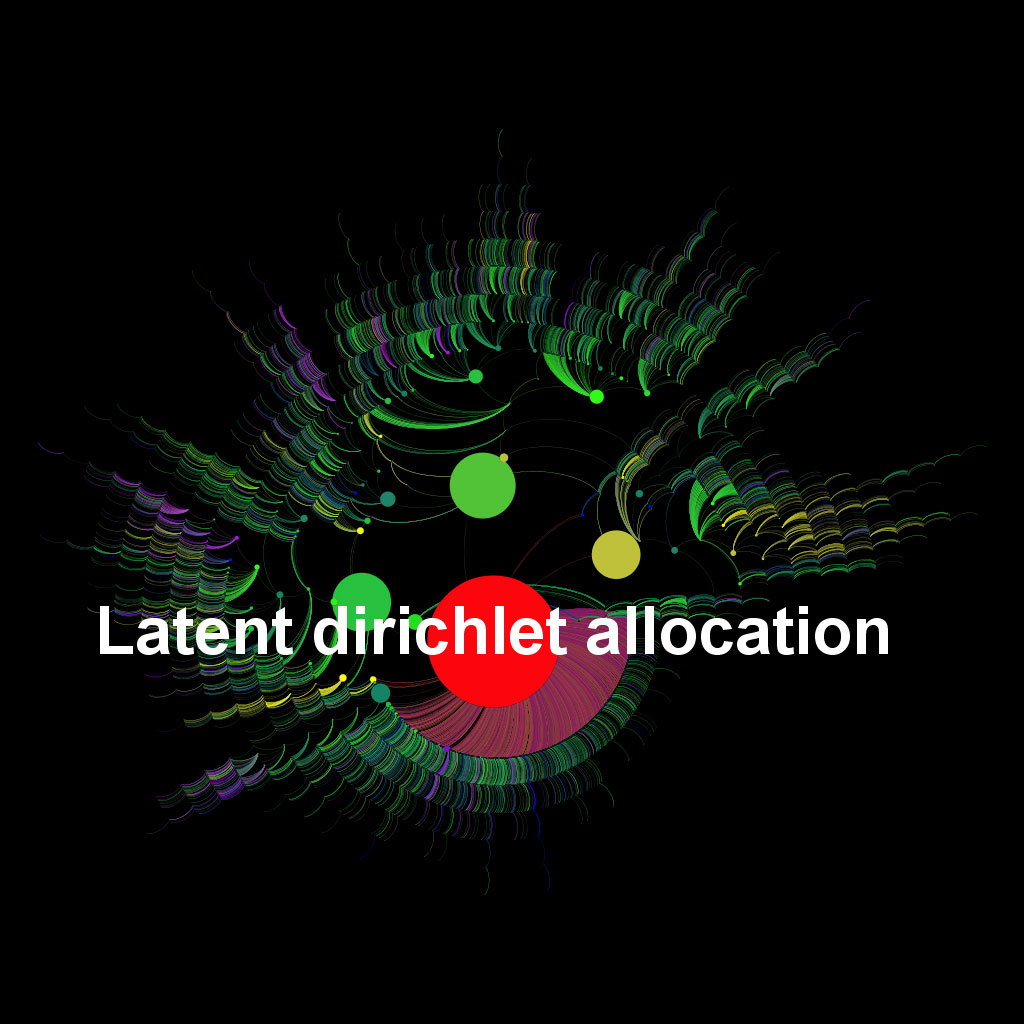}
\caption{}
\end{subfigure}
\begin{subfigure}{0.45\textwidth}
\centering
\includegraphics[width=0.9\linewidth]{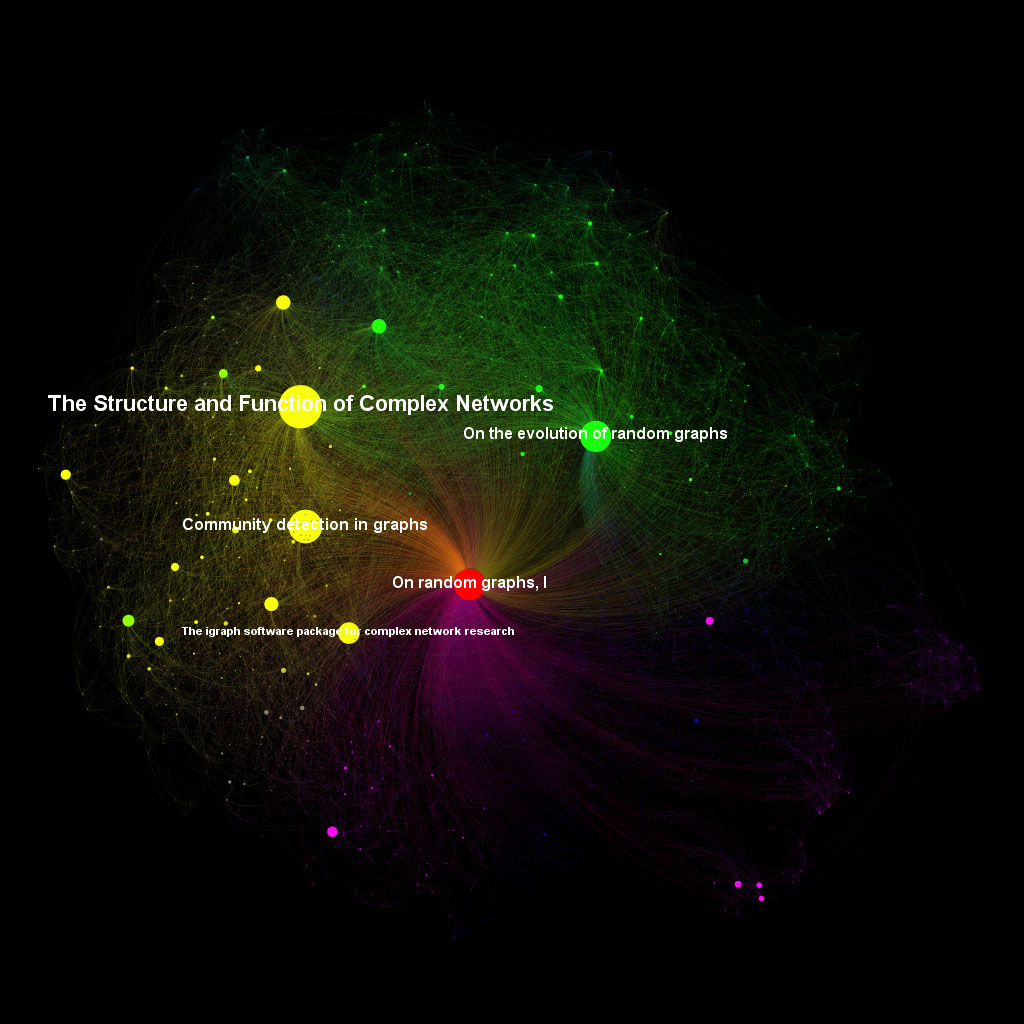}
\caption{}
\end{subfigure}
\begin{subfigure}{0.45\textwidth}
\centering
\includegraphics[width=0.9\linewidth]{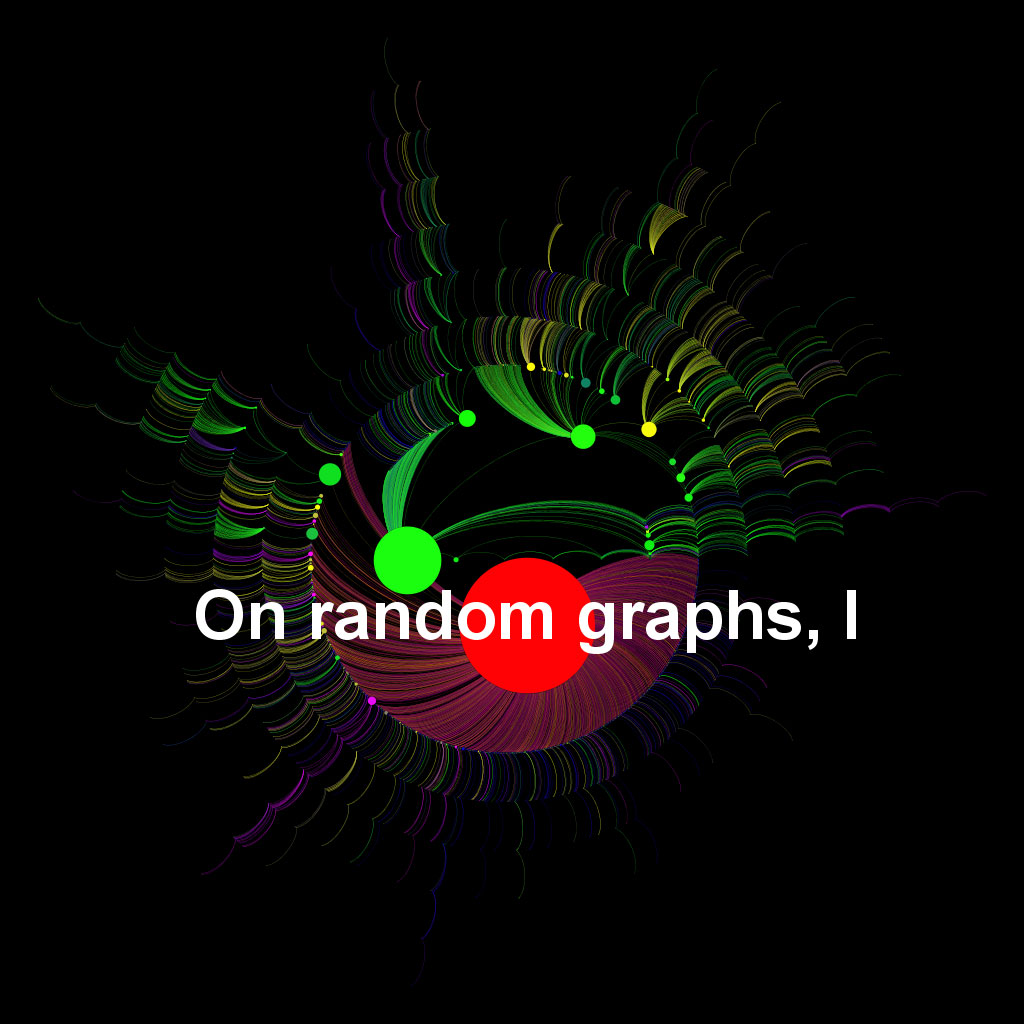}
\caption{}
\end{subfigure}
\caption{}
\label{fig:ima_1}
\end{figure}

\noindent The first step to answer this question is to define scientific topic and then to find an appropriate way to describe it. A scientific topic is in fact a complex network comprising of articles that have similar research interests. As citation is able to display the interaction among articles, we can thus define and represent a scientific topic by its citation network. By retrieving and integrating academic data from renowned databases including but not limited to DBLP, arXiv, Elsevier and Springer, we identified 47310 articles that have gained over 1000 citations and have had a non-trivial influence within their research fields. These articles were published between 1800 and 2019 and their research interests cover 294 domains in 16 disciplines: History, Computer science, Environmental science, Geology, Psychology, Mathematics, Physics, Materials science, Philosophy, Biology, Medicine, Sociology, Art, Economics, Chemistry and Political science. Some of them created new topics while others made major breakthroughs in existing fields. Their immense contribution and inspiration to subsequent researches has made them each a leader in their field of research. To this end, we refer to these papers as pioneering works and define a scientific topic led by each to be a citation network that consists of the pioneering work, child papers, which are all the articles that directly cite the pioneering work, and all the citations among them. We visualize our scientific topics with a graph that we call galaxy map. Galaxy map not only highlights the most influential child papers along with the pioneering work, but also does a preliminary clustering within the topic (Fig. \ref{fig:ima_1}(a,c,e)). We find that while some pioneering works still have an overwhelming impact in the scientific topics they founded, quite a few have several child papers who have established an authority comparable or even greater than themselves. Furthermore, in some of our examples, these prominent child papers seem to have transformed the original topic into multiple new topics (Fig. \ref{fig:ima_1}(e)). Much as galaxy map gives a nice overview of scientific topic's current status, the temporal evolution of scientific topic needs to be further depicted. With this regard, we go beyond the galaxy map representation and dig deeper into the topic citation network for a more intuitive perception of topic's flourishing dynamics.\\

\noindent Since we interpret scientific topics through their citation pattern, topic evolution is reflected by the development of topic citation network. Complicated academic citation networks are springing up all across the science community as a result of the explosive research activity growth, both in and across disciplines, and the prevalence of larger teams \cite{team-assembly-mechanism}$^{,}$\cite{large-team-develop}. The representation and characterization of complex network has attracted a huge amount of efforts, among which an appeal to statistical thermodynamics stands out as a principled school of thought \cite{network-thermo-complexity}. Some studies at the beginning of this century reveal the intimate connections between thermodynamic quantities and complex network dynamics \cite{stat-mechanics-centrality}. Recently, more literature has succeeded in characterizing natural networks \cite{thermo-natural-network}, neuron networks \cite{thermo-network-neuron} and biological networks \cite{thermo-cellular-network} through thermodynamic approaches. In particular, thermodynamic temperature is able to capture critical events in evolving networks \cite{thermodym-evolving-network}. These prior works inspired us in that heat corresponds with popularity and moreover, temperature quantifies partly our body feelings of weather. It would be most direct and intuitive if we could `feel’ topic vigor in the same way as we perceive the weather. Motivated by this thought, we try to depict the flourishing and perishing of scientific topics by measuring their knowledge temperature, a quantity designed to portrait topic impact and popularity evolution by leveraging the rich structural information hidden in citation networks.\\

\begin{figure}[htbp]
 \begin{subfigure}{\textwidth}
 \centering
 \includegraphics[width = \textwidth]{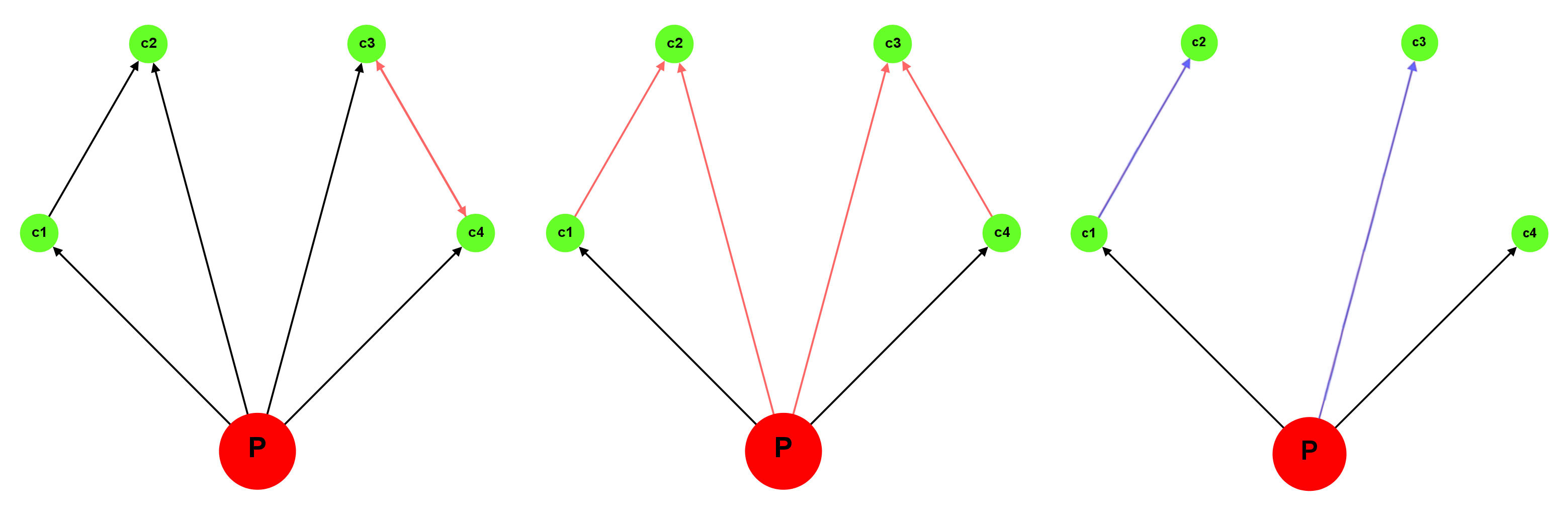}
 \caption{}
 \end{subfigure}

 \begin{subfigure}{\textwidth}
 \includegraphics[width = \textwidth]{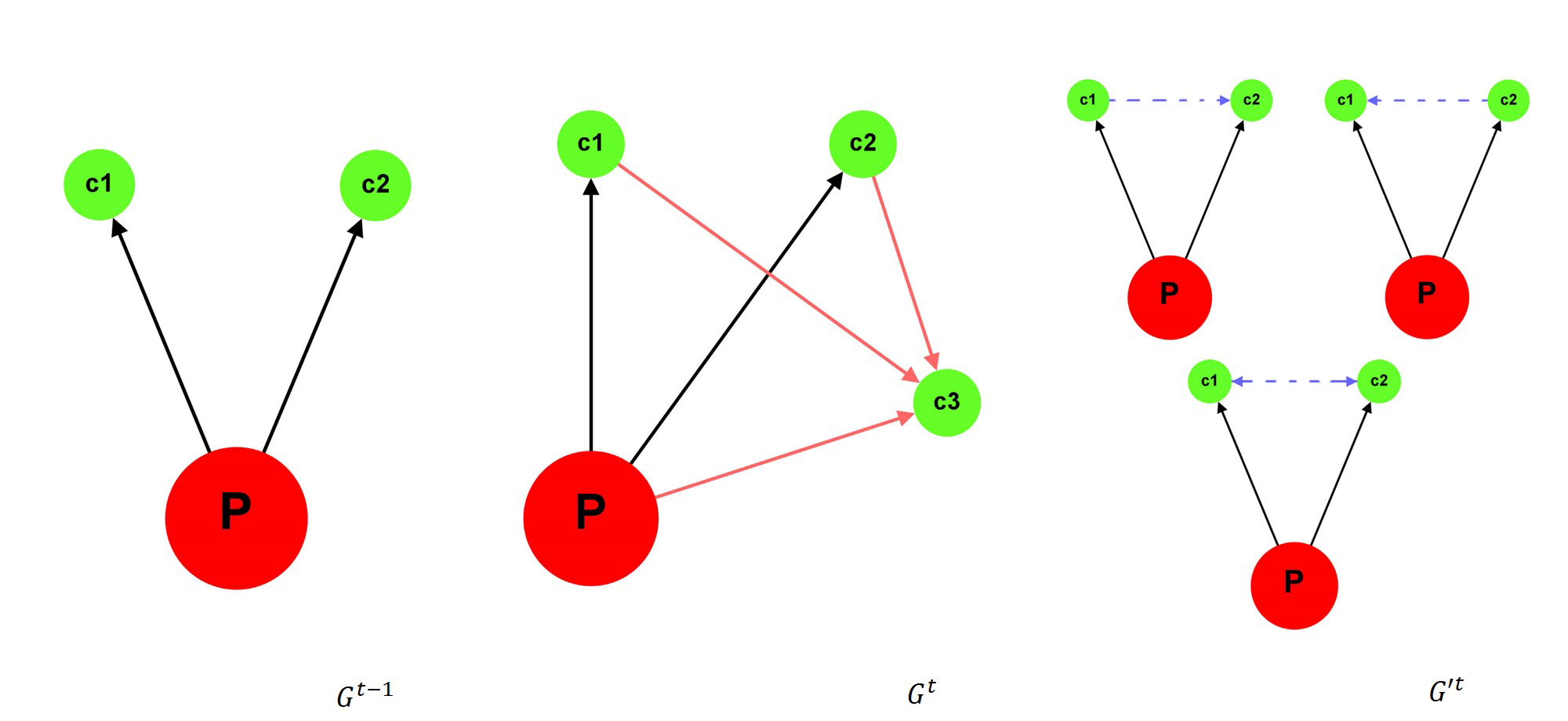}
 \caption{}
 \end{subfigure}
\caption{}
\label{fig:ima_4}
\end{figure}

\noindent Knowledge temperature depends on 3 factors: the evolution of topic size, the evolution of topic knowledge quantity and the advancement of knowledge structure. As knowledge is a sublimation of information and duplicated information is no longer valuable to knowledge generation, measuring knowledge quantity boils down to evaluating the volume of non-overlapped, or useful information. The latter, however, can be estimated by examining paper similarity, which essentially involves determining citation significance. As for knowledge structure, it is also closely related to the question whether a citation is important for an article. Therefore, in order to address the key issue in knowledge temperature conception: citation importance judgement, we extracted skeleton tree for each topic (Fig. \ref{fig:ima_1}(b,d,f)). Skeleton tree provides a more lucid topic representation than galaxy map and accentuates the most essential idea inheritance within the topic by preserving the most valuable citation for every child paper. In particular, we are able to answer 2 fundamental questions by tracing down a path in skeleton tree: from what thought an idea is greatly inspired and what new idea it has directly inspired. From another perspective, skeleton tree demonstrates certain clustering effect in its leaves as it puts intimately related articles together. We employed graph embedding techniques to extract topic skeleton tree. We first measured the importance of every citation in the topic based on structural information and then simplified topic citation network in 2 steps: firstly, remove the loops in the citation network and secondly, leave out relatively unimportant citations while ensuring the global connectivity (Fig. \ref{fig:ima_4}(a)). Because the extraction process involves a thorough investigation into citation network structure, topic skeleton tree serves as an indispensable tool for our knowledge temperature design and for the heat distribution visualization within the topic.\\

\noindent We evaluated topic knowledge temperature from 2 aspects: topic growth and recent structural change in topic knowledge. Our core idea is to make an analogy between topic citation network $G^t$ and ideal gas. At timestamp $t$, we define topic knowledge temperature $T^t$ as:
\begin{equation}
    T^t = T_{growth}^t + T_{structure}^t
\end{equation}
where $T_{growth}^t$ measures knowledge increment and $T_{structure}^t$ estimates the magnitude of changes in knowledge structure between 2 consecutive timestamps. \\

\noindent We initialized $T_{growth}^t$ by combining 2 ideal gas's internal energy expressions and updated $T_{growth}^t$ via ideal gas state equation, $PV = nRT$, under the assumption that $G^t$'s expansion is an isobaric process. With pressure $P$ being invariant and $R$ being constant, the variation of $T_{growth}^t$ is governed by the dynamics of topic mass $n_t$ and topic volume $V_t$. From a macroscopic view of information and knowledge, $n_t$ measures the total amount of overlapped information whereas $V_t$ represents the total amount of information. A simple qualitative analysis shows that $T_{growth}^t$ increases when topics succeed in accumulating distinct, or useful information, the knowledge source for the future. Intuitively, promising topics are able to attract a steady or even growing inflow of new information. On the contrary, staggering topics consume more useful information than they receive and their potential eventually drops. A rising $T_{growth}^t$ indicates an increasingly solid and rich knowledge base and thus reflects a topic's growing impact. Furthermore, an accelerating increase in $T_{growth}^t$ suggests a topic's greater capability in useful information collection and thus its faster gain in fame. \\

\noindent Inspired by the temperature design in prior work \cite{thermodym-evolving-network}$^{,}$\cite{thermodym-network-graph-poly}, we computed $T_{structure}^t$ between every two adjacent timestamps by making an analogy between $G^t$'s evolution and an isochoric process. The analogy is legitimate as long as the node number is fixed, which unfortunately does not hold for $G^t$. In order to solve this issue, we designed a graph shrinking algorithm that transforms the newcomers from timestamp $t-1$ and $t$ into virtual citations among nodes in $G^{t-1}$ (Fig. \ref{fig:ima_4}(b)). We defined $T_{structure}^t$ as the average structural change brought by a node in $G^t$:
\begin{equation}
    T_{structure}^t = \left|\frac{\frac{dU^t}{dS^t}}{|V^t|}\right| = \left|\frac{\frac{{U}'^t-U^{t-1}}{{S}'^t-S^{t-1}}}{|V^t|}\right|
\end{equation}
where $S^{t-1}$, ${S}'^t$ are the von Neumann entropy \cite{vonNeumann-entropy} of $G^{t-1}$  and ${G'}^t$, the weighted reduced graph of $G^t$ and $U^{t-1}$, ${U}'^t$ their internal energy. We approximated von Neumann entropy by node degree and set internal energy to be the number of edges for simplicity. Different from $T_{growth}^t$ which focuses more on continual knowledge increment, $T_{structure}^t$ is designed to capture recent critical events and hence assesses topic's short-term popularity.\\

\begin{figure}[htbp]

\begin{subfigure}{0.5\textwidth}
\centering
\includegraphics[width = 0.9\linewidth]{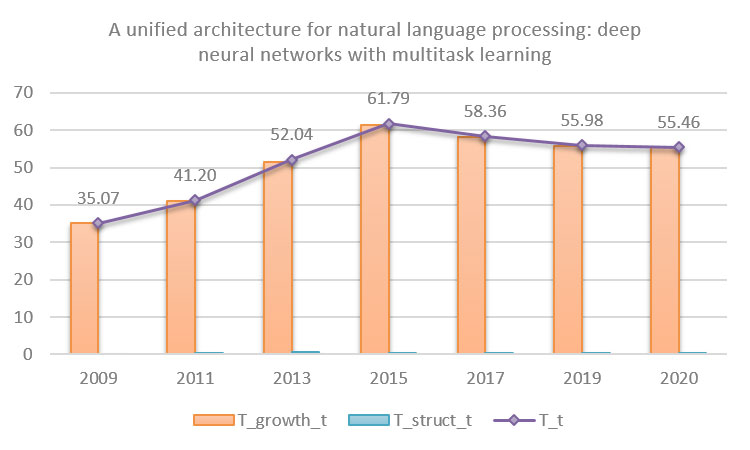}
\caption{}
\end{subfigure}
\begin{subfigure}{0.5\textwidth}
\centering
\includegraphics[width = 0.9 \linewidth]{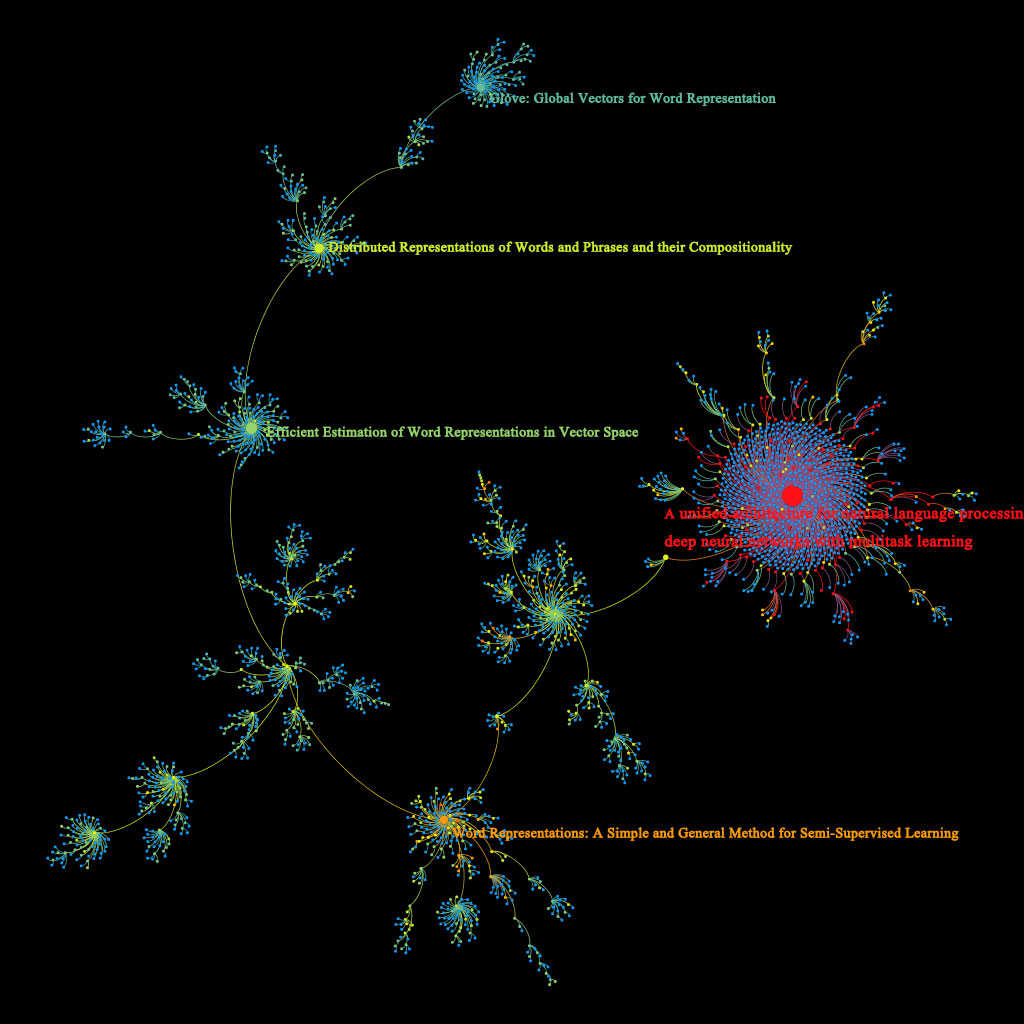}
\caption{}
\end{subfigure}

\begin{subfigure}{\textwidth}
\begin{minipage}[t]{0.33\linewidth}
\includegraphics[width = \linewidth]{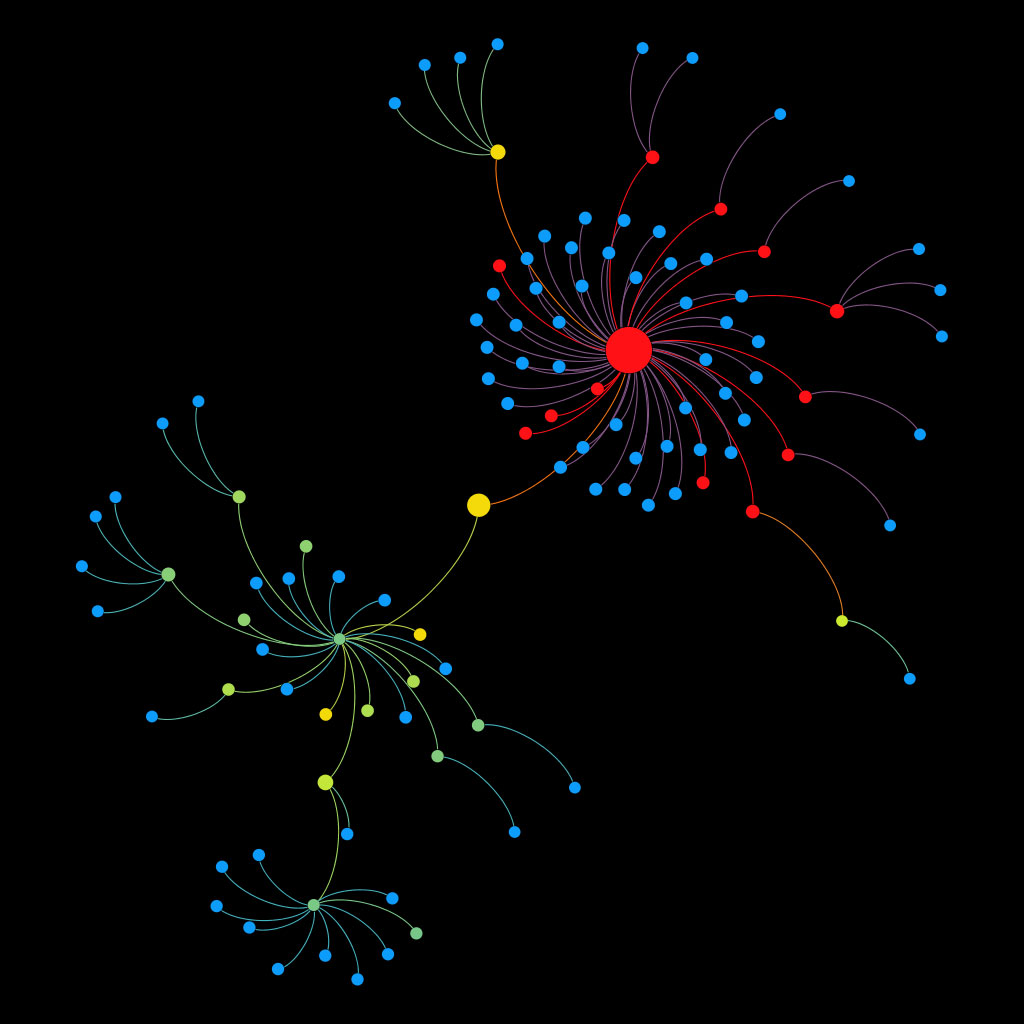}
\caption{}
\end{minipage}
\begin{minipage}[t]{0.33\linewidth}
\includegraphics[width = \linewidth]{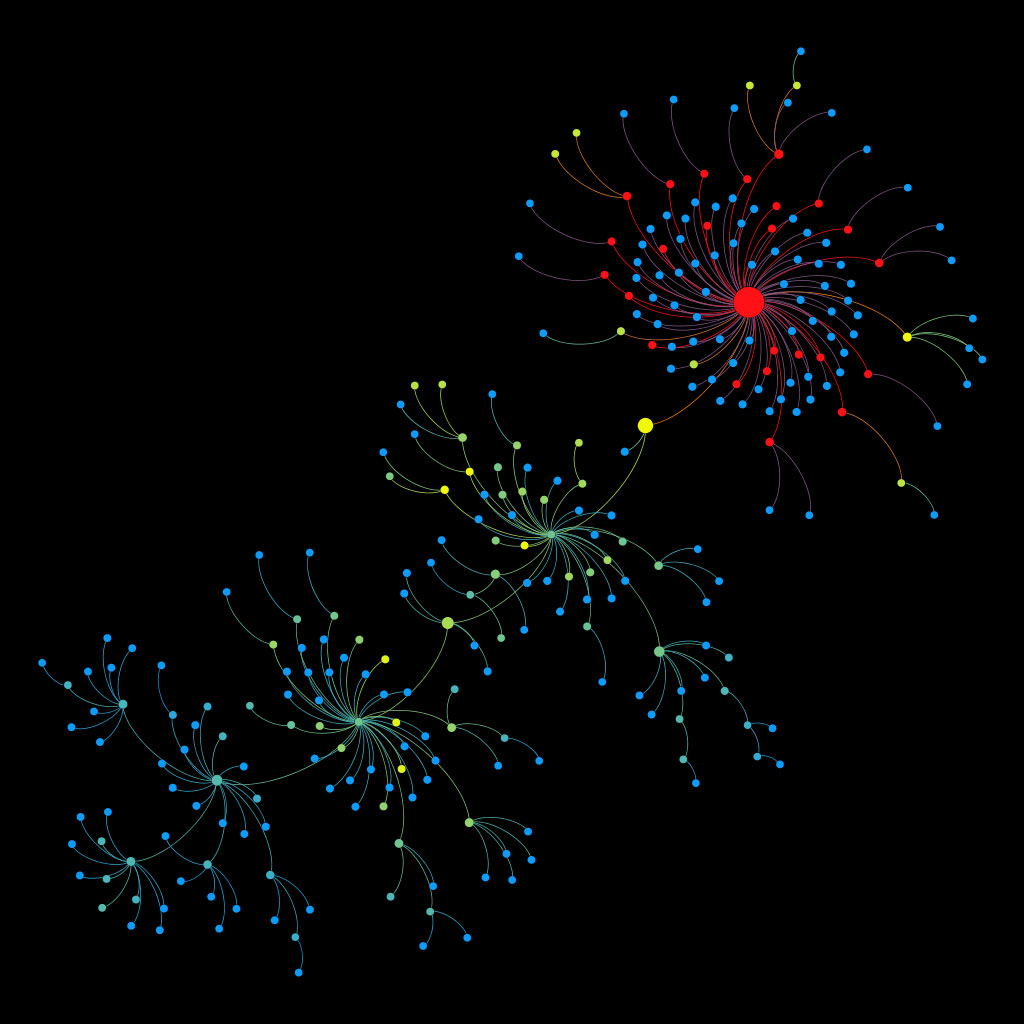}
\caption{}
\end{minipage}
\begin{minipage}[t]{0.33\linewidth}
\includegraphics[width = \linewidth]{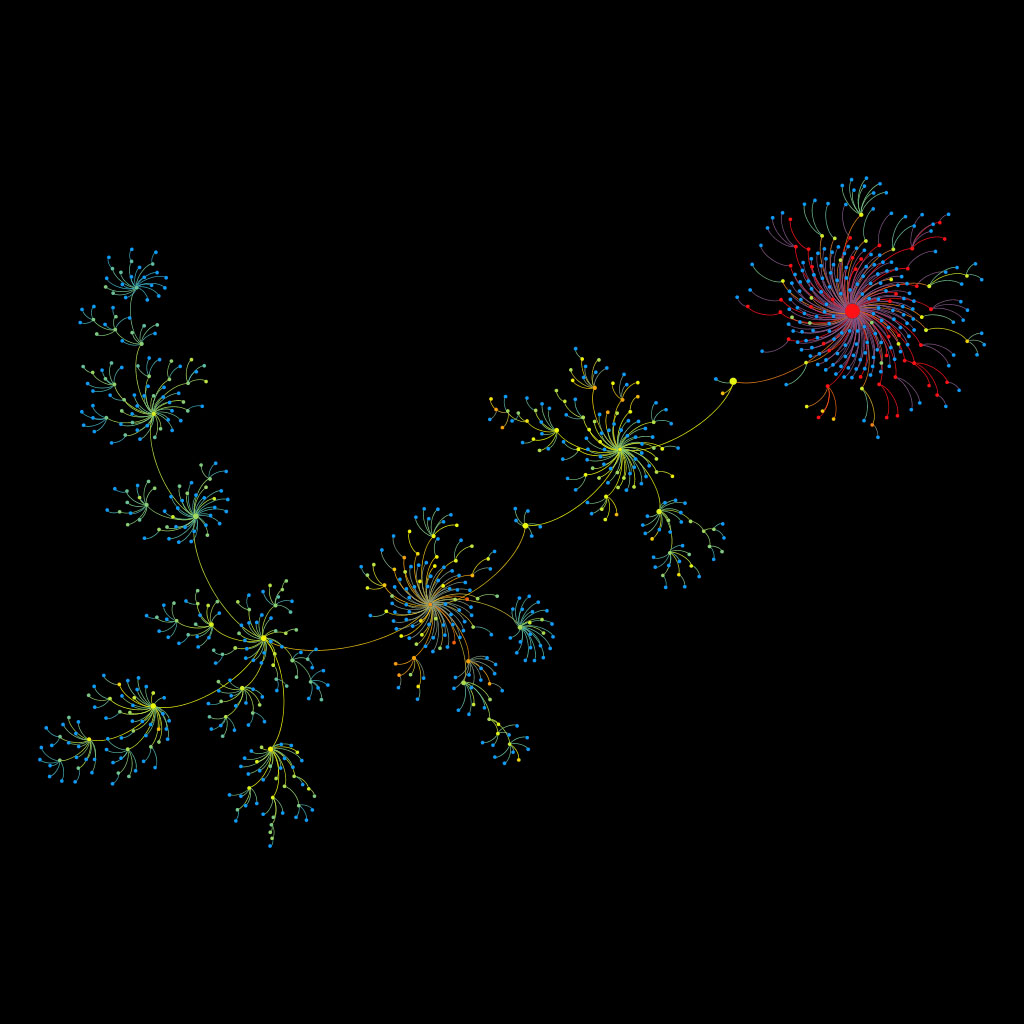}
\caption{}
\end{minipage}
\end{subfigure}

\begin{subfigure}{0.5\textwidth}
\centering
\includegraphics[width = 0.9\linewidth]{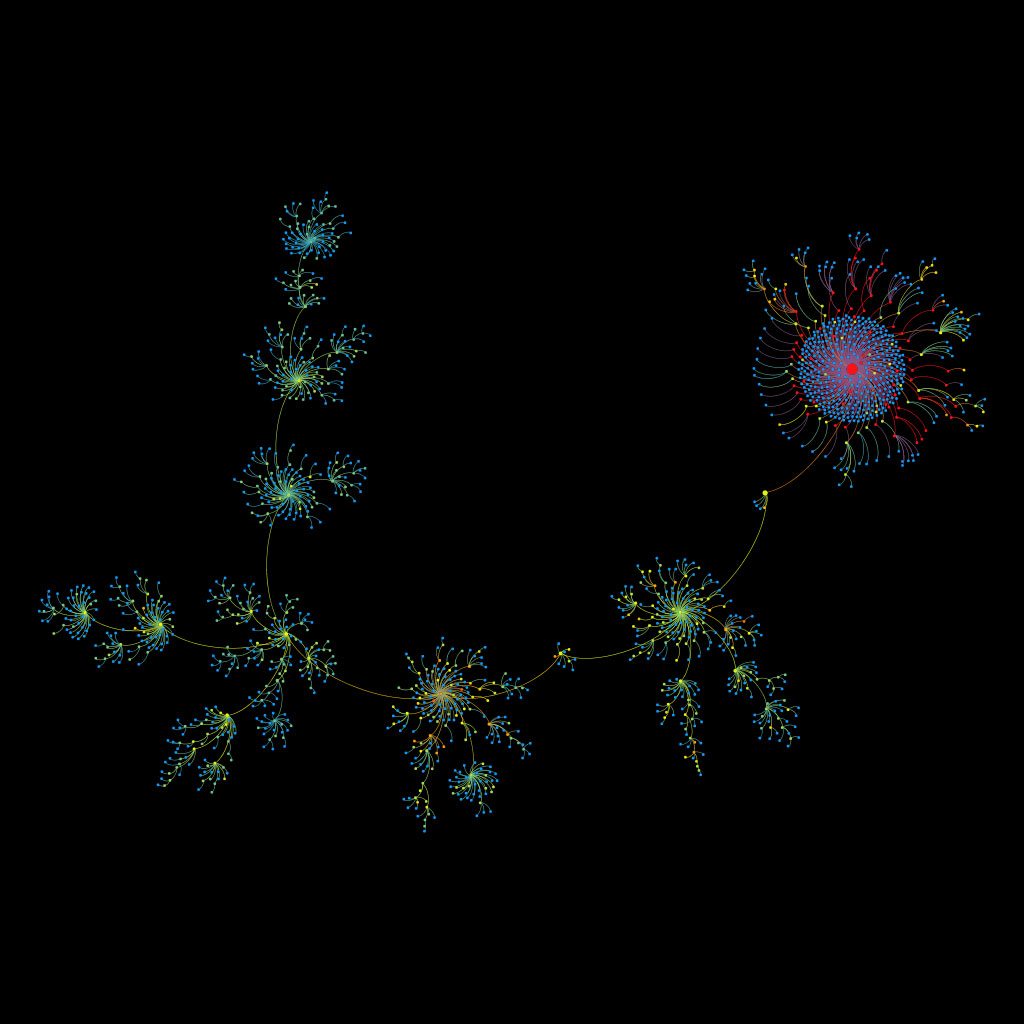}
\caption{}
\end{subfigure}
\begin{subfigure}{0.5\textwidth}
\centering
\includegraphics[width = 0.9\linewidth]{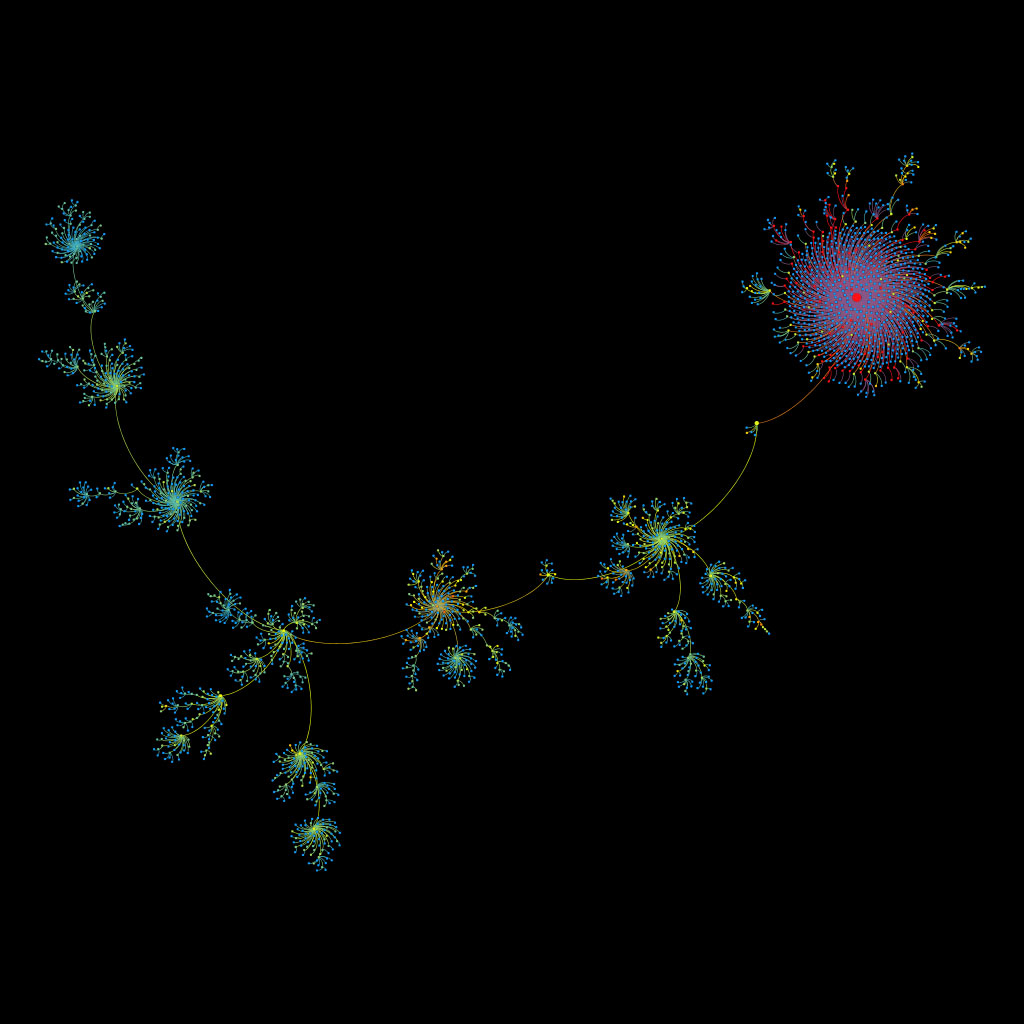}
\caption{}
\end{subfigure}

\caption{}
\label{fig:ima_2}
\end{figure}

\begin{figure}[htbp]

\begin{subfigure}{0.5\textwidth}
\centering
\includegraphics[width = 0.8\linewidth]{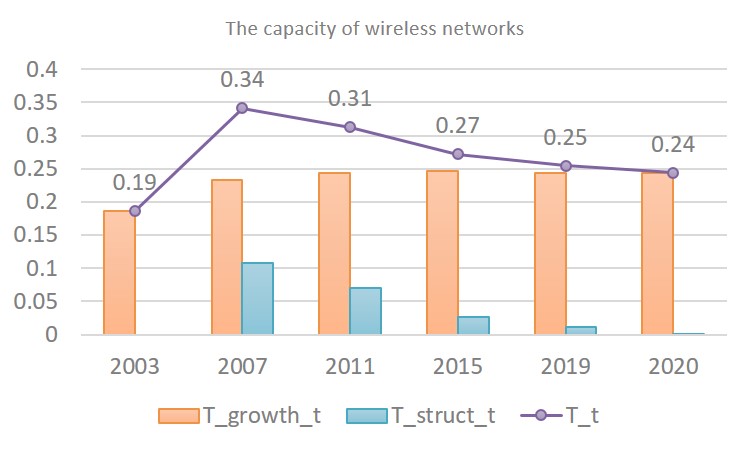}
\caption{}
\end{subfigure}
\begin{subfigure}{0.5\textwidth}
\centering
\includegraphics[width = 0.8\linewidth]{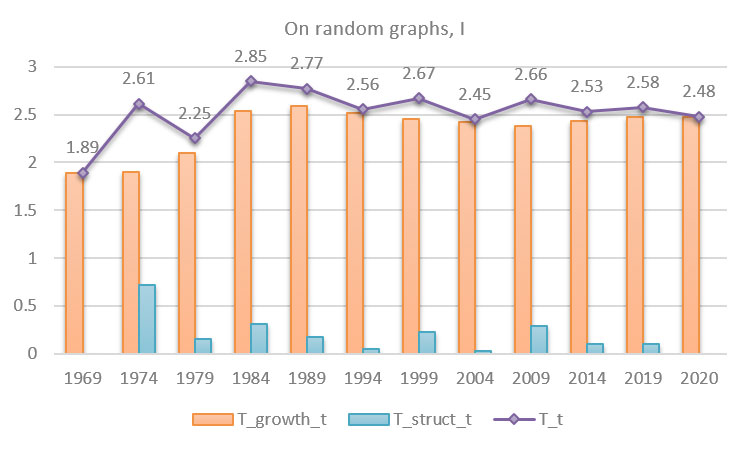}
\caption{}
\end{subfigure}

\begin{subfigure}{\textwidth}
\begin{minipage}[t]{0.5\linewidth}
\centering
\includegraphics[width = 0.9\linewidth]{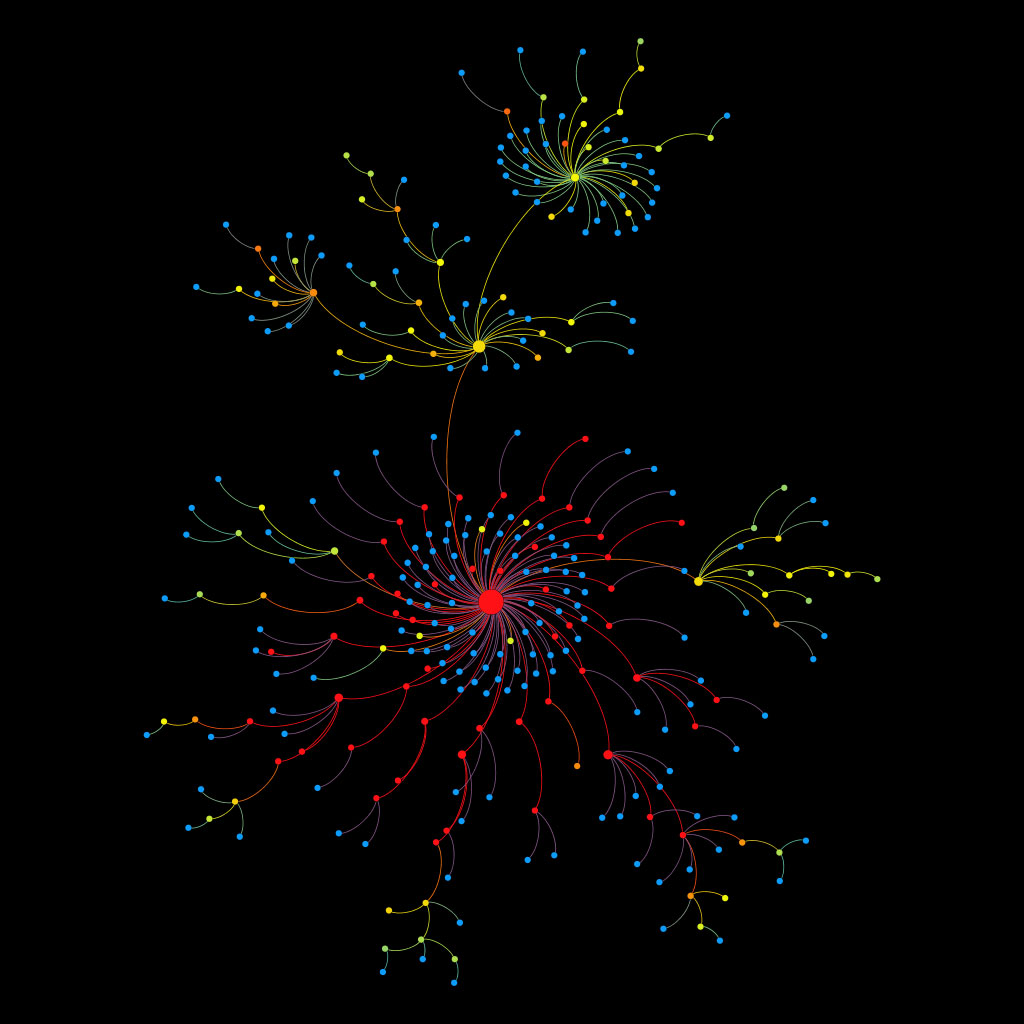}
\caption{}
\end{minipage}
\begin{minipage}[t]{0.5\linewidth}
\centering
\includegraphics[width = 0.9\linewidth]{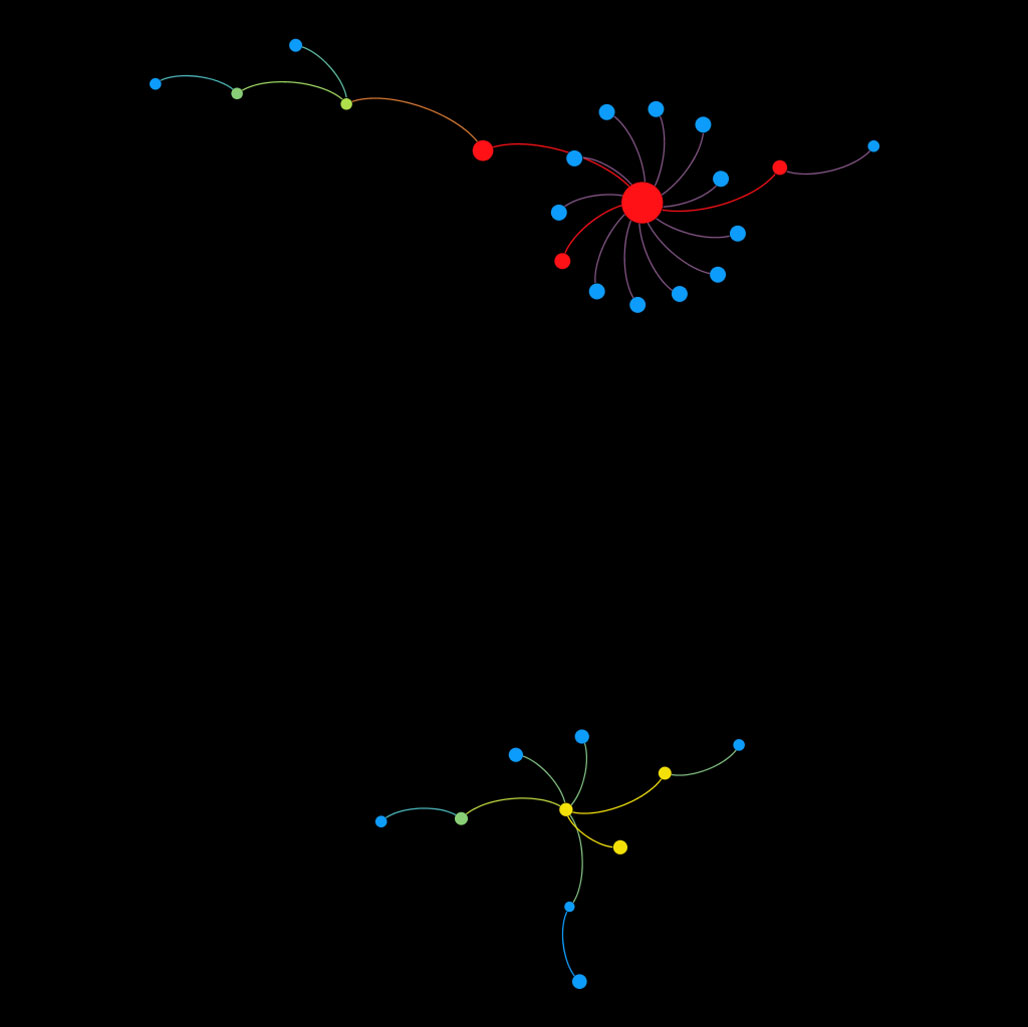}
\caption{}
\end{minipage}
\end{subfigure}

\begin{subfigure}{\textwidth}
\begin{minipage}[t]{0.5\linewidth}
\centering
\includegraphics[width = 0.9\linewidth]{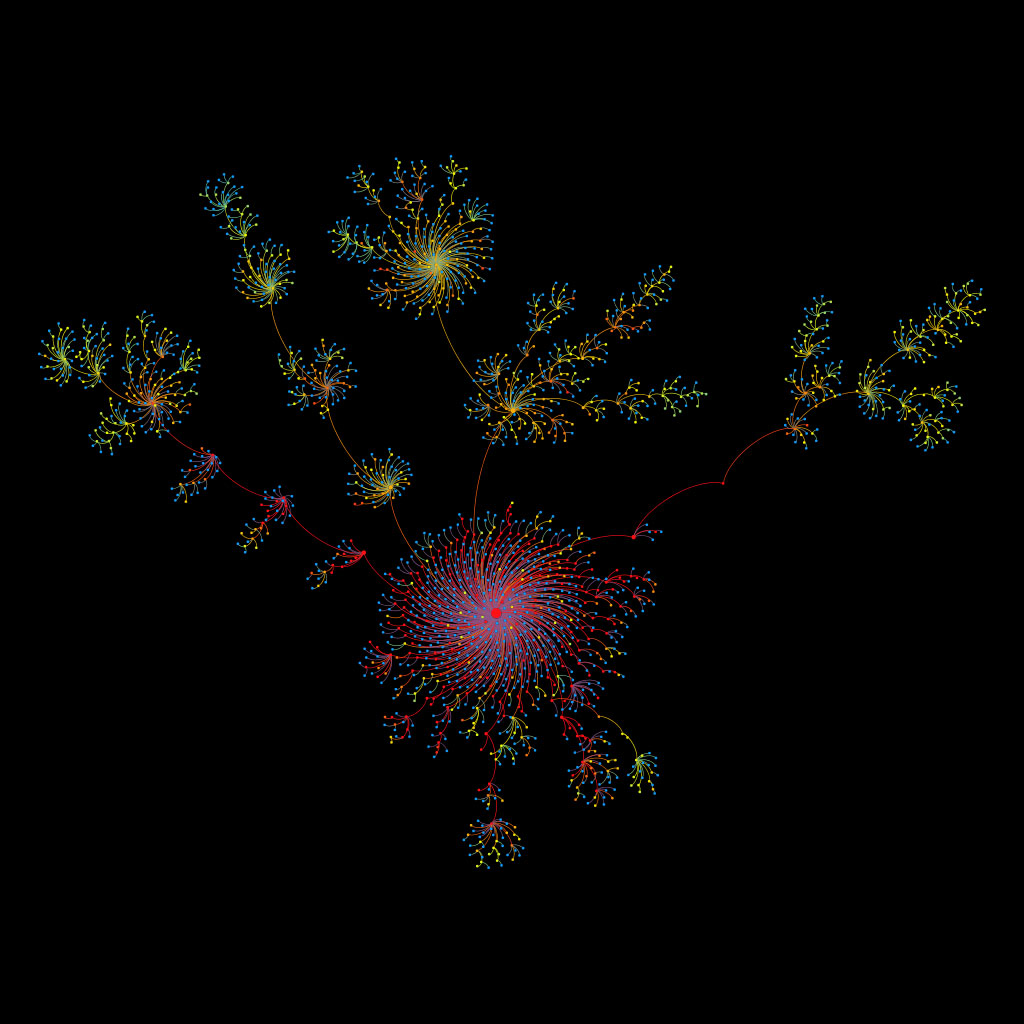}
\caption{}
\end{minipage}
\begin{minipage}[t]{0.5\linewidth}
\centering
\includegraphics[width = 0.9\linewidth]{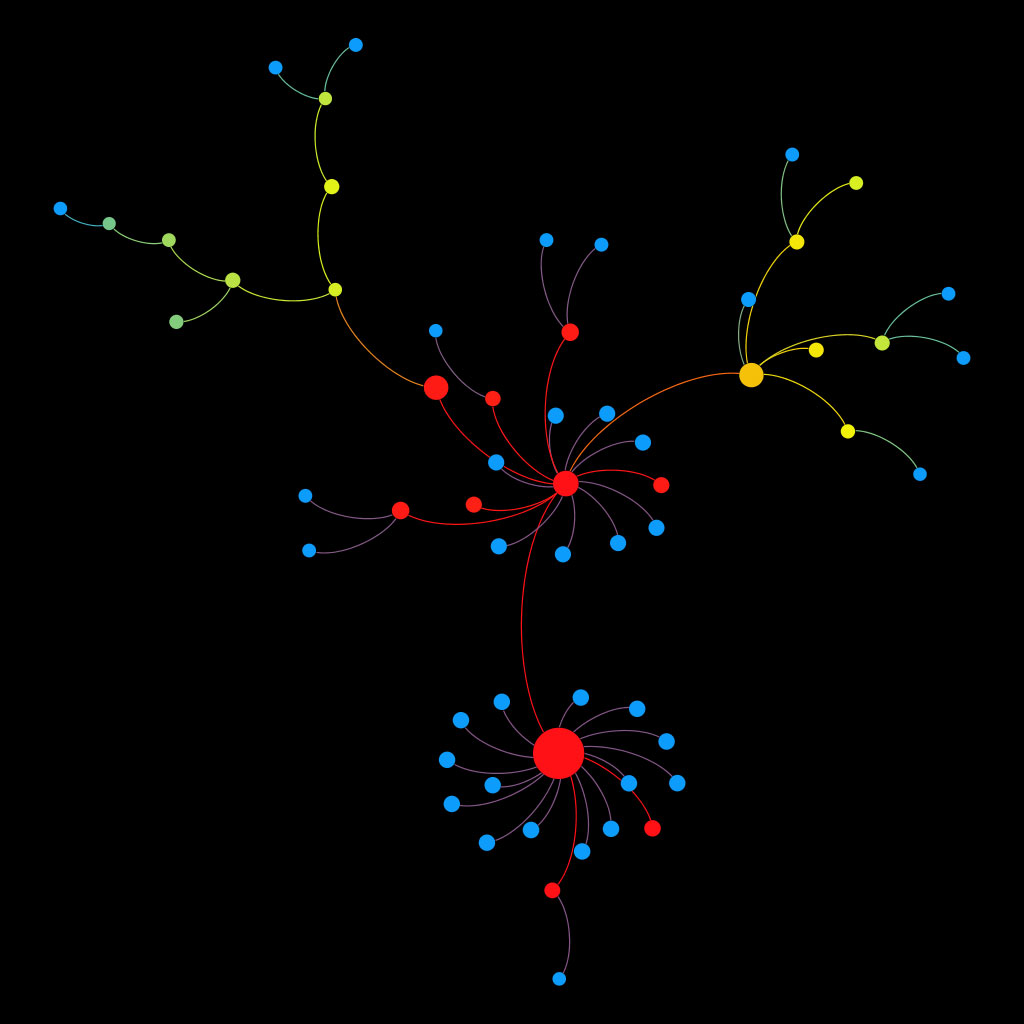}
\caption{}
\end{minipage}
\end{subfigure}

\caption{}
\label{fig:ima_3}
\end{figure}

\noindent Among all the topics, we identified 16 representative topics to conduct our knowledge temperature experiment. These articles were published between 1959 and 2014 and their research interests fall in domains including machine learning, wireless network, graph theory, biology and physics. These topics have sizes ranging from over 1000 articles and approximately 5000 citations to more than 31000 articles and nearly 200 thousand citations. We find that the temporal evolution of $T^t$ well depicts topic flourishing, with $T_{growth}^t$ quantifying knowledge accumulation and $T_{structure}^t$ reflecting knowledge structure shift. $T_{growth}^t$ varies smoothly and determines the overall trend of $T^t$ (Fig. \ref{fig:ima_2}(a)). A big rise in $T_{growth}^t$ correspond most often with a significant increase in topic size. Typically, during such periods, some child papers started to gain popularity and collect a non-trivial number of citations within the topic. They helped the pioneering work maintain the topic visibility \cite{long-term-sci-impact}$^{,}$\cite{aging-forget}. Their attractiveness to new ideas, added to that of the pioneering work, helped contribute to the enrichment of topic knowledge pool (Fig. \ref{fig:ima_2}(b)). A direct and visible consequence of this phenomenon is a fortification of existing knowledge structure, sometimes accompanied by a mild extension (Fig. \ref{fig:ima_2}(c-e)). Nonetheless, an ever-growing topic scale is not a guarantee for thriving periods. For instance, $T_{growth}^t$ of topic led by `Critical Power for Asymptotic Connectivity in Wireless Networks' has been on the decrease since 2011 despite a continuous size growth. This corresponds to the fact that almost all of the influential child papers within the topic were published no later than 2005. The lack of new, promising ideas and remarkable extensions to existing researches afterwards makes the topic lose community's attention and results in the topic's demise. As for topic led by `A unified architecture for natural language processing: deep neural networks with multitask learning', its decline in $T_{growth}^t$ since 2015 is somewhat atypical. The decrease is owing to the emergence of popular child papers published between 2013 and 2014 that largely excel their parent. Child papers `Efficient Estimation of Word Representations in Vector Space', `Distributed Representations of Words and Phrases and their Compositionality' and `Glove: Global Vectors for Word Representation' have each attracted around 600 citations within the topic, while their total citations have all surpassed 8000, much greater than their antecedent whose citation count still remains below 3000. They have had such big achievements that they have become the authorities in the domain. Consequently, they have won over the attention of subsequent studies, which in turn affects the knowledge accumulation of the topic created by their parent paper. We observe that articles published after 2016 in the topic have not had a comparable development. This confirms partly the shadowing effect caused by the prominent child papers mentioned above. $T_{structure}^t$, unlike $T_{growth}^t$, can vary greatly over time. It usually accounts for important fluctuations of $T^t$ (Fig.\ref{fig:ima_3}(a,b)). A high $T_{structure}^t$ usually marks one of the following 2 events: the formation of sub-topics and the fusion of sub-topics. The first event is a consequence of the arrival of rising stars in the topic. These articles, later proven influential to the topic evolution, either introduce multiple research directions or contribute to the flourishing of a single novel research focus. The second event takes place when there is subsequent literature uniting prior works' research. More specifically, the sub-topic merge occurs when there appears some unusual citations where an old article cites a young one and that the young article is crucial to topic development (Fig. \ref{fig:ima_3} (b,d,f)). Both the emergence of a single non-trivial research focus and the sub-topic merge can cause an obvious spike in $T_{structure}^t$. For instance, topic led by `Neural Networks for Pattern Recognition' had a sudden $T_{structure}^t$ increment when child paper `A Tutorial on Support Vector Machines for Pattern Recognition' established a third sub-topic direction. In topic led by `On random graphs, I', prominent child paper `On the evolution of random graphs' fuses prior works' ideas and changed topic landscape. However, the heat bought by such critical events are ephemeral. In the long run, their impact on topic's life-cycle is eventually reflected by the knowledge accumulation process, which is quantified by $T_{growth}^t$. We note that influential child papers play an important role in both $T^t$'s components and thus is crucial to topic's thriving. However, the duration between their publication and their visible contribution varies a lot \cite{time-window-citation}. \\

\noindent Besides knowledge temperature, we can also feel topic vigor by examining its skeleton tree. In fact, the evolution of knowledge temperature is consistent with the development of skeleton tree. Its skeleton tree thrives when a topic gains popularity and fame. In times when $T_{growth}^t$ rises, skeleton tree grows increasingly sturdy as newly published papers enrich existing research branches (Fig. \ref{fig:ima_2}(c-e)). During periods when $T_{structure}^t$ soars, topics usually form new research focus thanks to some prominent child papers. The trend is visualized by the emergence of new non-trivial clusters or branches. Sometimes, lately developed research directions prove to be a big success and start to defy topic authorities by attracting most new articles' attention. In such cases, skeleton tree also manifests a gravity shift, with new branches and clusters developing much faster than the previously dominating ones (Fig. \ref{fig:ima_3}(a,c,e)). Finally, if the rise of $T_{structure}^t$ is due to sub-topic merge, separated parts of skeleton tree are connected together by a young article which later proved crucial to topic development (Fig. \ref{fig:ima_3}(b,d,f)). When a topic loses it appeal, its skeleton tree stagnates, just like its knowledge temperature (Fig. \ref{fig:ima_2}(f,g)). \\

\noindent We observe a rich variation in $T^t$'s dynamics as each topic exhibits a unique development pattern. We identify 4 distinct topic life-cycles: rising topic, rise-then-fall topic, awakened topic and rise-and-fall-cycle topic. Rising topics demonstrate overall a steady and lasting $T^t$ increase. They welcome rather intermittently their child papers that enjoy popularity within the topic. This ensures to some extent a stable knowledge increment. Rise-then-fall topics reach their peak at some point and then go downhill owing to the lack of new development of existing ideas, the absence of new study focus or the shadowing of their outstanding child papers. In addition, their expansion pace slows down during the cooling down phase. Awakened topics can have a mild development for a duration as long as 20 years before experiencing an influence surge. Their sudden flourishing is largely due to scientific communities' recent frenzy in certain domains, such as artificial intelligence. Rise-and-fall-cycle topics manifest a more complicated $T^t$ pattern. However, their rises and falls also match the global background, such as the introduction of the Internet, the booming of artificial intelligence and the prevalence of online social networks (Detailed discussion is in Supplementary Information section S3.1-S3.4). \\

\noindent How is heat distributed within a topic? To answer this question, we interpreted $T^t$ as the average temperature of $G^t$ and computed knowledge temperature for every article based on $T^t$. Node knowledge temperature gauges a work's relative popularity and impact within the topic at a certain moment. At each timestamp $t$, we assumed the hottest and coldest works and then employed the heat equation to propagate the heat across $G^t$. For a node $u$, its temperature change $\frac{dT_u}{di}$ is (we omit the superscript $t$ of node temperature in the equation):
\begin{equation}
    \frac{dT_u}{di} = \sum_{v=1}^{|V^t|}\widetilde{A_{vu}^t}(T_v - T_u)
\end{equation}
where $\widetilde{A_{vu}^t}$ is the thermal conductivity between node $v$ and node $u$. We set the pioneering article to be the hottest node (knowledge temperature = 1) and all the underdeveloped papers to be the coldest nodes (knowledge temperature = 0). We modelled heat propagation via idea inheritance and youngster's contribution to knowledge renaissance respectively by forward and backward iterations of the heat equation. The number of iteration $i$ depends on the average hops between 2 randomly selected nodes. Finally we performed a scaling by $T^t$. Node $u$'s knowledge temperature at timestamp $t$, $T_u^t$ is therefore:
\begin{equation}
    T_u^t = T_{u,std}^t\cdot \frac{T^t}{\overline{T_{std}^t}}
\end{equation}
where $T_{u,std}^t$ is $u$'s temperature and $\overline{T_{std}^t}$ the average temperature derived from the heat equation.\\

\noindent We visualized node knowledge temperature by skeleton tree. If we let alone the coldest papers, we observe a ubiquitous phenomenon: the closer an article is to the pioneering work, the hotter it tends to be. Node knowledge temperature decreases along paths in skeleton tree (Fig. \ref{fig:ima_3}(c-f)). Although pioneering work is the only known hottest node, we identify other heat sources, the majority of which are the centers of non-trivial clusters. Most heat sources happen to be among the most-cited child papers within a topic. They possess primarily intrinsic value. Their own research content contributes a lot to topic's survival and flourishing. Another type of heat source are articles situated between clusters. Such papers may not have made astonishing discoveries nor have attracted many followers, but it is their studies that have inspired some influential subsequent work. Their value lies essentially in the enlightenment.\\

\begin{figure}[htbp]
\begin{subfigure}{\textwidth}
\centering
\begin{minipage}[t]{0.21\linewidth}
\includegraphics[width = \linewidth]{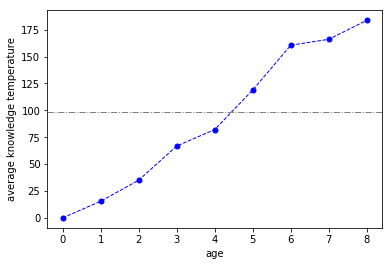}
\caption{}
\end{minipage}
\begin{minipage}[t]{0.21\linewidth}
\includegraphics[width = \linewidth]{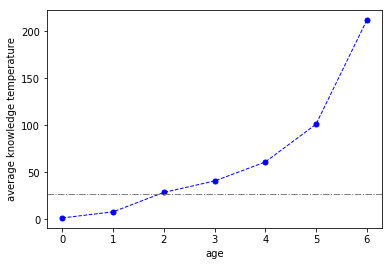}
\caption{}
\end{minipage}
\begin{minipage}[t]{0.21\linewidth}
\includegraphics[width = \linewidth]{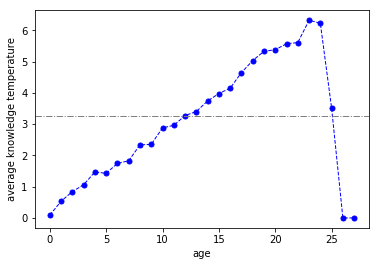}
\caption{}
\end{minipage}
\begin{minipage}[t]{0.21\linewidth}
\includegraphics[width = \linewidth]{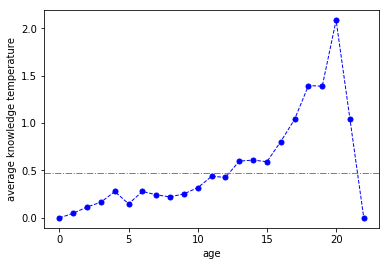}
\caption{}
\end{minipage}
\end{subfigure}

\begin{subfigure}{\textwidth}
\centering
\begin{minipage}[t]{0.21\linewidth}
\includegraphics[width = \linewidth]{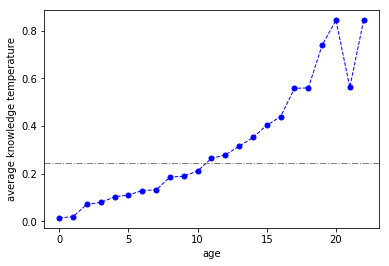}
\caption{}
\end{minipage}
\begin{minipage}[t]{0.21\linewidth}
\includegraphics[width = \linewidth]{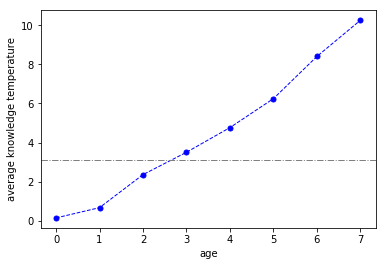}
\caption{}
\end{minipage}
\begin{minipage}[t]{0.21\linewidth}
\includegraphics[width = \linewidth]{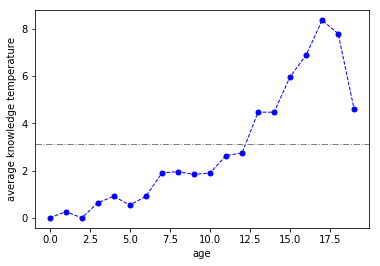}
\caption{}
\end{minipage}
\begin{minipage}[t]{0.21\linewidth}
\includegraphics[width = \linewidth]{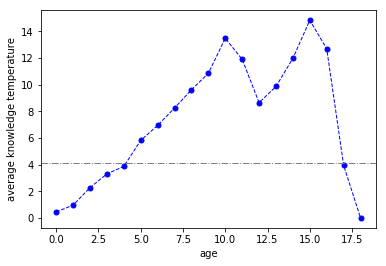}
\caption{}
\end{minipage}
\end{subfigure}

\begin{subfigure}{\textwidth}
\centering
\begin{minipage}[t]{0.21\linewidth}
\includegraphics[width = \linewidth]{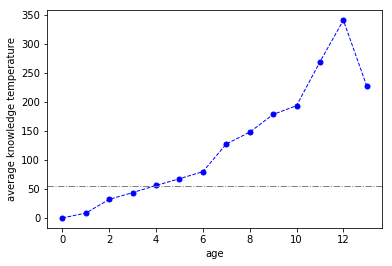}
\caption{}
\end{minipage}
\begin{minipage}[t]{0.21\linewidth}
\includegraphics[width = \linewidth]{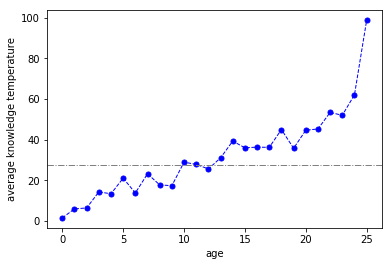}
\caption{}
\end{minipage}
\begin{minipage}[t]{0.21\linewidth}
\includegraphics[width = \linewidth]{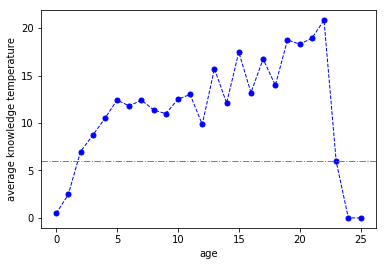}
\caption{}
\end{minipage}
\begin{minipage}[t]{0.21\linewidth}
\includegraphics[width = \linewidth]{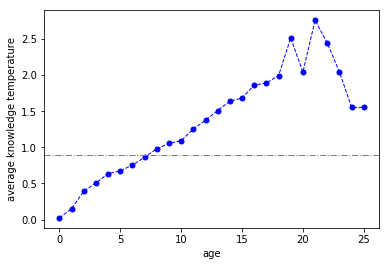}
\caption{}
\end{minipage}
\end{subfigure}

\begin{subfigure}{\textwidth}
\centering
\begin{minipage}[t]{0.21\linewidth}
\includegraphics[width = \linewidth]{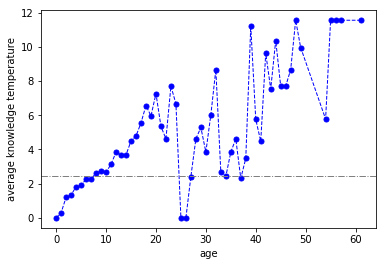}
\caption{}
\end{minipage}
\begin{minipage}[t]{0.21\linewidth}
\includegraphics[width = \linewidth]{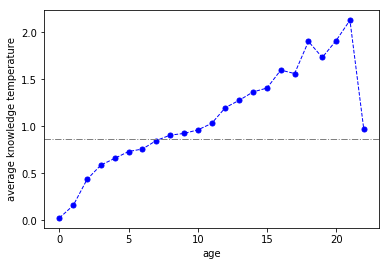}
\caption{}
\end{minipage}
\begin{minipage}[t]{0.21\linewidth}
\includegraphics[width = \linewidth]{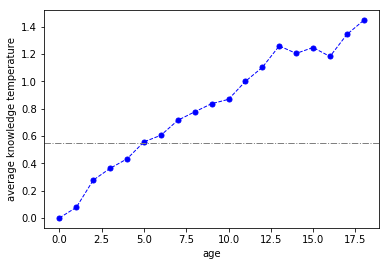}
\caption{}
\end{minipage}
\begin{minipage}[t]{0.21\linewidth}
\includegraphics[width = \linewidth]{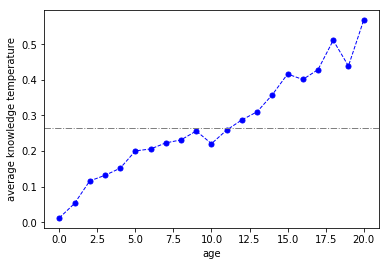}
\caption{}
\end{minipage}
\end{subfigure}
\caption{}
\label{fig:age_T}
\end{figure}

\noindent In an effort to better understand general heat distribution within topics, our preliminary observation prompted us to study the relation between node knowledge temperature and article age, as papers located in skeleton tree cores are parents or ancestors to papers on the periphery. We find that regardless of research themes, older papers indeed tend to have higher knowledge temperatures (Fig. \ref{fig:age_T}). Older papers take advantage of a longer time span and tend to better diffuse their ideas thanks to their numerous followers, a tendency in line with our intuition. Since we assume pioneering works possess the "hottest" knowledge, the gradual temperature decline well illustrates that idea inheritance and innovation are taking place simultaneously in every scientific topic. However, we observe a drop in average node knowledge temperature among the oldest papers in half of the topics. 2 phenomena can explain the anomaly. Some topics contain a tiny fraction of atypical citations where younger articles are cited by older papers or papers published at approximately the same time. When the younger articles happen to be pioneering works, the oldest papers are no longer the topic founders. They usually have inspired few or even no child papers in the topics. Consequently, they are among the coldest nodes. In rare cases, these papers inspired a certain quantity of works. But they remain "cold" owing to their relatively different research focus with that of the pioneering works even though they are connected to the latter. Their citations are more like peer bonds rather than a symbol of inspiration and idea inheritance. Such is the case for the pioneering work `Particle swarm optimization' and its peer and popular child paper `A new optimizer using particle swarm theory'.\\

\noindent Even if we let alone the cold old articles, the heat distribution is not that simple and monotonous. We observe in most topics that parent papers are not always hotter than its descendants. According to our design, node knowledge temperature is affected by 2 factors: the heat-level of its own research content and the promotion gained from its descendants. Therefore, a colder parent or ancestor is either due to its less prevalent ideas or a poor general performance of its children. This phenomenon implies that an important status within the topic does not necessarily bring much fame.\\

\noindent We further compared node knowledge temperature with in-topic citation count, a traditional article-level impact metrics, to get a better understanding of their similarities and differences (Fig. S49). We find a weak positive correlation between the two quantities among the best-cited papers in topics. In particular, we highlighted the most-cited child papers together with pioneering works on current skeleton trees. Most of them have a knowledge temperature above average as they are represented as yellow, orange or red nodes (current skeleton trees in Supplementary Information Fig. S3, S9, S15, S35 for example). However, there are exception. For instance, in topic led by `Particle swarm optimization', popular child paper `A new optimizer using particle swarm theory' (NOPST) is among the coldest despite the fact that it is the most influential child paper in terms of citation count (Fig. S35). NOPST was published in the same year as the pioneering work and it only cited the pioneering work. Its low temperature is due to its relatively different research focus with that of the pioneering work and an overall low heat level of its children. The latter is somehow also a consequence of the former, as the pioneering work has most prevalent idea. The focus difference is also reflected by their separation in the skeleton tree. \\

\noindent We also tracked the knowledge temperature evolution of relatively popular child papers within a topic and we find a similar phenomenon already observed at topic-level. While an article's own knowledge largely determines its heat level, child papers sometimes play a perceptible role in boosting or maintaining its popularity and impact. For example, in the topic led by paper `Bose-Einstein condensation in a gas of sodium atoms', article `Bose-Einstein condensation of exciton polaritons' has kept being hotter since its publication despite a global cooling since 2013 thanks to an above-average active development (Supplementary Information S3.2.7). Our finding is consistent with the research which demonstrates that papers need new citations to keep their visibility \cite{aging-forget}. Besides, in some topics, especially the one led by `Collective dynamics of ‘small-world’ networks', we frequently find that popular child papers were published in renowned journals such as Nature and Science (Supplementary Information section 3.4). Our observation accords with research which suggests a positive association between journal prestige and article high impact \cite{journal-pos-impact}. \\

\noindent Nonetheless, we find that several scientific topics are intimately connected. Some pioneering works occupy a primordial position in other topics' skeleton trees. Furthermore, these closely related topics manifest similar knowledge temperature dynamics. However, such similarity does not correspond very well with idea inheritance and development in some cases. For instance, paper `The capacity of wireless networks' (CMN) is the most successful child paper of the pioneering work `Critical Power for Asymptotic Connectivity in Wireless Networks'. It plays a crucial role in topic's prosperity (Fig. S12) by jointly inspiring one third of the topic members, most of which were published during the flourishing period. Besides, CMN surpassed and took over its predecessor to be the new authority in their domain in just a few years. Yet, according to their topic knowledge temperatures, it is the topic led by CMN that went downhill first. To this end, we wanted to design a mechanism that allows us to better capture the interactions among closely-connected topics. Following our skeleton tree notion, we were inspired by the nutrition transfer among real trees in a forest \cite{Wood-Wide-Web}. We hence treated scientific topics as trees and conceived a forest helping mechanism where thriving topics transfuse a small fraction of vigor to their dying siblings. The amount of shared energy depends on both the ages and the size of the topic group. When we compare topic knowledge temperatures before and after forest helping, we find that our helping mechanism regulates mildly the temperatures as if it took into account the "background popularity", average popularity of a bigger research topic to which the group belongs. Overall, forest helping slightly reduces the fluctuation of $T^t$ (Fig. S50).\\

\noindent In summary, we report a thermodynamic approach to depict the rise and fall of scientific topics. We design knowledge temperature, an intuitive and quantitative metrics to evaluate topic overall popularity and impact dynamics by fully leveraging the scale and structure dynamics of citation network through skeleton tree. A continuous streaming of useful information is the key to topics' prosperity in the long run, to which the arrival of eminent child papers contributes a lot. In the short term, critical events such as the merge and emergence of new sub-topic also boost topic's vigor. In addition, we also examine the heat diffusion within topics and discover that older articles generally have bigger chances to diffuse its ideas and thus enjoy a higher popularity within the topic. However, exceptions exists widely, suggesting that the positive correlation between heat-level and article's age and impact remains weak. Finally, we design a forest helping mechanism to better depict the idea inheritance and development among intimately-associated topics. Although knowledge temperature cannot directly be used as a scientific impact metrics, our study suggests a new possibility to quantify research impact in a most intuitive way.\\

\section*{Data Availability}

\noindent All code is available at https://github.com/drlisette/knowledge-temperature.\\

\noindent Data are available at https://github.com/drlisette/knowledge-temperature. Other related, relevant data are available from the corresponding author upon reasonable request.\\








\newpage

\clearpage

\bibliography{sample}

\section*{Author contributions statement}

\noindent L.F. conceived the idea to depict topic flourishing dynamics by thermodynamic temperature, checked model feasibility and summarised results. \\

\noindent D.L. designed the knowledge temperature model, did data visualization, conceived the experiments, analysed and summarised the results.\\

\noindent Q.L. processed the topic data and optimized the skeleton tree algorithm.\\

\noindent X.W. gave invaluable comments for paper writing.

\section*{Additional information}

\noindent \textbf{Competing interests}\\
\noindent The author(s) declare no competing interests.



\newpage

\section*{Figure captions}

\textbf{Figure 1: Comparison between galaxy map and topic skeleton tree.}
In galaxy map: Node size and title size are proportional to total citation count. Only the most-cited papers are labelled with titles. Node colour of pioneering work is red. Node colour of the other articles are determined by their positions under the ForceAltas layout algorithm. Nodes in the same cluster take a same colour (yellow, green, blue or pink). In topic skeleton tree: Node size (except pioneering work) is proportional to structure entropy. Pioneering work node is twice the maximum size of the child paper nodes. Node colour is the same as in galaxy map. Only pioneering work is labeled by its title. (a,b) Topic led by `Critical Power for Asymptotic Connectivity in Wireless Networks'. (a) Numerous child papers, especially `The capacity of wireless networks' and `HEED: a hybrid, energy-efficient, distributed clustering approach for ad hoc sensor networks', have outperformed the pioneering work. (b) After initial development, the topic has found two research focus. (c,d) Topic led by `Latent dirichlet allocation'. (c) The pioneering work has a dominant influence. (d) Three research directions have derived directly from the initial idea. (e,f) Topic led by `On random Graphs, I'. (e) Two influential child papers, `On the evolution of random graphs' and `The Structure and Function of Complex Networks' seem to split the topic into two parts. (f) The pioneering work has inspired in particular one school of thought. There is no significant division in topic's knowledge structure.\\

\noindent \textbf{Figure 2: Skeleton tree extraction and graph shrinking demo.} The red node labelled "P" represents the pioneering work. Green nodes are child papers. A directed edge from A to B represents "B cites A". (a) Skeleton tree extraction. From left to middle: loop cutting. Child papers $c3$ and $c4$ cites each other. We remove one of the two citations to get a tree structure. From middle to left: tree pruning. We remove redundant citations for every child paper so that it only keeps the most meaningful citation. (b) Graph shrinking for $T_{structure}^t$ computation. Graph shrinking process transforms the newly arrived articles into virtual citations among existing papers. For example, child paper $c3$ arrives between timestamp $t-1$ and $t$ and cites all papers in the topic. Its citations suggest that $c1$ and $c2$, disconnected in $G^{t-1}$, have certain connections in their research content. We remove $c3$ and add one or two virtual citations between $c1$ and $c2$ according to the general rule where the younger virtually cites the older. If $c1$ and $c2$ were published in the same year, they virtually cite each other in $G^t$'s shrinked counterpart, $G^{'t}$. \\

\noindent \textbf{Figure 3: Knowledge temperature (especially $T^t$ and $T_{growth}^t$) and skeleton tree evolution of topic led by `A unified architecture for natural language processing: deep neural networks with multitask learning'.} Nodes in skeleton tree are coloured according to their knowledge temperature, with red being the hottest, yellow being the average level and blue the coldest within the topic. Node size (except pioneering work) is proportional to (re-scaled) structure entropy \cite{structure-entropy}. Pioneering work node is twice the maximum size of the child paper nodes. (a) Knowledge temperature evolution. $T_{growth}^t$ dominates $T^t$. (b) Current topic skeleton tree. The pioneering work and 4 most top-cited papers within the topic are labelled by title. (c,d,e) Topic skeleton tree by the end of 2011, 2013 and 2015. The thriving period is characterized by a steady knowledge accumulation, depicted by a fast-growing skeleton tree where small new clusters emerge and existing branches become increasingly robust. (f,g) Topic skeleton tree by the end of 2017 and 2019. The stagnation period is reflected by a decelerating growth and an almost fixed tree shape.\\

\noindent \textbf{Figure 4: Knowledge temperature (especially $T^t$ and $T_{structure}^t$) and skeleton tree evolution of topics led by `The capacity of wireless networks' (CWN) and `On random graph, I' (RG).} Nodes in skeleton tree are coloured according to their knowledge temperature, with red being the hottest and blue the coldest within the topic. Node size (except pioneering work) is proportional to (re-scaled) structure entropy. Pioneering work node is twice the maximum size of the child paper nodes. (a,b) Knowledge temperature evolution. $T_{structure}^t$ accounts for $T^t$'s fluctuations. (c,e) Skeleton tree of the topic led by CWN by the end of 2003 and 2007. Advancements are visible in all directions. In particular, the gravity shift in the tree implies the emergence of new research focus, which in turn yields a soar in $T_{structure}^t$. (d,f) Skeleton tree of the topic led by RG by the end of 1979 and 1984. Article `On the evolution of random graphs' published in 1984 fuses the previously separated parts due to an atypical citation from an older article 'On the existence of a factor of degree one of a connected random graph'. The merge in topic knowledge structure pushed up $T_{structure}^t$ during that period. \\

\noindent \textbf{Figure 5: Relation between article age and node knowledge temperature for 16 topics.} Article age = 2020 - year of publication. Grey dotted horizontal line marks the topic knowledge temperature (average level) in 2020. (a) Topic led by `Regulatory T Cells: Mechanisms of Differentiation and Function'. (b) Topic led by `Empirical Evaluation of Gated Recurrent Neural Networks on Sequence Modeling'. (c) Topic led by `Neural networks for pattern recognition'. (d) Topic led by `Critical Power for Asymptotic Connectivity in Wireless Networks'. (e) Topic led by `The capacity of wireless networks'. (f) Topic led by `Efficient Estimation of Word Representations in Vector Space'. (g) Topic led by `Coverage problems in wireless ad-hoc sensor networks'. (h) Topic led by `A neural probabilistic language model'. (i) Topic led by `A unified architecture for natural language processing: deep neural networks with multitask learning'. (j) Topic led by `Bose-Einstein condensation in a gas of sodium atoms'. (k) Topic led by `Long short-term memory'. (l) Topic led by `Particle swarm optimization'. (m) Topic led by `On random graphs, I'. (n) Topic led by `Collective dynamics of ‘small-world’ networks'. (o) Topic led by `Latent dirichlet allocation'. (p) Topic led by `A FUNDAMENTAL RELATION BETWEEN SUPERMASSIVE BLACK HOLES AND THEIR HOST GALAXIES'.\\

\clearpage

\end{document}


\flushbottom
\maketitle
%
%

\tableofcontents

\section{Data Description}

We collected topic citation relations from academic databases including DBLP, arXiv, Elsevier and Springer. Each topic is led by an article that have had a profound influence in certain domains. We refer to these papers as pioneering papers or leading papers. A scientific topic includes a pioneering paper, all the articles that directly cites it and all the citations among them. We chose 16 topics among our dataset to conduct the knowledge temperature experiment. Pioneering paper information is listed in Table \ref{tab:tab_2} and topic size is listed in Table \ref{tab:tab_3}. Topics are ordered by publishing year.\\

\noindent Among 16 topics, we identify 3 topic groups, each containing 2 or 3 topics:
\begin{enumerate}
    \item wireless network group.\\ Group is jointly led by Critical Power for Asymptotic Connectivity in Wireless Networks and The capacity of wireless networks.
    \item RNN gated unit group.\\ Group is jointly led by Long short-term memory and Empirical Evaluation of Gated Recurrent Neural Networks on Sequence Modeling.
    \item word embedding group.\\ Group is jointly led by A unified architecture for natural language processing: deep neural networks with multitask learning, A neural probabilistic language model and Efficient Estimation of Word Representations in Vector Space.
\end{enumerate}

\begin{table}[htbp]
    \centering
    \begin{tabular}{ p{7cm}  p{1cm}  p{3.9cm}  p{3.9cm} }
        \hline
        leading paper & year & journal & conference series \\
        \hline
         On random graphs, I & 1959 & & \\

        Bose-Einstein condensation in a gas of sodium atoms & 1995 & Physical Review Letters & \\

        Particle swarm optimization & 1995 & & International Conference on Networks (ICON)\\

        Neural networks for pattern recognition & 1995 & Advances in Computers & \\

        Long short-term memory & 1997 & Neural Computation & \\

        Collective dynamics of ‘small-world’ networks & 1998 & Nature & \\

        Critical Power for Asymptotic Connectivity in Wireless Networks & 1999 &  & \\

        The capacity of wireless networks & 2000 & IEEE Transactions on Information Theory & \\

        A FUNDAMENTAL RELATION BETWEEN SUPERMASSIVE BLACK HOLES AND THEIR HOST GALAXIES & 2000 & The Astrophysical Journal & \\

        Coverage problems in wireless ad-hoc sensor networks & 2001 & & International Conference on Computer Communications (INFOCOM)\\

        Latent dirichlet allocation & 2003 & Journal of Machine Learning Research &\\

        A neural probabilistic language model & 2003 & Journal of Machine Learning Research & \\

        A unified architecture for natural language processing: deep neural networks with multitask learning & 2008 & & International Conference on Machine Learning (ICML)\\

        Regulatory T Cells: Mechanisms of Differentiation and Function & 2012 & Annual Review of Immunology & \\

        Efficient Estimation of Word Representations in Vector Space & 2013 &  & International Conference on Learning Representations (ICLR)\\

        Empirical Evaluation of Gated Recurrent Neural Networks on Sequence Modeling & 2014 & arXiv: Neural and Evolutionary Computing &\\
        \hline
    \end{tabular}
    \caption{Pioneering Paper Information}
    \label{tab:tab_2}
\end{table}

\newpage

\begin{table}[htbp]
    \centering
    \begin{tabular}{ p{12.8cm}  p{1.5cm}  p{1.5cm} }
        \hline
        leading paper & node num. & edge num. \\
        \hline
        On random graphs, I & 5389 & 17098 \\

        Bose-Einstein condensation in a gas of sodium atoms & 2338 & 9171 \\

        Particle swarm optimization & 31800 & 183341 \\

        Neural networks for pattern recognition & 17046 & 42748 \\

        Long short-term memory & 16777 & 98553 \\

        Collective dynamics of ‘small-world’ networks & 25548 & 206646 \\

        Critical Power for Asymptotic Connectivity in Wireless Networks & 1078 & 4998 \\

        The capacity of wireless networks & 7644 & 51788 \\

        A FUNDAMENTAL RELATION BETWEEN SUPERMASSIVE BLACK HOLES AND THEIR HOST GALAXIES & 2432 & 34120 \\

        Coverage problems in wireless ad-hoc sensor networks & 1546 & 8865 \\

        Latent dirichlet allocation & 18813 & 114969\\

        A neural probabilistic language model & 3265 & 22912 \\

        A unified architecture for natural language processing: deep neural networks with multitask learning & 2733 & 13855 \\

        Regulatory T Cells: Mechanisms of Differentiation and Function & 1381 & 4190 \\

        Efficient Estimation of Word Representations in Vector Space & 8133 & 36219 \\

        Empirical Evaluation of Gated Recurrent Neural Networks on Sequence Modeling & 2282 & 4675\\
        \hline
    \end{tabular}
    \caption{Topic Overview}
    \label{tab:tab_3}
\end{table}

\section{Model}

Our core idea is to treat citation network $G^t = (V^t, E^t)$ as a thermodynamic system, more specifically, ideal gas. $G^t$ is a directed graph whose nodes consist of a pioneering paper and all the articles that directly cites it and whose edges are the citations among them. Its adjacency matrix $A^t$ is defined as:
$$A_{uv}^t = \begin{cases} 1 & \text{$v$ cites $u$} \\
0 & \text{otherwise}
\end{cases}$$

\noindent As knowledge temperature relies on some quantities defined in skeleton tree extraction and knowledge entropy computation, we would like to organise our model description in the following order: we present first the construction of skeleton tree, then we define knowledge entropy. Next, we unfold our topic knowledge temperature design and at last we elaborate on node knowledge temperature.

\subsection{Topic Skeleton Tree}
Skeleton tree illustrates the knowledge structure of a topic. Its evolution reveals a topic's development pattern. The extraction of skeleton tree is essentially a process to reduce a graph to a tree. We note $G^t$'s skeleton tree ${Tree}^t = (V_T^t, E_T^t)$. For notation simplicity, we omit superscript $t$ for variables that appear in the rest of this subsection. There are altogether 3 steps in ${Tree}^t$'s construction:
\begin{enumerate}
    \item We perform node embedding and compute distance matrix $EmbedDist$ that shows the node pair-wise distance in embedding space.
    \item We derive matrix $DiffIdx$ based on $EmbedDist$ to measure the difference between every node pair. Vector $ReductionIdx$, a node score which serves to judge the citation importance, is computed afterwards. We rely on $ReductionIdx$ to prune $G^t$ in the following step.
    \item We reduce $G^t$ to ${Tree}^t$ by removing less important references while ensuring the overall connectivity. The significance of a citation is determined by the similarity of 2 papers, which is assessed through their reduction indices. The process involves loop cutting and tree pruning. In ${Tree}^t$, every node except the root, which is exactly the pioneering node, has at most one citation.
\end{enumerate}

\noindent We start by slightly modifying adjacency matrix $A$ by adding a self-loop to the pioneering work. This is for the convenience of spectral decomposition. Then, we compute out-degree matrix $D$ and normalized Laplacian matrix $\widetilde{L} = D^{-\frac{1}{2}}(D-A)D^{-\frac{1}{2}}$. $D$ is a diagonal matrix, with diagonal entries equal to the out-degree, or practically speaking the in-topic citation count of each node. We next perform a full spectral decomposition of $\widetilde{L}$. The eigenvectors are our node embeddings and $EmbedDist$ is a distance matrix with entry $EmbedDist_{u,v} = \parallel {eigenvector}_u - {eigenvector}_v \parallel_2$.\\

\noindent Now we proceed to compute difference matrix $DiffIdx$. For node pair $(u,v)$, we define their difference index $DiffIdx_{u,v}$ as:
$$DiffIdx_{u,v} = \sum_{v_{parent}}d_{u,v_{parent}}$$
$v_{parent}$s are the predecessors of $v$ and $d_{u,v_{parent}}$ is the shortest weighted path between $u$ and $v_{parent}$:
$$d_{u,v_{parent}} = \begin{cases} \sum_{(i,j)\in path} EmbedDist_{i,j} & \text{if there exists a path between $u$ and $v_{parent}$} \\
MaxDist \times avgStep & \text{otherwise}
\end{cases}
$$
$MaxDist$ is the biggest distance between two connected nodes, $MaxDist = \max_{(a,b)\in E^t}(EmbedDist_{a,b}A_{a,b})$ and $avgStep$ is the average hop number of all shortest paths between any two reachable nodes. $DiffIdx$ gauges the difference between $u$ and $v$ by involving works that inspire $v$. If $u$ and $v_{parent}$ is reachable from each other, it suggests that there is some degree of similarity in their ideas or research topics and thus we represent their distance by shortest path's weight. Else, we model their correlation by a long imaginary path of $avgStep$ hops and step length of $MaxDist$. Therefore, the greater $DiffIdx_{u,v}$ is, the more different $u$ and $v$ are. \\

\noindent For a node $u$, its reduction index $ReductionIdx_u$ is defined as the sum of its difference indices:
$$ReductionIdx_{u} = \sum_{v\in V^t\setminus u}DiffIdx_{u,v}$$
Vector $ReductionIdx$ helps to determine the importance of citations. A citation between two articles with similar reduction indices is considered more valuable than one between two papers with different reduction indices.\\

\noindent We are now ready to extract topic skeleton tree. The first step is to find and cut loops in $G^t$. We cut a loop by removing the least important edge (its extremities have the most different reduction indices). Nonetheless, we try to ensure that the edge we cut is not the last citation left for some node so as to preserve overall connectivity as much as possible. After loop cutting, we obtain a tree. The second step is to remove redundant citations in the tree. Recall that we only keep one citation for every node except the root in ${Tree}^t$. Fig. 2(a) illustrates the whole process with a toy example.\\

\subsection{Structure Entropy}
We adopt structure entropy$^{32}$ to determine the node size in the skeleton tree visualisation. Structure entropy measures the uncertainty of the tree structure if node $u$ is absent. Consequently, it makes sense to evaluate the importance of a paper to knowledge passing within the topic by structure entropy. For a node $u$ other than the root, its structure entropy $S_u^t$ is defined as:
$$S_u^t = -\frac{g_{T,u}^t}{2|E_T^t|}\log\frac{V_{T,u}^t}{V_{T,u_{parent}}^t} $$
$g_{T,u}^t$ is the cut size of the sub-tree ${Tree}_u^t$ whose root is $u$. It is the sum of the degree of nodes in ${Tree}_u^t$ in ${Tree}^t$. $E_T^t$ is the edge set of skeleton tree. $V_{T,u}^t$ is the number of nodes ${Tree}_u^t$ contains (the sum of out-degrees of ${Tree}_u^t$) and $V_{T,u_{parent}}^t$ the number of nodes $Tree_{u_{parent}}^t$ has.\\

\noindent The term before log measures the importance of ${Tree}_u^t$ to the whole skeleton tree and the log part describes the uncertainty of ${Tree}_u^t$ with respect to its parent sub-tree.\\

\noindent Structure entropy of the entire topic, $S^t$, is defined as the sum of node structure entropy:
$$S^t = \sum_{u\in T^t, u\neq root} S_u^t = -\sum_{u\in T^t, u\neq root}{\frac{g_{T,u}^t}{2|E_T^t|}\log\frac{V_{T,u}^t}{V_{T,u_{parent}}^t}}$$
.

\subsection{Topic Knowledge Temperature}
Topic knowledge temperature $T^t$ is defined as:
$$T^t = T_{growth}^t + T_{structure}^t$$
where $T_{growth}^t$ measures knowledge increment and $T_{structure}^t$ estimates the degree of latest structural changes in topic's knowledge framework.

\subsubsection{$T_{growth}^t$}
We initialise $T_{growth}^t$ by combining the 2 expressions of ideal gas's internal energy $U$:
$$ U = cnT $$
$$ U = ke^{\frac{S}{cn}}V^{-\frac{R}{c}}n^{\frac{R+c}{c}} $$
where $S$ is entropy, $n$ is substance amount (number of moles), $V$ is volume, $R$ is ideal gas constant, $c$ is heat capacity and $k$ adjusting coefficient.\\

\noindent As a result, $T_{growth}^0$ writes:
$$T_{growth}^0 = k e^{\frac{S_0}{cn_0}} {\left(\frac{n_0}{V_0}\right)}^{\frac{R}{c}}$$
where $S_0$ is the initial structure entropy of the topic , $n_0$ initial topic mass, $V_0$ initial topic volume, $k$ coefficient to be determined and $R$ and $c$ two constants.\\

\noindent  Next, we model $G^t$'s evolution as an isobaric process of ideal gas. Hence, according to the ideal gas state equation $PV=nRT$, by fixing pressure $P$, $T_{growth}^t$ is updated by the following expression:
$$T_{growth}^t = T_{growth}^{t-1} \frac{n_{t-1}}{n_t}\frac{V_t}{V_{t-1}}$$\\

\noindent We set topic volume $V_t$ to be the node number: $V_t = |V^t|$ and topic mass $n_t$ as $n_t = |V^t| - {UsefulInfo}^t$. Topic structure entropy $S_t$ is derived in the previous subsection, $S_t = S^t$. \\

\noindent ${UsefulInfo}^t$ is based on $DiffIdx$ in skeleton tree extraction:
$${UsefulInfo}^t = \sum_{(u,v)\in {Tree}^t} \frac{DiffIdx_{u,v}} {{\max_{(a,b)\in {Tree}^t} DiffIdx_{a,b}}}$$

\noindent Nevertheless, we would like to finish this part with a qualitative analysis of $T_{growth}^t$'s dynamics from a macroscopic view of information and knowledge. Knowledge originates from information, but information and knowledge have different characteristics. Information is only valuable for one time. Duplicate information does not create any additional value, thus cannot be used to create knowledge. Knowledge is like an understanding and a refinement of information. It is always valuable. Normally speaking we cannot have too much knowledge.\\

\noindent Bearing the interplay of knowledge and information in mind, we are now ready to interpret the symbolic meaning of volume $V_t$ and mass $n_t$. $V_t$ represents the total amount of information possessed by a topic at timestamp $t$. $UsefulInfo$ signifies the amount of useful information and thus $n_t$ symbolises the total amount of overlapped, or used information. We assume that each paper carries one unit of information. Yet we derive useful information edge by edge. This is because in a skeleton tree, all articles except the pioneering paper only have one citation, and if article $u$ and its 'parent' ('child') article have drastically different $DiffIdx$s, they are likely to have distinct research contents. In this case, therefore, even if one of them has completely overlapped content with some other article(s) , we can still roughly determine one unit of new information. \\

\noindent From the update rule of $T_{growth}^t$, we distinguish 3 cases (suppose $G^t$ always expands, thus $V_t$ always increases):
\begin{enumerate}
    \item $T_{growth}^t$ will not change if $V_t$ and $n_t$ have identical increase rate during the last period.
    \item $T_{growth}^t$ will decrease if $n_t$ increases faster than and $V_t$ over the last period.
    \item $T_{growth}^t$ will increase if $V_t$ increases faster than and $n_t$ over the last period.
\end{enumerate}
$T_{growth}^t$ goes up when the quantity of total information grows faster than the amount of duplicate information. Note that $V_t - n_t = {UsefulInfo}^t$, $T_{growth}^t$ rises when there is an accelerated increase in useful information. The more abundant useful information is, the bigger possibility for a topic to create new knowledge in the future and the greater potential a topic is. Otherwise, the topic "consumes" information faster than its information capital accumulation. If the tendency continues, it will have less information reserve for knowledge generation in the future. Its growth potential declines and eventually it 'dies'. Therefore, $T_{growth}^t$ reflects both how smoothly the knowledge accumulation goes and how promising the topic is at timestamp $t$. As knowledge enrichment eventually brings about scientific impact, $T_{growth}^t$ illustrates the long-term cumulative impact of a topic.

\subsubsection{$T_{structure}^t$}
For a thermodynamic system with freedom to vary its volume, temperature and pressure, the variation in internal energy $dU$ is given by $dU = TdS - PdV + mdn$, where $T$ is the temperature, $P$ the pressure, $dV$ the volume change,
$m$ the particle mass and $dn$ the change in the number of particles$^{26}$. The temperature $T$ for an evolving network with fixed node number can be derived as $T = \frac{dU}{dS}$ $^{25}$. It has been proved that with appropriate thermodynamic representations and some approximations, this relation is able to detect the critical events in a dynamic network$^{25,26}$.\\

\noindent Inspired by the above literature, we define $T_{structure}^t$ as:
$$T_{structure}^t = \left|\frac{\frac{dU^t}{dS^t}}{|V^t|}\right| = \left|\frac{\frac{{U}'^t-U^{t-1}}{{S}'^t-S^{t-1}}}{|V^t|}\right|$$
where $S^{t-1}$, ${S}'^t$ are the von Neumann entropy of $G^{t-1}$ and ${G'}^t$ and $U^{t-1}$, ${U}'^t$ the internal energy. ${G'}^t$ is a weighted reduced graph of $G^t$. It has all the nodes and edges of $G^{t-1}$. Besides, ${G'}^t$ contains virtual citations deduced from the new nodes coming between timestamp $t-1$ and timestamp $t$. Intuitively, $T_{structure}^t$ can be interpreted as the average structural change brought by an article in $G^t$. \\

\noindent The transformation from $G^t$ to ${G'}^t$ boils down to 2 tasks: remove new nodes and add virtual citations when possible. The edge weight of a real citation is 1. For every new node $x$, we distinguish 2 cases:
\begin{itemize}
    \item If $x$ has only 1 parent node $p_x$, then remove $x$. If $x$ has child node(s) $c_x$, connect it (them) to $x$'s unique parent node and set the edge weight $A_{p_xc_x} = \frac{1}{2}A_{xc_x}$.  Intuitively, since $x$ only cites 1 paper, its arrival cannot give us extra information about whether any of the node pair in $G^{t-1}$ that don't have a citation between them shares some of their research content.
    \item If $x$ has multiple parent nodes, find all its "youngest" ancestor nodes in $G^{t-1}$. If a parent node $p_x$ is in $G^{t-1}$, then $p_x$ is already a "youngest" ancestor node. Else, iteratively find $p_x$'s predecessors until they are in $G^{t-1}$. Note $x$'s youngest ancestor nodes in $G^{t-1}$ ($a_1, a_2,...,a_m$). Next, for each ancestor pair ($a_i$,$a_j$) between which there is no edge in $G^{t-1}$, add a directed virtual link according to their publishing year $y_i, y_j$ (note $A$ the real-time adjacency matrix, $m$ the total number of $x$'s youngest ancestor nodes):
    \begin{itemize}
        \item If $y_i < y_j$, add a directed weighted edge from $a_i$ to $a_j$ of weight $\frac{2\cdot\sum_{p_x}A_{p_xx}}{m(m-1)}$. The new edge means "$a_j$ virtually cites $a_i$".
        \item If $y_i > y_j$, add a directed weighted edge from $a_j$ to $a_i$ of weight $\frac{2\cdot\sum_{p_x}A_{p_xx}}{m(m-1)}$. The new edge means "$a_j$ virtually cites $a_i$".
        \item If $y_i = y_j$, add a bidirectional weighted edge between $a_i$ and $a_j$ of weight $\frac{\sum_{p_x}A_{p_xx}}{m(m-1)}$. The new edge means "$a_j$, $a_i$ virtually cites each other".
    \end{itemize}
\end{itemize}

\noindent Fig. 2(b) illustrates a simple graph shrinking case.\\

\noindent In case of a duplicate virtual link, we discard it. In order words, we always keep the first virtual link added between a node pair. Remove $x$ after adding all possible virtual links. Intuitively, since $x$ cites several papers, we can guess that these papers are somehow loosely connected to one another even if there is no direct citations among them. That is why we add virtual citations of weight less than 1.\\

\noindent We set $U^{t-1}$, ${U}'^t$ to be the sum of edge weight. As an authentic citation has a weight of 1, $U^{t-1}$ reduces to the number of edges $U^{t-1} = |V^{t-1}|$. Therefore, if we note ${A}'^t$ and ${E'}^t$ the adjacency matrix and the edge set of ${G'}^t$ respectively, $${U}'^t - U^{t-1} = \sum\limits_{(u,v)\in {E'}^t\setminus E^{t-1}} {A}'^{t}_{uv}$$ \\

\noindent We approximate $S^t$ and ${S}'^t$ by node degree. The von Neumann entropy for a directed graph is the sum of the von Neumann entropy of its strongly connected (SC) components$^{27}$:
$$S =\sum_{SC} S_{SC}$$
Now assume the strong connectivity and we extend the entropy computation for unweighted directed graph$^{25,27}$ to that for a weighted directed graph $G = (V, E)$. First define some notations:\\

\noindent Bidirectional edge set $E_{bd}$:  $$E_{bd} = \{(u,v)|(u,v)\in E \,\text{and}\, (v,u)\in E \}$$

\noindent Adjacency matrix $A$:
$$A_{uv} = \begin{cases}
w_{uv} & \text{if} (u,v)\in E\\
0 & \text{otherwise}
\end{cases}$$

\noindent In-degree and out-degree of node $u$:
$$d_u^{in} = \sum_{v\in V}A_{vu}\qquad d_u^{out} = \sum_{v\in V}A_{uv}$$

\noindent Transition matrix $P$:
$$P_{uv} = \begin{cases}
\frac{A_{uv}}{d_u^{out}} & \text{if} (u,v)\in E\\
0 & \text{otherwise}
\end{cases}$$

\noindent Normalized Laplacian matrix $\tilde{L}$:
$$\tilde{L} = \begin{cases}
1 & u=v, d_v^{out} \neq 0\\
-\frac{A_{uv}}{\sqrt{d_u^{out}}\sqrt{d_v^{out}}} & u\neq v,(u,v)\in E\\
0 & \text{otherwise}
\end{cases}$$

\noindent We note $\tilde{\lambda_s}$ normalized Laplacian eigenvalue and $\phi$ unique left eigenvector of transition matrix $P$.

\noindent The von Neumann entropy of $G$ is the
Shannon entropy associated with the normalized Laplacian eigenvalues.By adopting the quadratic approximation
to the Shannon entropy (i.e. $ - x\ln x \approx x(1 - x)$), we have$^{27}$ $$ \begin{aligned} S & = -\Sigma_{s=1}^{|V|}\frac{\tilde{\lambda_s}}{|V|}\ln\frac{\tilde{\lambda_s}}{|V|}\\
& = \Sigma_{s=1}^{|V|}\frac{\tilde{\lambda_s}}{|V|}(1-\frac{\tilde{\lambda_s}}{|V|})\\
& = \frac{tr(\tilde{L})}{|V|} - \frac{tr(\tilde{L}^2)}{{|V|}^2}\\
& = 1 - \frac{tr(\tilde{L}^2)}{{|V|}^2}
\end{aligned} $$

\noindent Now we expand the equation $tr(\tilde{L}^2) = |V|+\frac{1}{2}(tr(P^2)+tr(P\Phi^{-1}P^T\Phi))$ $^{27}$for $G$:
$$tr(P^2) = \Sigma_{u\in V}\Sigma_{v \in V}P_{uv}P_{vu} = \Sigma_{(u,v)\in E_{bd}}\frac{A_{uv}A_{vu}}{d_u^{out}d_v^{out}}$$
$$\begin{aligned} tr(P\Phi^{-1}P^T\Phi) & = \Sigma_{u\in V}\Sigma_{v \in V}{P_{uv}}^2\frac{\phi(u)}{\phi(v)}\\
& = \Sigma_{(u,v)\in E}\frac{\phi(u)}{\phi(v)}\cdot\frac{{A_{uv}}^2}{d_u^{out}} \end{aligned}$$

\noindent Combine the simplifications together and we have an approximation of $G$'s entropy:
$$S = 1-\frac{1}{|V|}-\frac{1}{2{|V|}^2}\left(\Sigma_{(u,v)\in E_{bd}}\frac{A_{uv}A_{vu}}{d_u^{out}d_v^{out}} + \Sigma_{(u,v)\in E}\frac{\phi(u)}{\phi(v)}\cdot\frac{{A_{uv}}^2}{d_u^{out}}\right)$$

\noindent Finally, we obtain $S^t$ and ${S}'^t$:
$$S^t = \sum_{SC} S_{SC}^t = \sum_{SC} 1-\frac{1}{|V_{SC}|}-\frac{1}{2{|V_{SC}|}^2}\left(\Sigma_{(u,v)\in E_{SC,bd}}\frac{1}{d_u^{out}d_v^{out}} + \Sigma_{(u,v)\in E}\frac{1}{d_u^{out}}\cdot\frac{\phi(u)}{\phi(v)}\right)$$
$${S}'^t = \sum_{SC} {{S}'}_{SC}^t = \sum_{SC} 1-\frac{1}{|{{V}'}_{SC}|}-\frac{1}{2{|{{V}'}_{SC}|}^2}\left(\Sigma_{(u,v)\in {{E}'}_{SC,bd}}\frac{{{{A}'}_{SC}}_{uv}{{{A}'}_{SC}}_{vu}}{d_u^{out}d_v^{out}} + \Sigma_{(u,v)\in E}\frac{\phi(u)}{\phi(v)}\cdot\frac{{{{{A}'}_{SC}}_{uv}}^2}{d_u^{out}}\right)$$

\subsection{Node Knowledge Temperature}
We employ the heat equation to compute node knowledge temperature. For a node $u$, its temperature change $\frac{dT_u^t}{dt}$ is:
$$\frac{dT_u}{dt} = \sum_{i=1}^{|V^t|}\widetilde{A_{iu}^t}(T_i - T_u)$$
where $\widetilde{A_{iu}^t}$ is defined as:
$$\widetilde{A_{iu}^t} = A_{iu}^t\cdot\left(0.5+\frac{{DiffIdx}_{i,u} - \min_{(a,b)\in E^t} {DiffIdx}_{a,b}}{\max_{(a,b)\in E^t} {DiffIdx}_{a,b} - \min_{(a,b)\in E^t} {DiffIdx}_{a,b} }\right)$$
$DiffIdx$ is defined previously in subsection topic skeleton tree. $\widetilde{A_{iu}}$ is the thermal conductivity between node $i$ and node $u$. \\


\noindent Before the heat diffusion, we need to fix the temperature of certain nodes and to precise the number of iteration of the heat equation. We assume that the pioneering work is the hottest and all the inactive papers are the coldest. An article $u$ is considered inactive if either of the following criteria is met:
\begin{enumerate}
    \item $u$ does not have any citation until timestamp $t$
    \item If $u$ joins in the topic before timestamp $t-1$ and $u$ does not have any new citations between timestamp $t-1$ and timestamp $t$.
\end{enumerate}

\noindent We first diffuse heat backward by transposing the adjacency matrix $\widetilde{A}$ for 1 iteration, then forward for $\lfloor avgStep \rfloor$ iterations. $avgStep$, defined during skeleton tree extraction, can be interpreted as the average hops between 2 random nodes in $G^t$. Backward propagation models the popularity gain in idea thanks to the newcomers and forward propagation models the heat diffusion due to the inheritance of topic knowledge. \\

\noindent We obtain node knowledge temperature ranging from 0 to 1 after applying the heat equation. The last step is to scale node knowledge temperature by topic knowledge temperature. Note $T_{u,std}^t$ and $T_u^t$ node $u$'s temperatures before and after the scaling and $\overline{T_{std}^t}$ the average node knowledge temperature before the scaling, we have
$$T_u^t = T_{u,std}^t\cdot \frac{T^t}{\overline{T_{std}^t}}$$

\subsection{Forest Helping}
Forest helping is designed for a group of similar topics. Through this mechanism, thriving topics "transfuse" a small part of their energy to other stagnant sister topics. The helping does not change the total energy of topic group:
$$\sum_{j=1}^K cn_j^tT_j^t = \sum_{j=1}^K cn_j^tT_{j,forest}^t$$
where $K$ is the number of topics in a group and $T_{j, forest}^t$ the average temperature of topic $j$ after the helping.\\

\noindent If all topics in the group are hotter than last period, no helping takes place. Else, all of the topics with a rising knowledge temperature help the rest.  \\

\noindent We model the probability that "a thriving topic is willing to help others" follows a beta distribution $B(1, \sum_{j=1}^K a_j)$, $a_j$ being topic age. Beta distribution varies from 0 to 1, which corresponds with option "not help" and option "help with all I have". We assume a prosperous topic will give an amount of energy equal to the expectation of the distribution. Hence, at time $t$, the energy that a topic gives away is proportional to its own knowledge temperature and is inversely proportional to the ages of the entire group:
$$\Delta E = cn^t\frac{1}{1+\sum_{j=1}^K a_j^t}T^t$$

\noindent The energy received by each topic in need of help is proportional to its node number. Therefore, they have an identical increase in their knowledge temperatures:
$$\delta T = \frac{\Delta E}{\sum_{j}n_j^t} $$

\noindent As topics mature, their initially close connection in thoughts will wear off by time. Consequently, the amount of energy transmitted through forest helping will decrease.

\section{Experiments}
We first present our results and analysis for individual topic, next discuss the forest helping results for topic group. Note that most of the data for 2020 only cover the first 2 months, therefore the latest temperature is not definite. The data in the tables are rounded to 3 decimal places. We set two constants in $T_{growth}^t$'s calculation as $R=8,c=1$. For topics with more than 5000 articles, the coefficient $k=10$ in $T_{growth}^0$'s computation. Else, $k=100$.\\

\noindent In this section, we refer to "popular child papers" as the child papers with high in-topic citations unless explicitly specified. Child papers with titles in topic's current skeleton tree are the ones with the highest in-topic citations, whereas the highlighted child papers in galaxy maps are the ones that has won the most total citation counts.\\

\noindent Based on the evolution of knowledge temperature, we classify topics into 4 categories: rising topic, rise-and-fall topic, awakened topic and rise-fall-cycle topic. Among 16 topics, 9 follow a rise-then-fall pattern, with their knowledge temperature reaching record high shortly after birth. 3 topics have been almost always on the rise until today. 2 topics have waited a long time before being recognised and having a surge in knowledge temperature. We refer to them as awakened topics. The rest exhibit a periodic knowledge temperature variation characterised by multiple up-down cycles.\\

\subsection{Rising Topics}
\subsubsection{Regulatory T Cells: Mechanisms of Differentiation and Function}

\begin{figure}[htbp]
\centering
\begin{subfigure}[t]{0.6\linewidth}
\includegraphics[width=\linewidth]{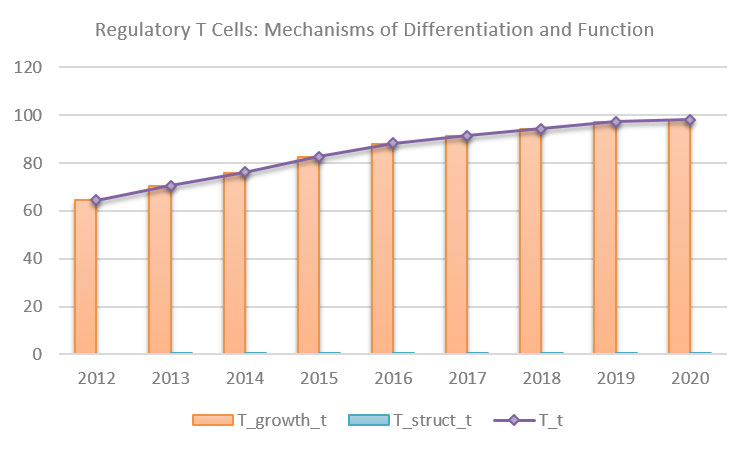}
\end{subfigure}

\begin{subfigure}[t]{\linewidth}
\centering
\begin{tabular}{ccccccccc}
\hline
year & $|V^t|$ & $|E^t|$ & $n_t$ & $V_t$ & ${UsefulInfo}^t$ & $T_{growth}^t$ & $T_{struct}^t$ & $T^t$\\
\hline
2012 & 49 & 58 & 46.15 & 49 & 2.85 & 64.466 &   & 64.466 \\

2013 & 241 & 346 & 207.527 & 241 & 33.473 & 70.51 & 0.058 & 70.567 \\

2014 & 460 & 773 & 367.669 & 460 & 92.331 & 75.964 & 0.09 & 76.053 \\

2015 & 659 & 1321 & 484.941 & 659 & 174.059 & 82.509 & 0.046 & 82.556 \\

2016 & 841 & 1949 & 579.149 & 841 & 261.851 & 88.168 & 0.043 & 88.211 \\

2017 & 1027 & 2633 & 682.316 & 1027 & 344.684 & 91.388 & 0.022 & 91.411 \\

2018 & 1199 & 3334 & 771.581 & 1199 & 427.419 & 94.35 & 0.024 & 94.375 \\

2019 & 1356 & 4053 & 845.96 & 1356 & 510.04 & 97.323 & 0.016 & 97.339 \\

2020 & 1381 & 4190 & 854.519 & 1381 & 526.481 & 98.125 & 0.004 & 98.129 \\
\hline
\end{tabular}
\end{subfigure}

\caption{Regulatory T cells: topic statistics and knowledge temperature evolution}
\label{fig:3209_chart}
\end{figure}

The topic has been thriving ever since its birth in 2012 (Fig. \ref{fig:3209_chart}). It has a very stable annual growth of $T^t$ and $T_{growth}^t$, which corresponds with its seemingly uniform publishing rhythm: an annual publication count always over 10\% of the total size between 2013 and 2019. In addition, popular child papers came at a steady speed during 2012 and 2015. They have helped maintain a stable knowledge accumulation.\\

\begin{figure}[htbp]
\begin{subfigure}{\textwidth}
\begin{minipage}[t]{0.5\linewidth}
\includegraphics[width = \linewidth]{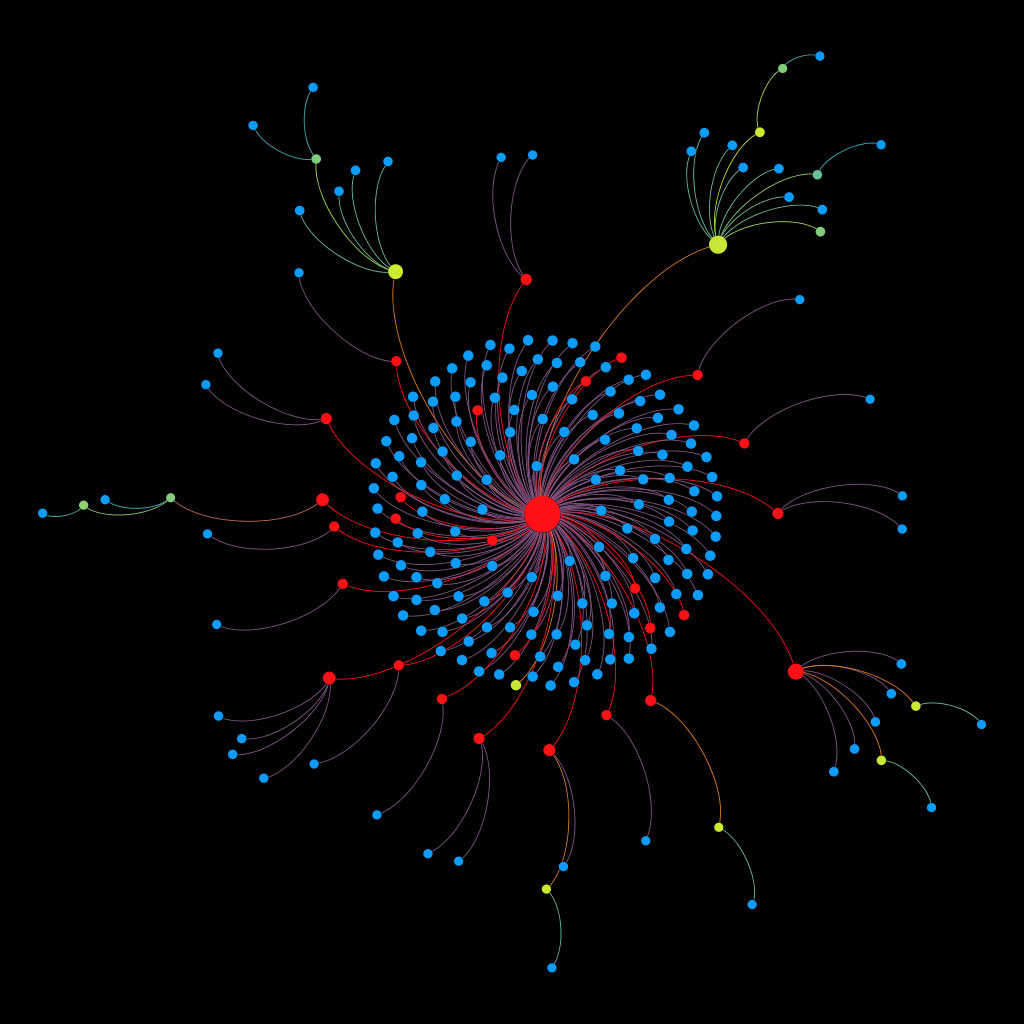}
\caption{Skeleton tree until 2013}
\end{minipage}
\begin{minipage}[t]{0.5\linewidth}
\includegraphics[width = \linewidth]{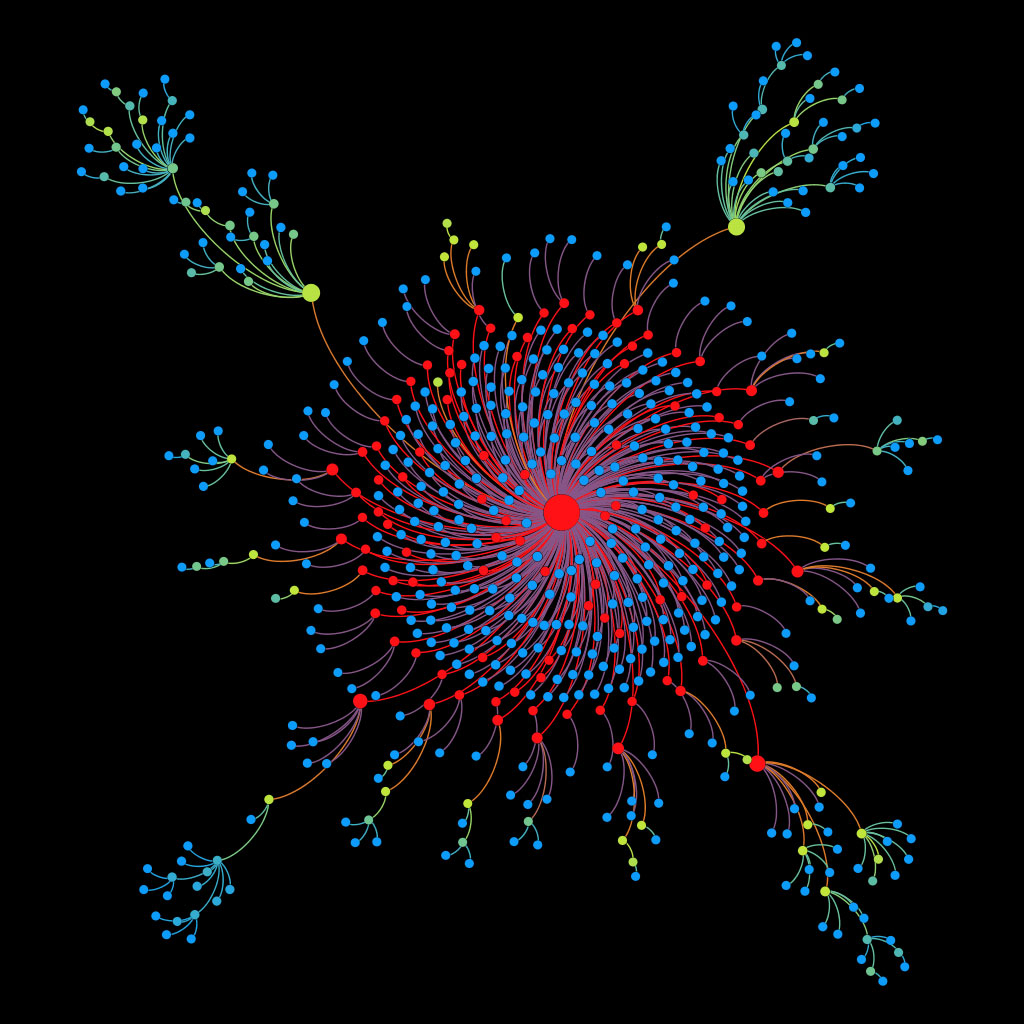}
\caption{Skeleton tree until 2015}
\end{minipage}
\end{subfigure}
\vspace{2mm}
\begin{subfigure}{\textwidth}
\begin{minipage}[t]{0.5\linewidth}
\includegraphics[width = \linewidth]{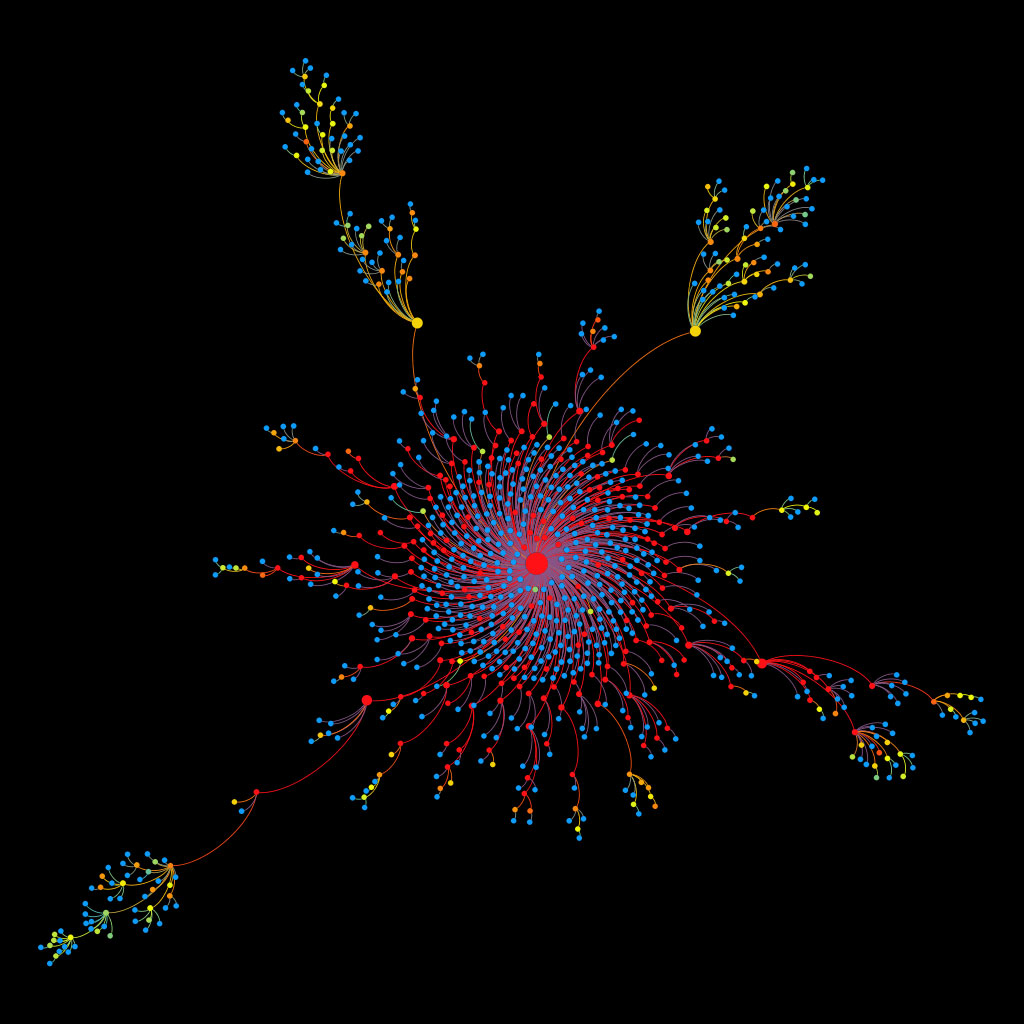}
\caption{Skeleton tree until 2017}
\end{minipage}
\begin{minipage}[t]{0.5\linewidth}
\includegraphics[width = \linewidth]{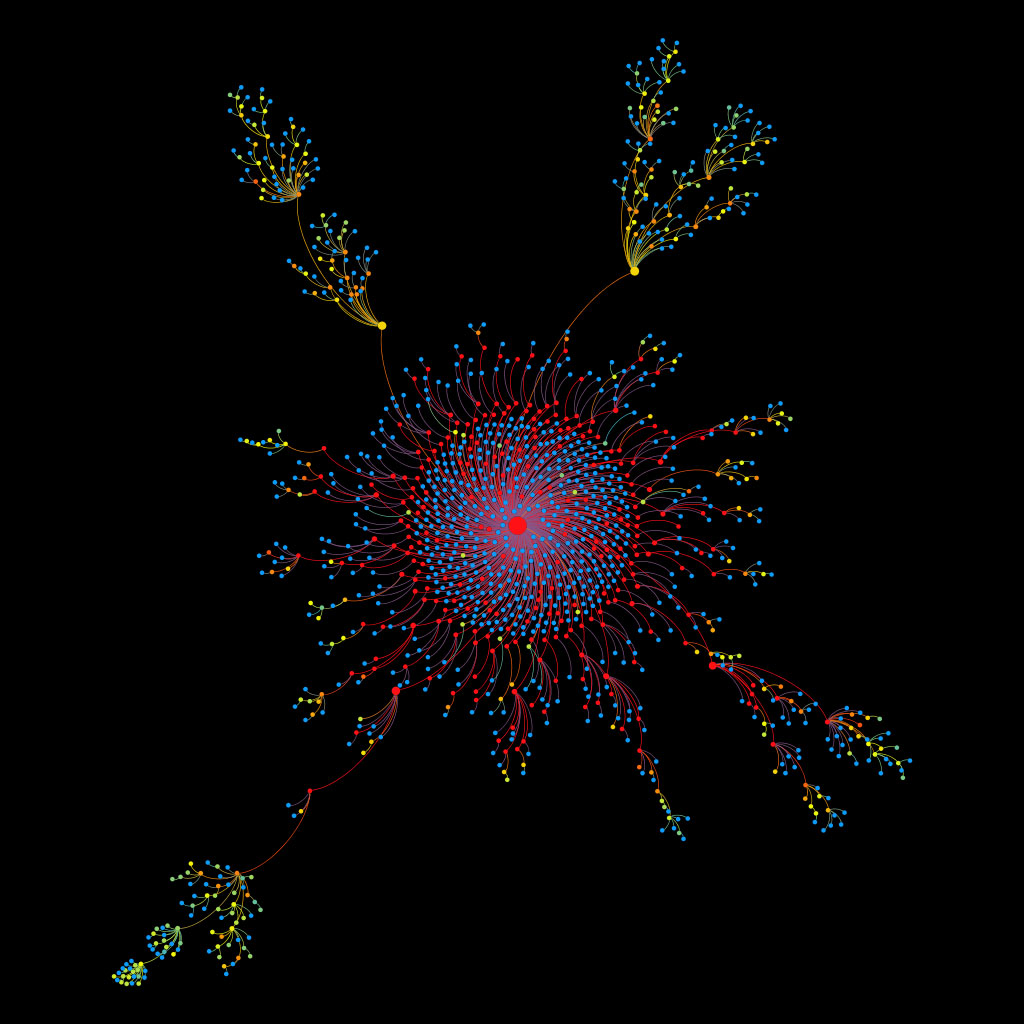}
\caption{Skeleton tree until 2019}
\end{minipage}
\end{subfigure}
\caption{Regular T Cells: Skeleton tree evolution}
\label{fig:3209-tree_evo}
\end{figure}

\noindent $T_{structure}^t$ remains tiny, suggesting that this topic has a gradual knowledge structure progression and has not experienced a sudden short-term impact gain. Indeed, although we observe constant visible development in skeleton tree, we don't see any disruptive changes in the overall structure (Fig. \ref{fig:3209-tree_evo}). Under the leadership of several popular child papers, the topic have been succeeded in developing some sub-directions, as is reflected by the fact that multiple non-trivial branches have been gradually growing out of the central cluster led by the pioneering work. Yet so far the pioneering paper remains the absolute topic center. Moreover, tiny twigs are forming around the center at a seemingly uniform speed, which may be a good sign for more novel research focus. The vigor of skeleton tree shows again the topic's slowly yet firmly rising popularity and impact.\\

\begin{figure}
    \centering
    \begin{subfigure}{\linewidth}
    \begin{minipage}[t]{0.55\textwidth}
    \centering
    \includegraphics[width = 0.9 \linewidth]{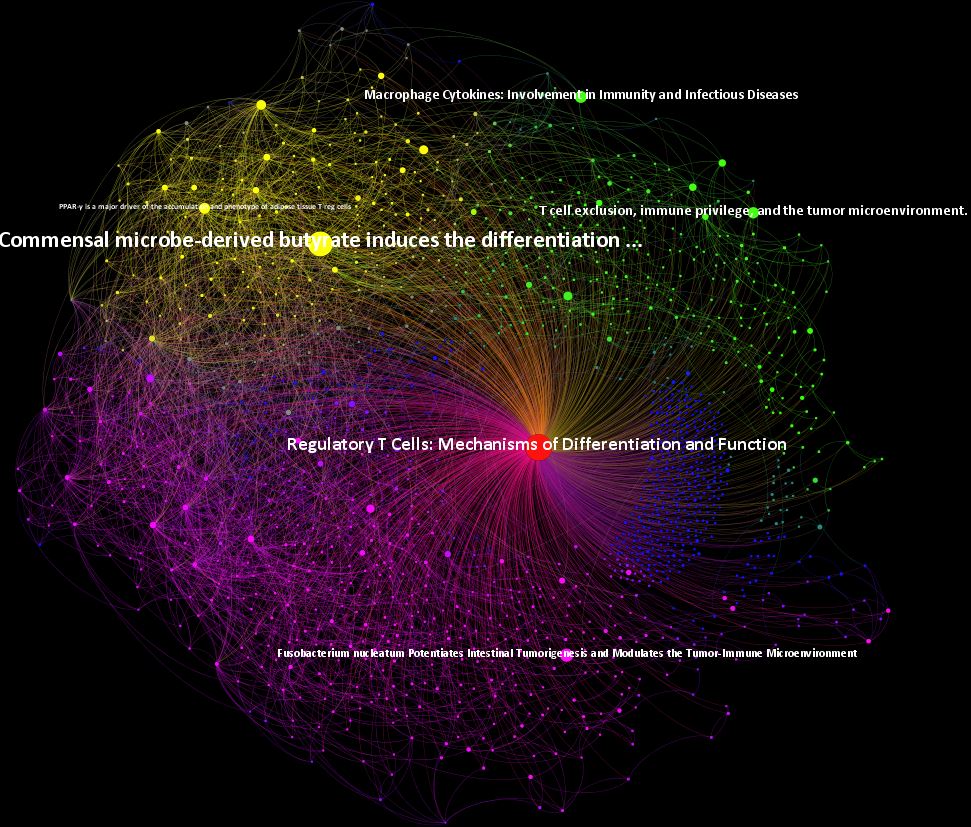}
    \end{minipage}
    \begin{minipage}[t]{0.45\textwidth}
    \centering
    \includegraphics[width = 0.935\linewidth]{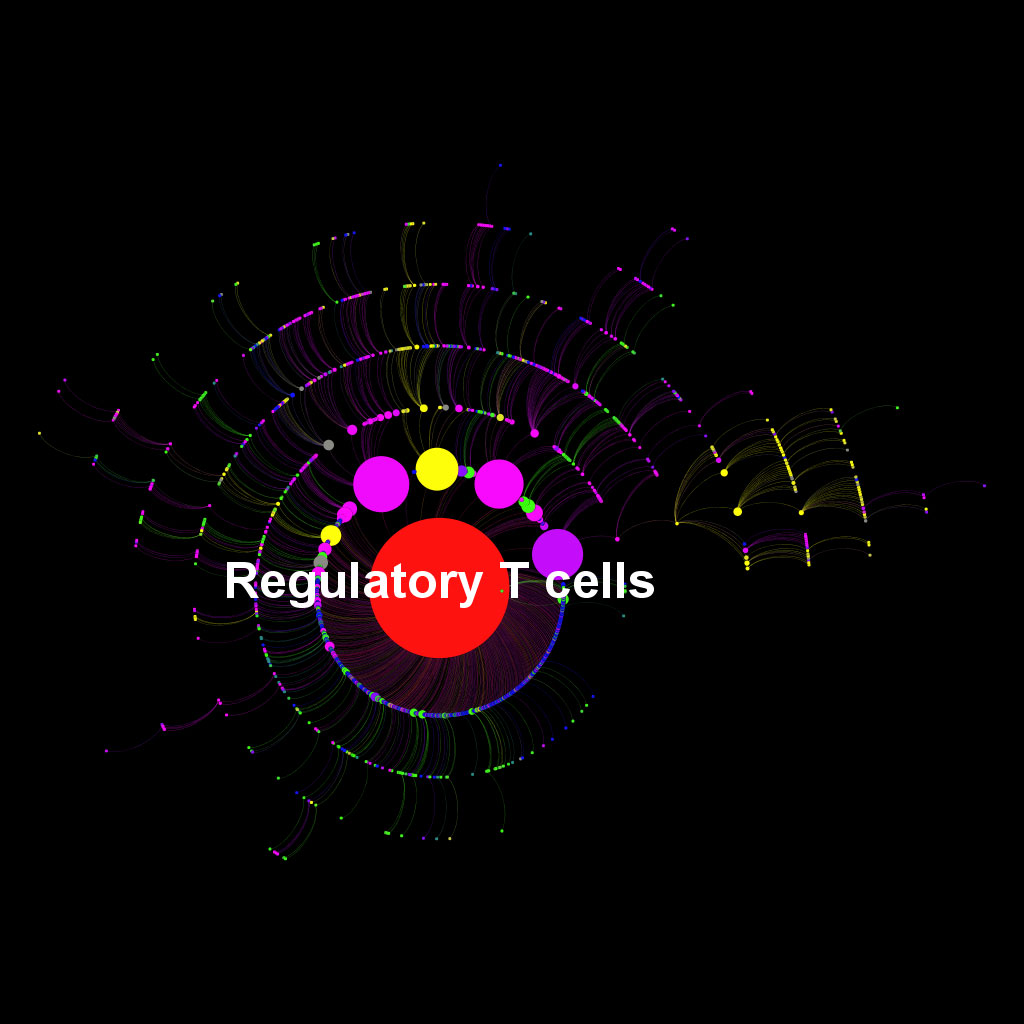}
    \end{minipage}
    \end{subfigure}

    \vspace{5mm}

    \begin{subfigure}{0.55\linewidth}
    \centering
    \includegraphics[width = \linewidth]{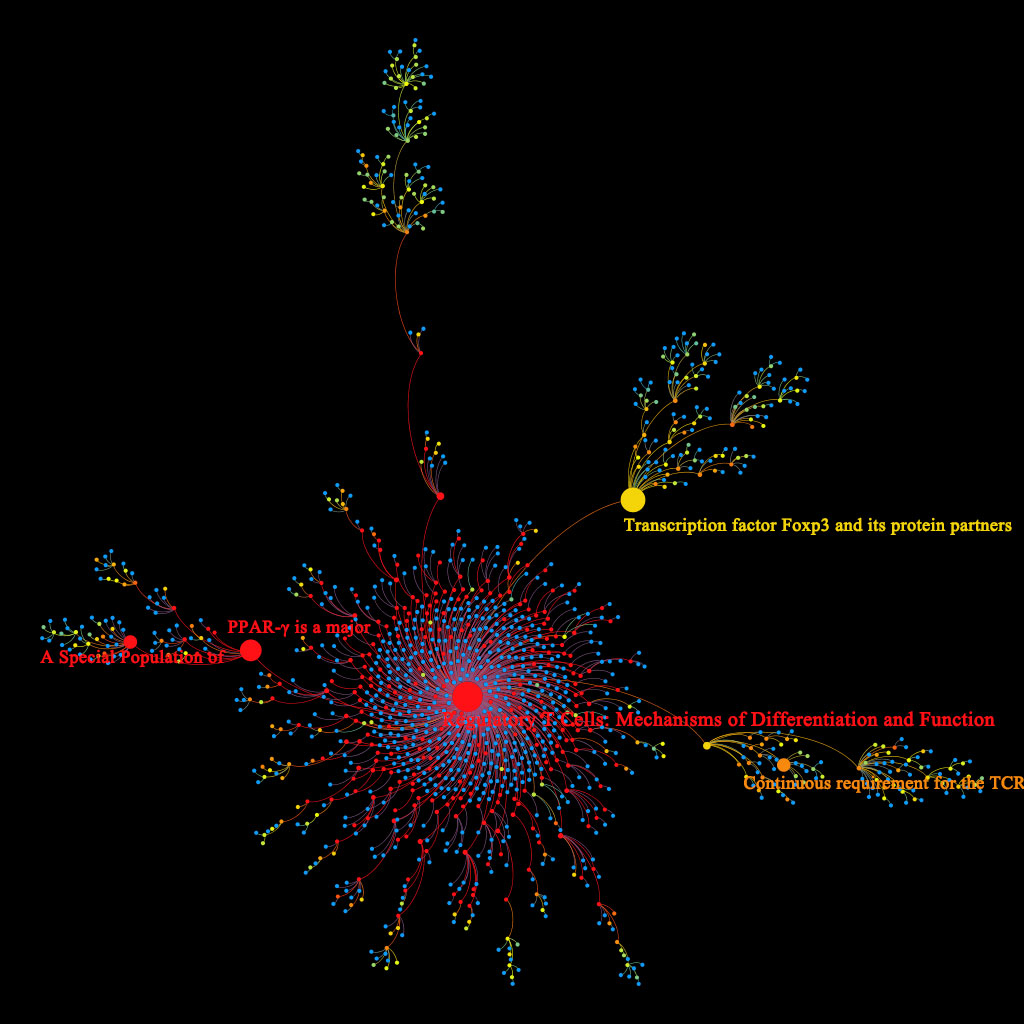}
    \end{subfigure}
    \begin{subfigure}{0.4\linewidth}
    \begin{minipage}[t]{\textwidth}
    \centering
    \includegraphics[width = 0.85\linewidth]{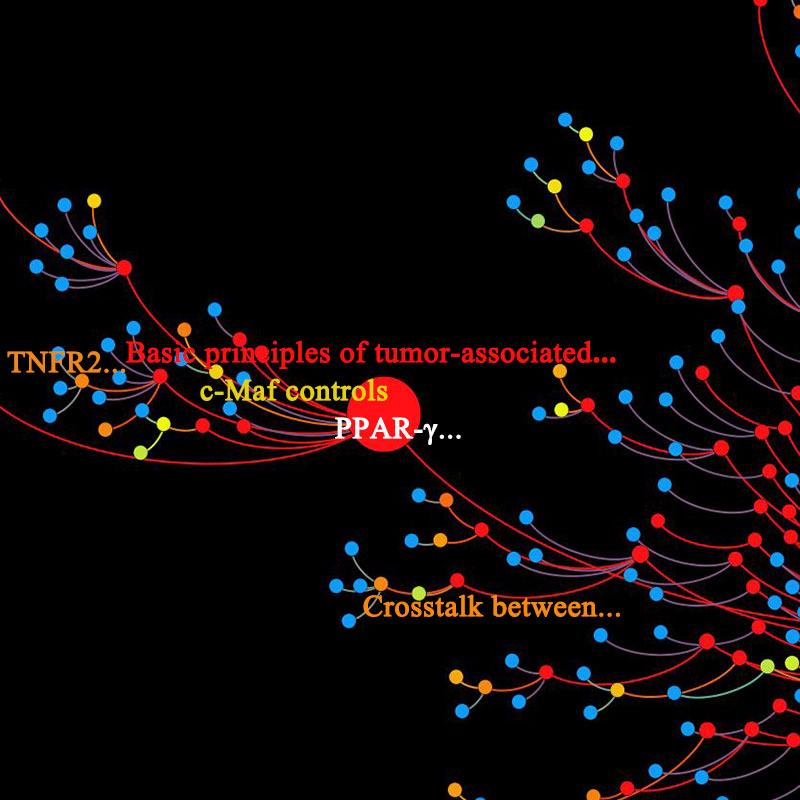}
    \end{minipage}
    \end{subfigure}
    \caption{Regular T Cells: Galaxy map, current skeleton tree and its regional zoom. Papers with more than 55 in-topic citations are labelled by title in the skeleton tree. Except the pioneering work, corresponding nodes' size is amplified by 3 times.}
    \label{fig:3209-2020}
\end{figure}

\noindent Now we closely examine its latest skeleton tree (Fig. \ref{fig:3209-2020}). Almost all the hottest articles surround the pioneering paper and node knowledge temperature decreases globally as the articles are located farther away from the pioneering paper. Note that the blue nodes that surround the pioneering work are articles with little development within the topic. If we let alone these coldest papers, the heat distribution fits the general rules "the older the hotter" (Fig. 5(a)) and "the more influential the hotter" (Fig. \ref{fig:citation_T}(a))). Nonetheless, there are exceptions. Age and citations are not guarantee for heat-level. For example, popular child paper `Transcription factor Foxp3 and its protein partners form a complex regulatory network' is colder than some of its child papers in the research branch it leads. The intrinsic difference of their research ideas, which is partly reflected by the average heat-level of their citations, causes the temperature difference. Besides, we also identify some young and hot articles. For example, 2 papers published in 2017, `TNFR2: A Novel Target for Cancer Immunotherapy' (TNFR2) and `Crosstalk between Regulatory T Cells and Tumor-Associated Dendritic Cells Negates Anti-tumor Immunity in Pancreatic Cancer' and 1 paper published in Nature Immunology in 2018, `c-Maf controls immune responses by regulating disease-specific gene networks and repressing IL-2 in CD4 + T cells' all have a knowledge temperature above average. All of them have already inspired several works. Their popularity not only manifests the boosting effect of new articles on original work, but also shows the lasting activity of this topic. Overall, these atypical examples suggest that the positive correlation between node knowledge temperature and age or pure impact in terms of citation statistics is weak.  \\

\noindent In particular, we find the knowledge temperature evolution of paper `Basic principles of tumor-associated regulatory T cell biology' (BPTRT), published in 2013 in journal \textit{Trends in Immunology} very interesting. This article is the parent paper of `TNFR2: A Novel Target for Cancer Immunotherapy' in 2020's skeleton tree. Its temperature dropped from 213.26 to around 170 between 2013 to 2016 despite the fact that it had new followers and that the whole topic went hotter during this period. By the end of the following year, its temperature skyrocketed to around 330. The sudden gain is the result of an accumulated influence during period 2013-2016 and the global heat diffusion owing to the topic's gradual development. Its temperature has mildly climbed up since 2016, which is in accordance with topic knowledge temperature dynamics. The arrival of its promising child, TNFR2. TNFR2 has helped keep BPTRT's heat-level with its own development. This example well illustrates child article's role in maintaining parent paper's popularity and impact. \\

\noindent We observe in addition certain clustering effect in the skeleton tree. For example, almost all direct children of paper `Pregnancy imprints regulatory memory that sustains anergy to fetal antigen' have similar research themes as itself (Table \ref{tab:3209-clustering}). This confirms the effectiveness of our skeleton tree extraction algorithm.

\begin{table}
    \centering
    \begin{tabular}{p{15cm} p{1cm}}
        \hline
        title & year\\
        \hline
         \textcolor{red}{Pregnancy} imprints regulatory memory that sustains anergy to fetal antigen predictions using deep neural networks & 2012 \\

         Mechanisms of T cell tolerance towards the allogeneic \textcolor{orange}{fetus} & 2013\\

        \textcolor{orange}{Pregnancy} Complications and Unlocking the Enigma of Fetal Tolerance Regulatory T Cells: New Keys for Further & 2014\\

        Regulatory T Cells: New Keys for Further Unlocking the Enigma of Fetal Tolerance and \textcolor{orange}{Pregnancy} Complications & 2014\\

        The immunology of \textcolor{orange}{pregnancy}: regulatory T cells control maternal immune tolerance toward the fetus & 2014\\

        Regulatory T Cells: Types, Generation and Function & 2014\\

        \textcolor{orange}{Daughter’s} Tolerance of \textcolor{orange}{Mom} Matters in Mate Choice & 2015\\

        Regulatory T cells in \textcolor{orange}{embryo} implantation and the immune response to \textcolor{orange}{pregnancy} & 2018\\

        Alloreactive fetal T cells promote uterine contractility in \textcolor{orange}{preterm labor} via IFN-$\gamma \rm{\,and\,} TNF-\alpha$ & 2018\\
        \hline
    \end{tabular}
    \caption{Regular T Cells: Clustering effect example. First line is the parent paper and the rest children.}
    \label{tab:3209-clustering}
\end{table}

\subsubsection{Empirical Evaluation of Gated Recurrent Neural Networks on Sequence Modeling}

\begin{figure}[htbp]
\centering
\begin{subfigure}[t]{0.6\linewidth}
\centering
\includegraphics[width=\linewidth]{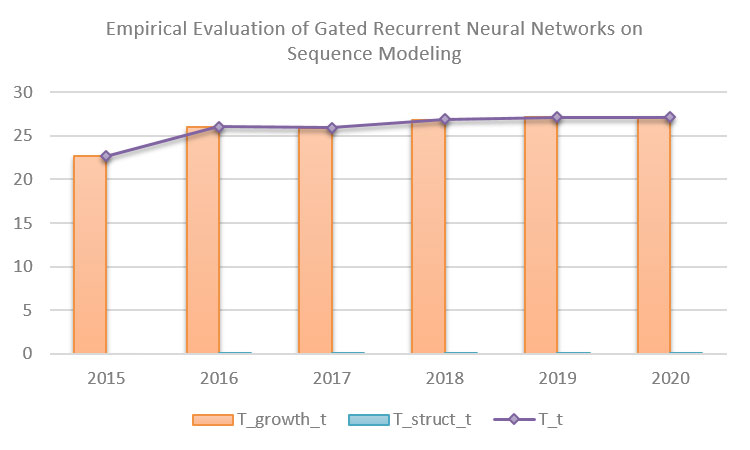}
\end{subfigure}

\begin{subfigure}[t]{\linewidth}
\centering
\begin{tabular}{ccccccccc}
\hline
year & $|V^t|$ & $|E^t|$ & $n_t$ & $V_t$ & ${UsefulInfo}^t$ & $T_{growth}^t$ & $T_{struct}^t$ & $T^t$\\
\hline
2015 & 53 & 76 & 43.823 & 53 & 9.178 & 22.686 &   & 22.686 \\

2016 & 295 & 514 & 212.885 & 295 & 82.115 & 25.994 & 0.03 & 26.024 \\

2017 & 749 & 1377 & 543.23 & 749 & 205.77 & 25.864 & 0.058 & 25.921 \\

2018 & 1328 & 2619 & 929.445 & 1328 & 398.555 & 26.802 & 0.085 & 26.887 \\

2019 & 2109 & 4287 & 1459.035 & 2109 & 649.965 & 27.115 & 0.034 & 27.149 \\

2020 & 2282 & 4675 & 1576.618 & 2282 & 705.382 & 27.151 & 0.011 & 27.162 \\
\hline
\end{tabular}
\end{subfigure}
\caption{GRU: topic statistics and knowledge temperature evolution}
\label{fig:168338164_chart}
\end{figure}

As is shown by the basic statistics and $T^t$, the topic is keeping popularity and steadily gaining impact (Fig.\ref{fig:168338164_chart}). Its popular child papers came at a steady speed during 2015 and 2017. Apart from enriching topic knowledge pool with their own ideas, they also attracted new researches' attention and thus have helped maintain a stable knowledge accumulation. The topic has been accelerating its expansion since 2017. It witnessed the biggest annual publication count in 2019. Yet as most child papers published no earlier than 2018 have had little development, the publication surge did not result in a significant uprise in $T^t$. \\

\begin{figure}[htbp]
\begin{subfigure}{\textwidth}
\begin{minipage}[t]{0.5\linewidth}
\includegraphics[width = \linewidth]{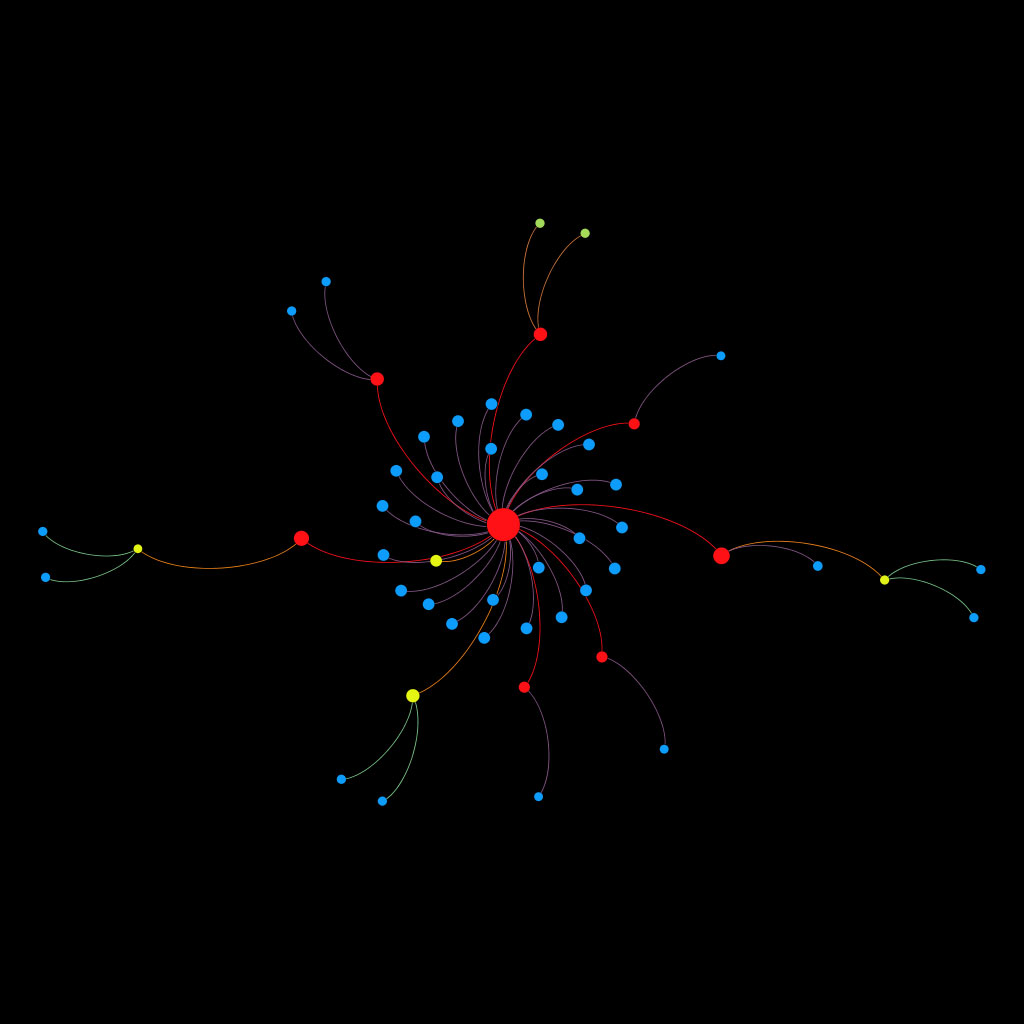}
\caption{Skeleton tree until 2015}
\end{minipage}
\begin{minipage}[t]{0.5\linewidth}
\includegraphics[width = \linewidth]{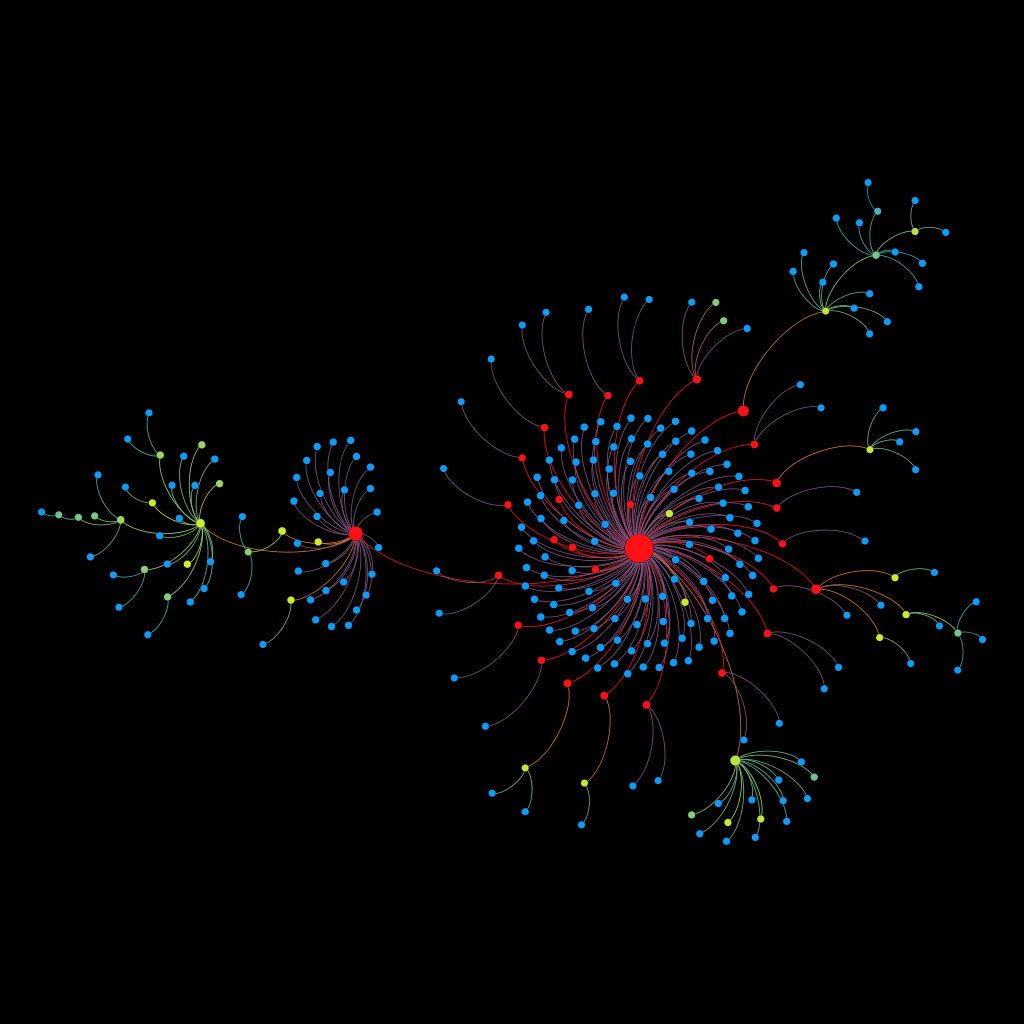}
\caption{Skeleton tree until 2016}
\end{minipage}
\end{subfigure}
\vspace{2mm}
\begin{subfigure}{\textwidth}
\begin{minipage}[t]{0.5\linewidth}
\includegraphics[width = \linewidth]{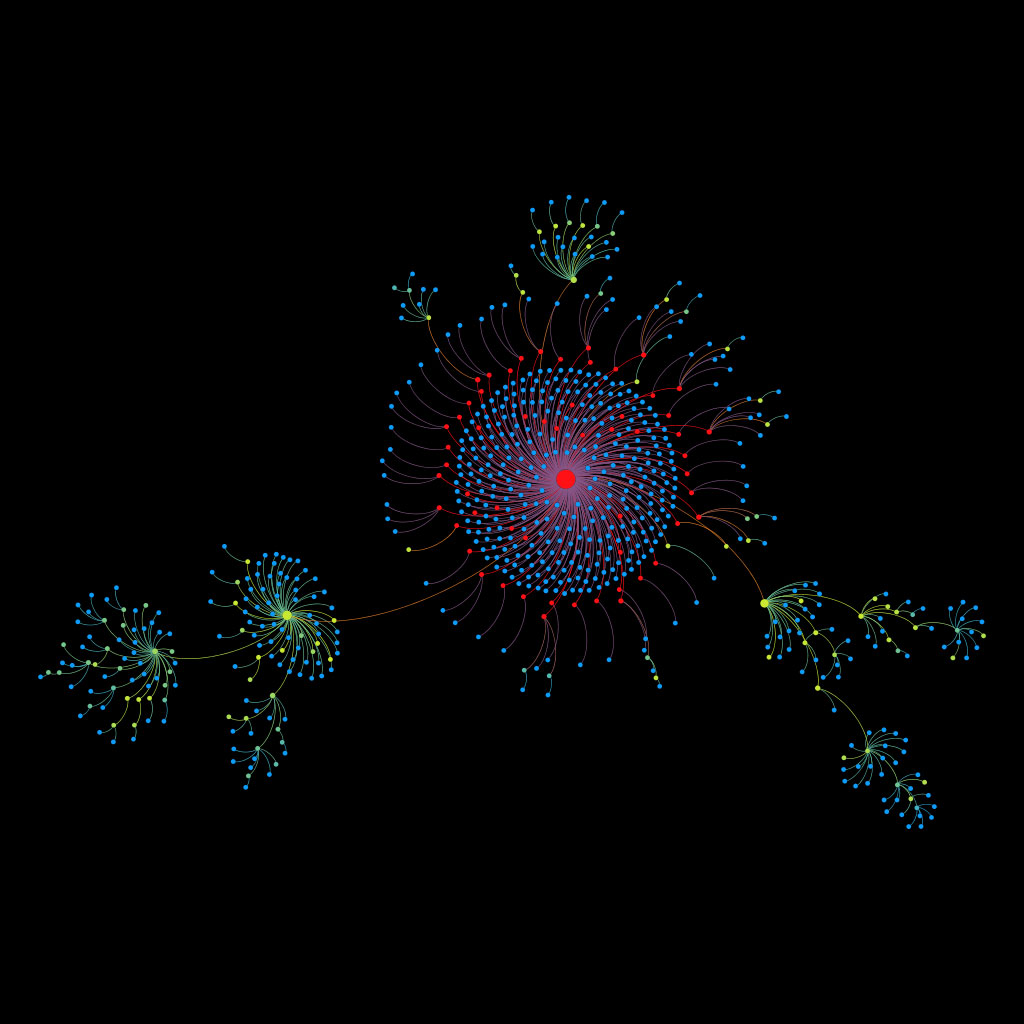}
\caption{Skeleton tree until 2017}
\end{minipage}
\begin{minipage}[t]{0.5\linewidth}
\includegraphics[width = \linewidth]{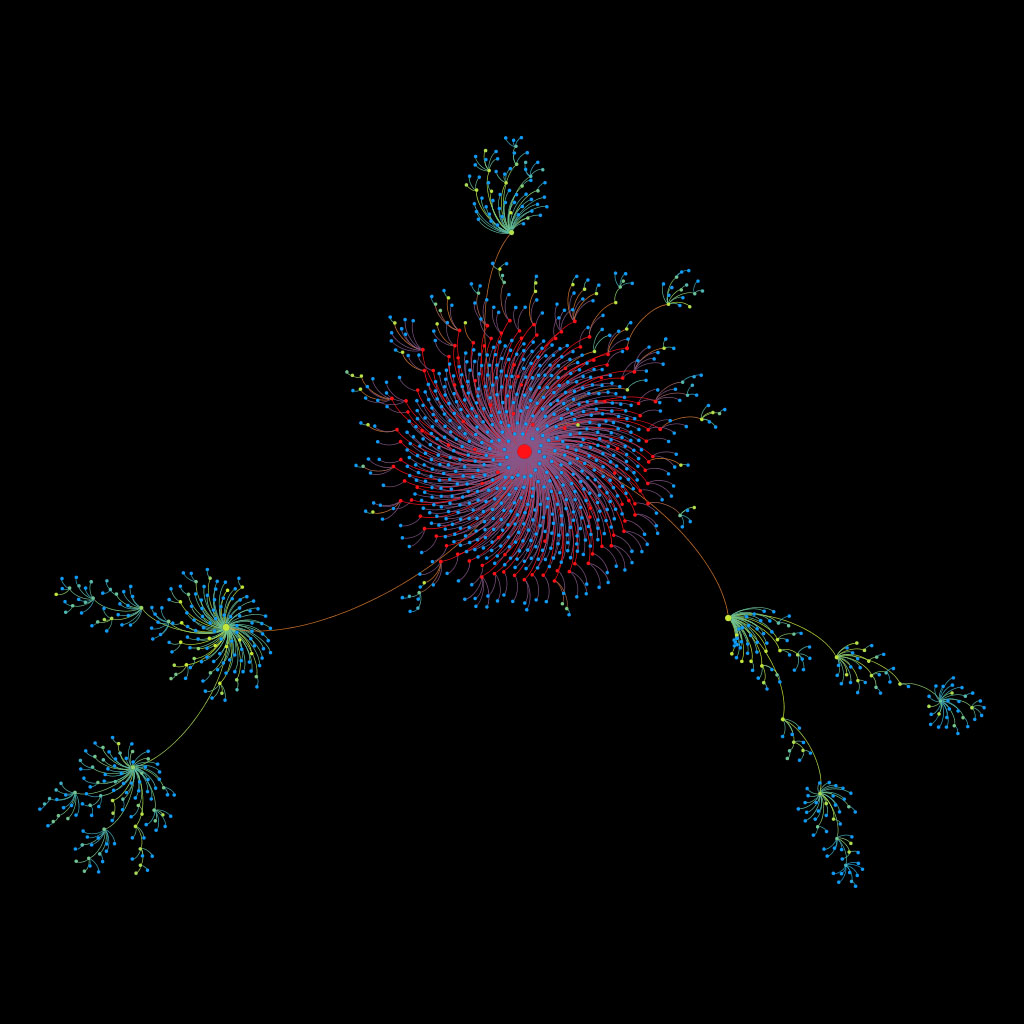}
\caption{Skeleton tree until 2018}
\end{minipage}
\end{subfigure}
\caption{GRU: Skeleton tree evolution}
\label{fig:168338164-tree_evo}
\end{figure}

\noindent $T_{structure}^t$ remains tiny compared to $T_{growth}^t$, suggesting that the topic has a gradual knowledge structure progression and has not experienced a sudden short-term impact gain. Indeed, although its skeleton tree has constant visible development (Fig. \ref{fig:168338164-tree_evo}), so far no child paper is able to defy the absolute authority of the pioneering paper, the center of the biggest cluster. Several popular child papers have each led a research sub-field in the topic, as is depicted by the small bundles extending from the central cluster. In particular,  popular child paper `LSTM: A Search Space Odyssey' in 2017 has inspired 2 schools of thoughts. The maturation of these newly emerged research directions accounts for a higher $T_{structure}^t$ in the first years of the topic. Overall, we observe a universal non-trivial growth in the skeleton tree. The vigor of skeleton tree shows again the slowly yet firmly increasing popularity and impact of this topic.\\

\begin{figure}
    \centering
    \begin{subfigure}{\linewidth}
    \begin{minipage}[t]{0.5\textwidth}
    \centering
    \includegraphics[width = 0.9\linewidth]{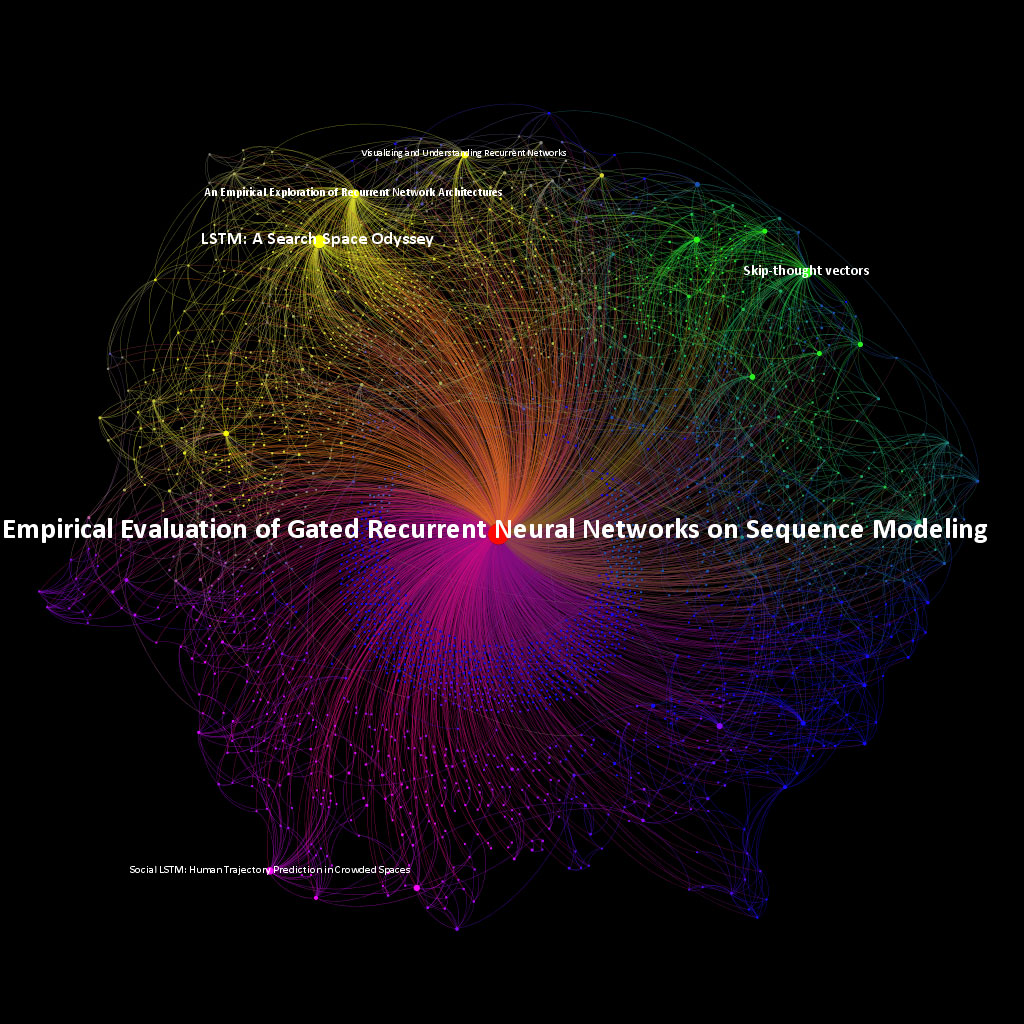}
    \end{minipage}
    \begin{minipage}[t]{0.5\textwidth}
    \centering
    \includegraphics[width = 0.9\linewidth]{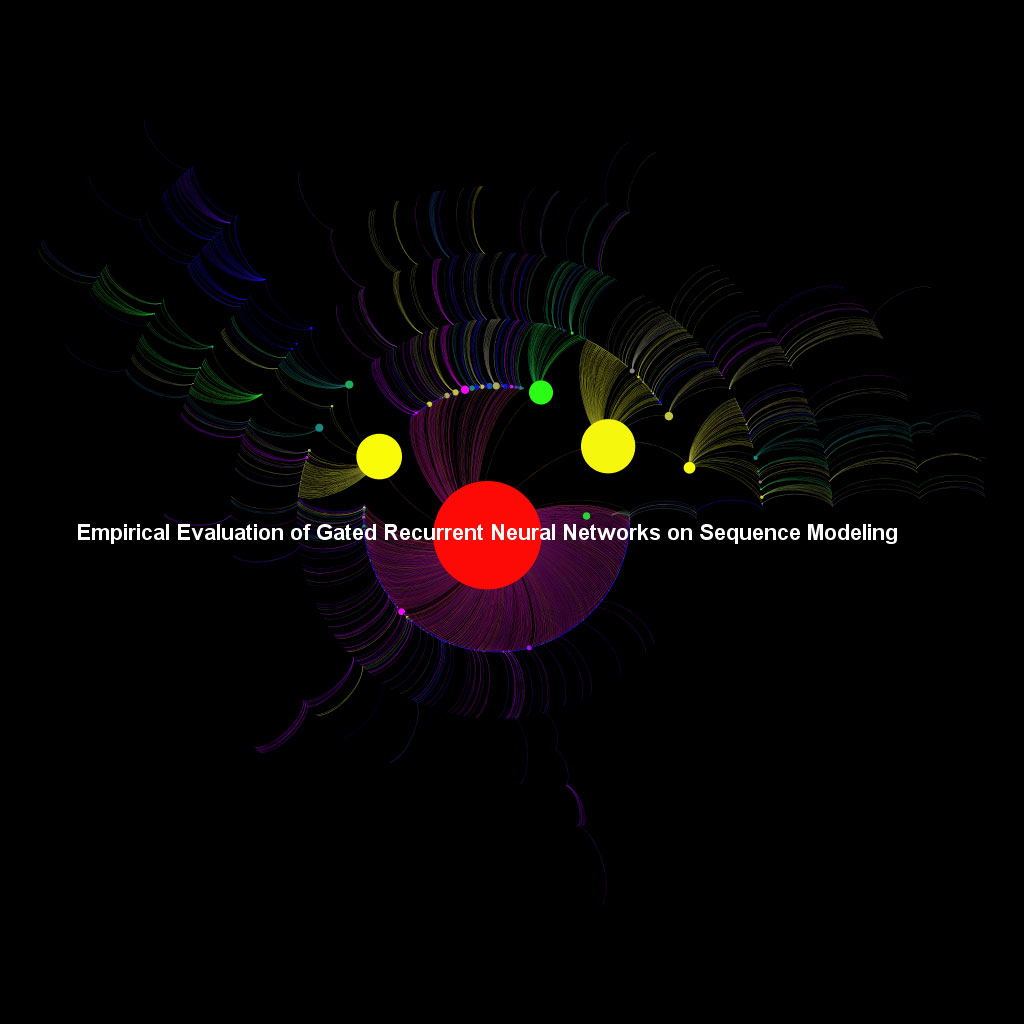}
    \end{minipage}
    \end{subfigure}

    \vspace{5mm}

    \begin{subfigure}{0.6\linewidth}
    \includegraphics[width = \linewidth]{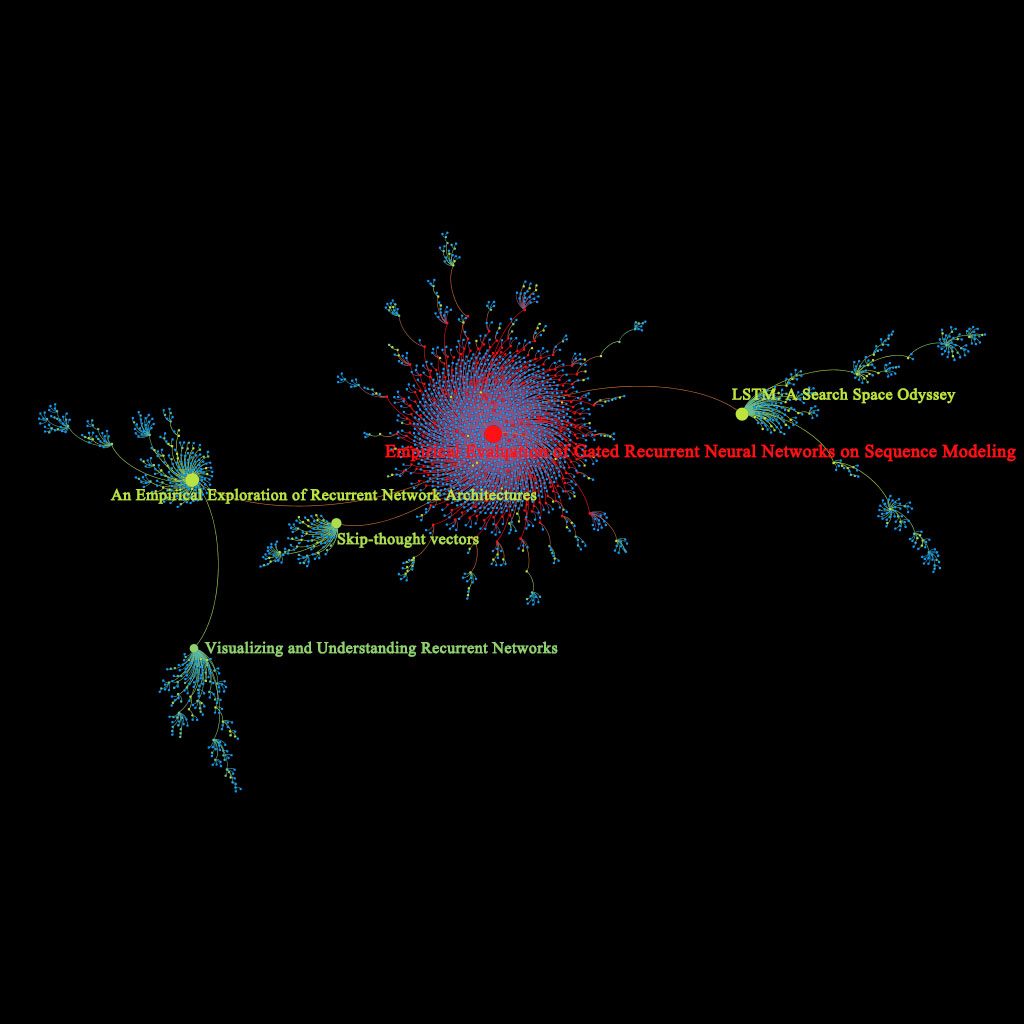}
    \end{subfigure}
    \caption{GRU: Galaxy map and current skeleton tree. Papers with more than 60 in-topic citations are labelled by title in the skeleton tree. Except the pioneering work, corresponding nodes' size is amplified by 3 times.}
    \label{fig:168338164-2020}
\end{figure}

\noindent Now we closely examine its latest skeleton tree (Fig. \ref{fig:168338164-2020}). The decrease in node knowledge temperature from root, the pioneering work, to leaves is obvious, which accords with the general rule "the older the hotter" (Fig. 5(b)). Note that the blue nodes that surround the pioneering work and popular child papers are articles with little development within the topic. In particular, the heat distribution is rather concentrated in old papers. This phenomenon is in line with our above observation that young child papers have little authority in the topic. The limited heat diffusion is also why most popular child papers have a node knowledge temperature no greater than average. This topic is quite young. It needs more time to fully explore the potential of new ideas and to trigger a thorough heat diffusion in its range. \\

\noindent In particular, we find the knowledge temperature evolution of the second most-cited paper `An Empirical Exploration of Recurrent Network Architectures', published in 2015 in journal \textit{International Conference on Machine Learning} very interesting (Fig. \ref{fig:168338164-2020}). This article became much hotter from 2015 to 2016 thanks to its numerous child papers. However, its temperature reduces by half from 182.578 to 89.19 the next year upon the arrival of the third most-cited paper `LSTM: A Search Space Odyssey', the leader of the right major branch in the skeleton tree (Fig. \ref{fig:168338164-tree_evo} (b,c)). Since then, its temperature has been slightly decreasing to around 80 in 2020. The sudden drop is a vivid illustration of the rivalry within the topic.  \\

\noindent We observe in addition certain clustering effect in the skeleton tree. For example, almost all direct children of paper `Machine Health Monitoring Using Local Feature-Based Gated Recurrent Unit Networks' study the industrial applications of gated recurrent unit network  (Table \ref{tab:168338164-clustering}). This illustrates the effectiveness of our skeleton tree extraction algorithm.

\begin{table}
    \centering
    \begin{tabular}{p{15cm} p{1cm}}
        \hline
        title & year\\
        \hline
         \textcolor{red}{Machine Health} Monitoring Using Local Feature-Based Gated Recurrent Unit Networks & 2018 \\

         Integrating Convolutional Neural Network and Gated Recurrent Unit for \textcolor{orange}{Hyperspectral Image} Spectral-Spatial Classification & 2018\\

        Comparison of Deep learning models on time series forecasting : a case study of \textcolor{orange}{Dissolved Oxygen Prediction} & 2019\\

        \textcolor{orange}{Anomaly Detection of Wind Turbine} Generator Based on Temporal Information & 2019\\

        \textcolor{orange}{Energy} price prediction based on independent component analysis and gated recurrent unit neural network & 2019\\

        Condition \textcolor{orange}{monitoring of wind turbines} based on spatio-temporal fusion of SCADA data by convolutional neural networks and gated recurrent units & 2019\\

        Intelligent \textcolor{orange}{Fault Diagnosis} of Rolling Bearing Using Adaptive Deep Gated Recurrent Unit & 2019\\

        \textcolor{orange}{Abnormality Diagnosis} Model for \textcolor{orange}{Nuclear Power Plants} Using Two-Stage Gated Recurrent Units & 2020\\

        \hline
    \end{tabular}
    \caption{GRU: Clustering effect example. First line is the parent paper and the rest children.}
    \label{tab:168338164-clustering}
\end{table}

\subsubsection{Neural networks for pattern recognition}
\begin{figure}[htbp]
\centering
\begin{subfigure}[t]{0.7\linewidth}
\centering
\includegraphics[width=\linewidth]{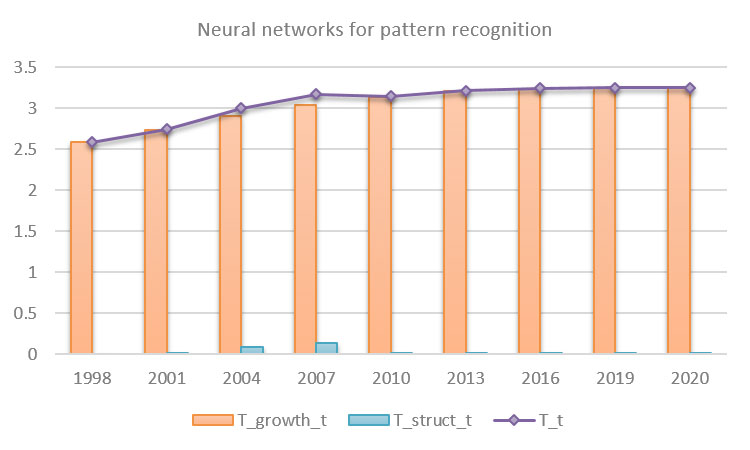}
\end{subfigure}

\begin{subfigure}[t]{\linewidth}
\centering
\begin{tabular}{ccccccccc}
\hline
year & $|V^t|$ & $|E^t|$ & $n_t$ & $V_t$ & ${UsefulInfo}^t$ & $T_{growth}^t$ & $T_{struct}^t$ & $T^t$\\
\hline
1998 & 586 & 848 & 494.418 & 586 & 91.582 & 2.583 &   & 2.583 \\

2001 & 2235 & 3764 & 1779.662 & 2235 & 455.338 & 2.737 & 0.008 & 2.745 \\

2004 & 4761 & 9302 & 3564.22 & 4761 & 1196.78 & 2.911 & 0.089 & 3 \\

2007 & 8058 & 17236 & 5785.872 & 8058 & 2272.128 & 3.035 & 0.137 & 3.172 \\

2010 & 11202 & 25723 & 7789.449 & 11202 & 3412.551 & 3.134 & 0.012 & 3.146 \\

2013 & 13788 & 33517 & 9374.213 & 13788 & 4413.787 & 3.205 & 0.005 & 3.21 \\

2016 & 15763 & 39120 & 10605.175 & 15763 & 5157.825 & 3.239 & 0.004 & 3.243 \\

2019 & 16927 & 42423 & 11352.527 & 16927 & 5574.473 & 3.249 & 0.001 & 3.25 \\

2020 & 17046 & 42748 & 11431.177 & 17046 & 5614.823 & 3.25 & 0 & 3.25 \\
\hline
\end{tabular}
\end{subfigure}
\caption{Pattern recognition: topic statistics and knowledge temperature evolution}
\label{fig:99188113_chart}
\end{figure}

\begin{figure}[htbp]
\begin{subfigure}{\textwidth}
\begin{minipage}[t]{0.33\linewidth}
\includegraphics[width = \linewidth]{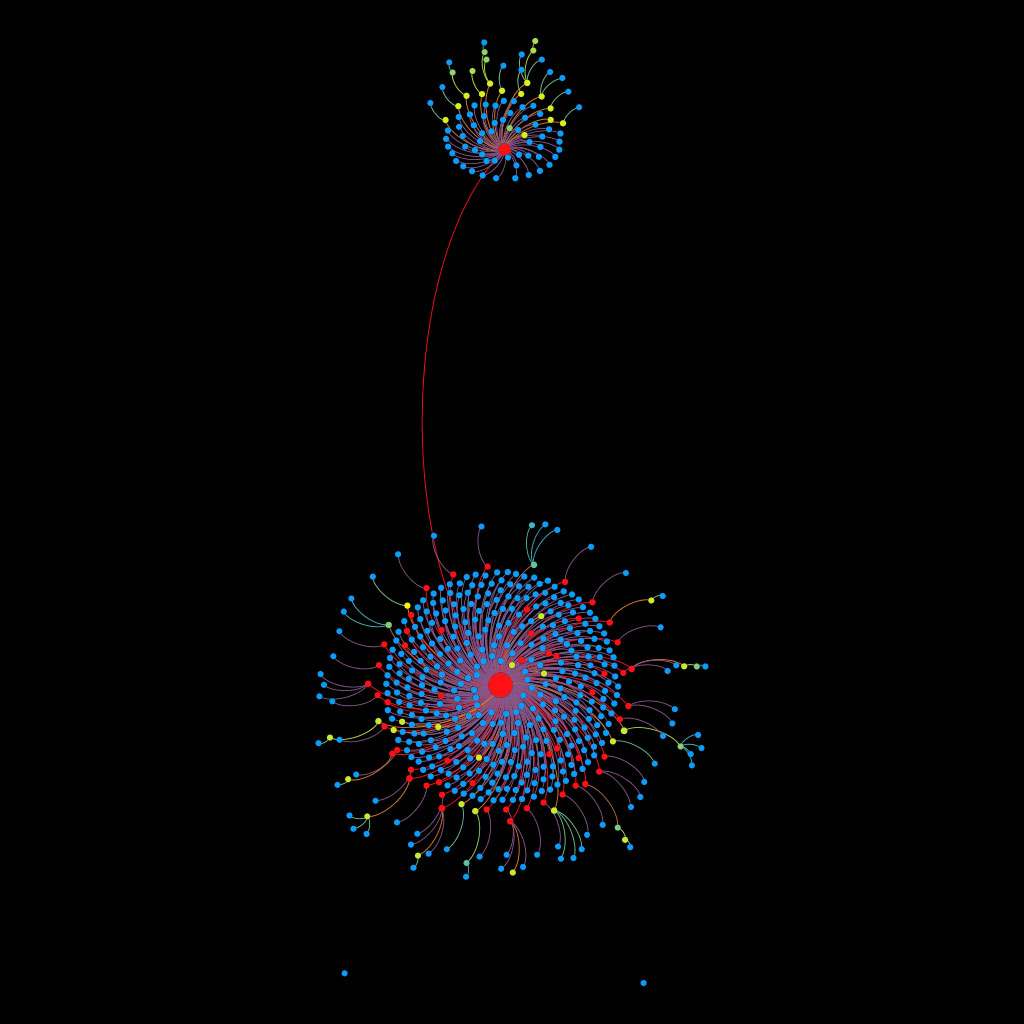}
\caption{Skeleton tree until 1998}
\end{minipage}
\begin{minipage}[t]{0.33\linewidth}
\includegraphics[width = \linewidth]{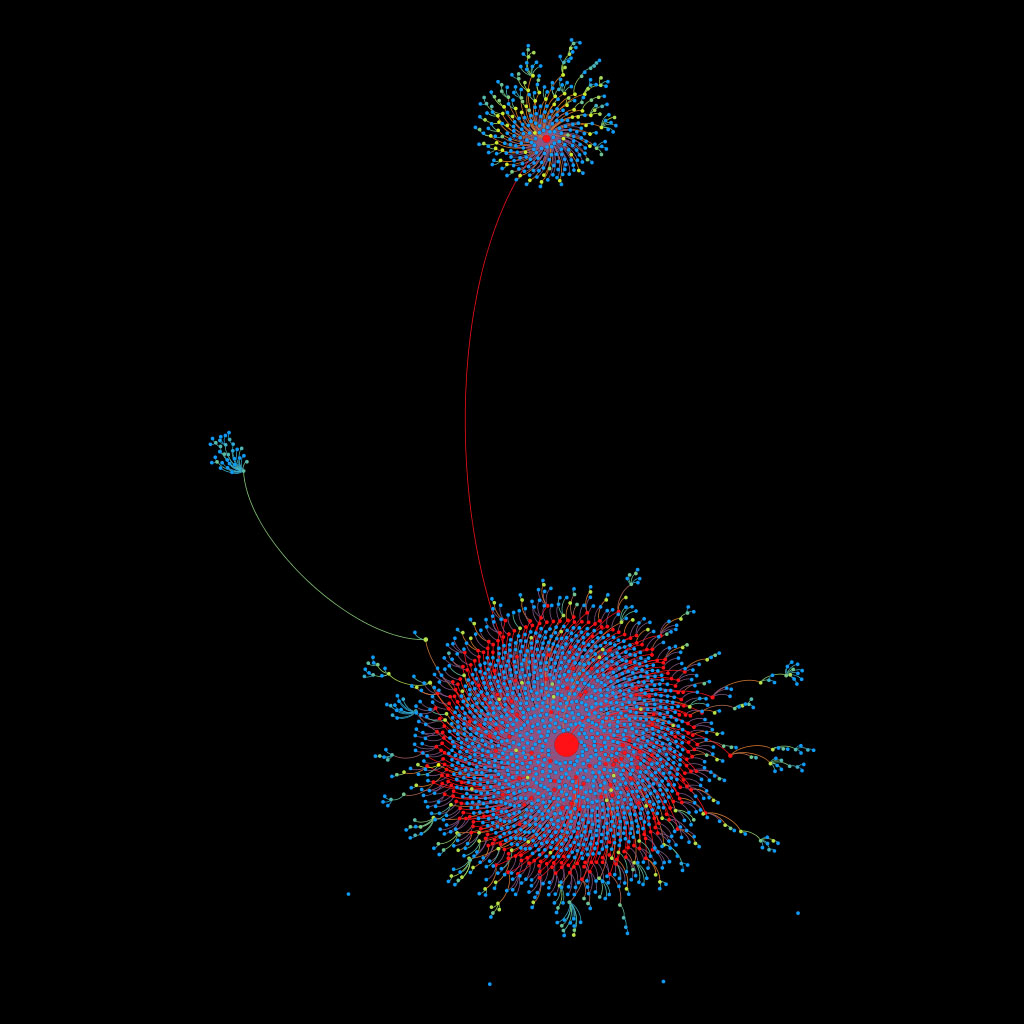}
\caption{Skeleton tree until 2001}
\end{minipage}
\begin{minipage}[t]{0.33\linewidth}
\includegraphics[width = \linewidth]{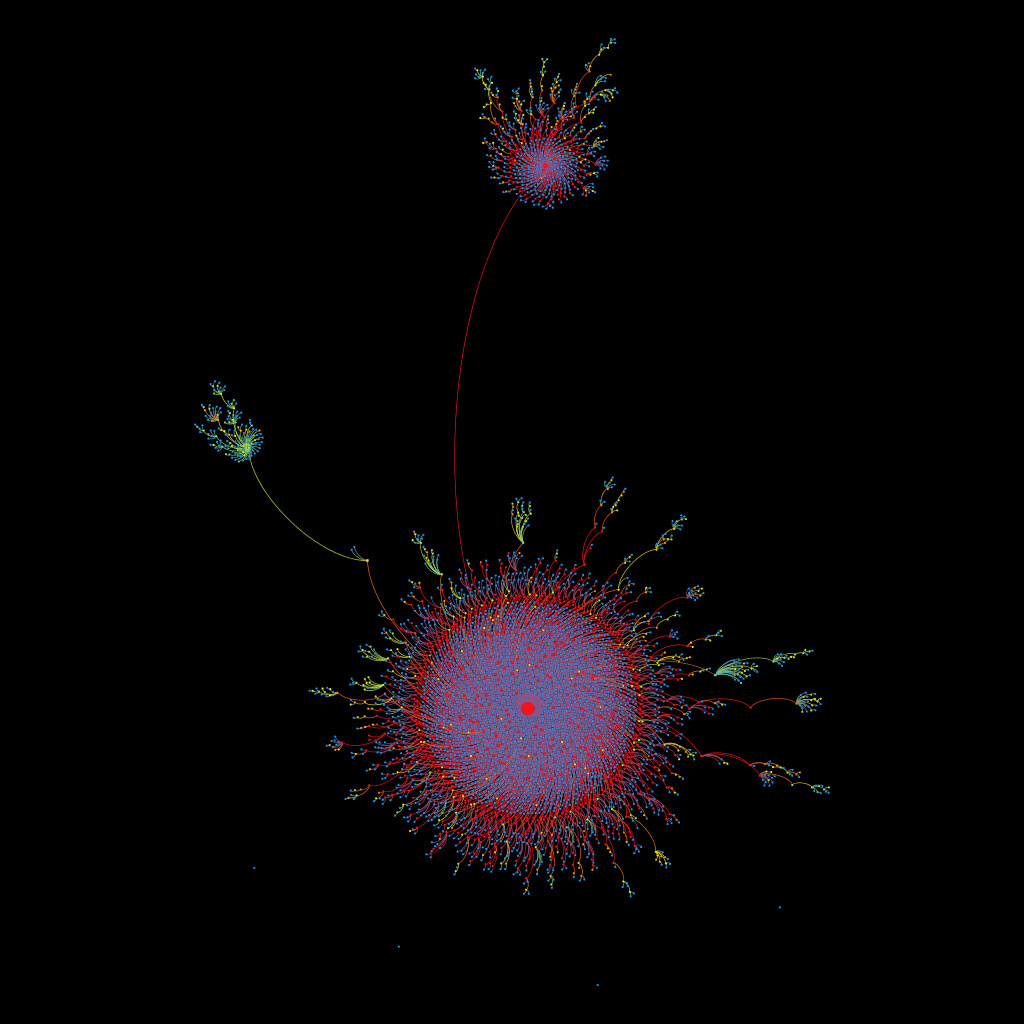}
\caption{Skeleton tree until 2004}
\end{minipage}
\end{subfigure}
\vspace{2mm}
\begin{subfigure}{\textwidth}
\begin{minipage}[t]{0.33\linewidth}
\includegraphics[width = \linewidth]{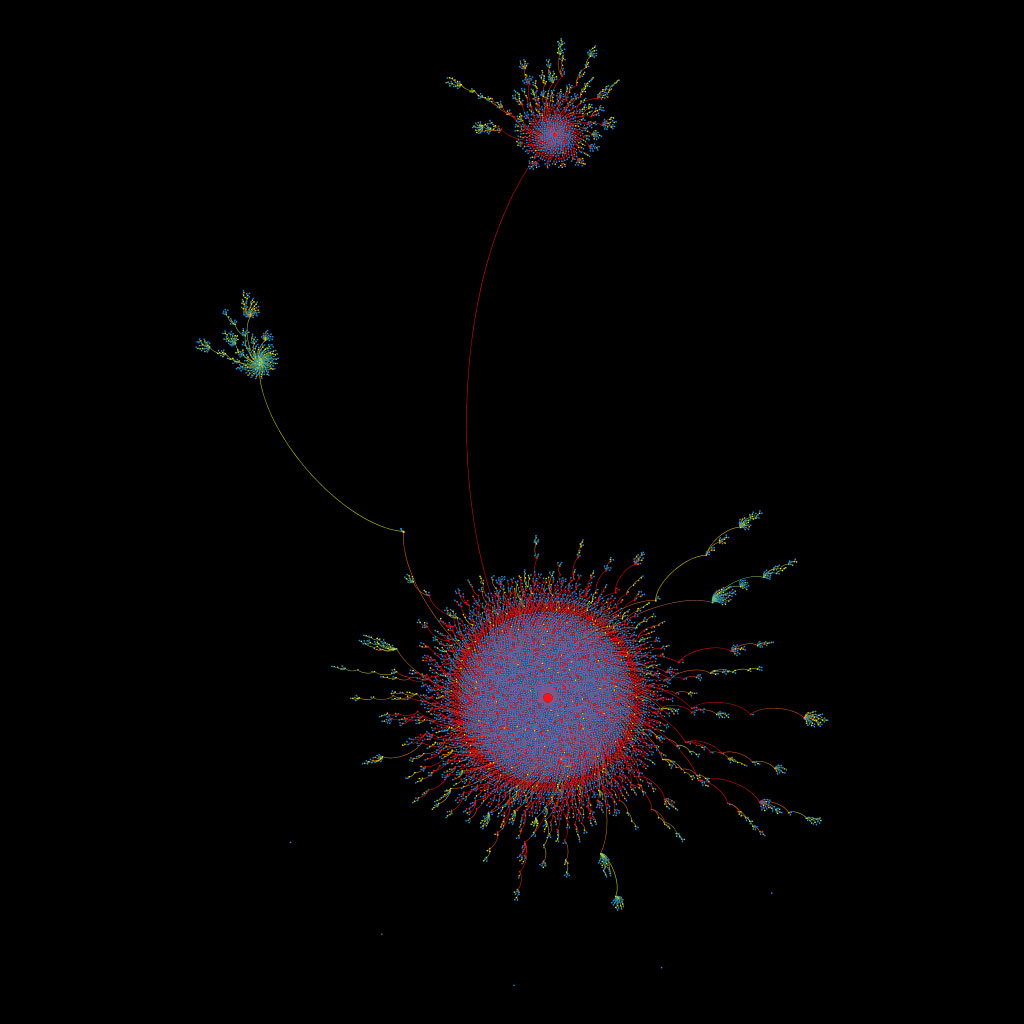}
\caption{Skeleton tree until 2007}
\end{minipage}
\begin{minipage}[t]{0.33\linewidth}
\includegraphics[width = \linewidth]{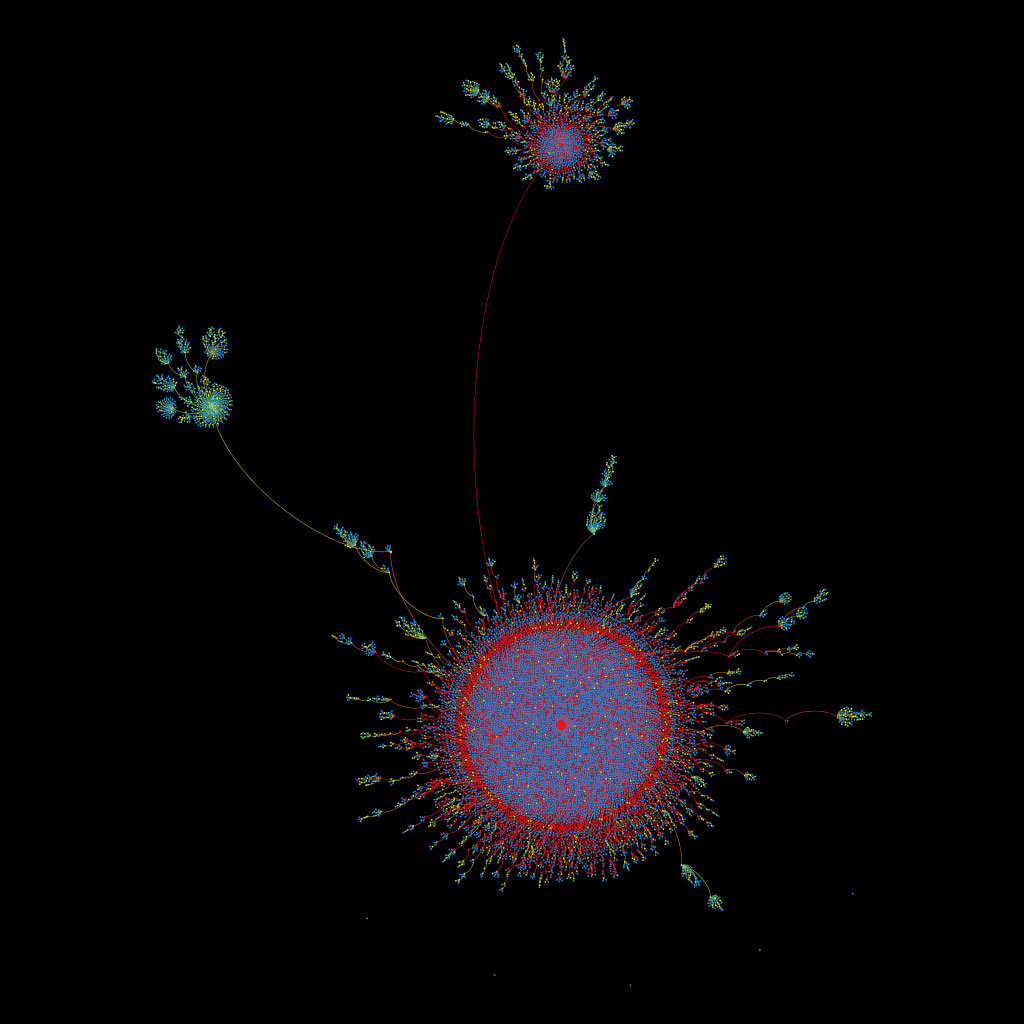}
\caption{Skeleton tree until 2010}
\end{minipage}
\begin{minipage}[t]{0.33\linewidth}
\includegraphics[width = \linewidth]{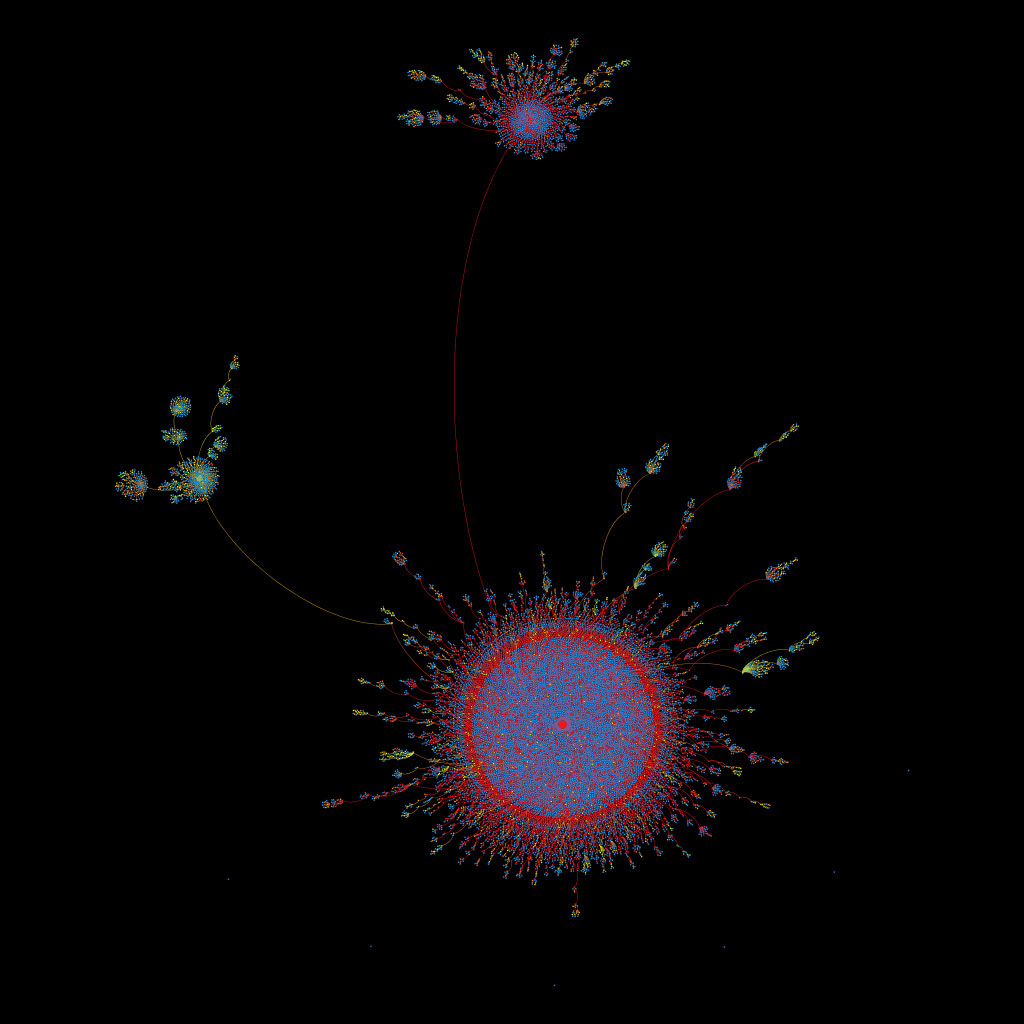}
\caption{Skeleton tree until 2016}
\end{minipage}
\end{subfigure}
\caption{Pattern recognition: Skeleton tree evolution}
\label{fig:99188113-tree_evo}
\end{figure}

\begin{figure}[htbp]
\centering
    \begin{subfigure}{\linewidth}
    \begin{minipage}[t]{0.5\textwidth}
    \centering
    \includegraphics[width = 0.9\linewidth]{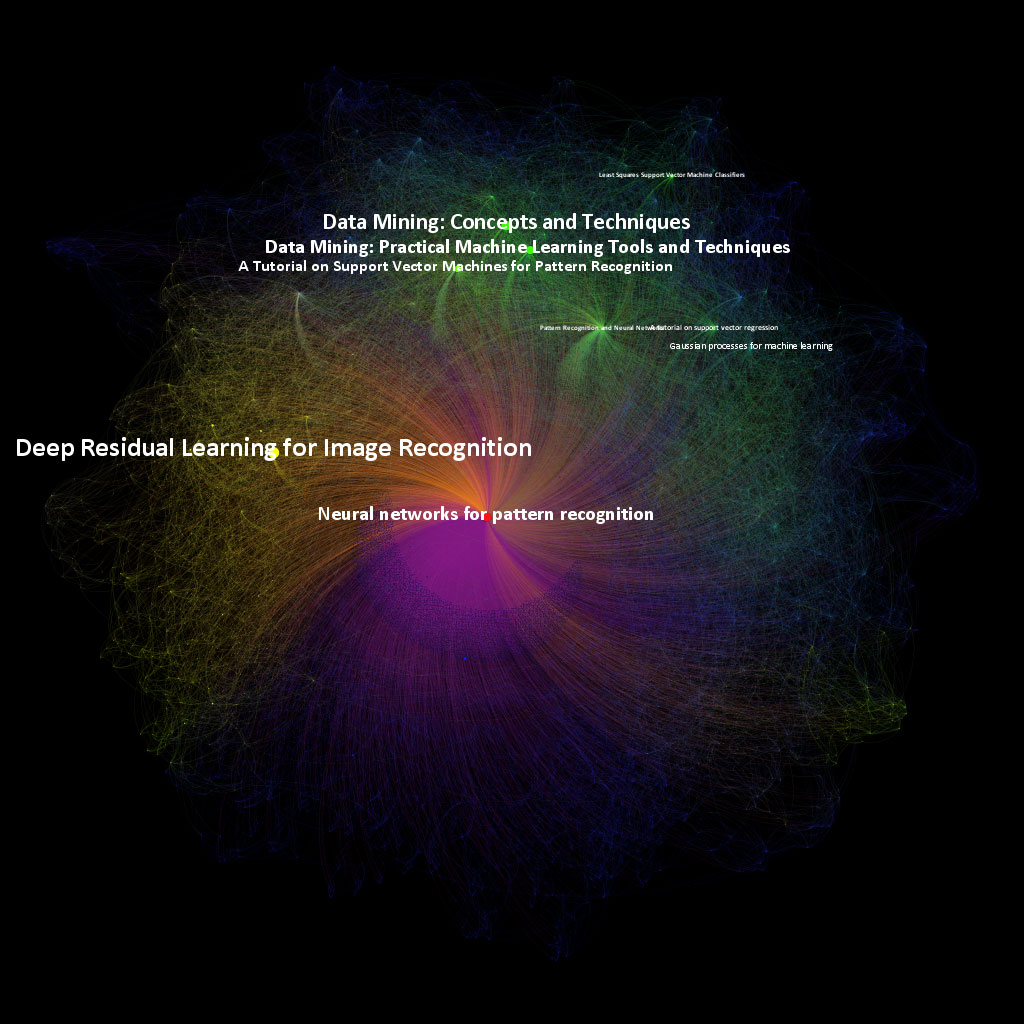}
    \end{minipage}
    \begin{minipage}[t]{0.5\textwidth}
    \centering
    \includegraphics[width = 0.9\linewidth]{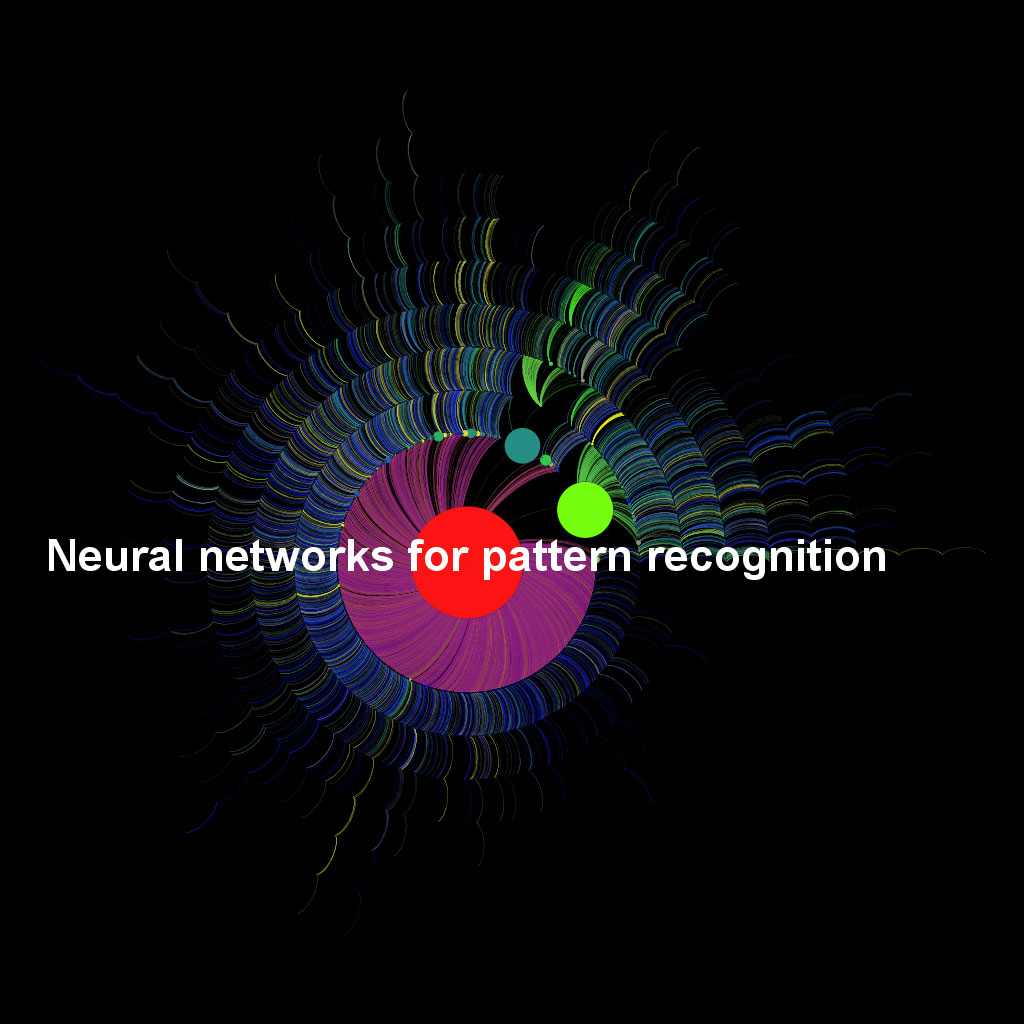}
    \end{minipage}
    \end{subfigure}

    \vspace{5mm}

    \begin{subfigure}{0.6\linewidth}
    \centering
    \includegraphics[width = \linewidth]{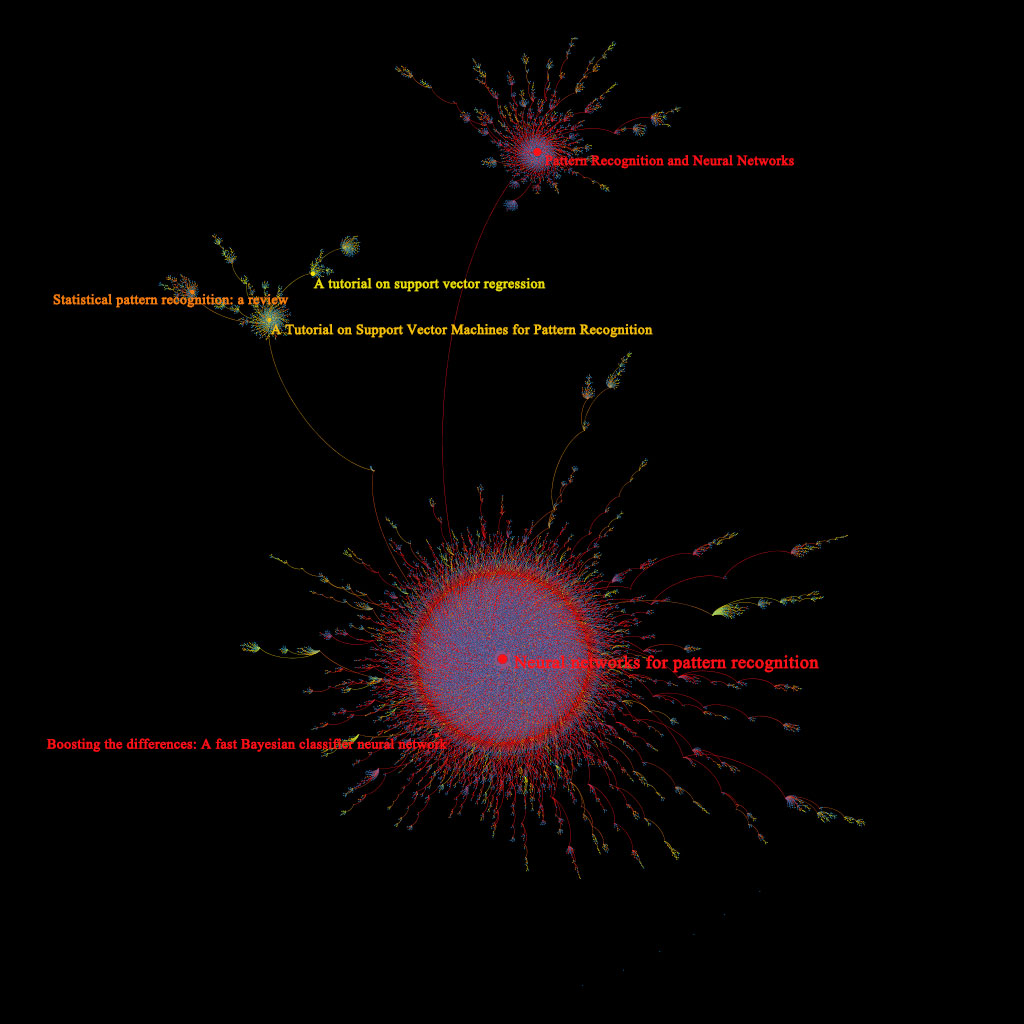}
    \end{subfigure}
    \begin{subfigure}{0.35\linewidth}
    \begin{minipage}[t]{\textwidth}
    \centering
    \includegraphics[width = 0.9\linewidth]{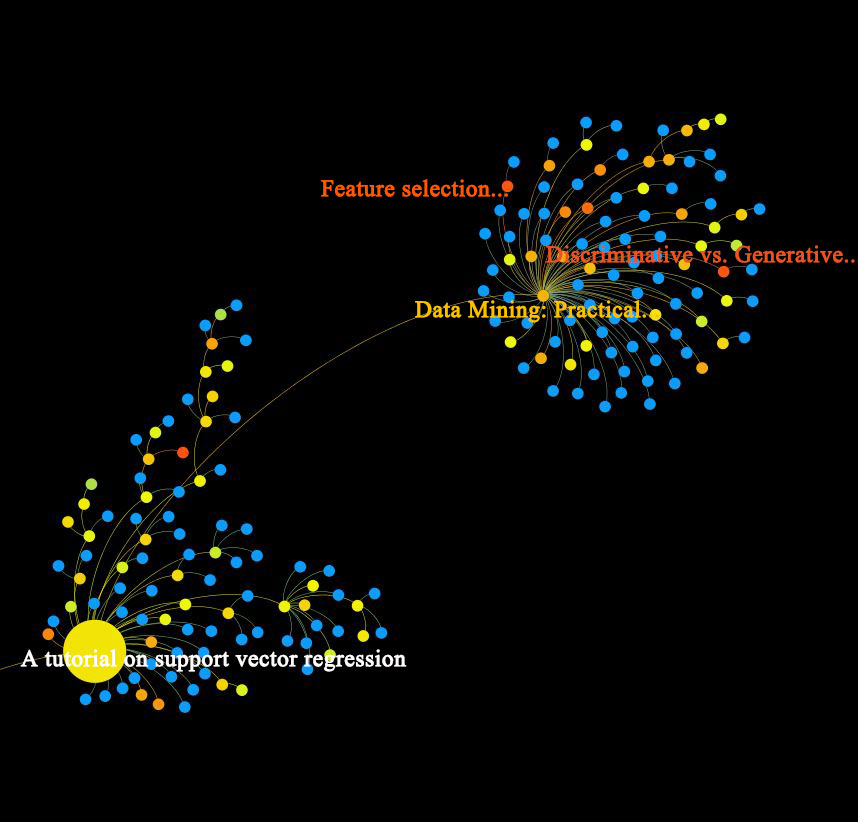}
    \end{minipage}
    \end{subfigure}
\caption{Pattern recognition: Galaxy map, current skeleton tree and its regional zoom. Papers with more than 230 in-topic citations are labelled by title in the skeleton tree. Except the pioneering work, corresponding nodes' size is amplified by 5 times.}
\label{fig:99188113-2020}
\end{figure}

The topic gained popularity and impact steadily in its first 10 years, as is shown by its increasing size and $T^t$ (Fig. \ref{fig:99188113_chart}). During this period, influential child papers within the topic, namely `Pattern Recognition and Neural Networks' (PRNN) published in 1996 and `A Tutorial on Support Vector Machines for Pattern Recognition' (SVMPR) published in 1998, shaped the skeleton tree altogether with the pioneering work. Their enrichment to topic knowledge structure accounts for a slightly higher $T_{structure}^t$ back then, which is manifested by the formation of 2 clusters in the skeleton tree (Fig. \ref{fig:99188113-tree_evo}). Yet the pioneering work is still the absolute authority in the topic. In particular, the cluster in the top is led by PRNN and the top-left small cluster surrounds SVMPR  (Fig. \ref{fig:99188113-2020}). Meanwhile, their arrival pushed up the $T_{growth}^t$ as they also enlarged knowledge base together with common descendants with the pioneering work. Afterwards, despite a constant increase in total size, topic's $T^t$ increment has slowed down. The popular child papers coming after 2000, namely `Boosting the differences: A fast Bayesian classifier neural network' published in 2000 ,`A tutorial on support vector regression' published in 2004 and `Data Mining: Concepts and Techniques' published in 2011 have mostly extended the sub-field led by SVMPR. Judging from skeleton tree, they have not contributed as much as their antecedent (Fig. \ref{fig:99188113-tree_evo}). As a result, the topic has been accumulating its knowledge and popularity much slower than before. Nonetheless, globally speaking, this is a rising topic.\\

\noindent Now we closely examine the interior of this topic. 20 years of development allows a full exploration of the mainstream ideas and a thorough heat diffusion within the topic (Fig. \ref{fig:99188113-tree_evo}). Today, the most popular child papers all have a node knowledge temperature above average (Fig. \ref{fig:99188113-2020}) and they serve as heat sources together with the pioneering work. As the articles are located farther away from them, node knowledge temperature decreases globally. Node knowledge temperature also drops evenly with article age (Fig. 5(c)). The drastic heat-level drop in biggest ages is due to the fact that the topic contains several articles published earlier than the pioneering work and these articles have few followers. Besides, the blue nodes that surround the pioneering work and the most popular child papers are papers with few or no in-topic citations. However, even if we let alone these oldest articles and the aforementioned papers with little subsequent development, the general rule "the older the hotter" is not robust. For example, article `Data Mining: Practical Machine Learning Tools and Techniques' (DM) published in 1999 is slightly hotter than its child papers `Discriminative vs. Generative Classifiers: An In-Depth Experimental Comparison using Cost Curves' (DGC) published in 2005 and `Feature selection and classification in multiple class datasets' (FSC) published in 2011. DM is coloured orange while DGC and FSC are coloured orange-red. This is due to the intrinsic difference of their content, which is reflected by their distinct citations. This example also suggests that the general rule "the more influential the hotter" is weak (Fig. \ref{fig:citation_T} (c)). \\

\noindent We observe in addition certain clustering effect in the skeleton tree. For example, all child papers of `Selection of input parameters to model direct solar irradiance by using artificial neural networks' study the topic's application in energy radiation (Table \ref{tab:99188113-clustering}). This confirms the effectiveness of our skeleton tree extraction algorithm.\\

\begin{table}
    \centering
    \begin{tabular}{p{15cm} p{1cm}}
        \hline
        title & year\\
        \hline
       Selection of input parameters to model direct \textcolor{red}{solar irradiance} by using artificial neural networks & 2004 \\

        Estimation of Surface \textcolor{orange}{Solar Radiation} with Artificial Neural Networks & 2008\\

        Improvement of temperature-based ANN models for \textcolor{orange}{solar radiation} estimation through exogenous data assistance & 2011\\

         Splitting Global \textcolor{orange}{Solar Radiation} into Diffuse and Direct Normal Fractions Using Artificial Neural Networks & 2012\\

        Prediction of daily global \textcolor{orange}{solar irradiation} data using Bayesian neural network: A comparative study & 2012\\

        Assessment of ANN and SVM models for estimating normal direct \textcolor{orange}{irradiation} (Hb) & 2016\\
        \hline
    \end{tabular}

    \caption{Pattern recognition: Clustering effect example. First line is the parent paper and the rest children.}
    \label{tab:99188113-clustering}
\end{table}

\subsection{Rise-then-fall Topics}
\subsubsection{Critical Power for Asymptotic Connectivity in Wireless Networks}

\begin{figure}[htbp]
\centering
\begin{subfigure}[t]{0.5\linewidth}
\centering
\includegraphics[width=\linewidth]{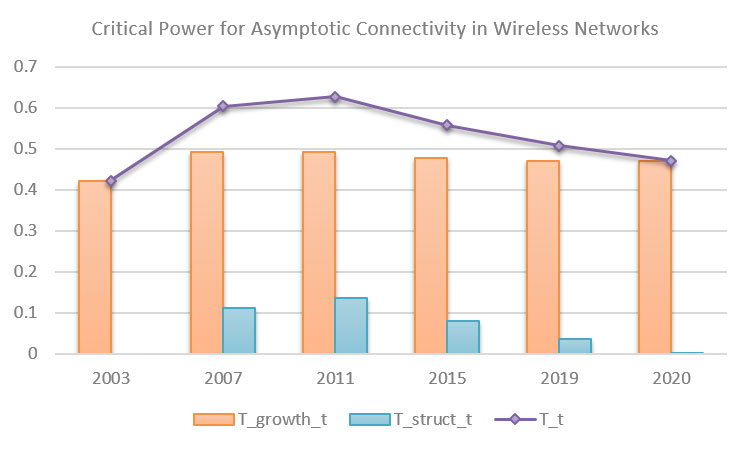}
\end{subfigure}

\begin{subfigure}[t]{0.7\linewidth}
\centering
\begin{tabular}{ccccccccc}
\hline
year & $|V^t|$ & $|E^t|$ & $n_t$ & $V_t$ & ${UsefulInfo}^t$ & $T_{growth}^t$ & $T_{struct}^t$ & $T^t$\\
\hline
2003 & 83 & 236 & 41.688 & 83 & 41.312 & 0.422 &   & 0.422 \\

2007 & 412 & 1697 & 177.643 & 412 & 234.357 & 0.492 & 0.111 & 0.603 \\

2011 & 783 & 3514 & 337.789 & 783 & 445.211 & 0.492 & 0.135 & 0.626 \\

2015 & 992 & 4607 & 440.339 & 992 & 551.661 & 0.478 & 0.079 & 0.557 \\

2019 & 1074 & 4984 & 484.238 & 1074 & 589.762 & 0.47 & 0.037 & 0.507 \\

2020 & 1078 & 4998 & 486.525 & 1078 & 591.475 & 0.47 & 0 & 0.47 \\
\hline
\end{tabular}
\end{subfigure}
\caption{Critical Power: topic statistics and knowledge temperature evolution}
\label{fig:62270017_chart}
\end{figure}

\begin{figure}[htbp]
\begin{subfigure}{\textwidth}
\begin{minipage}[t]{0.5\linewidth}
\includegraphics[width = \linewidth]{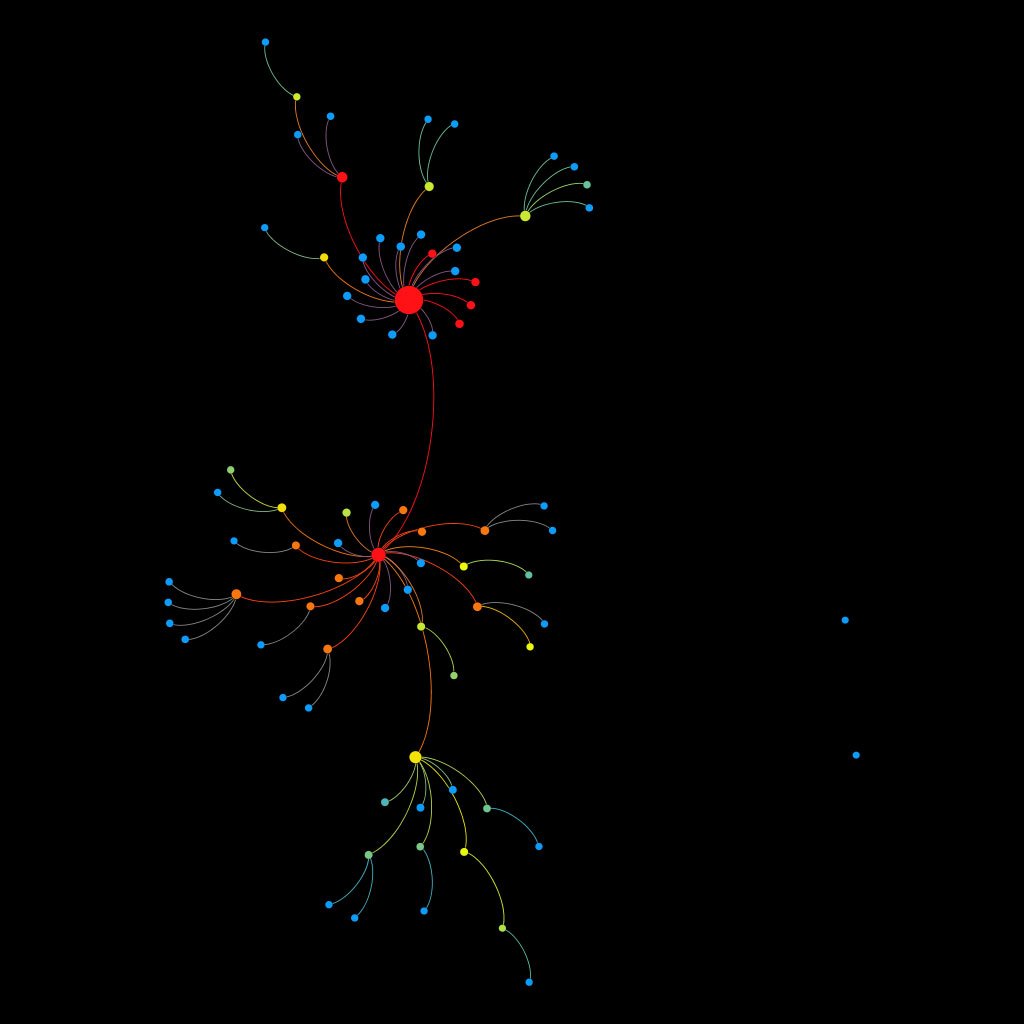}
\caption{Skeleton tree until 2003}
\end{minipage}
\begin{minipage}[t]{0.5\linewidth}
\includegraphics[width = \linewidth]{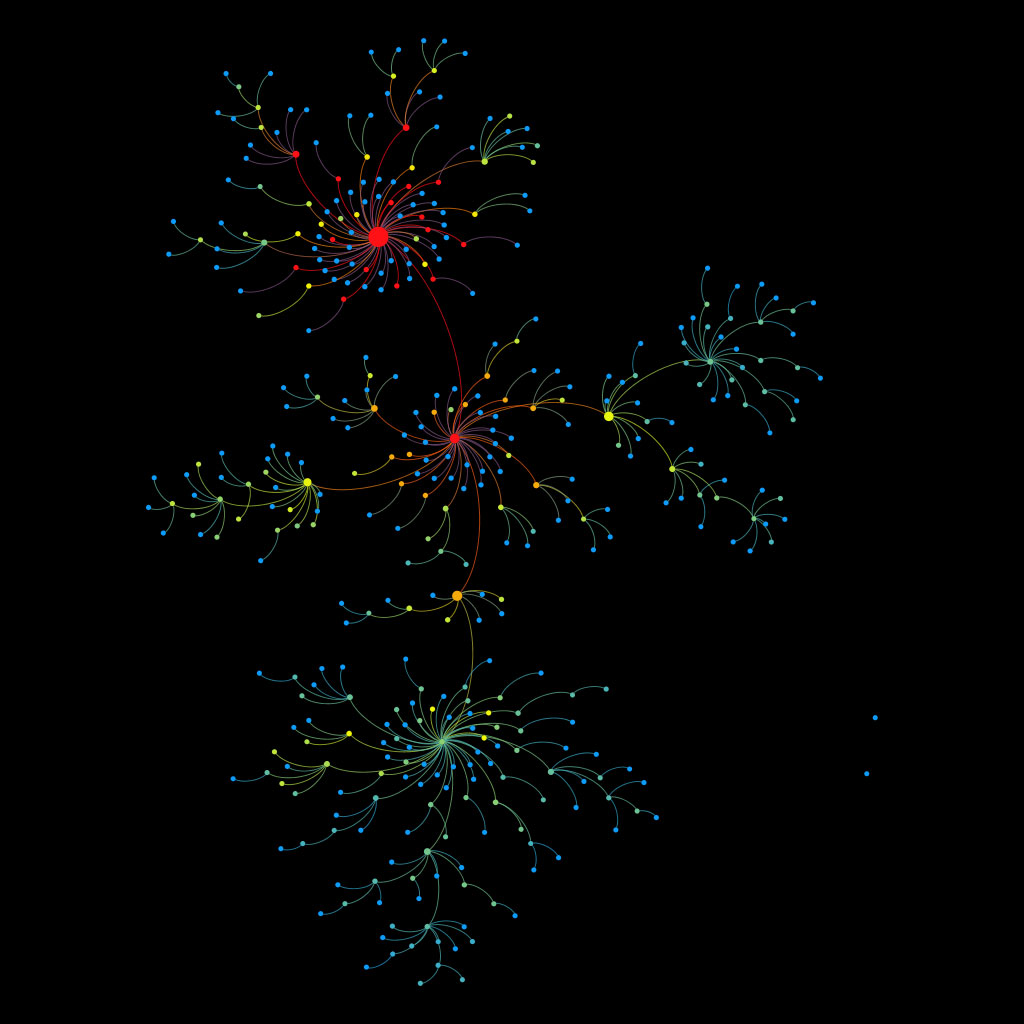}
\caption{Skeleton tree until 2007}
\end{minipage}
\end{subfigure}
\vspace{2mm}
\begin{subfigure}{\textwidth}
\begin{minipage}[t]{0.5\linewidth}
\includegraphics[width = \linewidth]{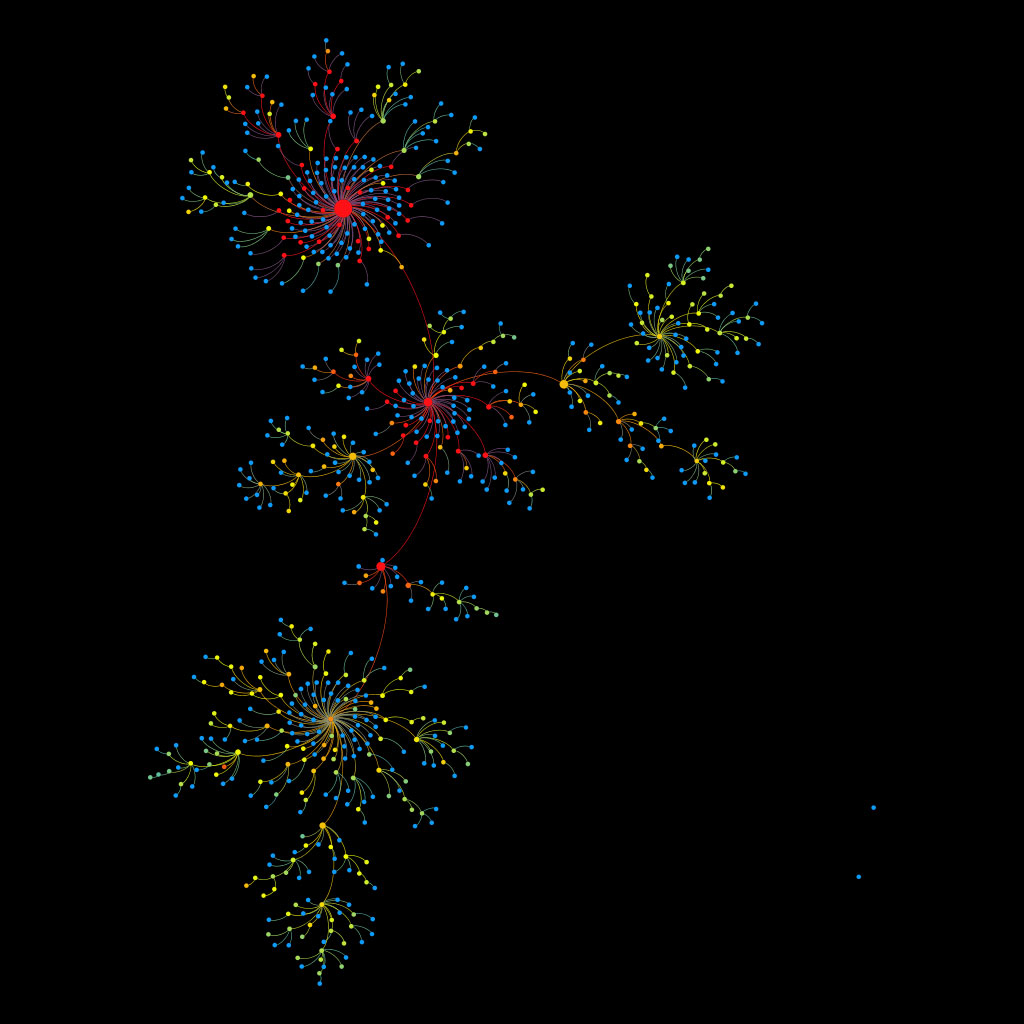}
\caption{Skeleton tree until 2011}
\end{minipage}
\begin{minipage}[t]{0.5\linewidth}
\includegraphics[width = \linewidth]{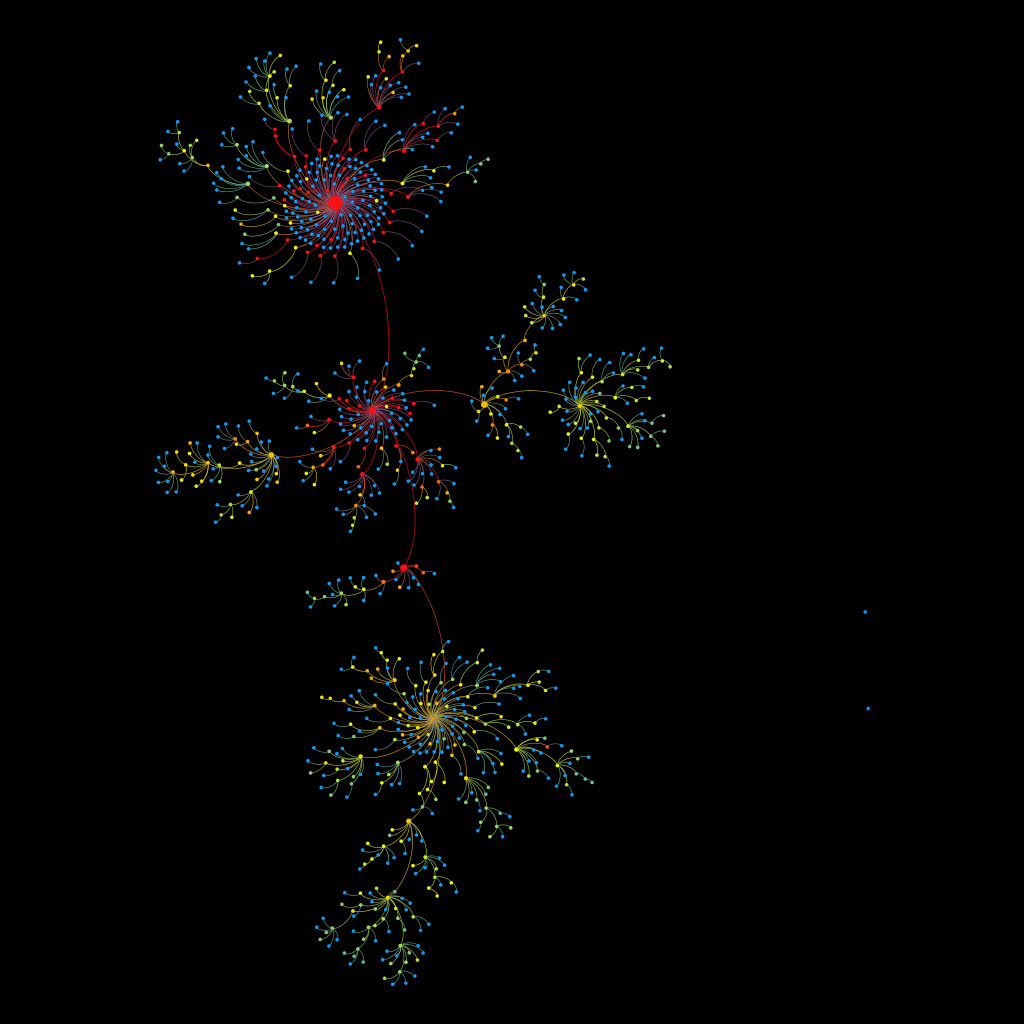}
\caption{Skeleton tree until 2015}
\end{minipage}
\end{subfigure}
\caption{Critical Power: Skeleton tree evolution}
\label{fig:62270017-tree_evo}
\end{figure}

\begin{figure}[htbp]
\centering
    \begin{subfigure}{\linewidth}
    \begin{minipage}[t]{0.5\textwidth}
    \centering
    \includegraphics[width = 0.9\linewidth]{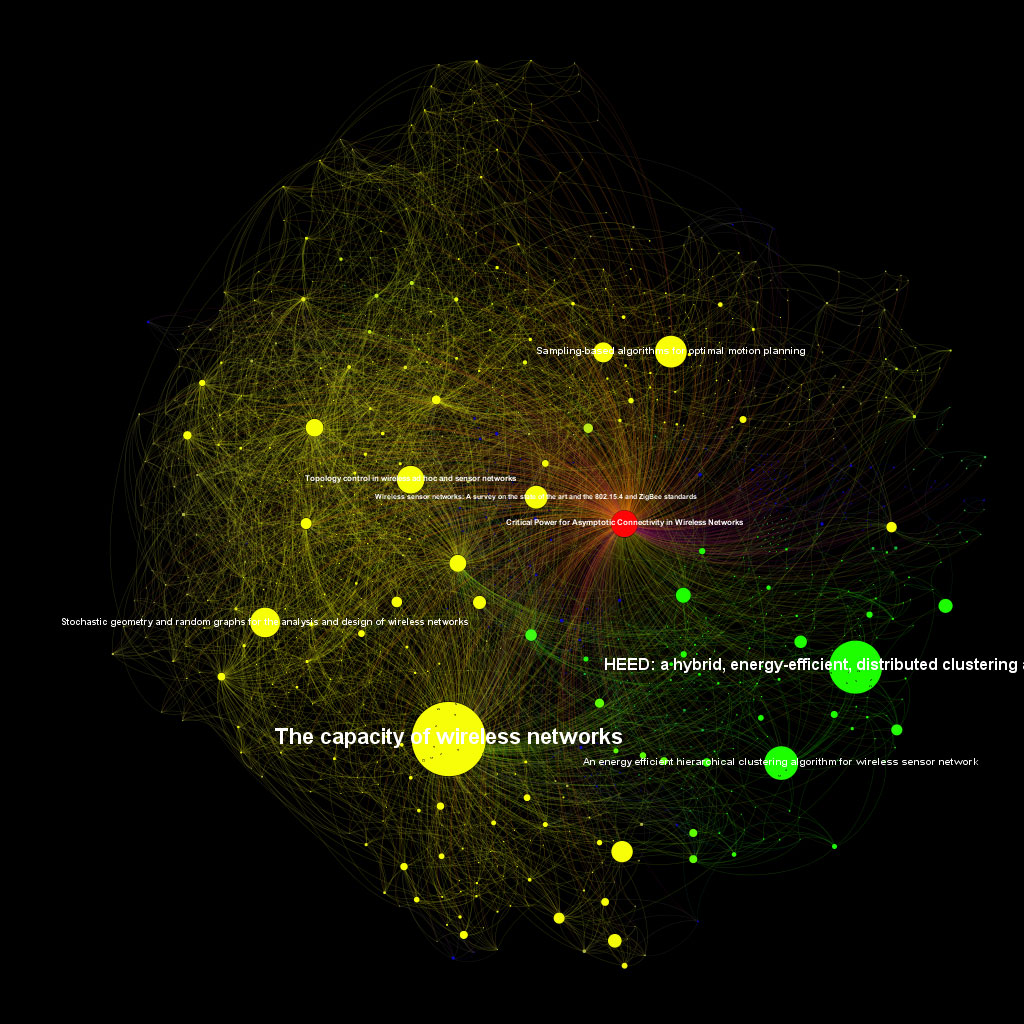}
    \end{minipage}
    \begin{minipage}[t]{0.5\textwidth}
    \centering
    \includegraphics[width = 0.9\linewidth]{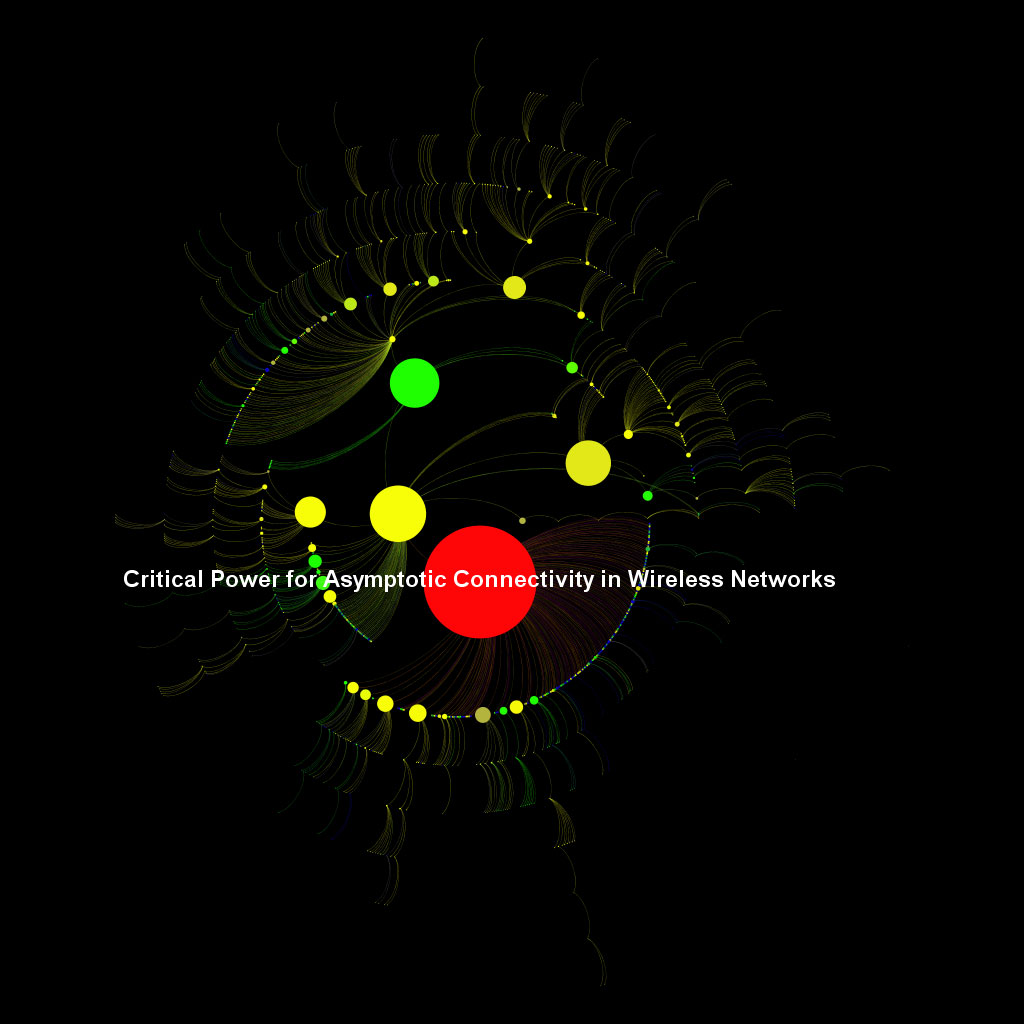}
    \end{minipage}

    \vspace{5mm}

    \end{subfigure}
    \begin{subfigure}{0.6\linewidth}
    \centering
    \includegraphics[width = \linewidth]{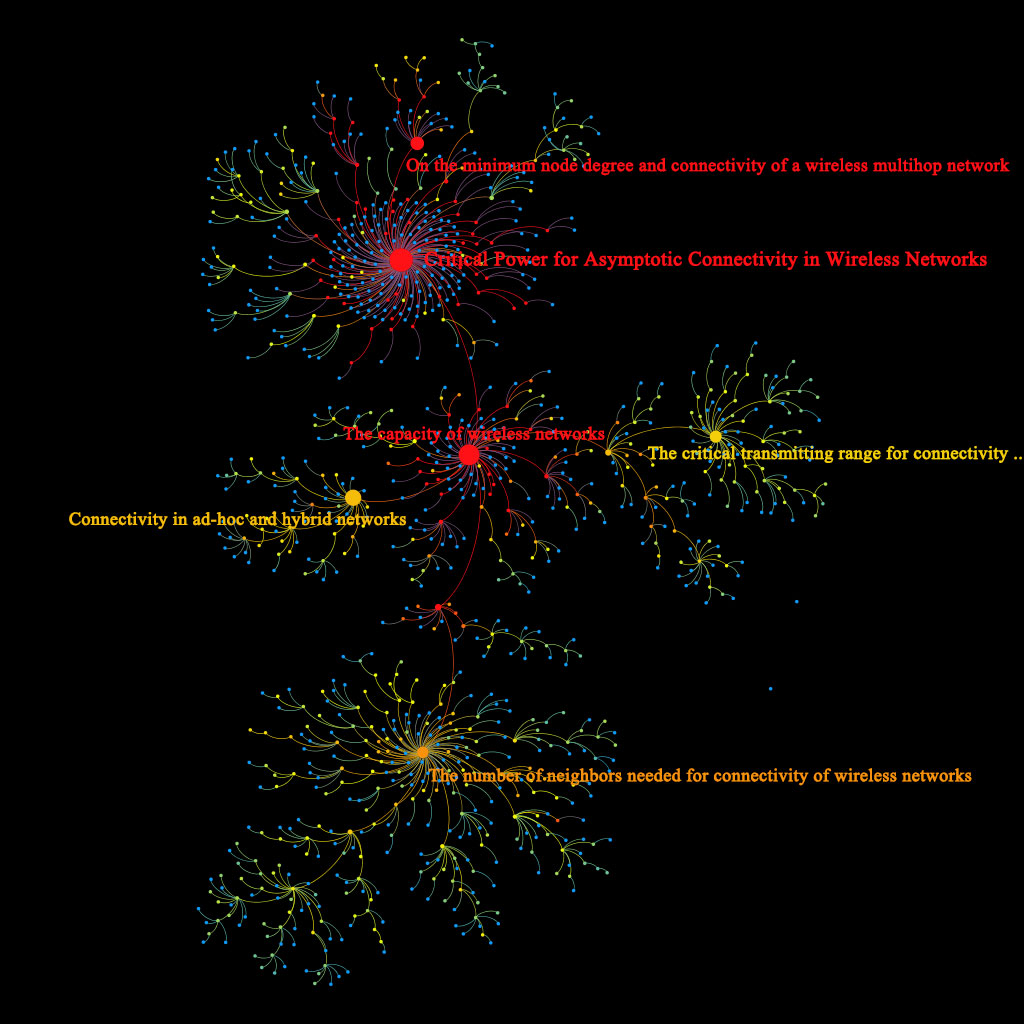}
    \end{subfigure}
    \begin{subfigure}{0.35\linewidth}
    \centering
    \includegraphics[width = 0.9 \linewidth]{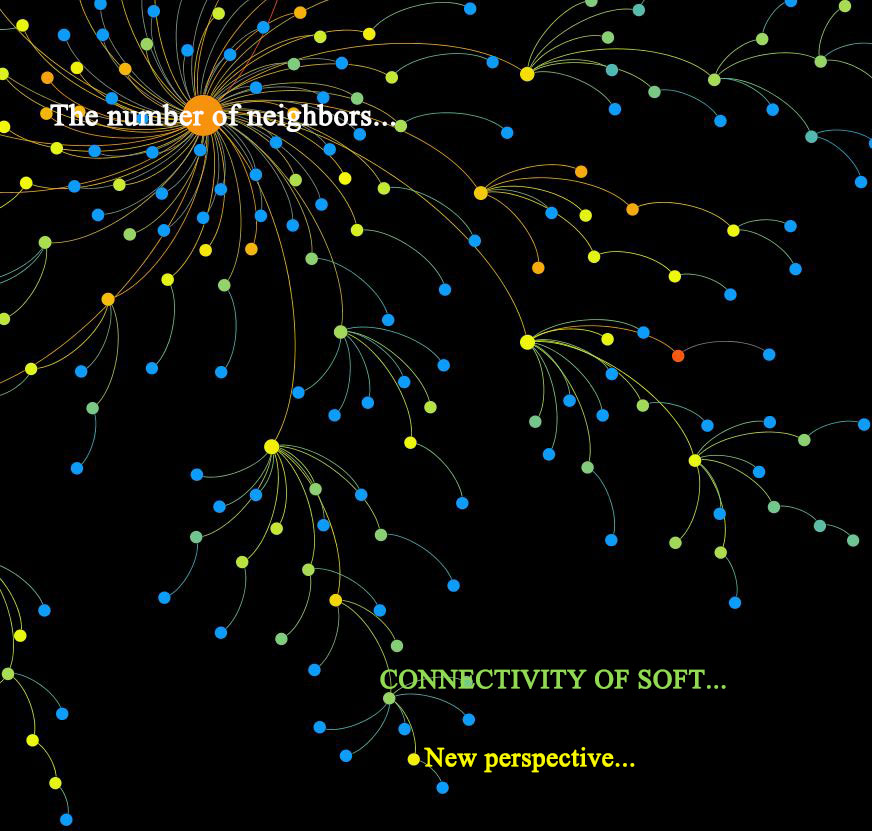}
    \end{subfigure}
\caption{Critical Power: Galaxy map, current skeleton tree and its regional zoom. Papers with more than 100 in-topic citations are labelled by title in the skeleton tree. Except the pioneering work, corresponding nodes' size is amplified by 3 times.}
\label{fig:62270017-2020}
\end{figure}

As is shown by the basic statistics and $T_{growth}^t$, the topic reached its peak around 2011 (Fig. \ref{fig:62270017_chart}). The decline in scale growth and $T_{growth}^t$ is obvious afterwards. The majority of popular child papers were published no later than 2004. They pushed up $T_{growth}^t$ with their new ideas and contributed to the flourishing before 2010. In particular, popular child papers `The capacity of wireless networks' published in 2000 and `The number of neighbors needed for connectivity of wireless networks' published in 2004 each leads a non-trivial research sub-direction, demonstrated as clusters in the skeleton tree (Fig. \ref{fig:62270017-2020}). Their substantial extension to the topic knowledge structure is additionally illustrated by a high $T_{structure}^t$ in the early days. However, the glory did not last for long. After 2010, the continuous lack of young influential child papers gradually resulted in a decreasing topic visibility and thus a shrinking inflow of useful information, its knowledge source. The trend is also reflected in the stagnation of skeleton tree. While we are still able to detect some development on the periphery of all 3 clusters from 2007 to 2011, the skeleton tree seems to take a definitive form after 2011. The snapshots look almost identical (Fig. \ref{fig:62270017-tree_evo}). Consequently, both $T_{growth}^t$ and $T_{structure}^t$ have plunged. After 10 years of golden age, the topic is now perishing.\\

\noindent Now we closely examine the heat distribution within the topic (Fig. \ref{fig:62270017-2020}). We observe a quick heat diffusion during the flourishing period (Fig. \ref{fig:62270017-tree_evo}(b,c)). Now heat diffusion is complete as popular child papers all have a knowledge temperature above average and the child papers published during the golden period are relatively hot in general (Fig. 5(d)). An obvious exception lies in the oldest child papers. Their low average temperature is because they were published at the same time or earlier than the pioneering work and they have few or no followers. Besides the pioneering work, popular child paper `The capacity of wireless networks' is also a heat source within the topic. As articles are located farther away from them, they gradually cool down. The blue nodes that surround the pioneering work and the popular child paper `The capacity of wireless networks' in central clusters are papers with few or no in-topic followers. However, the general rules "the older the hotter" and "the more influential the hotter" (Fig. \ref{fig:citation_T}(d)) are not robust. For instance, paper `New perspective on sampling-based motion planning via random geometric graphs' (SBMP) published in 2018 is hotter than its parent, `CONNECTIVITY OF SOFT RANDOM GEOMETRIC GRAPHS' (CSRG), an article published in 2016. SBMP has an average knowledge temperature while CSRG has a temperature below average. This can be mainly attributed to their different research focus, which is reflected by their distinct citations and citations' average heat-level. Another reason may be that even though CSRG has had a much better development, the dozen articles it has inspired have gained little popularity and impact, thus they do not help boost CSRG's status. \\

\noindent We find article `Power Control in Ad-Hoc Networks: Theory, Architecture, Algorithm and Implementation of the COMPOW Protocol' particularly interesting. It is not a cluster center, nor does it have many articles around, yet it has a big structure entropy and a highest knowledge temperature. We think this is due to its strategic position, right between 2 clusters respectively led by `The capacity of wireless networks' and `The number of neighbors needed for connectivity of wireless networks'. The article itself may not have a big impact, but it has inspired a handful of influential literature. Its value lies in enlightenment.\\

\noindent We observe in addition certain clustering effect in the skeleton tree (Table \ref{tab:62270017-clustering}). For example, almost all child papers of `CONNECTIVITY OF SOFT RANDOM GEOMETRIC GRAPHS' have similar research themes as itself. This confirms the effectiveness of our skeleton tree extraction algorithm.\\

\begin{table}
    \centering
    \begin{tabular}{cc}
        \hline
        title & year\\
        \hline
         CONNECTIVITY OF SOFT \textcolor{red}{RANDOM GEOMETRIC GRAPHS} & 2016 \\

         Isolation and Connectivity in \textcolor{orange}{Random Geometric Graphs} with Self-similar Intensity Measures & 2018\\

        On Resilience and Connectivity of Secure Wireless Sensor Networks Under Node Capture Attacks & 2017\\

        New perspective on sampling-based motion planning via \textcolor{orange}{random geometric graphs} & 2018\\
        \hline
    \end{tabular}
    \caption{Critical Power: Clustering effect example. First line is the parent paper and the rest children.}
    \label{tab:62270017-clustering}
\end{table}

\subsubsection{The capacity of wireless networks}

\begin{figure}[htbp]
\centering

\begin{subfigure}[t]{0.6\linewidth}
\centering
\includegraphics[width=\linewidth]{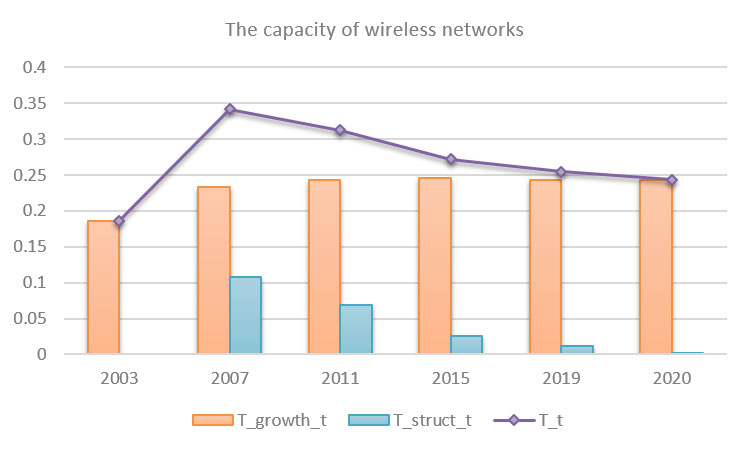}
\end{subfigure}

\begin{subfigure}[t]{\linewidth}
\centering
\begin{tabular}{ccccccccc}
\hline
year & $|V^t|$ & $|E^t|$ & $n_t$ & $V_t$ & ${UsefulInfo}^t$ & $T_{growth}^t$ & $T_{struct}^t$ & $T^t$\\
\hline
2003 & 325 & 860 & 197.224 & 325 & 127.776 & 0.186 &   & 0.186 \\

2007 & 2220 & 10999 & 1076.03 & 2220 & 1143.97 & 0.233 & 0.108 & 0.342 \\

2011 & 4956 & 30466 & 2302.961 & 4956 & 2653.039 & 0.243 & 0.07 & 0.313 \\

2015 & 6867 & 46263 & 3152.523 & 6867 & 3714.477 & 0.246 & 0.026 & 0.272 \\

2019 & 7621 & 51667 & 3535.877 & 7621 & 4085.123 & 0.244 & 0.011 & 0.255 \\

2020 & 7644 & 51789 & 3546.091 & 7644 & 4097.909 & 0.244 & 0 & 0.244 \\
\hline
\end{tabular}
\end{subfigure}
\caption{Capacity Wireless Network: topic statistics and knowledge temperature evolution}
\label{fig:438420345_chart}
\end{figure}

\begin{figure}[htbp]
\begin{subfigure}{\textwidth}
\begin{minipage}[t]{0.5\linewidth}
\includegraphics[width = \linewidth]{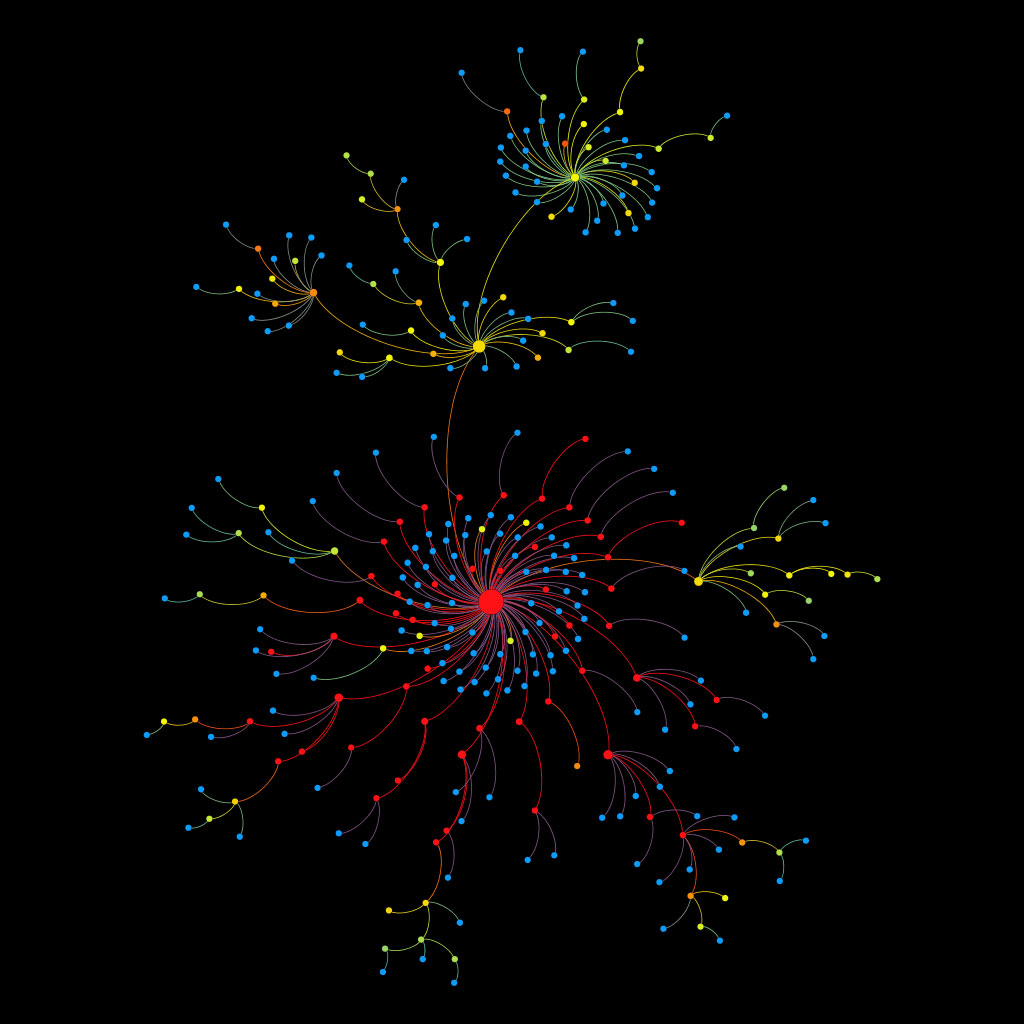}
\caption{Skeleton tree until 2003}
\end{minipage}
\begin{minipage}[t]{0.5\linewidth}
\includegraphics[width = \linewidth]{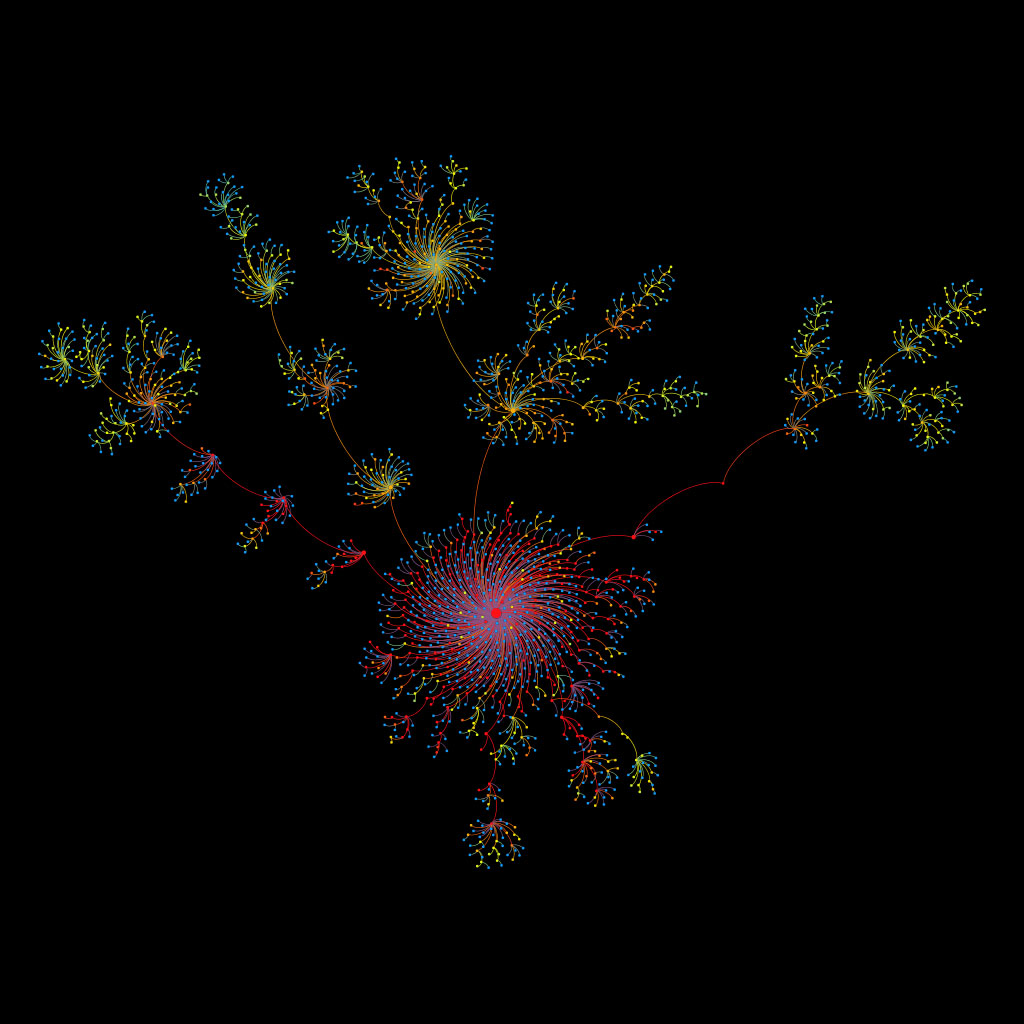}
\caption{Skeleton tree until 2007}
\end{minipage}
\end{subfigure}
\vspace{2mm}
\begin{subfigure}{\textwidth}
\begin{minipage}[t]{0.5\linewidth}
\includegraphics[width = \linewidth]{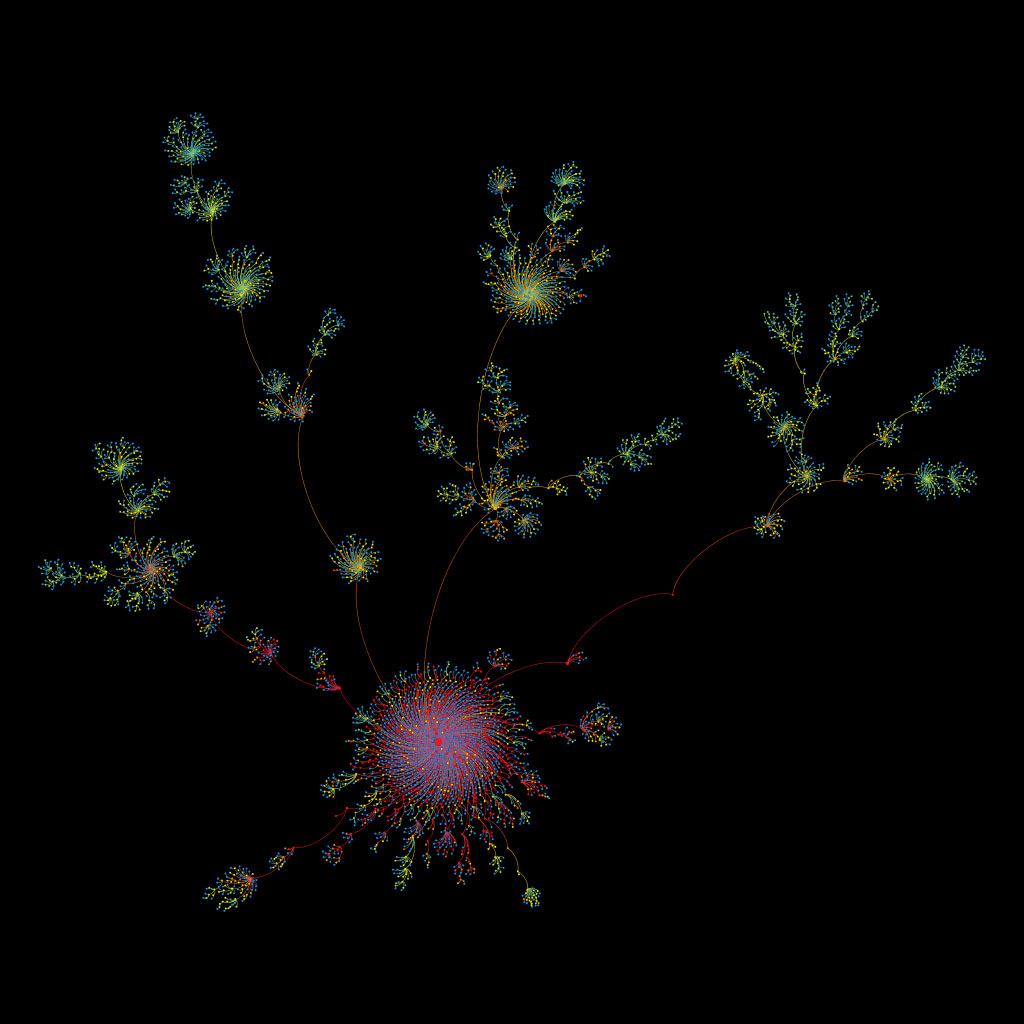}
\caption{Skeleton tree until 2011}
\end{minipage}
\begin{minipage}[t]{0.5\linewidth}
\includegraphics[width = \linewidth]{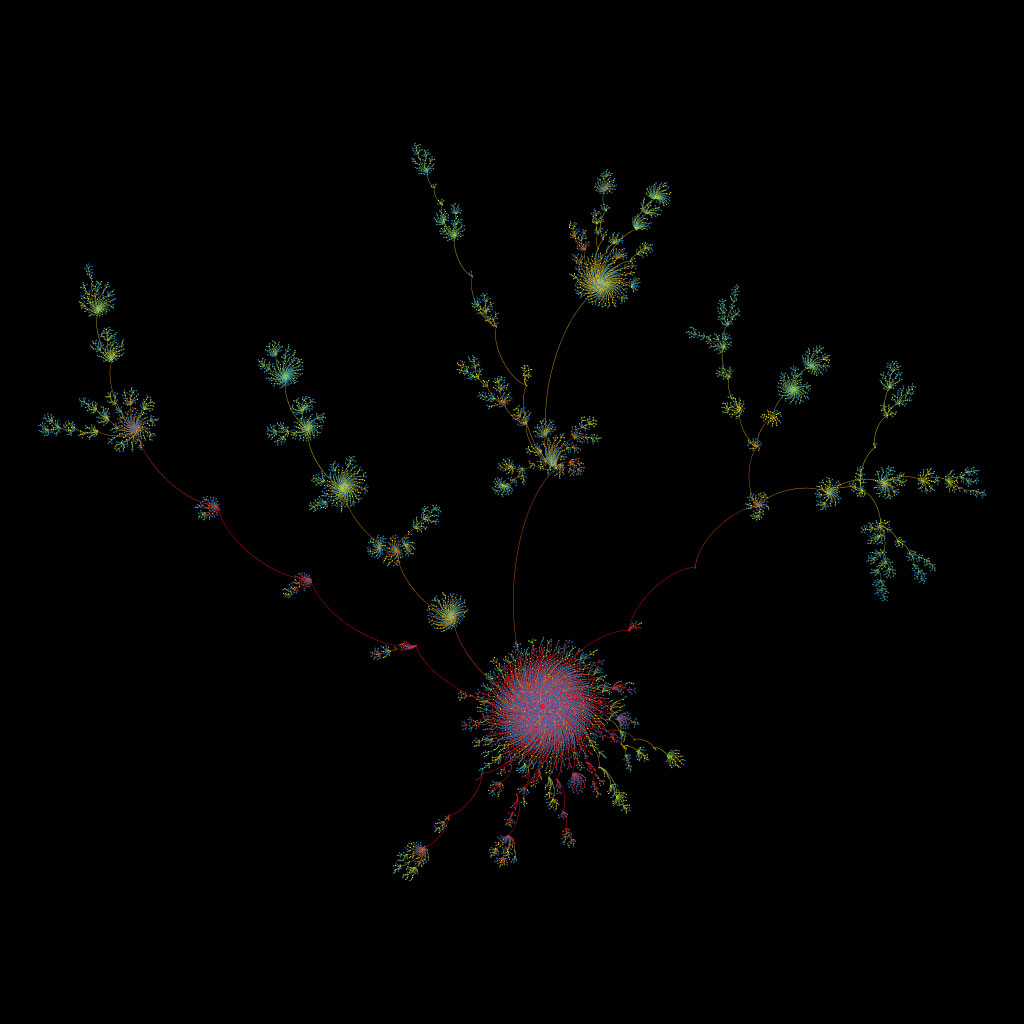}
\caption{Skeleton tree until 2015}
\end{minipage}
\end{subfigure}
\caption{Capacity Wireless Network: Skeleton tree evolution}
\label{fig:438420345-tree_evo}
\end{figure}

As is shown by $T^t$, the topic reached its peak at some time around 2007 (Fig. \ref{fig:438420345_chart}). The batch of popular child papers arriving between 2001 and 2004, namely `Capacity of Ad hoc wireless networks', `Mobility increases the capacity of ad-hoc wireless networks', `A network information theory for wireless communication: scaling laws and optimal operation' and `Impact of interference on multi-hop wireless network performance', largely enriched the topic knowledge base by inspiring several research sub-fields, as is reflected by the significant structure advancement in skeleton tree from 2003 to 2007 (Fig. \ref{fig:438420345-tree_evo}). As a result, we observe a soar both in $T_{growth}^t$ and $T_{structure}^t$. Popular child papers continued to come until 2007. But the younger ones did not cause a stir as much. Only 1 of them has made visible contribution to knowledge structure evolution: `Closing the Gap in the Capacity of Wireless Networks Via Percolation Theory' published in 2007 opened up a new research focus and led to the end division of a major branch in the skeleton tree by 2011. The decreasing exposure gained by its child papers and a decelerating evolution in knowledge pattern caused $T_{structure}^t$ to drop after 2007. But the residual attractiveness continued to draw a abundant quantity of "new blood" and ensured the rise in $T_{growth}^t$ for a while longer. After 2011, despite a continuous size expansion and a steady knowledge accumulation, the topic has been gradually phased out due to an overall mediocre development of child papers published after 2009. The wear-off of the community's focus is illustrated by an immediate drop in $T_{structure}^t$ in 2015, which also accounts for the down trend of $T^t$. Correspondingly, we observe fewer remarkable changes in skeleton tree during this period. While the cooling-down is mainly due to attention loss before 2015, recent temperature drop is caused by knowledge supply shortage. The focus loss has eventually resulted in diminishing publications and affected its long-term knowledge accumulation. To sum up, after around 10 years of glory, the topic is now going downhill.\\


\begin{figure}[htbp]
\centering
    \begin{subfigure}{\textwidth}
    \begin{minipage}[t]{0.5\textwidth}
    \centering
    \includegraphics[width = 0.9\linewidth]{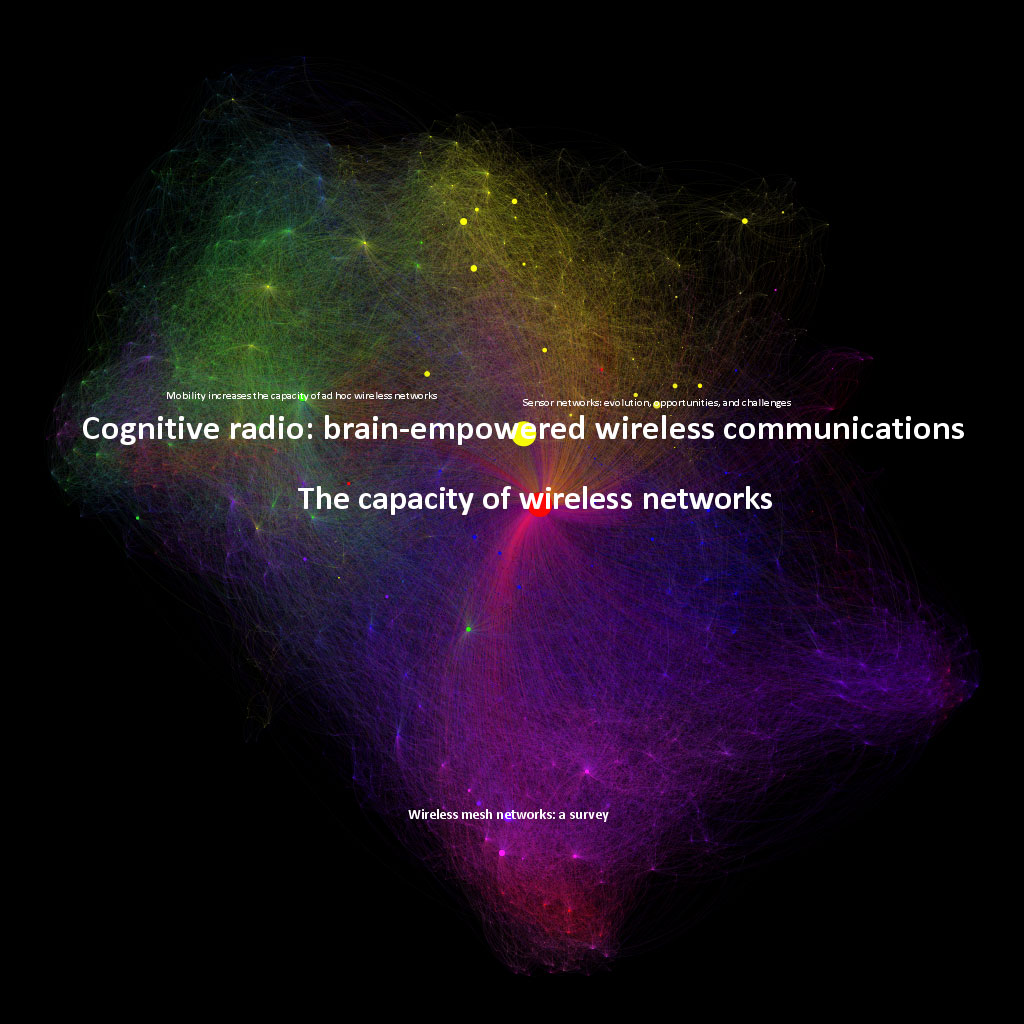}
    \end{minipage}
    \begin{minipage}[t]{0.5\textwidth}
    \centering
    \includegraphics[width = 0.9\linewidth]{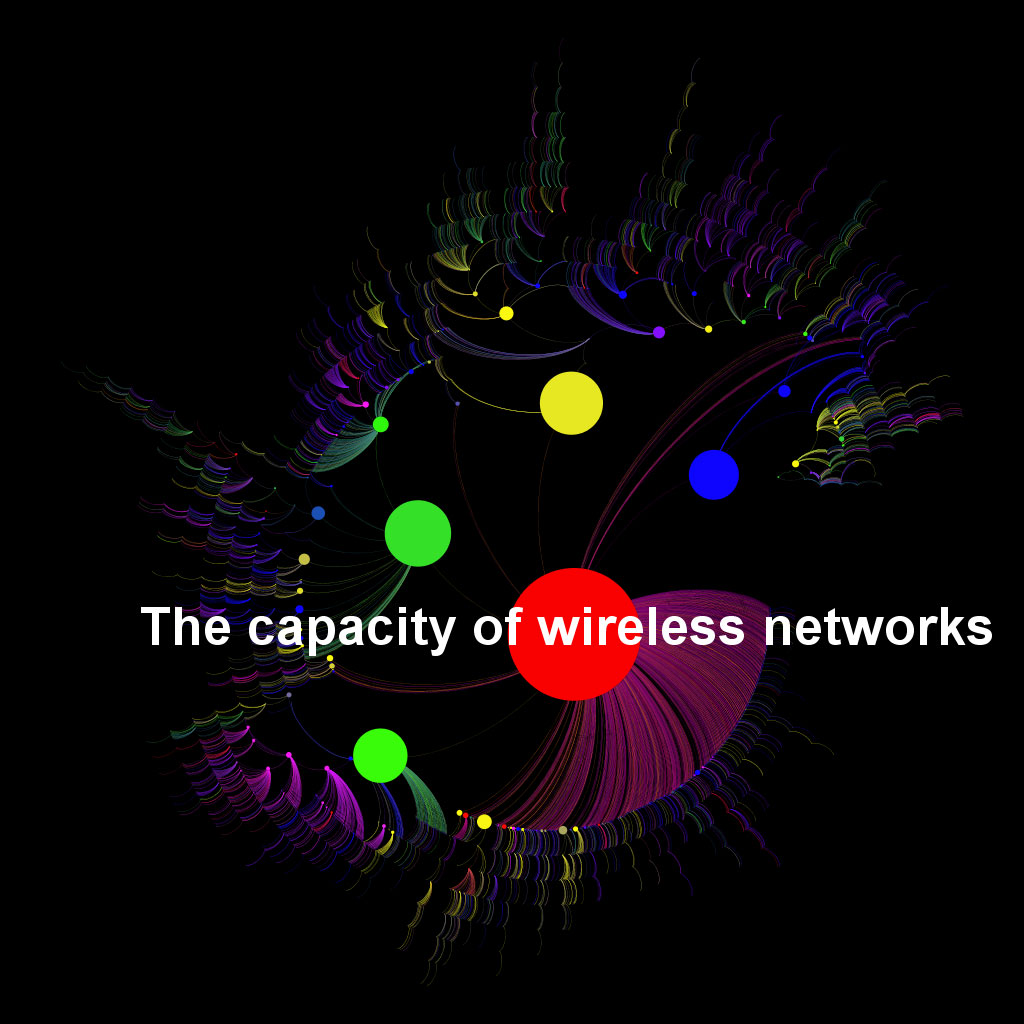}
    \end{minipage}
    \end{subfigure}

    \vspace{5mm}

   \begin{subfigure}{0.6\textwidth}
   \centering
   \includegraphics[width = \linewidth]{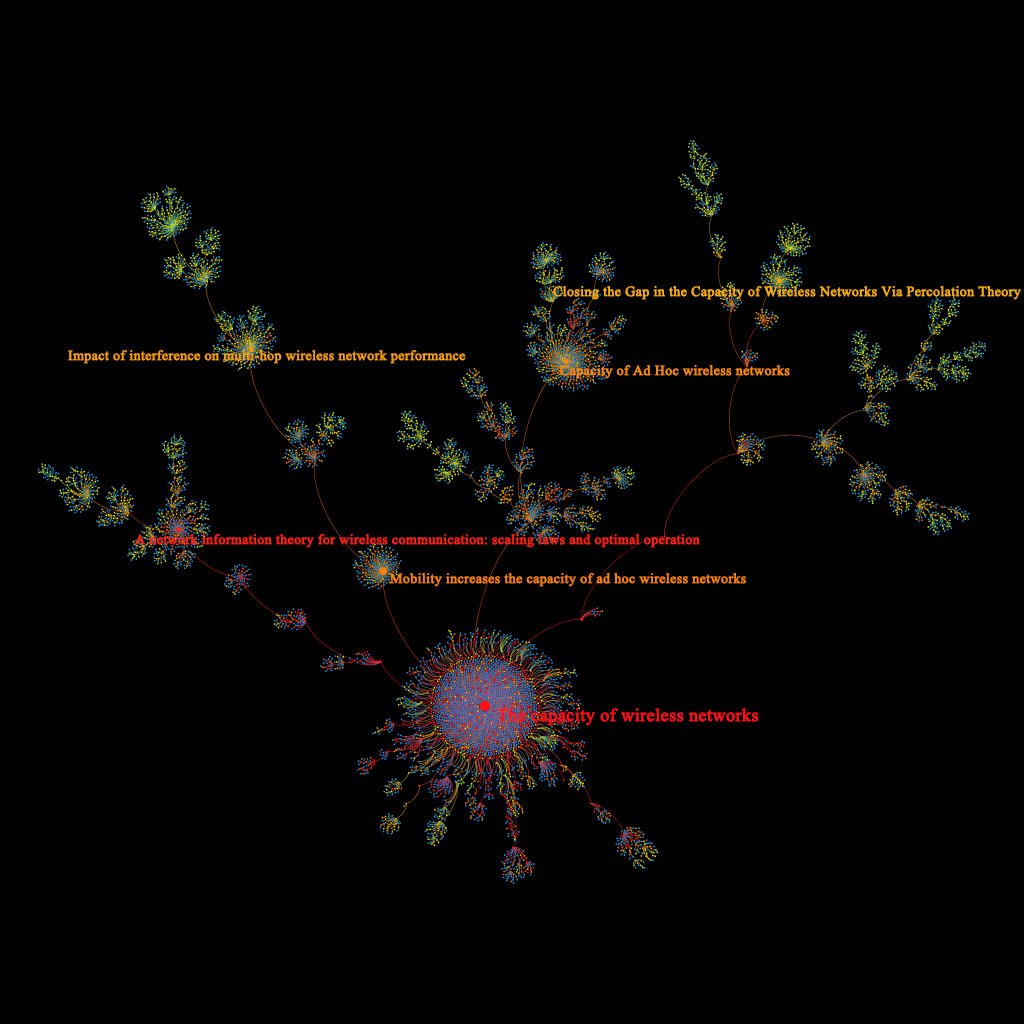}
   \end{subfigure}
   \begin{subfigure}{0.35\textwidth}
   \centering
   \includegraphics[width = 0.9\linewidth]{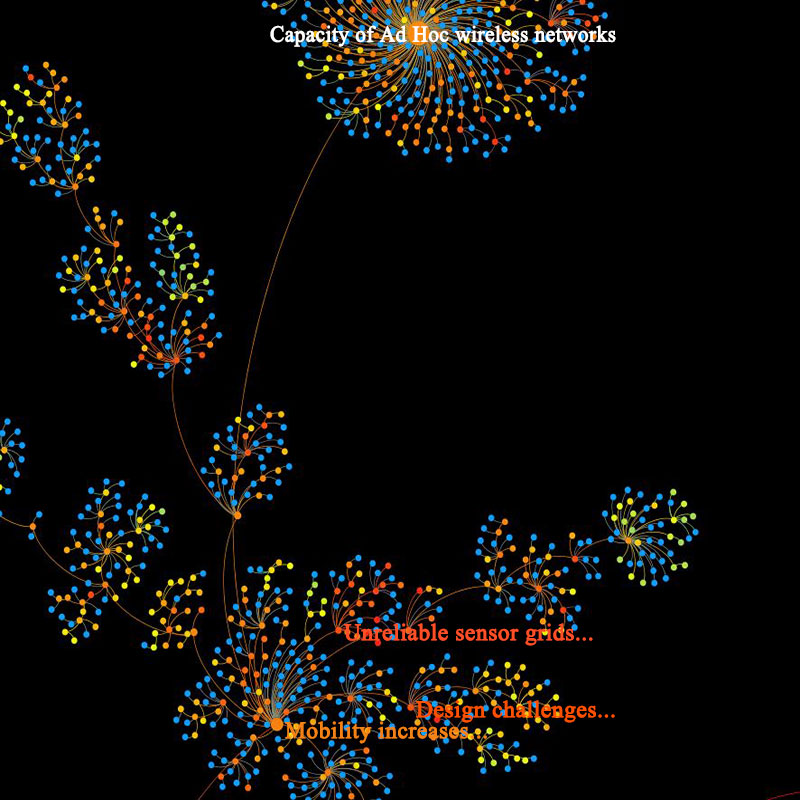}
   \end{subfigure}
\caption{Capacity Wireless Network: Galaxy map, current skeleton tree and its regional zoom. Papers with more than 500 in-topic citations are labelled by title in the skeleton tree. Except the pioneering work, corresponding nodes' size is amplified by 3 times.}
\label{fig:438420345-2020}
\end{figure}

\noindent Now we probe into the topic and closely examine the heat distribution in its latest skeleton tree (Fig. \ref{fig:438420345-2020}). After 20 years of development, the heat diffusion is nearly completed as popular child papers all have a knowledge temperature above average and the child papers published in the first 10 years are relatively hot in general (Fig. 5(e)). The popular child papers and the pioneering work are the multiple heat sources within the topic. If we let alone the blue nodes surrounding the pioneering work and popular child papers, which are papers with few or without any in-topic citations, it is clear that node knowledge temperature decreases globally as the articles are located farther away from them. However, there are exceptions to general rules "the more influential the hotter" (Fig. \ref{fig:citation_T}(e)) and "the older the hotter". For example, paper `Mobility increases the capacity of ad-hoc wireless networks' (MAWN) published in 2001, which is at the junction between the central cluster and a principal branch, is slightly colder than 2 of its children: `Design challenges for energy-constrained ad hoc wireless networks' (DCAWN) published in 2002 and `Unreliable sensor grids: coverage, connectivity and diameter' (USG) published in 2003. MAWN is coloured orange while DCAWN and USG are coloured orange-red and red. The main reason of this uncommon phenomenon is their different research focus, which is reflected by their distinct citations and citations' average heat-level. Another reason may be that even though MAWN has inspired much more child papers, few of its numerous followers have so far achieved remarkable development, hence their limited boosting effect. \\

\noindent We observe in addition certain clustering effect in the skeleton tree (Table \ref{tab:438420345-clustering}). For example, almost all child papers of `A Delay-Efficient Algorithm for Data Aggregation in Multihop Wireless Sensor Networks' have similar research themes as itself. This proves the effectiveness of our skeleton tree extraction algorithm.\\

\begin{table}
    \centering
    \begin{tabular}{p{15cm} p{1cm}}
        \hline
        title & year\\
        \hline
        A \textcolor{red}{Delay}-Efficient Algorithm for \textcolor{red}{Data Aggregation} in Multihop Wireless Sensor Networks & 2011 \\

         In-Network Estimation with \textcolor{orange}{Delay} Constraints in Wireless Sensor Networks & 2013\\

       Estimate \textcolor{orange}{Aggregation} with \textcolor{orange}{Delay} Constraints in Multihop Wireless Sensor Networks & 2011\\

        Genetic Local Search for Conflict-Free Minimum-\textcolor{orange}{Latency} \textcolor{orange}{Aggregation} Scheduling in Wireless Sensor Networks & 2018\\

        Interference-Fault Free \textcolor{orange}{Data Aggregation} in Tree-Based WSNs & 2016\\

        GLS and VNS Based Heuristics for Conflict-Free Minimum-\textcolor{orange}{Latency} \textcolor{orange}{Aggregation} Scheduling in WSN. & 2019\\

        \textcolor{orange}{Data Aggregation} Scheduling Algorithms in Wireless Sensor Networks: Solutions and Challenges & 2014\\

        Efficient scheduling for periodic \textcolor{orange}{aggregation} queries in multihop sensor networks & 2012\\

        Layer-Based Data \textcolor{orange}{Aggregation} and Performance Analysis in Wireless Sensor Networks & 2013\\

        Neither Shortest Path Nor Dominating Set: \textcolor{orange}{Aggregation} Scheduling by Greedy Growing Tree in Multihop Wireless Sensor Networks & 2011\\

        Composite interference mapping model for Interference Fault-Free Transmission in WSN & 2015 \\

        Weighted fairness guaranteed \textcolor{orange}{data aggregation} scheduling algorithm in wireless sensor networks & 2012\\

        A fuzzy-rule-based packet reproduction routing for sensor networks & 2018\\
        \hline
    \end{tabular}
    \caption{Capacity Wireless Network: Clustering effect example. First line is the parent paper and the rest children.}
    \label{tab:438420345-clustering}
\end{table}

\subsubsection{Efficient Estimation of Word Representations in Vector Space}

\begin{figure}[htbp]
\centering
\begin{subfigure}[t]{0.6\linewidth}
\includegraphics[width=\linewidth]{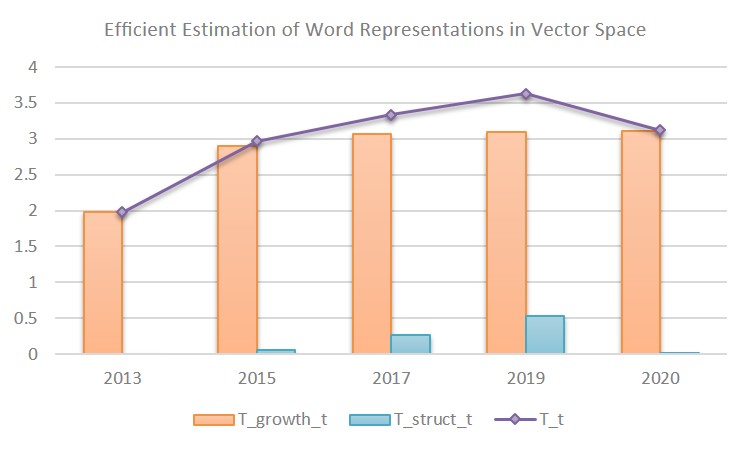}
\end{subfigure}

\begin{subfigure}[t]{\linewidth}
\centering
\begin{tabular}{ccccccccc}
\hline
year & $|V^t|$ & $|E^t|$ & $n_t$ & $V_t$ & ${UsefulInfo}^t$ & $T_{growth}^t$ & $T_{struct}^t$ & $T^t$\\
\hline
2013 & 29 & 42 & 23.5 & 29 & 5.5 & 1.978 &  & 1.978\\

2015 & 1197 & 4014 & 660.232 & 1197 & 536.768 & 2.91 & 0.061  & 2.967 \\

2017 & 4136 & 16798 & 2159.275 & 4136 & 1976.725 & 3.07 & 0.268 & 3.338 \\

2019 & 7736 & 34285 & 3999.585 & 7736 & 3736.415 & 3.1 & 0.53 & 3.63 \\

2020 & 8133 & 36219 & 4199.586 & 8133 & 3933.414 & 3.104 & 0.015 & 3.119 \\
\hline
\end{tabular}
\end{subfigure}
\caption{Efficient word representation: topic statistics and knowledge temperature evolution}
\label{fig:372720438_chart}
\end{figure}

\begin{figure}[htbp]
\begin{subfigure}{\textwidth}
\begin{minipage}[t]{0.5\linewidth}
\includegraphics[width = \linewidth]{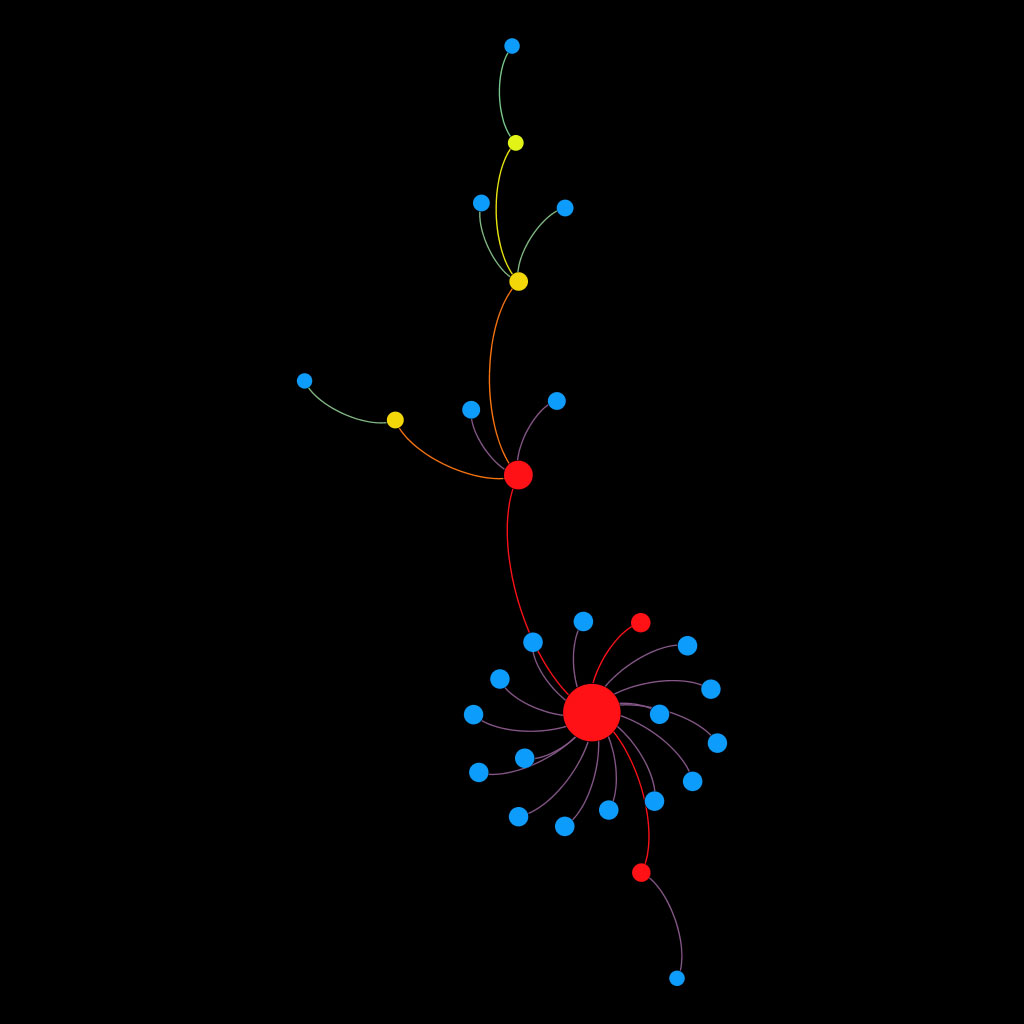}
\caption{Skeleton tree until 2013}
\end{minipage}
\begin{minipage}[t]{0.5\linewidth}
\includegraphics[width = \linewidth]{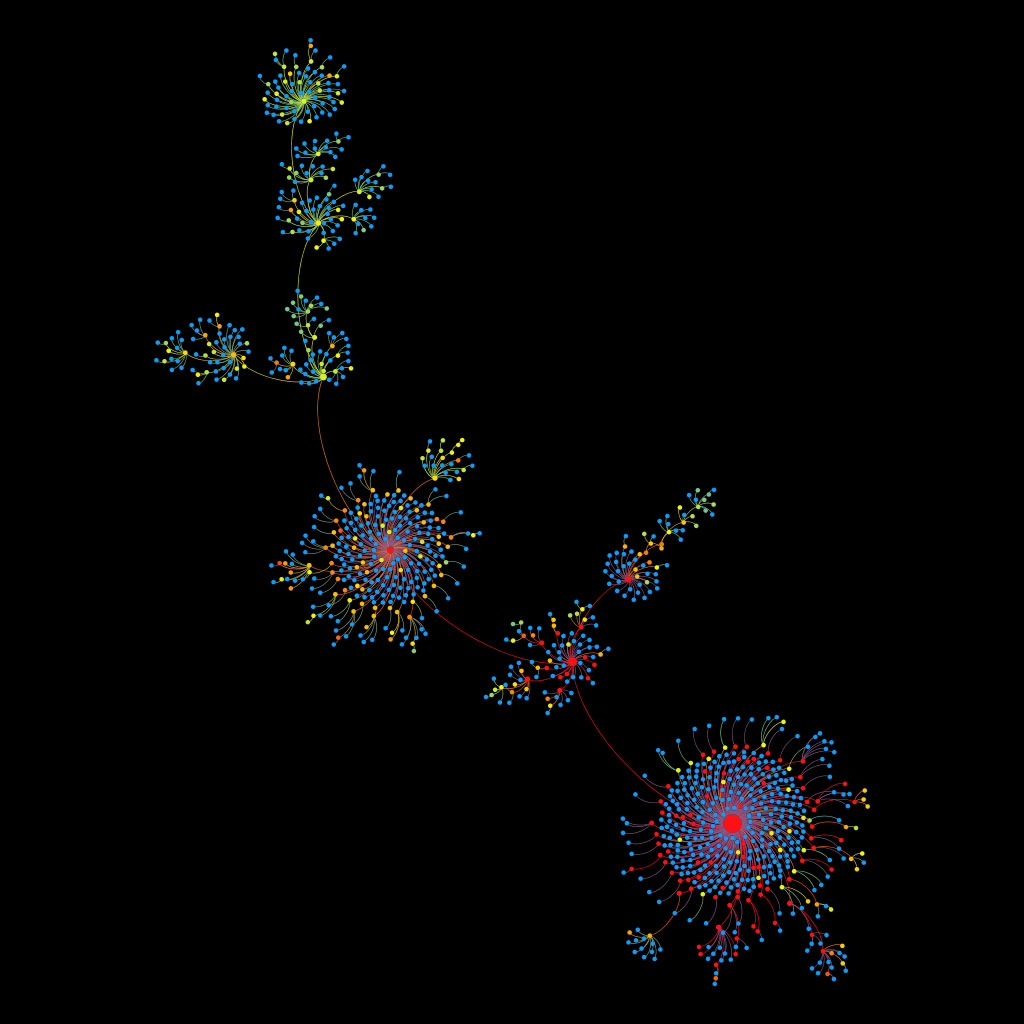}
\caption{Skeleton tree until 2015}
\end{minipage}
\end{subfigure}
\vspace{2mm}
\begin{subfigure}{\textwidth}
\begin{minipage}[t]{0.5\linewidth}
\includegraphics[width = \linewidth]{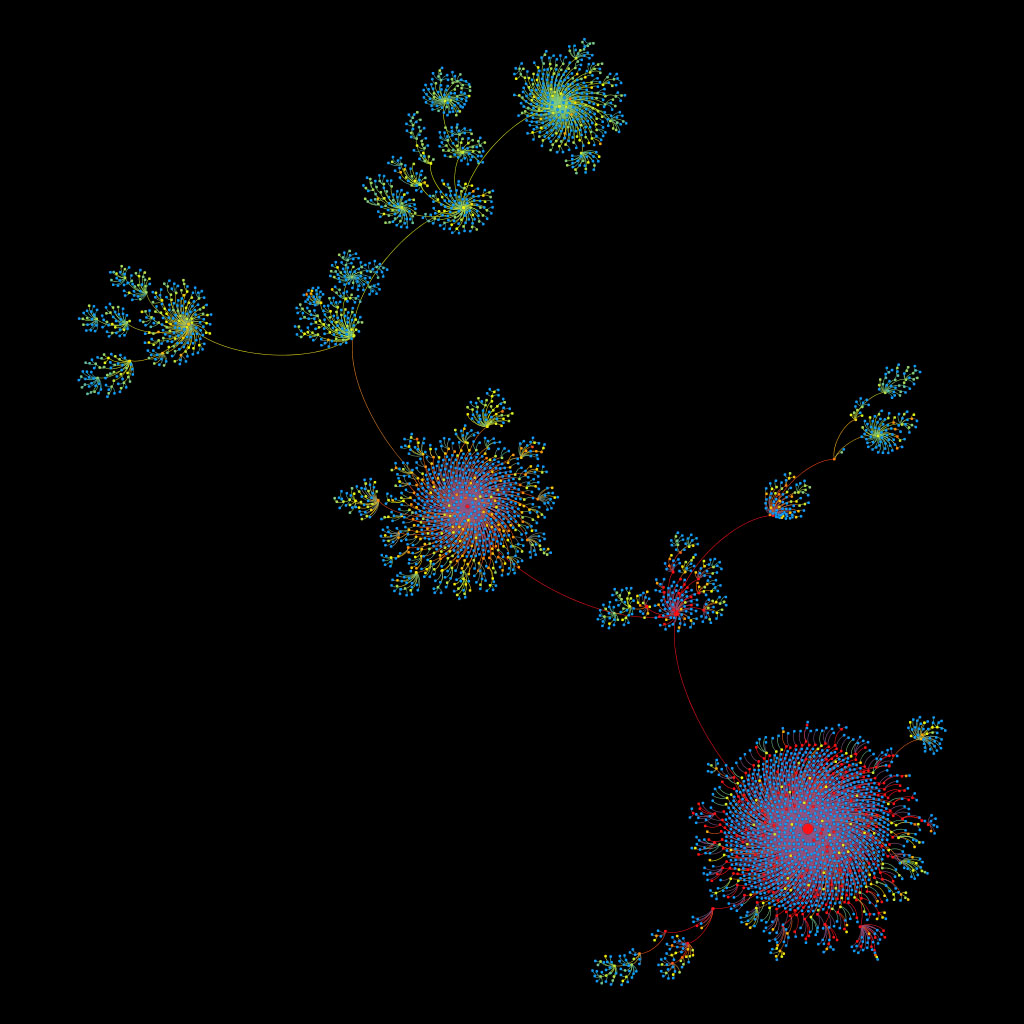}
\caption{Skeleton tree until 2017}
\end{minipage}
\begin{minipage}[t]{0.5\linewidth}
\includegraphics[width = \linewidth]{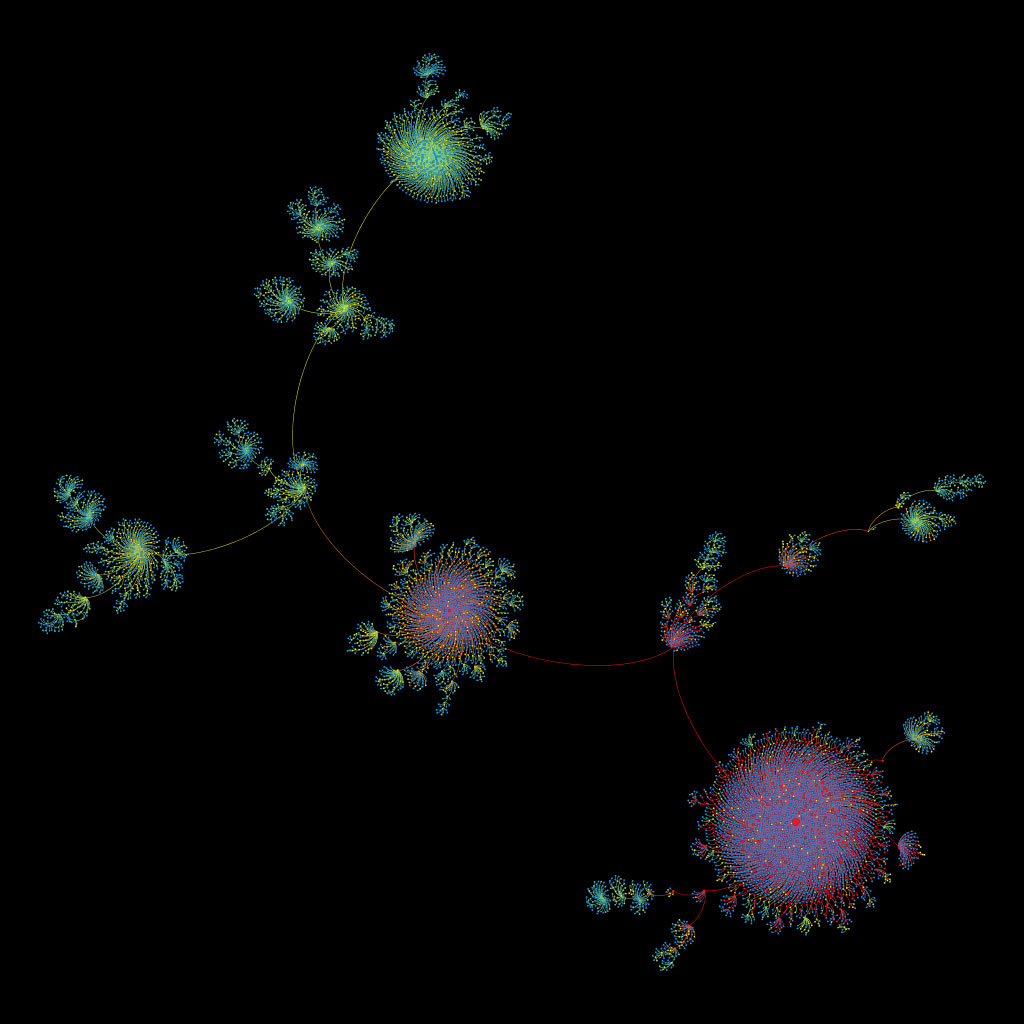}
\caption{Skeleton tree until 2019}
\end{minipage}
\end{subfigure}
\caption{Efficient word representation: Skeleton tree evolution}
\label{fig:372720438-tree_evo}
\end{figure}

The popularity and impact gain in the first years is mainly due to a fast accumulation of useful information. By the end of 2013, 2 influential child papers, `Linguistic Regularities in Continuous Space Word Representations' (LRCSWR) and `Distributed Representations of Words and Phrases and their Compositionality' (DRWPC) had formed the fundamentals of topic knowledge structure. LRCSWR is the red node in the middle of the then skeleton tree and its child, DRWPC, is represented by the yellow-green node above itself (Fig. \ref{fig:372720438-tree_evo}(a)). During the next 2 years, the topic expanded quickly thanks to the substantial development of all 3 papers. DRWPC emerged as the second topic center following the pioneering work (Fig. \ref{fig:372720438-tree_evo}(b)). In addition, DRWPC helped extending topic knowledge structure by inspiring a new research direction. This research branch later proved to be a novel research focus. Starting from 2016, owing to a multidimensional development  the topic has been maintaining a knowledge reserve quantity corresponding to its size, which is reflected by its steady $T_{growth}^t$ (Fig. \ref{fig:372720438_chart}). More importantly, the research branch that emerged by the end of 2015 has developed into 2 new non-trivial research directions due to the popularity rise in 2 child papers published in 2014: `Glove: Global Vectors for Word Representation' (Glove)  and `Distributed Representations of Sentences and Documents' (DRSD). They brought new knowledge, attracted the attention of the latest research attention, and catalysed an accelerated topic
 knowledge structure evolution, which is captured by a rising $T_{structure}^t$. This year, there has not been any significant new trend so far. Therefore, the topic cools down a bit due to a  $T_{structure}^t$ drop. Unless the topic succeeds in "breeding" some new focus or having some breakthrough to existing sub-topics in the near future, it starts to go downhill after 6 years of thriving. \\

\begin{figure}
    \centering
    \begin{subfigure}{\linewidth}
    \begin{minipage}[t]{0.55\textwidth}
    \centering
    \includegraphics[width = \linewidth]{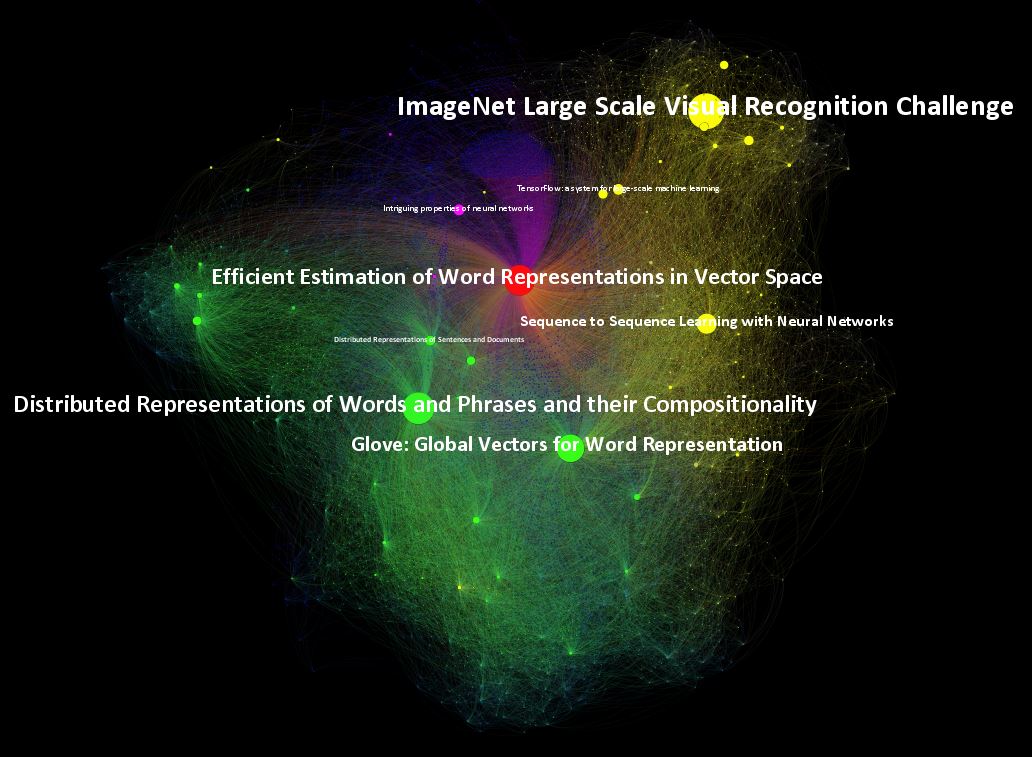}
    \end{minipage}
    \begin{minipage}[t]{0.45\textwidth}
    \centering
    \includegraphics[width =0.9\linewidth]{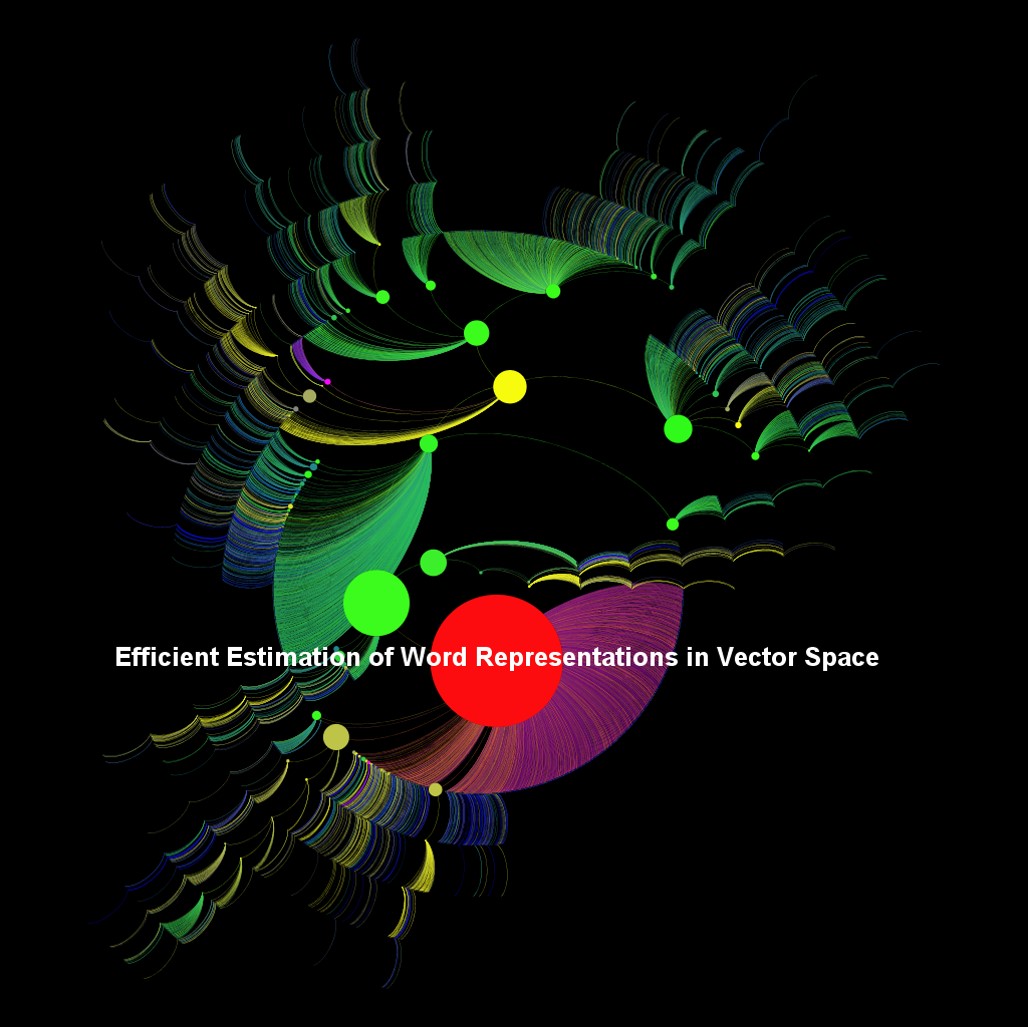}
    \end{minipage}
    \end{subfigure}

    \vspace{5mm}

    \begin{subfigure}{0.6\linewidth}
    \includegraphics[width = \linewidth]{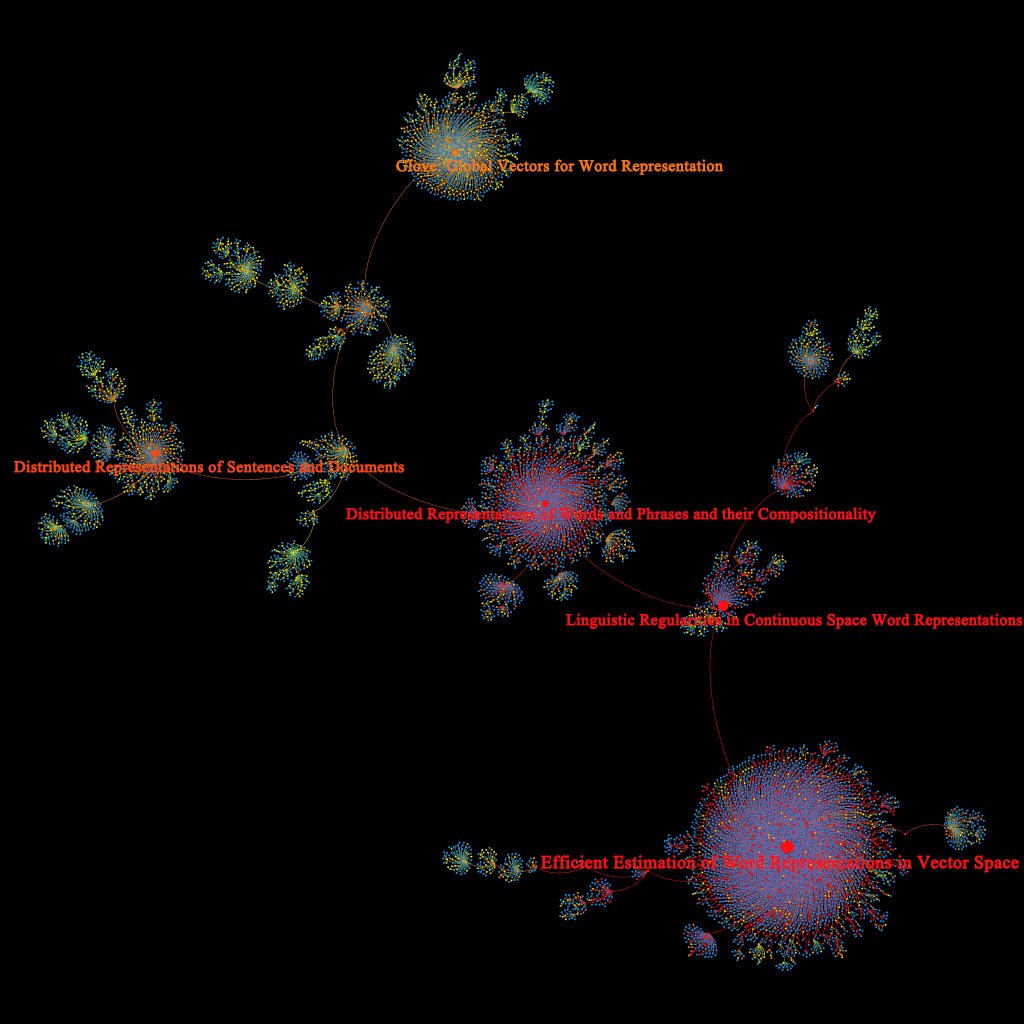}
    \end{subfigure}
    \caption{Efficient word representation: Galaxy map and current skeleton tree. Papers with more than 700 in-topic citations are labelled by title in the skeleton tree. Except the pioneering work, corresponding nodes' size is amplified by 3 times.}
    \label{fig:372720438-2020}
\end{figure}

\noindent Now we probe into the topic and closely examine the heat distribution in its latest skeleton tree (Fig. \ref{fig:372720438-2020}). The topic's fast development accompanies a continuous heat diffusion. The older popular child papers has become the hottest since 2015 and the younger ones, namely DRSD and Glove, has recently evolved into topic's new heat sources. It is clear that node knowledge temperature decreases globally as the articles are located farther away from them. This phenomenon fits the general rule "the older the hotter" (Fig. 5(f)) and "the more influential the hotter"  (Fig. \ref{fig:citation_T}(f)). Note that the blue nodes that surround the pioneering work and popular child papers in central parts are papers with few or without any in-topic citations.\\

\noindent We observe in addition certain clustering effect in the skeleton tree (Table \ref{tab:372720438-clustering}). For example, in current skeleton tree, all child papers of `Sentiment Embeddings with Applications to Sentiment Analysis' published in journal \textit{IEEE Transactions on Knowledge and Data Engineering} in 2016 specialize in sentiment analysis. This proves the effectiveness of our skeleton tree extraction algorithm.\\

\begin{table}
    \centering
    \begin{tabular}{p{15cm} p{1cm}}
        \hline
        title & year\\
        \hline
        Sentiment Embeddings with Applications to \textcolor{red}{Sentiment Analysis} & 2016 \\

        Deep Learning Adaptation with Word Embeddings for \textcolor{orange}{Sentiment Analysis} on Online Course Reviews & 2020\\

        Learning Word Representations for \textcolor{orange}{ Sentiment Analysis} & 2017\\

        Improving Aspect-Based \textcolor{orange}{Sentiment Analysis} via Aligning Aspect Embedding & 2019\\

        Attention-based long short-term memory network using sentiment lexicon embedding for aspect-level \textcolor{orange}{sentiment analysis} in Korean & 2019\\

        Deep Learning for Aspect-Based \textcolor{orange}{Sentiment Analysis}: A Comparative Review & 2019\\

        An efficient preprocessing method for supervised \textcolor{orange}{sentiment analysis} by converting sentences to numerical vectors: a twitter case study & 2019\\

        Deep learning for \textcolor{orange}{sentiment analysis}: A survey & 2018\\

        Deep Learning in \textcolor{orange}{Sentiment Analysis} & 2018\\

        \textcolor{orange}{Sentiment analysis} using deep learning approaches: an overview & 2020\\
        \hline
    \end{tabular}
    \caption{Efficient word representation: Clustering effect example. First line is the parent paper and the rest children.}
    \label{tab:372720438-clustering}
\end{table}

\subsubsection{Coverage problems in wireless ad-hoc sensor networks}

This topic reached its peak around 2010 thanks to a surge in $T_{structure}^t$. Most of its popular child papers were published by the end of 2006. Among them, the older ones laid the foundation of multiple research sub-directions and the younger ones further developed these new research branches. For instance, papers `Unreliable sensor grids: coverage, connectivity and diameter' and `Sensor placement for grid coverage under imprecise detections' published in 2002 and 2003 extended primarily the idea of the pioneering work. They formed the 2 big branches surrounding the central cluster in skeleton tree by 2007 (Fig. \ref{fig:344180001-tree_evo}(a), \ref{fig:344180001-2020}). Paper `The coverage problem in a wireless sensor network' (CPWS) published in 2005, however, created a second smaller cluster by furthering the study of his predecessor `Localized algorithms in wireless ad-hoc networks: location discovery and sensor exposure' (LAWAN) published in 2001. Other popular papers published between 2005 and 2006 were split into 2 parties, one group supporting the growth in central cluster led by the pioneering work, the other group enriching the newer cluster built essentially by CPWS. As a result, we observe non-trivial growth in every corner of the skeleton tree during 2007 and 2010 (Fig. \ref{fig:344180001-tree_evo}(b)). Nonetheless, along with the multidimensional flourishing, the knowledge structure started its gravity redistribution due to the maturation of the research sub-directions. This silent transformation is captured by the high $T_{structure}^t$ around 2010.
The aforementioned popular child papers as well as their inspirations for future works also make great contributions to the knowledge accumulation. They helped push up $T_{growth}^t$ until 2010. Afterwards, the topic experienced first an absence of promising child papers and then a decline in useful information supply due to its decelerated expansion. Consequently, $T_{growth}^t$ has stagnated. The skeleton tree has unsurprisingly lost its vigor during this period (Fig. \ref{fig:344180001-tree_evo} (c,d)). To sum up, this topic, after a rapid development in its early days, demonstrates now a decreasing activity and a diminishing popularity and impact.\\

\noindent The topic's skeleton tree is a bit special in that it is comprised of 2 parts. The separation is due to the isolation of LAWAN from the pioneering work. LAWAN cites both the pioneering work and `Dynamic fine-grained localization in Ad-Hoc networks of sensors' (DLANS). Because of a closer relation between LAWAN and DLANS, its connection to the pioneering work is cut off in skeleton tree extraction. A similar reason caused the separation of DLANS and the pioneering work. LAWAN, along with several intimately related papers, is thus completely separated from the pioneering work. They form a mini bundle beside the central cluster in 2004 skeleton tree. Shortly after, the arrival of popular child paper, CPWS, largely developed this tiny bundle and turned it into the big aggregation under the central cluster (Fig. \ref{fig:344180001-tree_evo}).\\

\begin{figure}[htbp]
\centering
\begin{subfigure}[t]{0.6\linewidth}
\includegraphics[width=\linewidth]{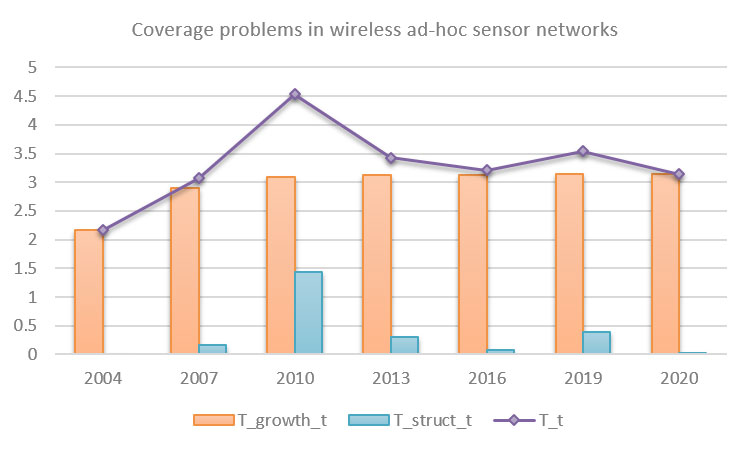}
\end{subfigure}

\begin{subfigure}[t]{\linewidth}
\centering
\begin{tabular}{ccccccccc}
\hline
year & $|V^t|$ & $|E^t|$ & $n_t$ & $V_t$ & ${UsefulInfo}^t$ & $T_{growth}^t$ & $T_{struct}^t$ & $T^t$\\
\hline
2004 & 146 & 337 & 90.189 & 146 & 55.811 & 2.166 &   & 2.166 \\

2007 & 542 & 2420 & 249.982 & 542 & 292.018 & 2.902 & 0.17 & 3.072 \\

2010 & 972 & 5118 & 420.165 & 972 & 551.835 & 3.096 & 1.44 & 4.536 \\

2013 & 1313 & 7325 & 562.623 & 1313 & 750.377 & 3.123 & 0.308 & 3.431 \\

2016 & 1490 & 8460 & 637.376 & 1490 & 852.624 & 3.129 & 0.08 & 3.209 \\

2019 & 1544 & 8846 & 657.378 & 1544 & 886.622 & 3.143 & 0.396 & 3.539 \\

2020 & 1546 & 8865 & 658.507 & 1546 & 887.493 & 3.142 & 0.001 & 3.143 \\
\hline
\end{tabular}
\end{subfigure}
\end{figure}

\begin{figure}[htbp]
\begin{subfigure}{\textwidth}
\begin{minipage}[t]{0.5\linewidth}
\includegraphics[width = \linewidth]{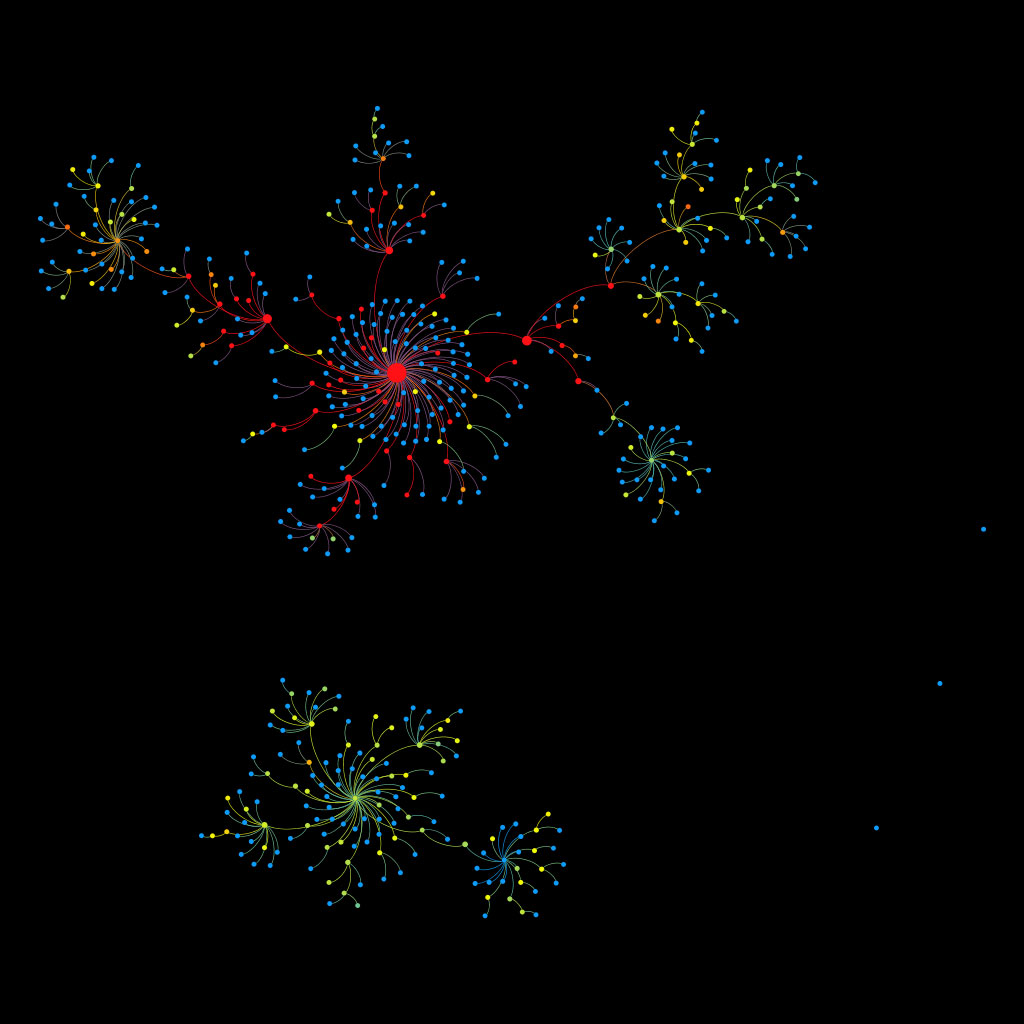}
\caption{Skeleton tree until 2007}
\end{minipage}
\begin{minipage}[t]{0.5\linewidth}
\includegraphics[width = \linewidth]{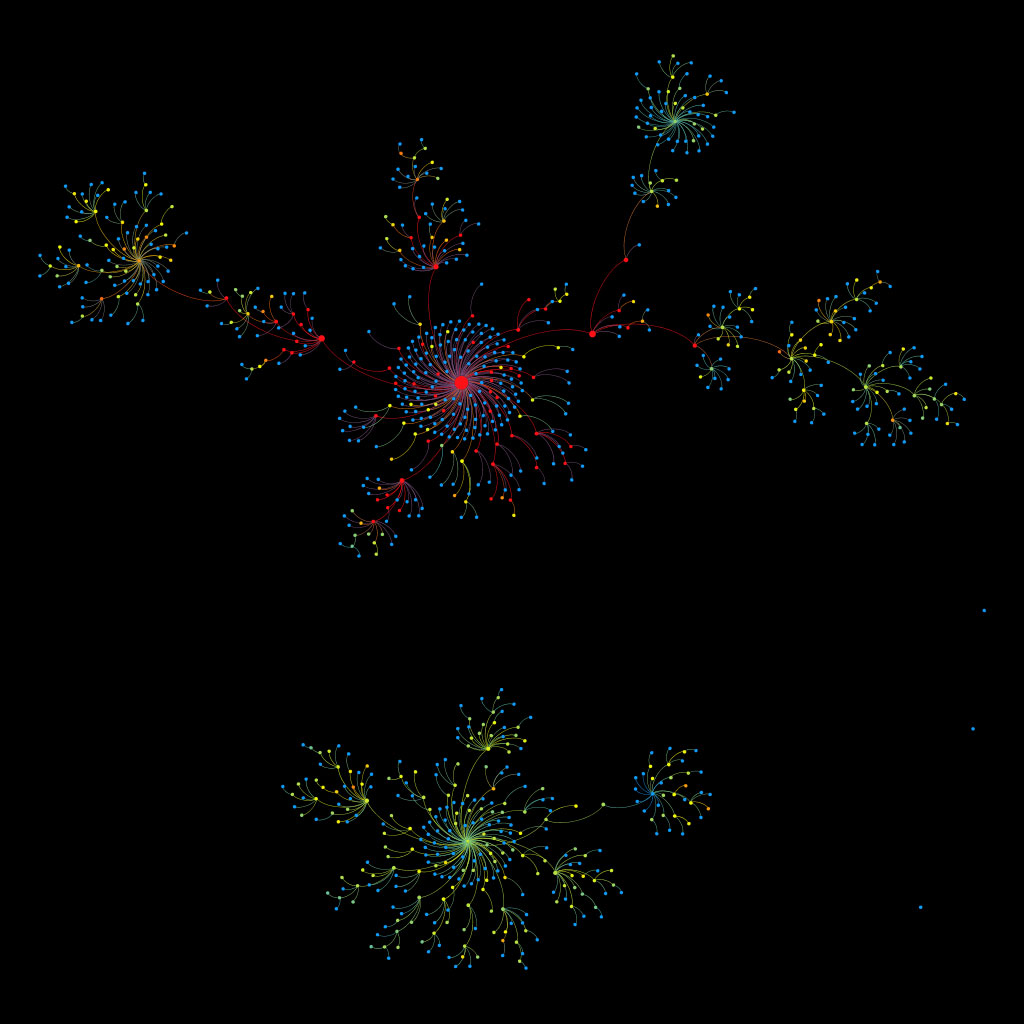}
\caption{Skeleton tree until 2010}
\end{minipage}
\end{subfigure}
\vspace{2mm}
\begin{subfigure}{\textwidth}
\begin{minipage}[t]{0.5\linewidth}
\includegraphics[width = \linewidth]{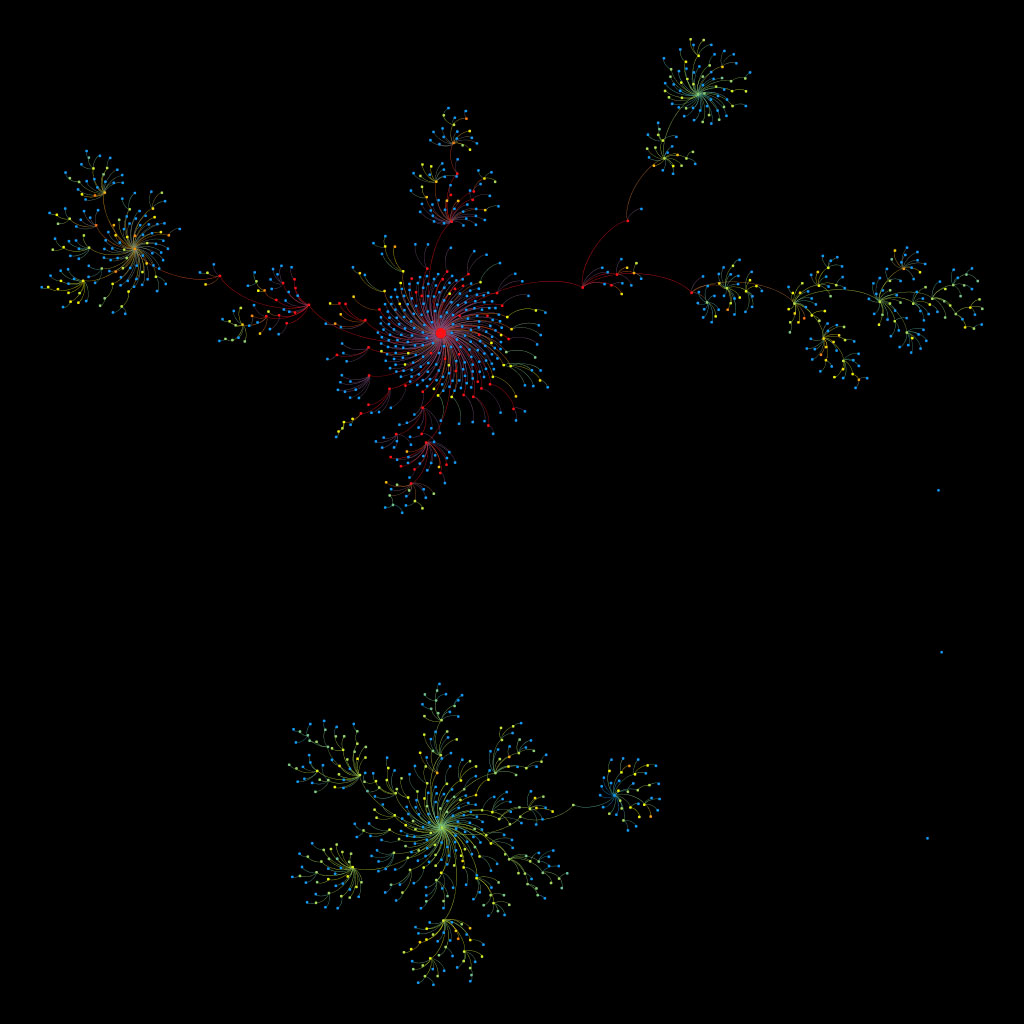}
\caption{Skeleton tree until 2013}
\end{minipage}
\begin{minipage}[t]{0.5\linewidth}
\includegraphics[width = \linewidth]{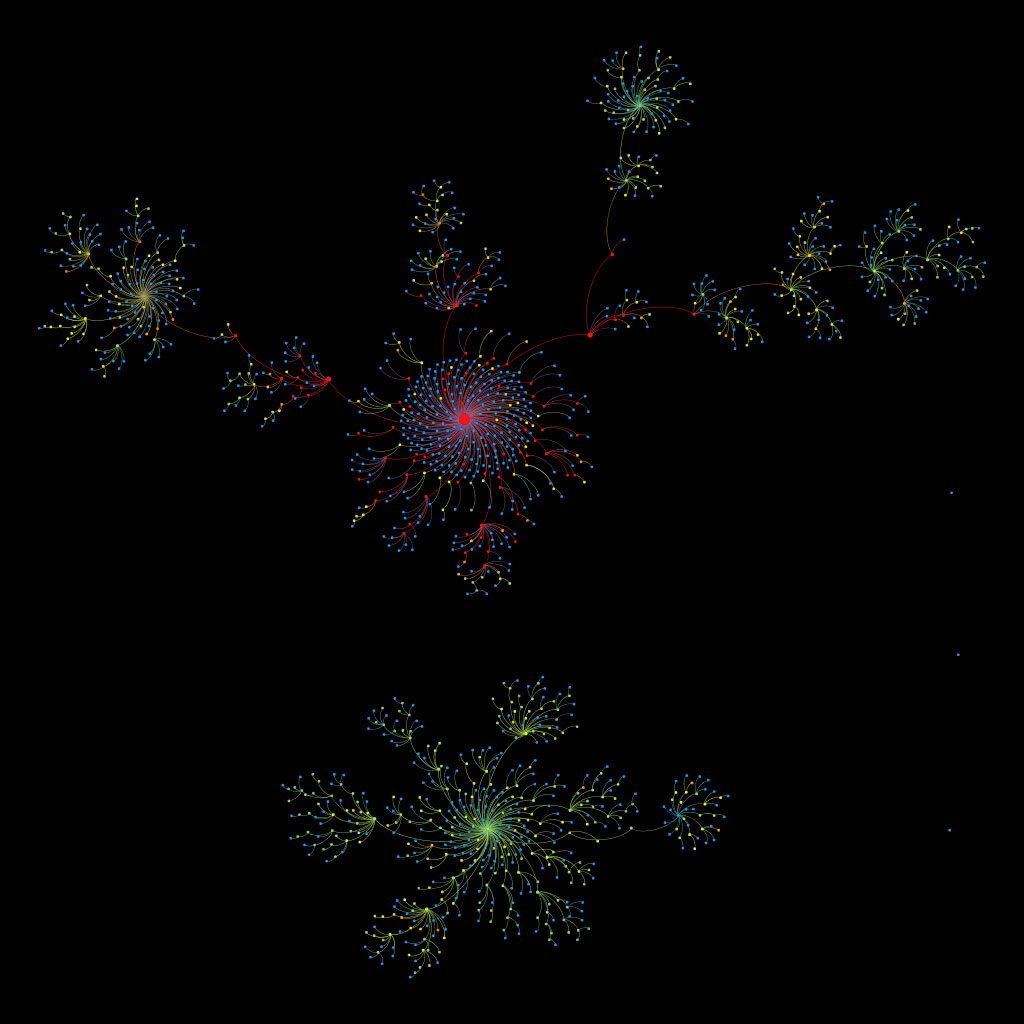}
\caption{Skeleton tree until 2016}
\end{minipage}
\end{subfigure}
\caption{Coverage problems: Skeleton tree evolution}
\label{fig:344180001-tree_evo}
\end{figure}

\noindent Now we closely examine the heat distribution within the topic (Fig. \ref{fig:344180001-2020}). After 19 years of development, the heat diffusion is nearly completed as most popular child papers have a knowledge temperature above average and the child papers published during the flourishing period are relatively hot in general (Fig. 5(g)). Half of the most popular child papers serve as heat sources and node knowledge temperature decreases globally as the articles are located farther away from them. This corresponds with the general rule "the older the hotter". Yet as several papers published at the same time as the pioneering work either have had few development or have not been cited by any recent works, they are the coldest and thus bring down the average knowledge of the oldest articles. In addition, the blue nodes that surround the pioneering work and popular child papers are papers with few or without any in-topic followers. However, we still find exceptions even if we let alone the oldest papers. Paper `Minimal and maximal exposure path algorithms for wireless embedded sensor networks' (MMEPA) published in 2003 is colder than, for instance, its child `Smart Path-Finding with Local Information in a Sensory Field' published in 2006 and `An Algorithm for Target Traversing Based on Local Voronoi Diagram' published in 2007. These 2 child papers are represented as orange nodes yet the MMEPA is a green node. This is mainly due to their relatively different research focus as most of their in-topic citations do not overlap with one another. Another reason may be that even though MMEPA has inspired much more child papers, few of them have achieved remarkable development, hence their limited boosting effect. In addition, this counter example also suggests that the general rule "the more influential the hotter" is very weak in this topic (Fig. \ref{fig:citation_T}(g)). \\

\begin{figure}[htbp]
    \centering
    \begin{subfigure}{\linewidth}
    \begin{minipage}[t]{0.5\textwidth}
    \centering
    \includegraphics[width = \linewidth]{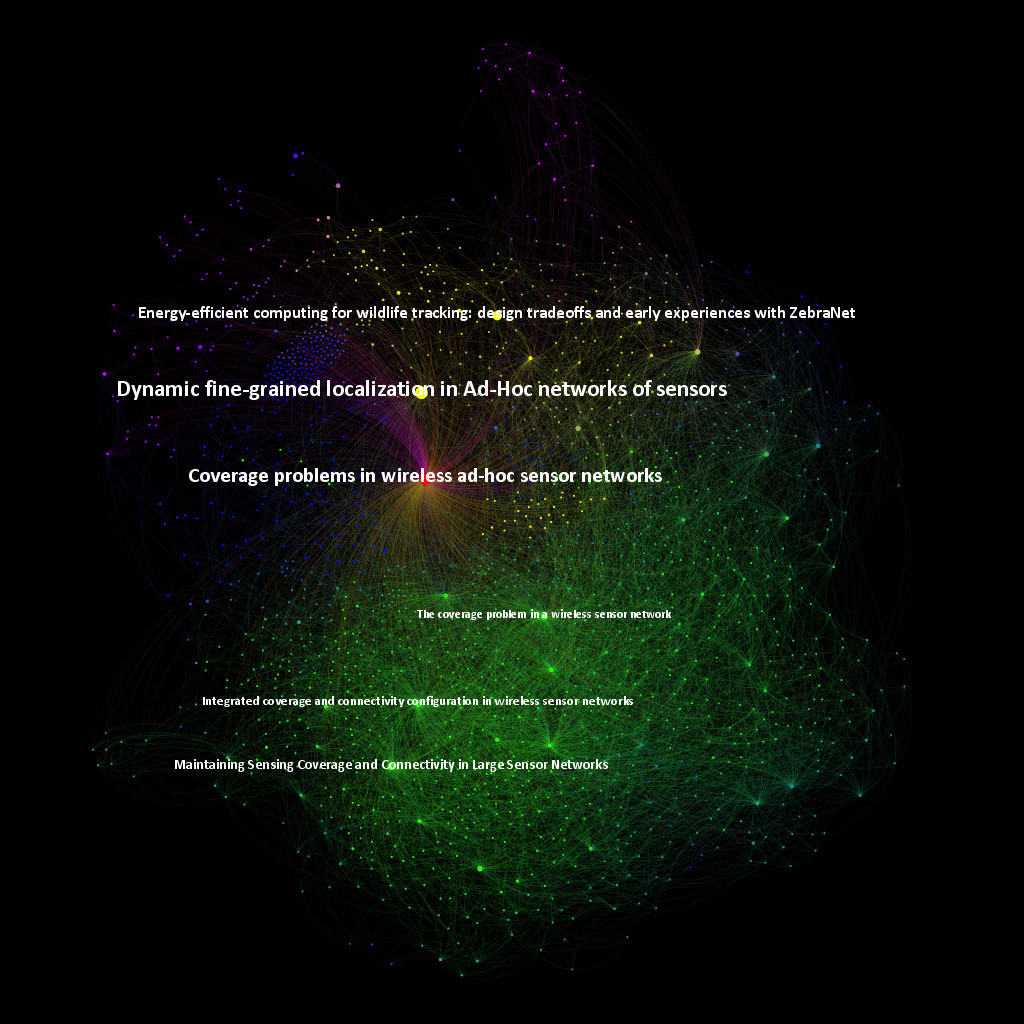}
     \end{minipage}
     \begin{minipage}[t]{0.5\textwidth}
    \centering
    \includegraphics[width = \linewidth]{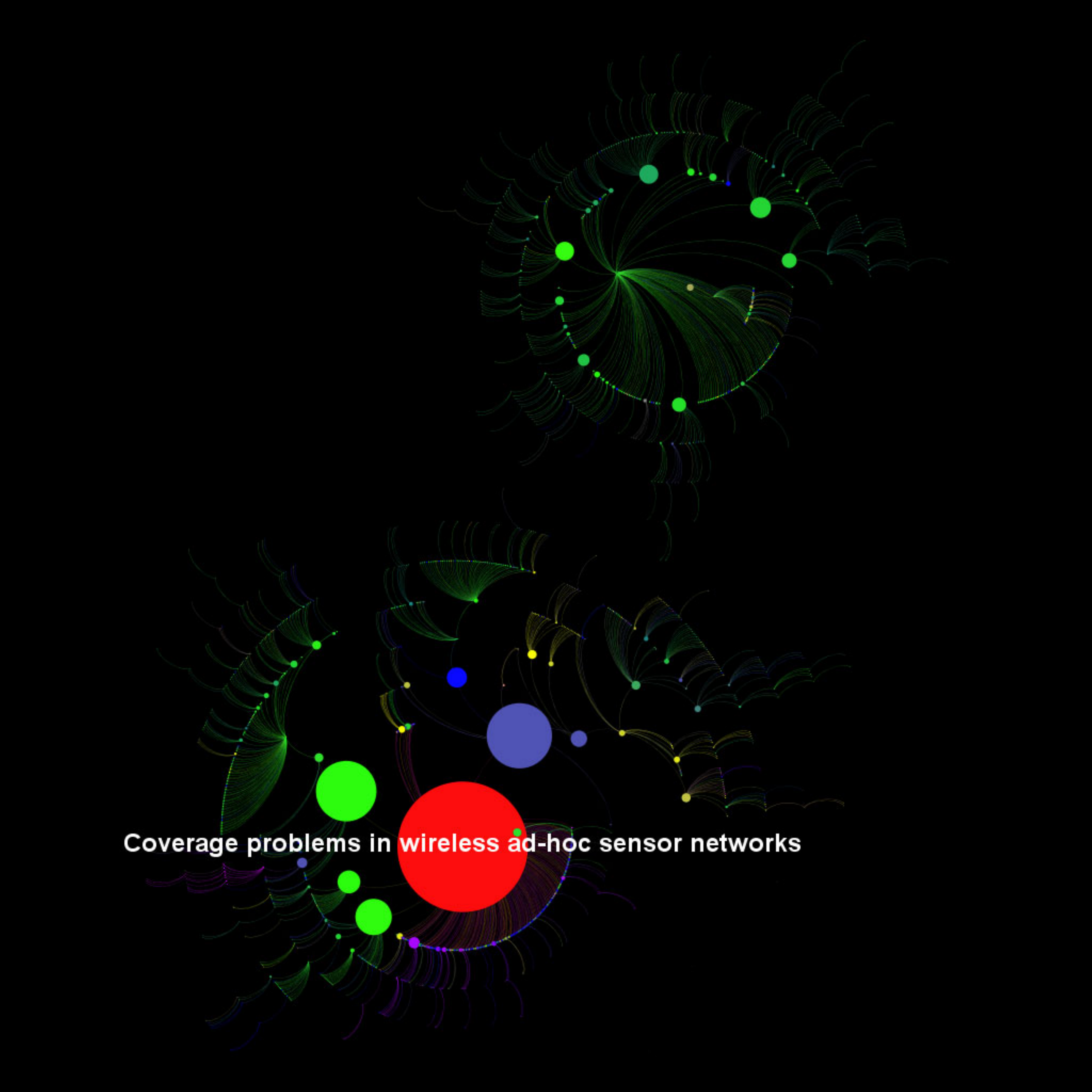}
     \end{minipage}

    \vspace{5mm}

    \end{subfigure}
    \begin{subfigure}{0.6\linewidth}
    \includegraphics[width = \linewidth]{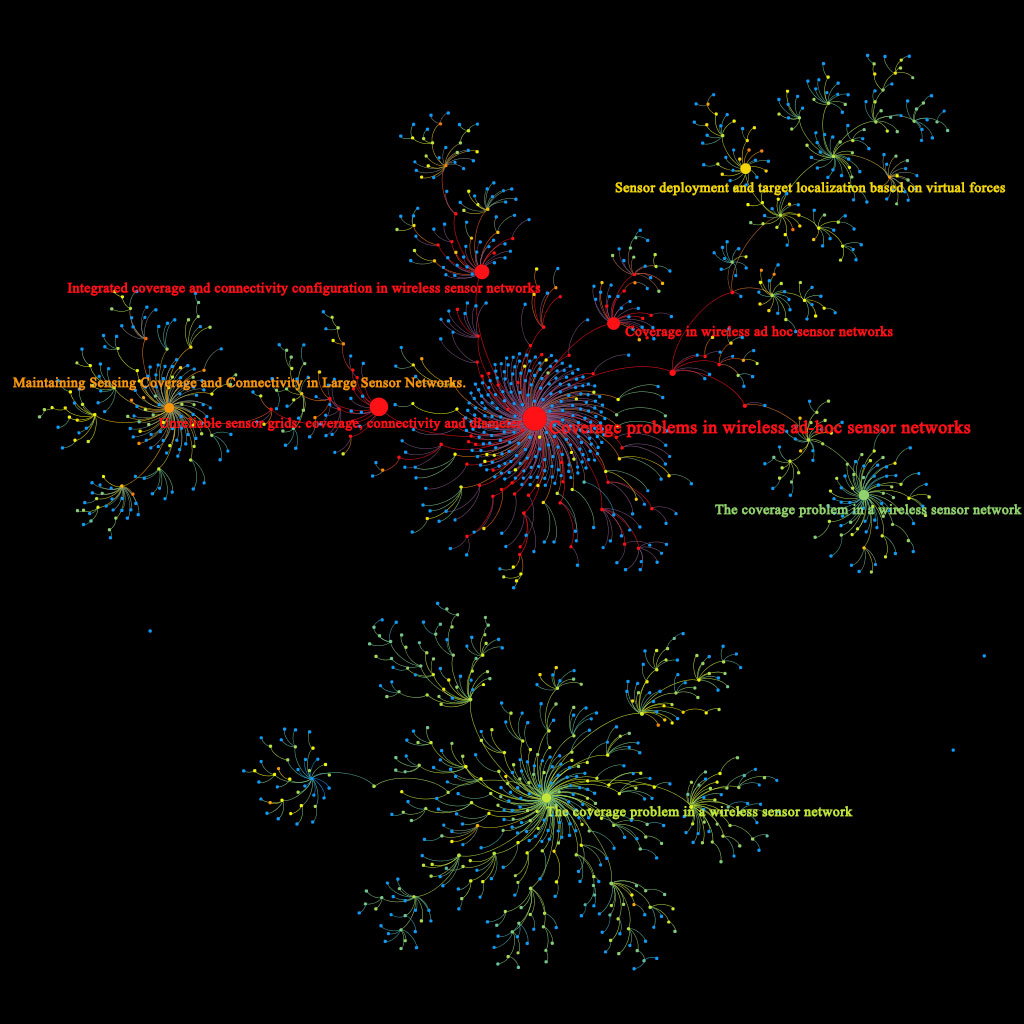}
    \end{subfigure}
    \begin{subfigure}{0.35\linewidth}
    \centering
    \includegraphics[width = 0.9\linewidth]{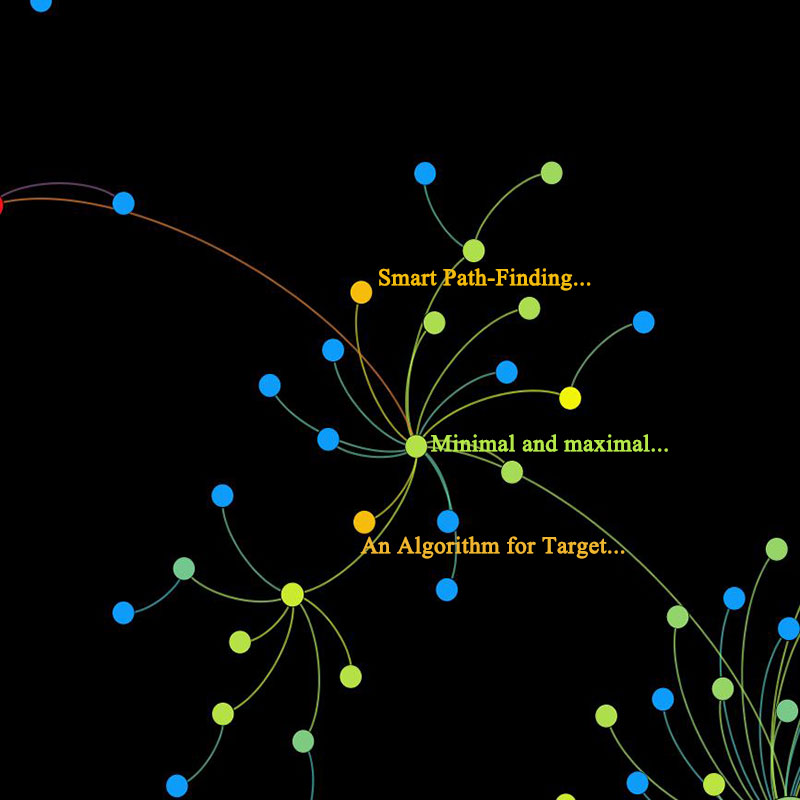}
    \end{subfigure}
    \caption{Coverage problems: Galaxy map, current skeleton tree and its regional zoom. Papers with more than 150 in-topic citations are labelled by title in the skeleton tree. Except the pioneering work, corresponding nodes' size is amplified by 3 times.}
    \label{fig:344180001-2020}
\end{figure}

\subsubsection{A neural probabilistic language model}
\begin{figure}[htbp]
\centering
\begin{subfigure}[t]{0.6\linewidth}
\includegraphics[width=\linewidth]{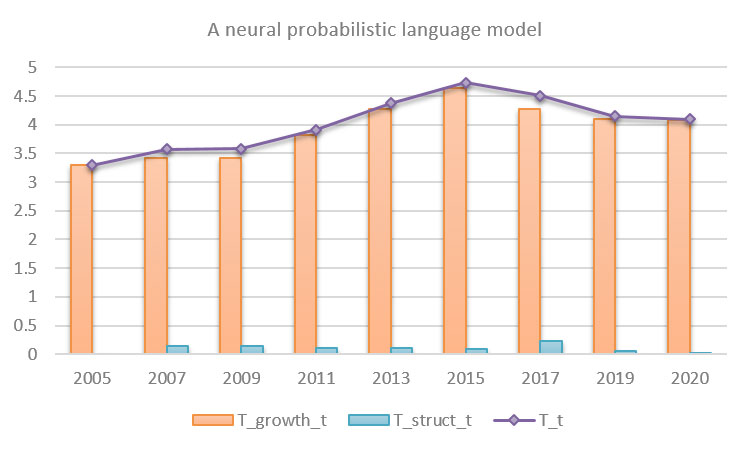}
\end{subfigure}

\begin{subfigure}[t]{\linewidth}
\centering
\begin{tabular}{ccccccccc}
\hline
year & $|V^t|$ & $|E^t|$ & $n_t$ & $V_t$ & ${UsefulInfo}^t$ & $T_{growth}^t$ & $T_{struct}^t$ & $T^t$\\
\hline
2005 & 25 & 53 & 16.204 & 25 & 8.796 & 3.297 &   & 3.297 \\

2007 & 61 & 172 & 38.068 & 61 & 22.932 & 3.424 & 0.151 & 3.575 \\

2009 & 103 & 305 & 64.249 & 103 & 38.751 & 3.426 & 0.151 & 3.577 \\

2011 & 173 & 649 & 96.959 & 173 & 76.041 & 3.813 & 0.101 & 3.914 \\

2013 & 341 & 1852 & 170.71 & 341 & 170.29 & 4.269 & 0.101 & 4.37 \\

2015 & 1050 & 8050 & 483.016 & 1050 & 566.984 & 4.646 & 0.085 & 4.731 \\

2017 & 2179 & 15992 & 1090.162 & 2179 & 1088.838 & 4.271 & 0.229 & 4.5 \\

2019 & 3157 & 22213 & 1648.768 & 3157 & 1508.232 & 4.092 & 0.053 & 4.145 \\

2020 & 3265 & 22912 & 1711.825 & 3265 & 1553.175 & 4.076 & 0.015 & 4.091 \\
\hline
\end{tabular}
\end{subfigure}
\caption{Neural language model: topic statistics and knowledge temperature evolution}
\label{fig:256500874_chart}
\end{figure}

\begin{figure}[htbp]
\begin{subfigure}{\textwidth}
\begin{minipage}[t]{0.5\linewidth}
\includegraphics[width = \linewidth]{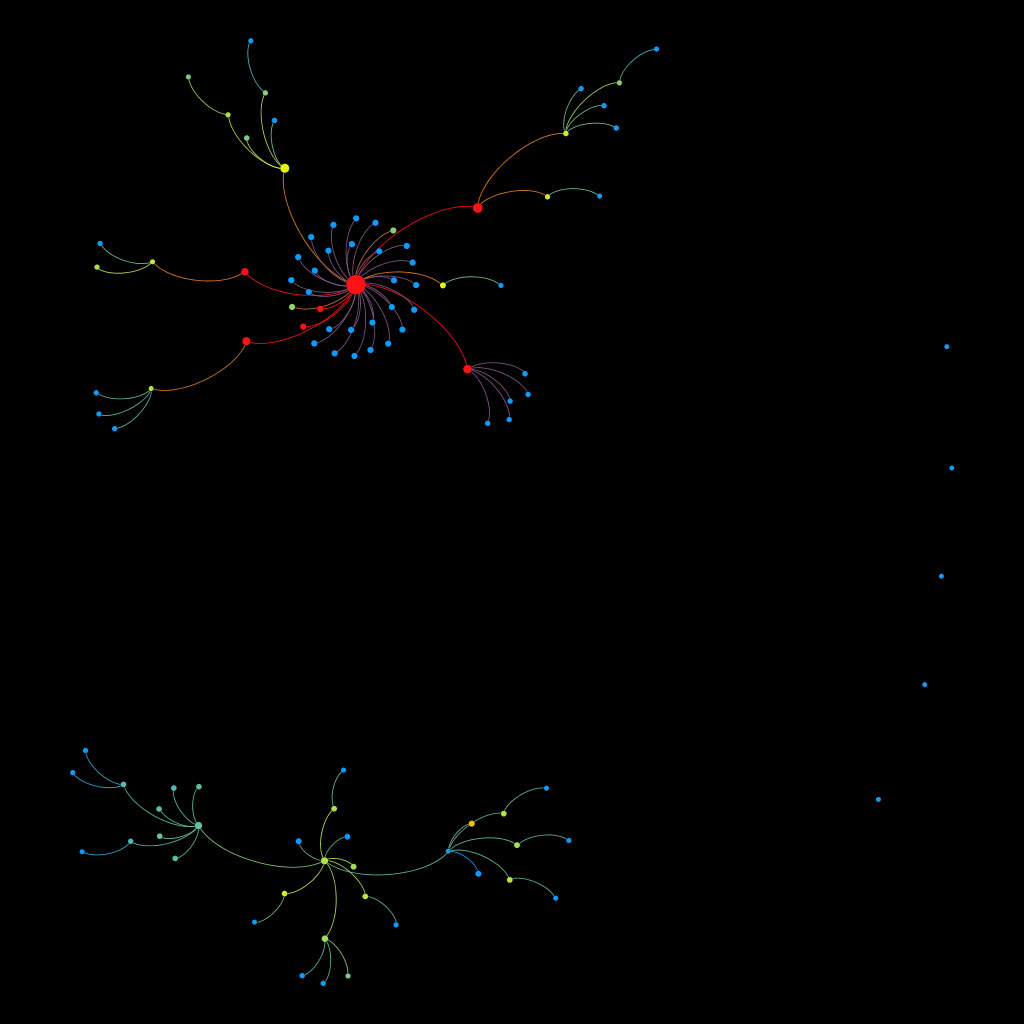}
\caption{Skeleton tree until 2009}
\end{minipage}
\begin{minipage}[t]{0.5\linewidth}
\includegraphics[width = \linewidth]{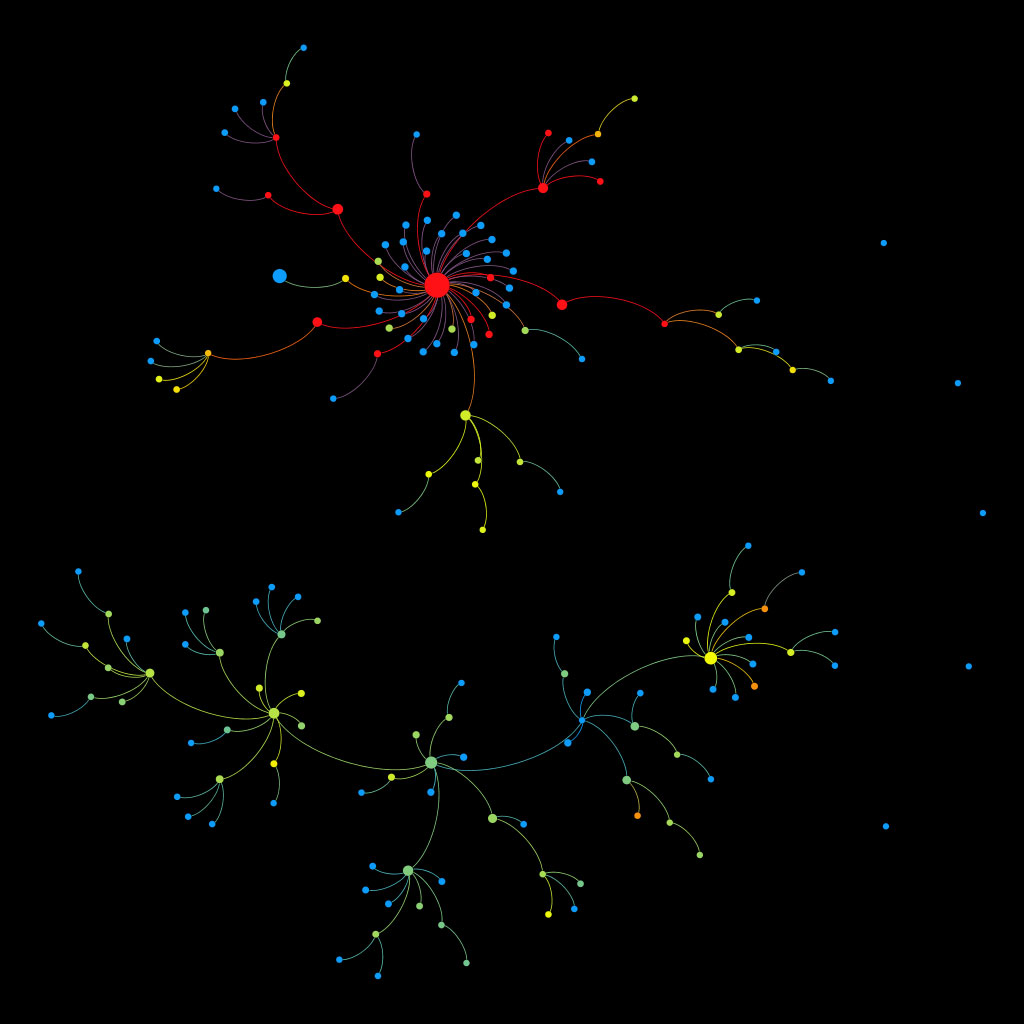}
\caption{Skeleton tree until 2011}
\end{minipage}
\end{subfigure}
\vspace{2mm}
\begin{subfigure}{\textwidth}
\begin{minipage}[t]{0.5\linewidth}
\includegraphics[width = \linewidth]{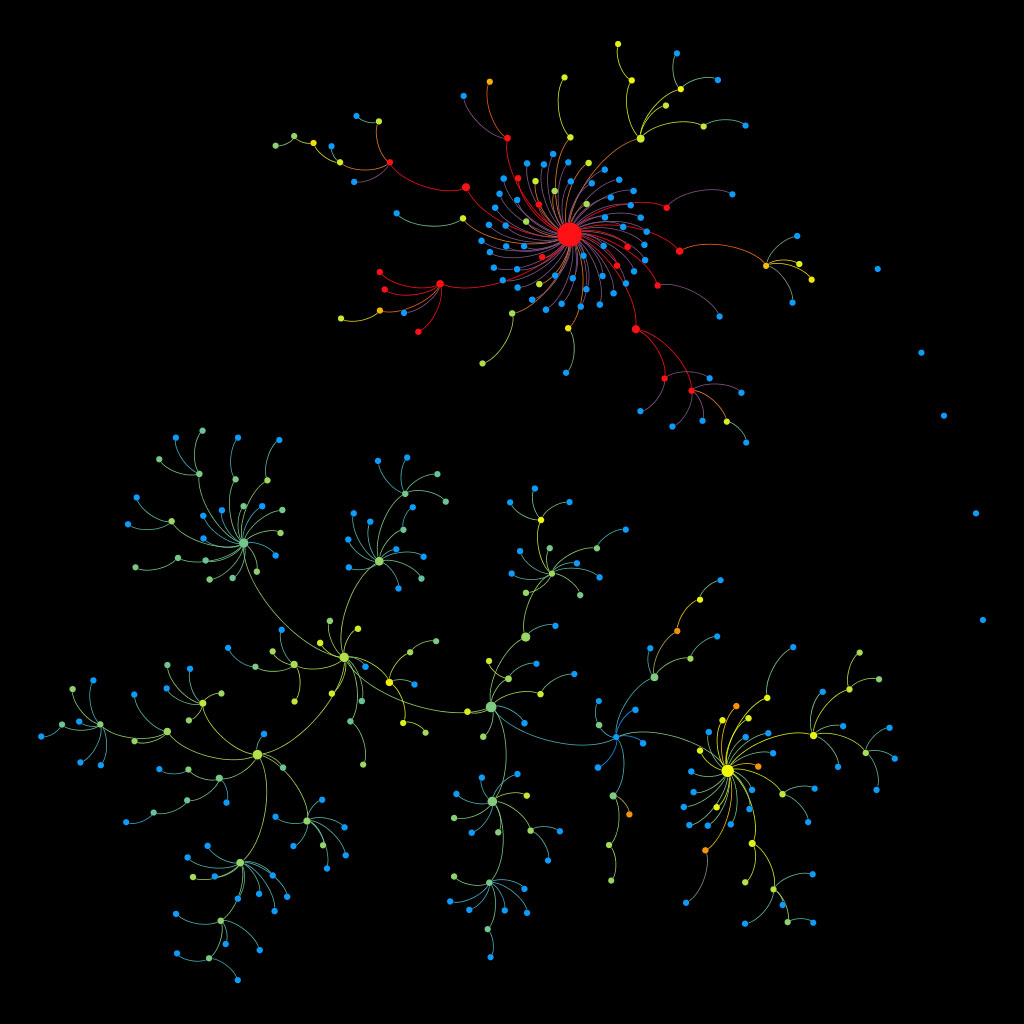}
\caption{Skeleton tree until 2013}
\end{minipage}
\begin{minipage}[t]{0.5\linewidth}
\includegraphics[width = \linewidth]{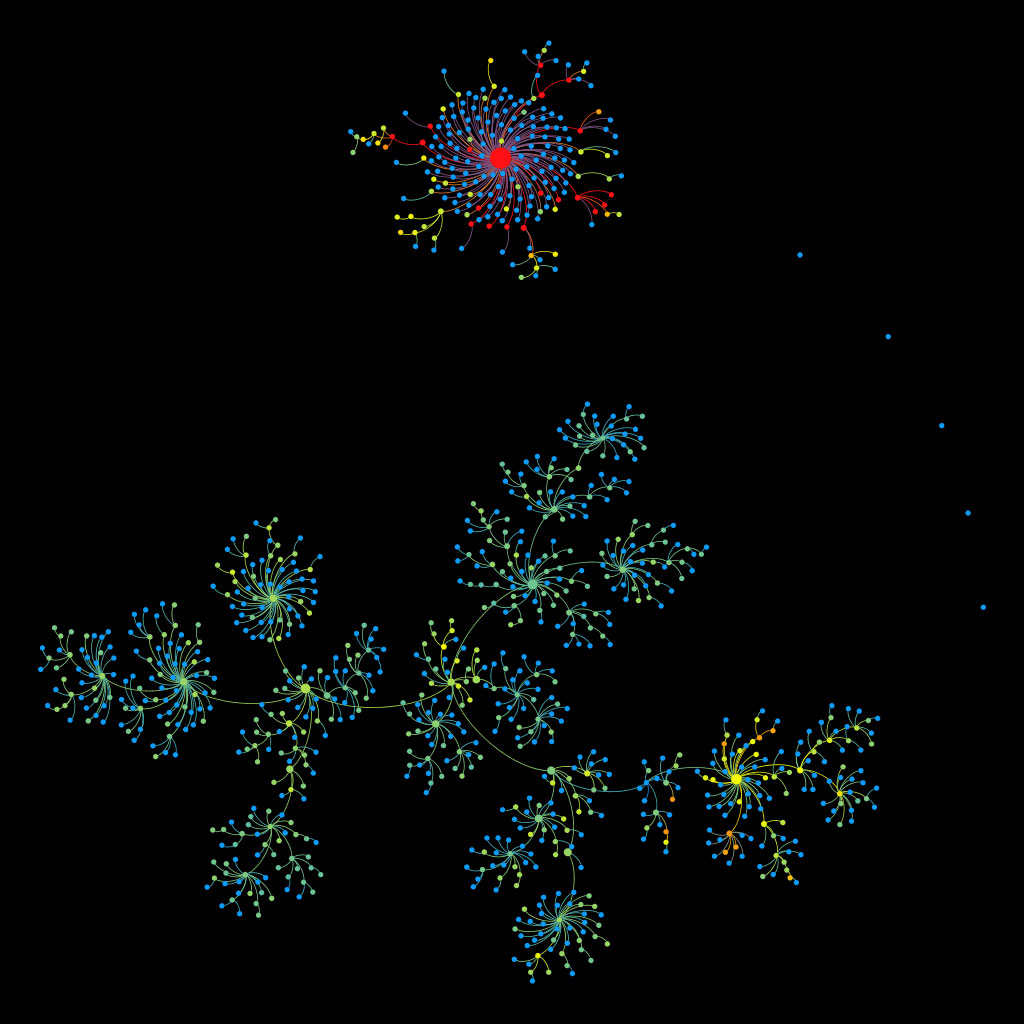}
\caption{Skeleton tree until 2015}
\end{minipage}
\end{subfigure}
\caption{Neural language model: Skeleton tree evolution}
\label{fig:256500874-tree_evo}
\end{figure}

\begin{figure}[htbp]
    \centering
    \begin{subfigure}{\linewidth}
    \begin{minipage}[t]{0.55\textwidth}
    \centering
    \includegraphics[width =0.9 \linewidth]{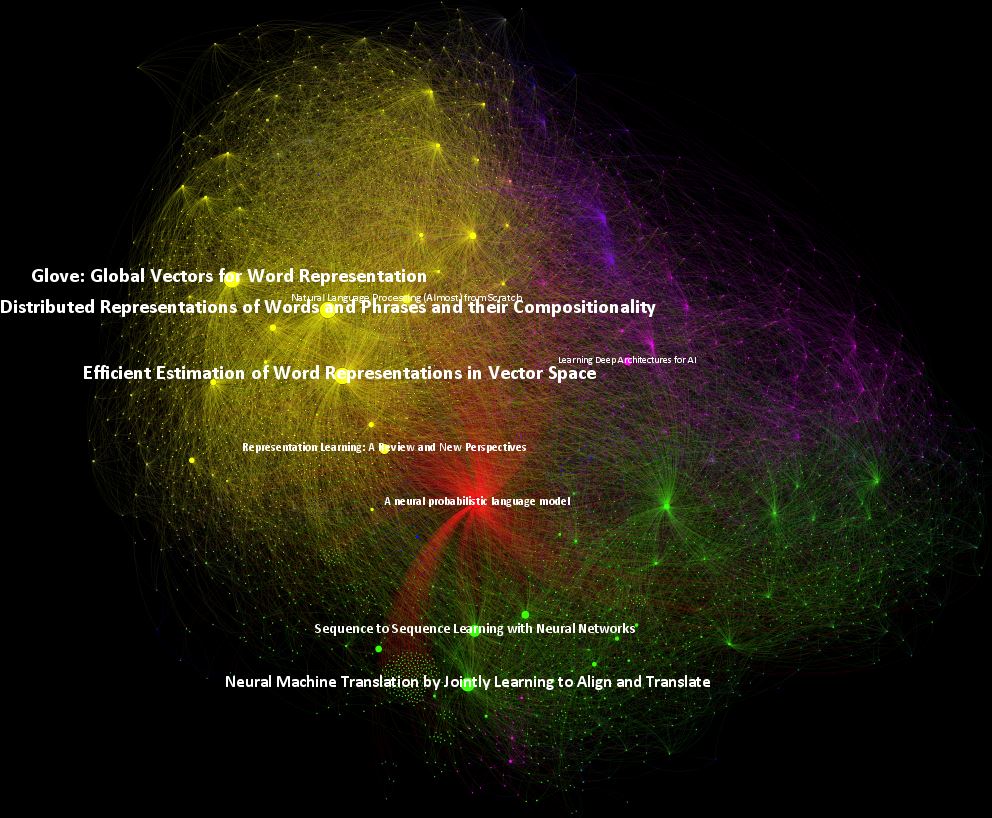}
    \end{minipage}
    \begin{minipage}[t]{0.45\textwidth}
    \centering
    \includegraphics[width =0.9 \linewidth]{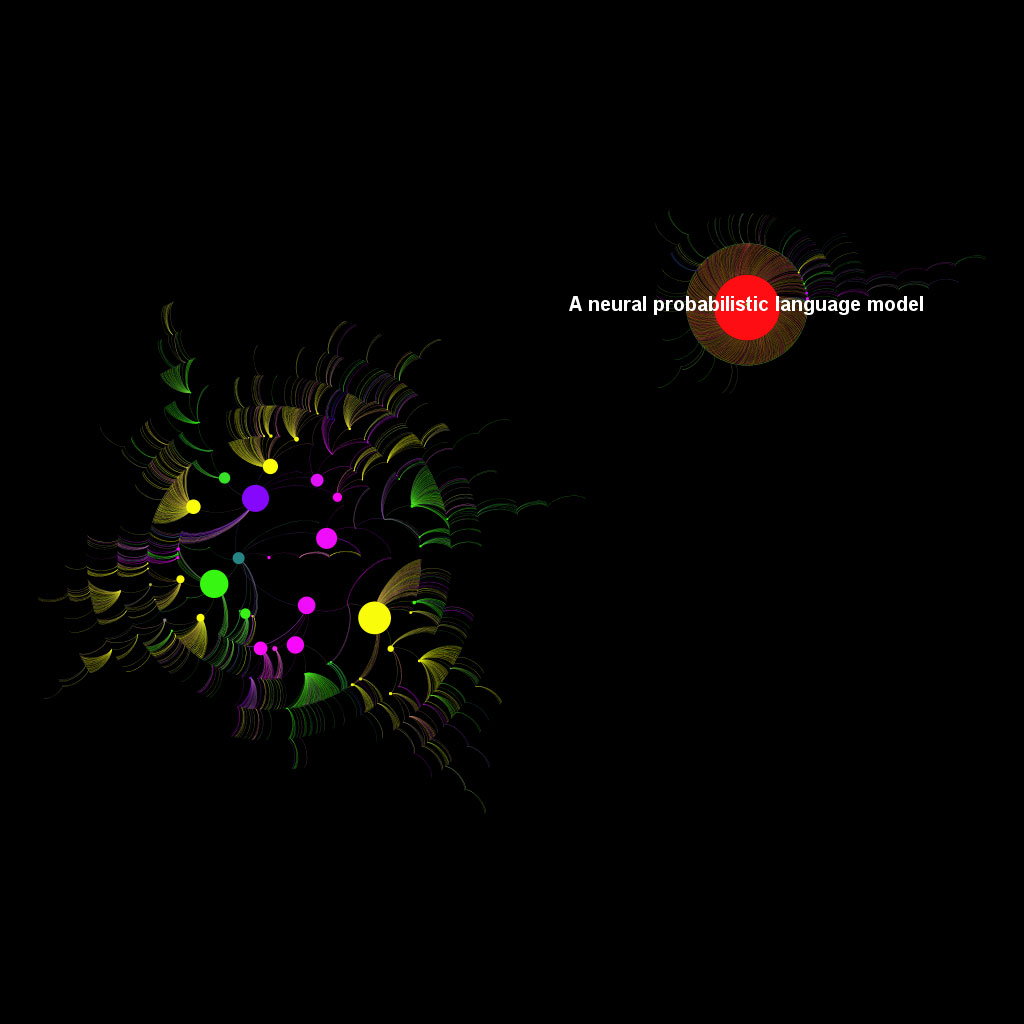}
    \end{minipage}
    \end{subfigure}

    \vspace{5mm}

    \begin{subfigure}{0.6\linewidth}
    \includegraphics[width = \linewidth]{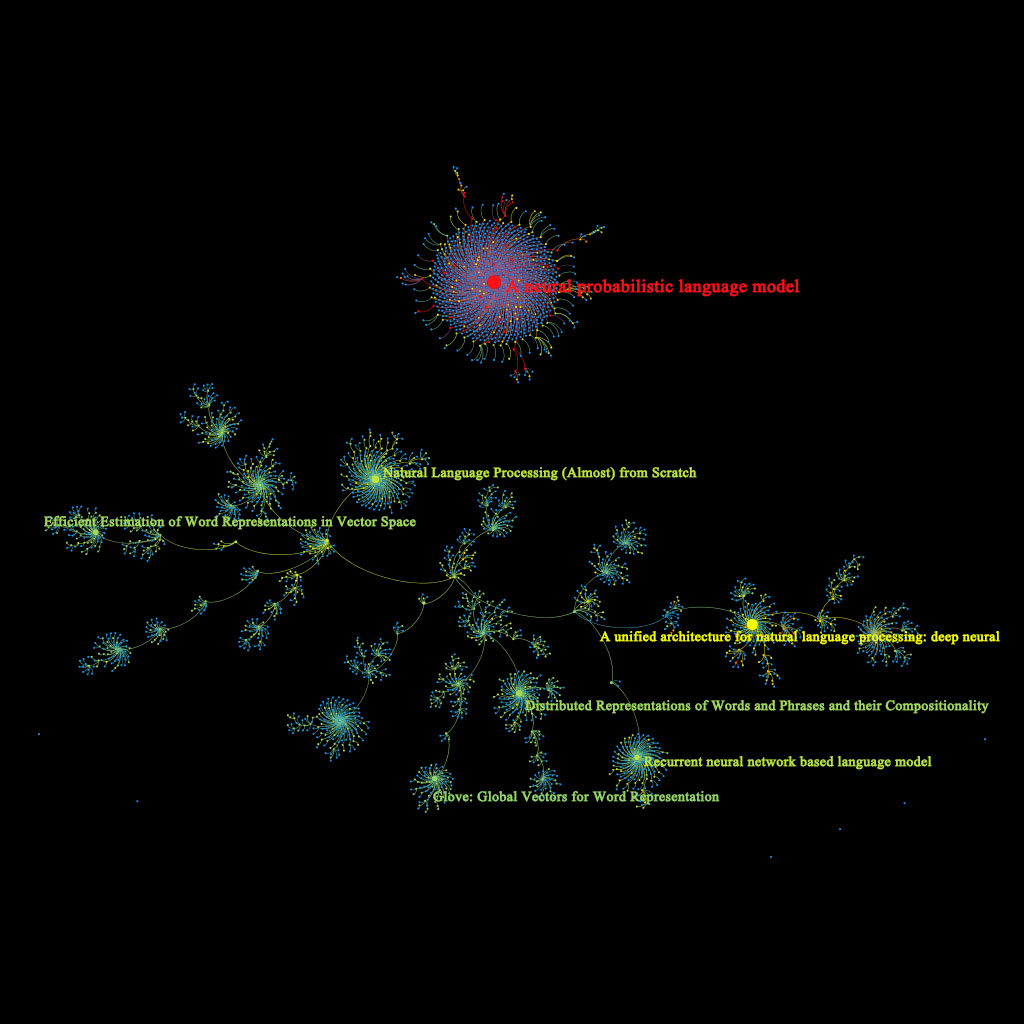}
    \end{subfigure}
    \begin{subfigure}{0.35\linewidth}
    \centering
    \includegraphics[width =0.9 \linewidth]{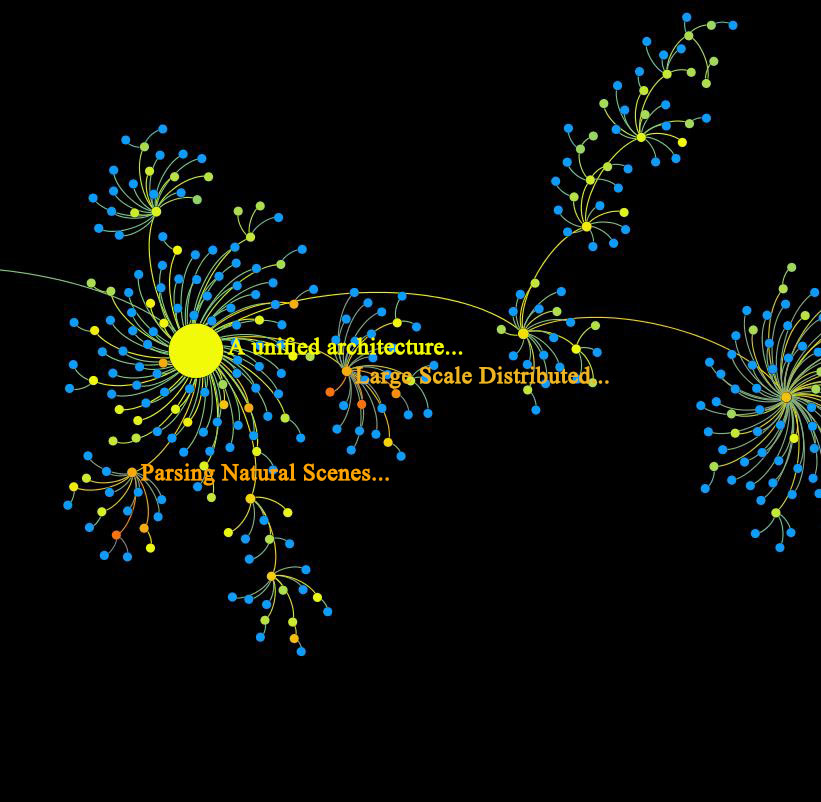}
    \end{subfigure}
    \caption{Neural language model: Galaxy map, current skeleton tree and its regional zoom. Papers with more than 600 in-topic citations are labelled by title in the skeleton tree. Except the pioneering work, corresponding nodes' size is amplified by 3 times.}
    \label{fig:256500874-2020}
\end{figure}

Unlike many topics that welcome the majority of their popular child papers shortly after their birth, this topic waited for a long time. Most of its prominent child papers came during 2010 and 2014. Their arrival opened up new research sub-fields (Fig. \ref{fig:256500874-2020}) and infused much vigor and new knowledge to the topic, which strongly boosted $T_{growth}^t$ during 2011 and 2015 (Fig. \ref{fig:256500874_chart}). Although the topic continued to grow fast after 2015, few child papers stood out and none has created new research focus so far. As a result, the knowledge accumulation process is affected by the overall quality slump and the topic started to cool down owing to the lack of new outstanding ideas. In terms of knowledge structure evolution, the topic manifests a smooth and steady progress (Fig. \ref{fig:256500874-tree_evo}).  Since the arrival of popular child papers is quite evenly spanned over 2010 and 2014, their contribution to the thriving is more reflected as knowledge and impact accumulation than a short-term popularity gain. To conclude, after a recent boom thanks to its popular child papers, the topic is now going downhill.\\

\noindent The skeleton tree is a bit special because it is made up of 2 parts. This is due to the separation of paper `Connectionist language modeling for large vocabulary continuous speech recognition' (CLM) from the pioneering work, the only citation CLM has within the topic. In fact, CLM was published a bit earlier than the pioneering work, therefore its relation with the pioneering work may not be tight. This results in the edge cutting during skeleton tree extraction. CLM later inspired `Efficient training of large neural networks for language modeling', whose work turned out to have a greater influence on the aforementioned popular child papers than that of the pioneering work. That is why skeleton tree finally takes a separated form. \\

\noindent Now we closely examine the current heat distribution with its latest skeleton tree (Fig. \ref{fig:256500874-2020}). The pioneering work remains the only heat source in the topic and almost all of the most popular child papers have a knowledge temperature below average. Although they have indeed vitalized the topic, more importantly they themselves have proposed novel ideas that made them overshadow the pioneering work and become the new authorities in the domain (Fig. \ref{fig:256500874-2020} galaxy map). The relatively loose connection to the core topic idea has resulted in their low knowledge temperature. Their "coolness" is also the reason that the cluster they are in is much colder than the one led by the pioneering work. Overall, we observe the general rule "the older the hotter" (Fig. 5(h)). The blue nodes that surround the pioneering work and popular child papers are papers with few or without any in-topic citations. Node knowledge temperature decrease is clear as we walk down the paths in skeleton tree. However, there are exceptions. Hit paper `A unified architecture for natural language processing: deep neural networks with multitask learning' (UANLP) published in 2008 is colder than, for instance, its well-developed child `Large Scale Distributed Deep Networks' published in 2012 and `Parsing Natural Scenes and Natural Language with Recursive Neural Networks' published in 2011. These 2 child papers are represented as orange nodes yet the UANLP is a yellow node. Their temperature difference lies mainly in their research focus reflected by their citation patterns. Although these 2 child papers both have a few followers in the latest skeleton tree, they are still less popular than their parent in terms of idea diffusion. This counter example also illustrates that the general rules "the more influential the hotter" is very weak in the topic (Fig. \ref{fig:citation_T}(h)). \\

\noindent We observe in addition certain clustering effect in the skeleton tree (Table \ref{tab:256500874-clustering}). For example, all child papers of `Road2Vec: Measuring Traffic Interactions in Urban Road System from Massive Travel Routes' have a research interest related to geographic relation. This confirms the effectiveness of our skeleton tree extraction algorithm. In addition, this small bundle is very younger, hence their research interest may be among the latest trends.\\

\begin{table}
    \centering
    \begin{tabular}{p{15cm} p{1cm}}
        \hline
        title & year\\
        \hline
        Road2Vec: Measuring Traffic Interactions in \textcolor{red}{Urban Road} System from Massive Travel Routes & 2017 \\

         Knowledge Embedding with \textcolor{orange}{Geospatial Distance} Restriction for Geographic Knowledge Graph Completion & 2019\\

         A \textcolor{orange}{regionalization} method for clustering and partitioning based on \textcolor{orange}{trajectories} from NLP perspective & 2019\\

         From Motion Activity to \textcolor{orange}{Geo-Embeddings}: Generating and Exploring Vector Representations of Locations, Traces and Visitors through Large-Scale Mobility Data & 2019\\

         Detecting \textcolor{orange}{geo-relation} phrases from web texts for triplet extraction of geographic knowledge: a context-enhanced method & 2019 \\
        \hline
    \end{tabular}
    \caption{Neural language model: Clustering effect example. First line is the parent paper and the rest children.}
    \label{tab:256500874-clustering}
\end{table}

\subsubsection{A unified architecture for natural language processing: deep neural networks with multitask learning}

\begin{figure}[htbp]
\centering
\begin{subfigure}[t]{0.7\linewidth}
\centering
\includegraphics[width=\linewidth]{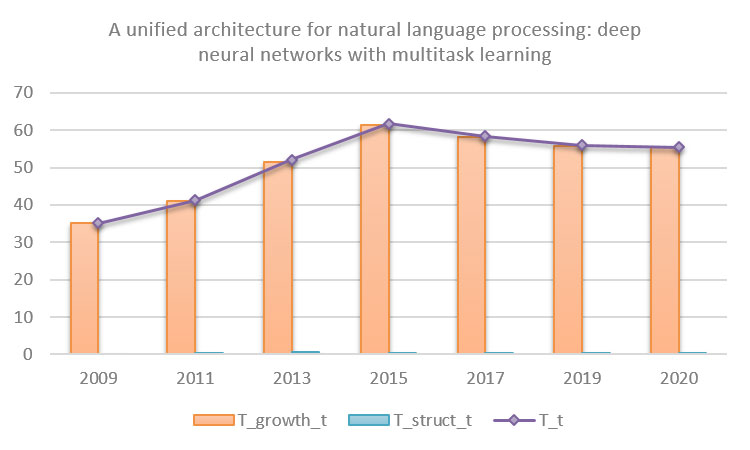}
\end{subfigure}

\begin{subfigure}[t]{\linewidth}
\centering
\begin{tabular}{ccccccccc}
\hline
year & $|V^t|$ & $|E^t|$ & $n_t$ & $V_t$ & ${UsefulInfo}^t$ & $T_{growth}^t$ & $T_{struct}^t$ & $T^t$\\
\hline
2009 & 24 & 36 & 20.883 & 24 & 3.117 & 35.066 &   & 35.066 \\

2011 & 113 & 236 & 83.927 & 113 & 29.073 & 41.081 & 0.117 & 41.199 \\

2013 & 291 & 1021 & 172.368 & 291 & 118.632 & 51.511 & 0.532 & 52.043 \\

2015 & 889 & 4818 & 441.78 & 889 & 447.22 & 61.399 & 0.39 & 61.79 \\

2017 & 1766 & 9451 & 926.156 & 1766 & 839.844 & 58.18 & 0.178 & 58.358 \\

2019 & 2640 & 13483 & 1441.288 & 2640 & 1198.712 & 55.888 & 0.087 & 55.976 \\

2020 & 2733 & 13855 & 1503.842 & 2733 & 1229.158 & 55.45 & 0.01 & 55.46 \\
\hline
\end{tabular}
\end{subfigure}
\caption{A unified architecture for NLP: topic statistics and knowledge temperature evolution}
\label{fig:223688399_chart}
\end{figure}

\begin{figure}[htbp]
\begin{subfigure}{\textwidth}
\begin{minipage}[t]{0.33\linewidth}
\includegraphics[width = \linewidth]{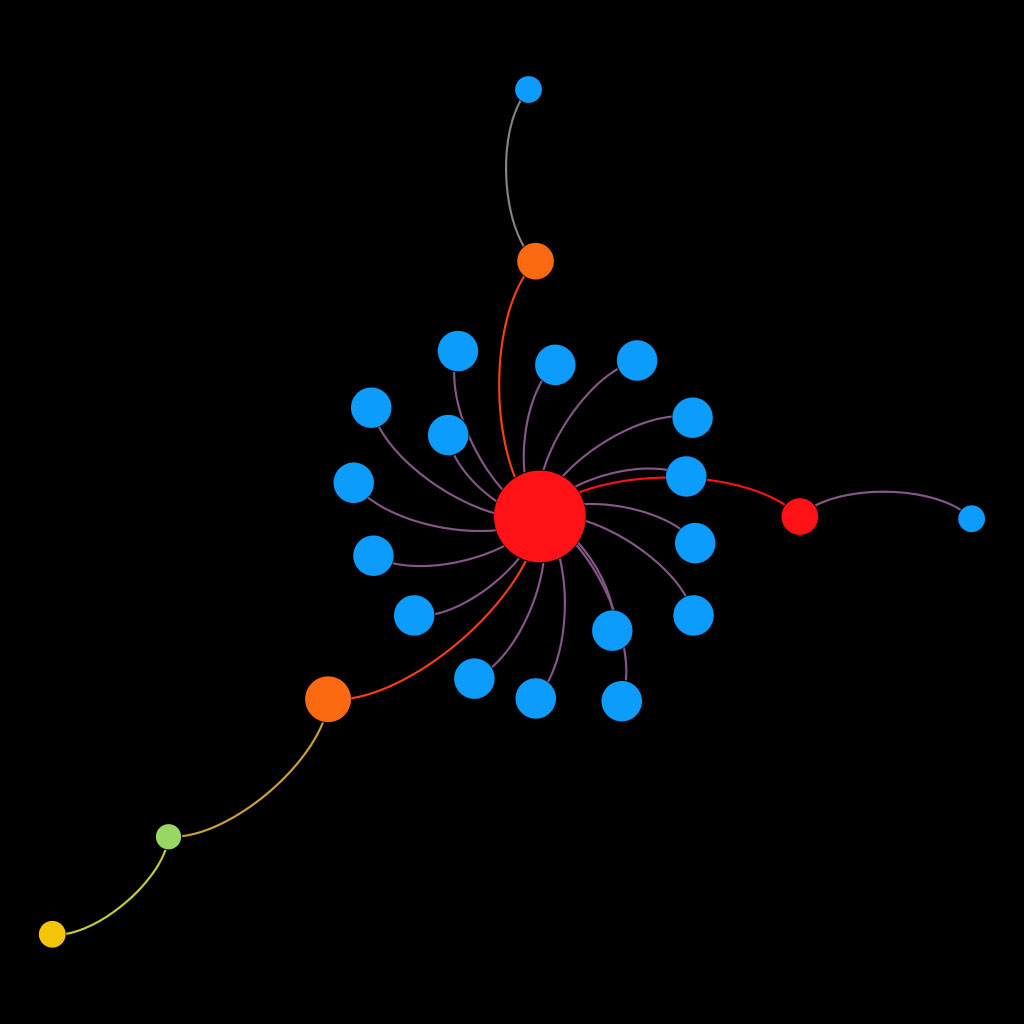}
\caption{Skeleton tree until 2009}
\end{minipage}
\begin{minipage}[t]{0.33\linewidth}
\includegraphics[width = \linewidth]{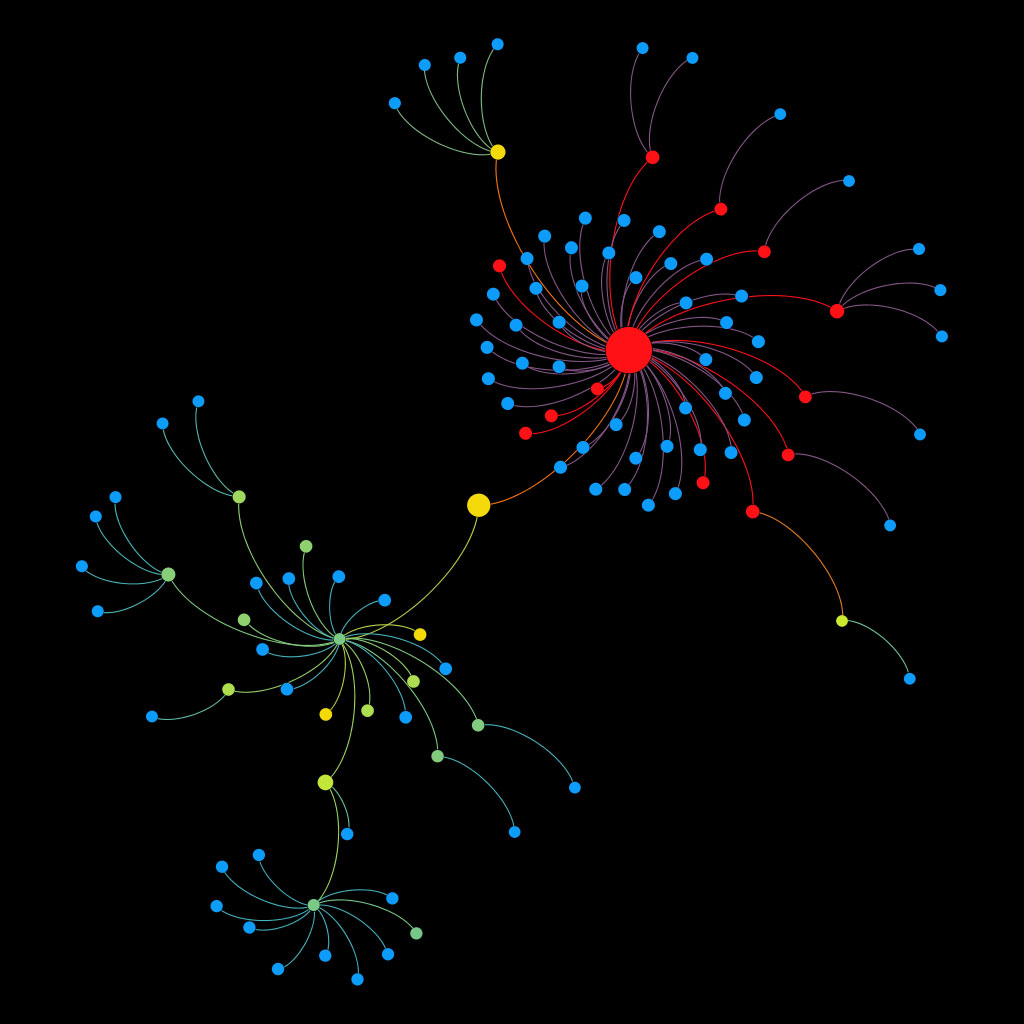}
\caption{Skeleton tree until 2011}
\end{minipage}
\begin{minipage}[t]{0.33\linewidth}
\includegraphics[width = \linewidth]{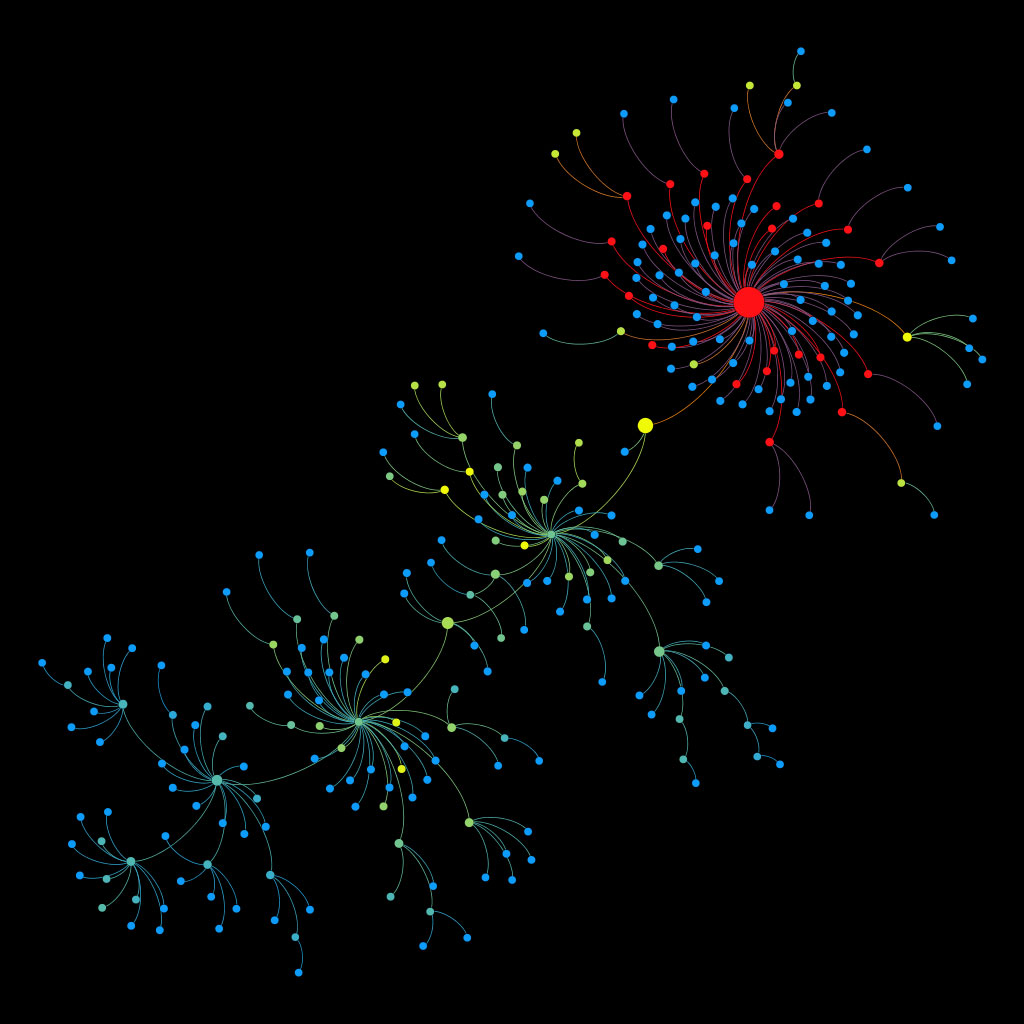}
\caption{Skeleton tree until 2013}
\end{minipage}
\end{subfigure}
\vspace{2mm}
\begin{subfigure}{\textwidth}
\begin{minipage}[t]{0.33\linewidth}
\includegraphics[width = \linewidth]{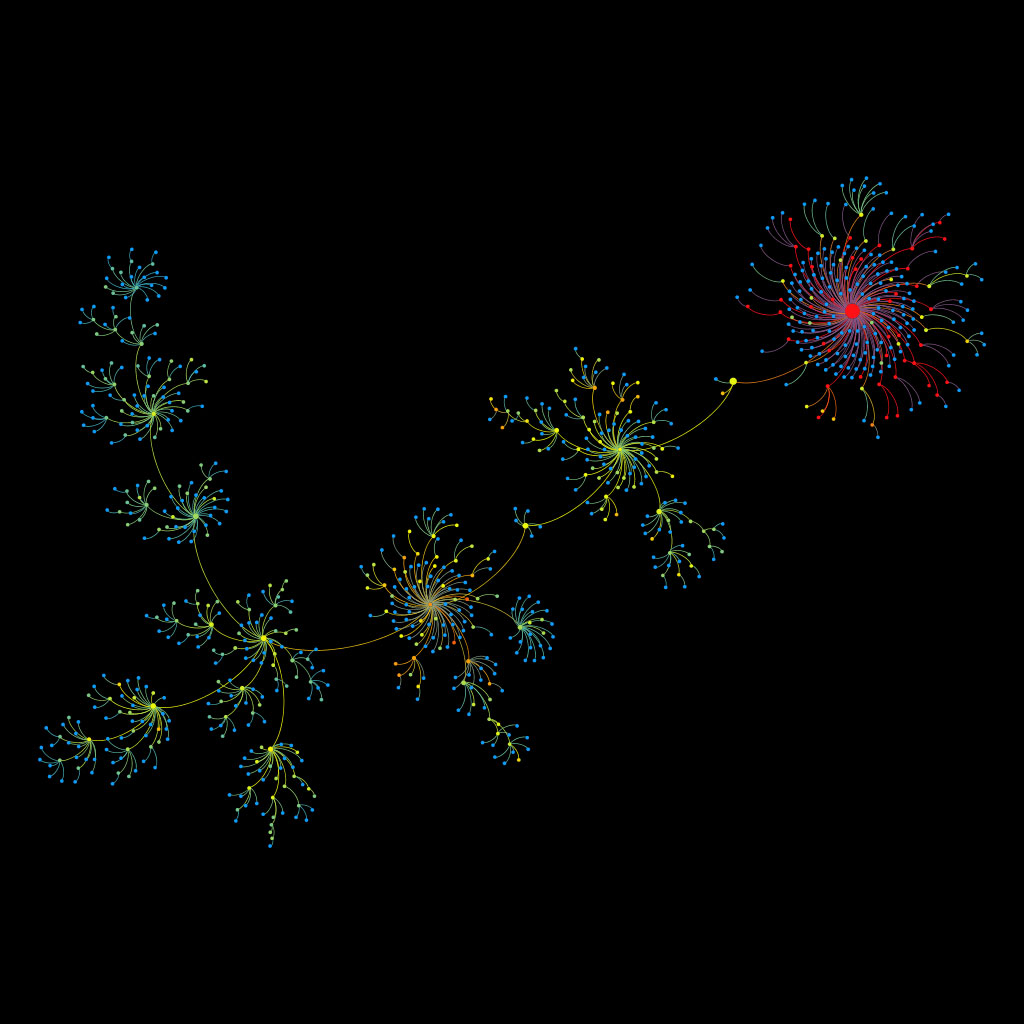}
\caption{Skeleton tree until 2015}
\end{minipage}
\begin{minipage}[t]{0.33\linewidth}
\includegraphics[width = \linewidth]{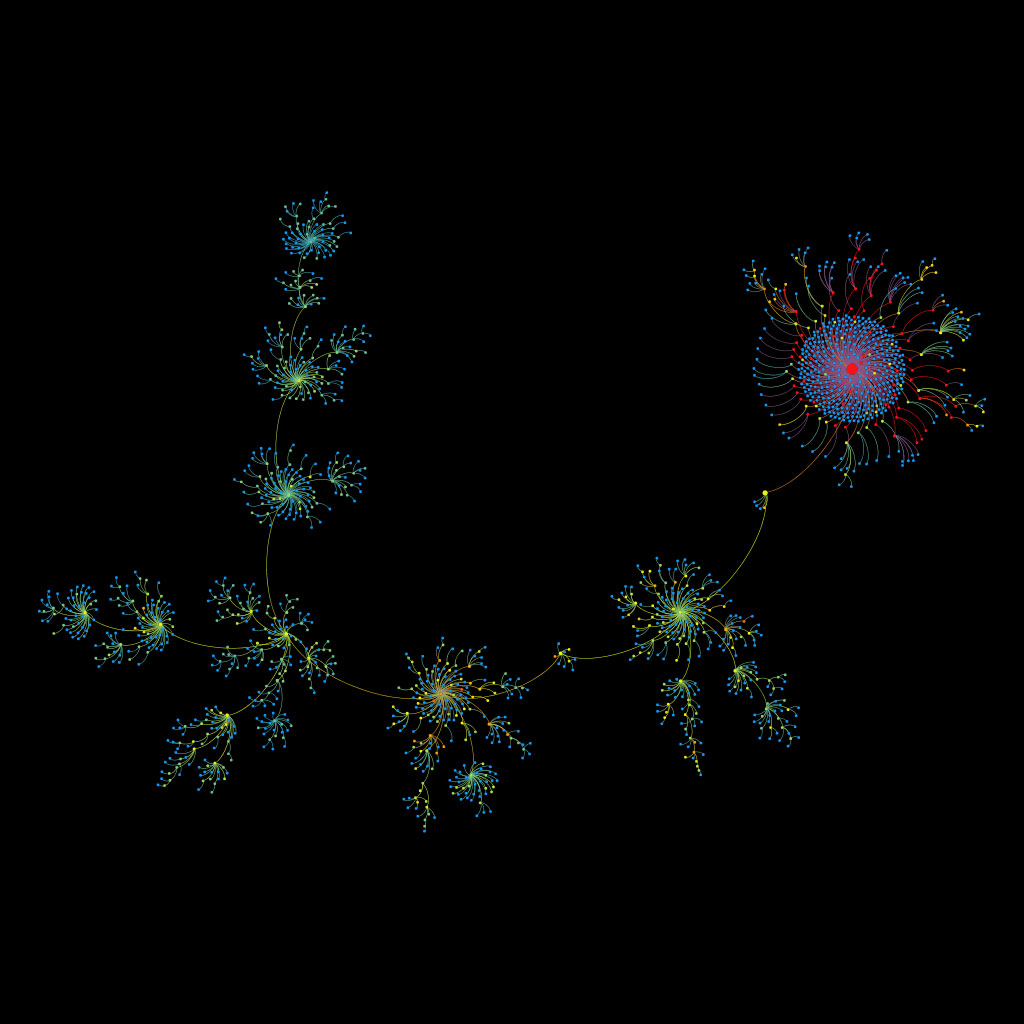}
\caption{Skeleton tree until 2017}
\end{minipage}
\begin{minipage}[t]{0.33\linewidth}
\includegraphics[width = \linewidth]{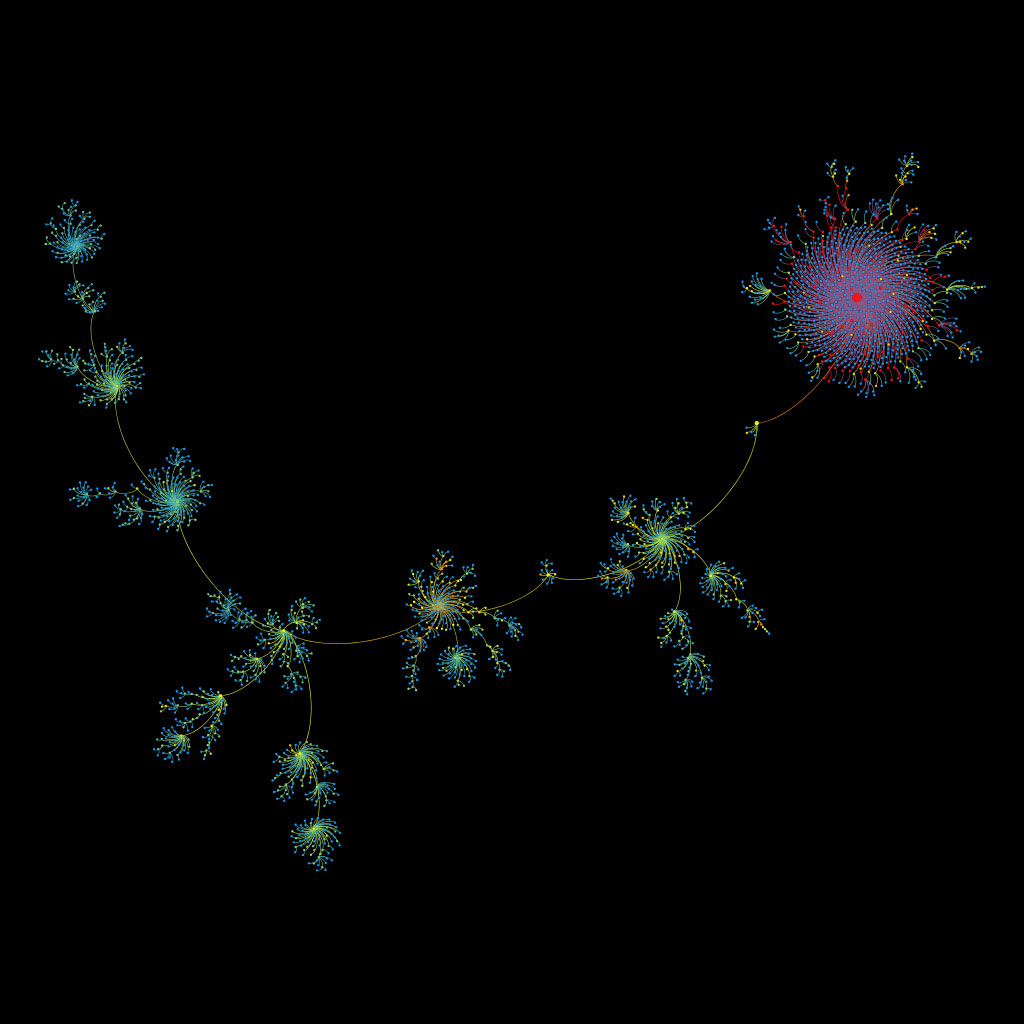}
\caption{Skeleton tree until 2019}
\end{minipage}
\end{subfigure}
\caption{A unified architecture for NLP: Skeleton tree evolution}
\label{fig:223688399-tree_evo}
\end{figure}

As is shown by $T_{growth}^t$ and $T^t$, the topic continuously gained fame between 2009 and 2015 (Fig. \ref{fig:223688399_chart}). Almost all of its most influential child papers were published during this period. After that, despite a steady size growth, the topic has gradually cooled down. This is because the majority of prominent child papers, namely `Efficient Estimation of Word Representations in Vector Space' (EEWRVS), `Distributed Representations of Words and Phrases and their Compositionality' (DRWPC) and `Word Representations: A Simple and General Method for Semi-Supervised Learning' (WRSSL), were published no later than 2013. They brought large amounts of new knowledge and, more importantly, attracted much immediate attention after their publication. By the end of 2015, these child papers, having collected a fair share of in-topic citations, had already become crucial members of the topic. Together with the pioneering work, they shaped topic knowledge (Fig. \ref{fig:223688399-tree_evo}(d)). Child papers published no earlier than 2016 enriched the ideas proposed by the aforementioned popular child papers (Fig. \ref{fig:223688399-tree_evo}(e,f)). Very few have had a significant subsequent development even though the topic has succeeded in attracting a stable stream of recent attention.  Therefore, the enrichment of knowledge base has slowed down and thus the knowledge temperature has slightly dropped. To sum up, the topic demonstrates a rise-then-fall dynamics.\\

\begin{figure}[htbp]
    \centering
    \begin{subfigure}{\linewidth}
    \begin{minipage}[t]{0.5\textwidth}
    \centering
    \includegraphics[width = 0.9 \linewidth]{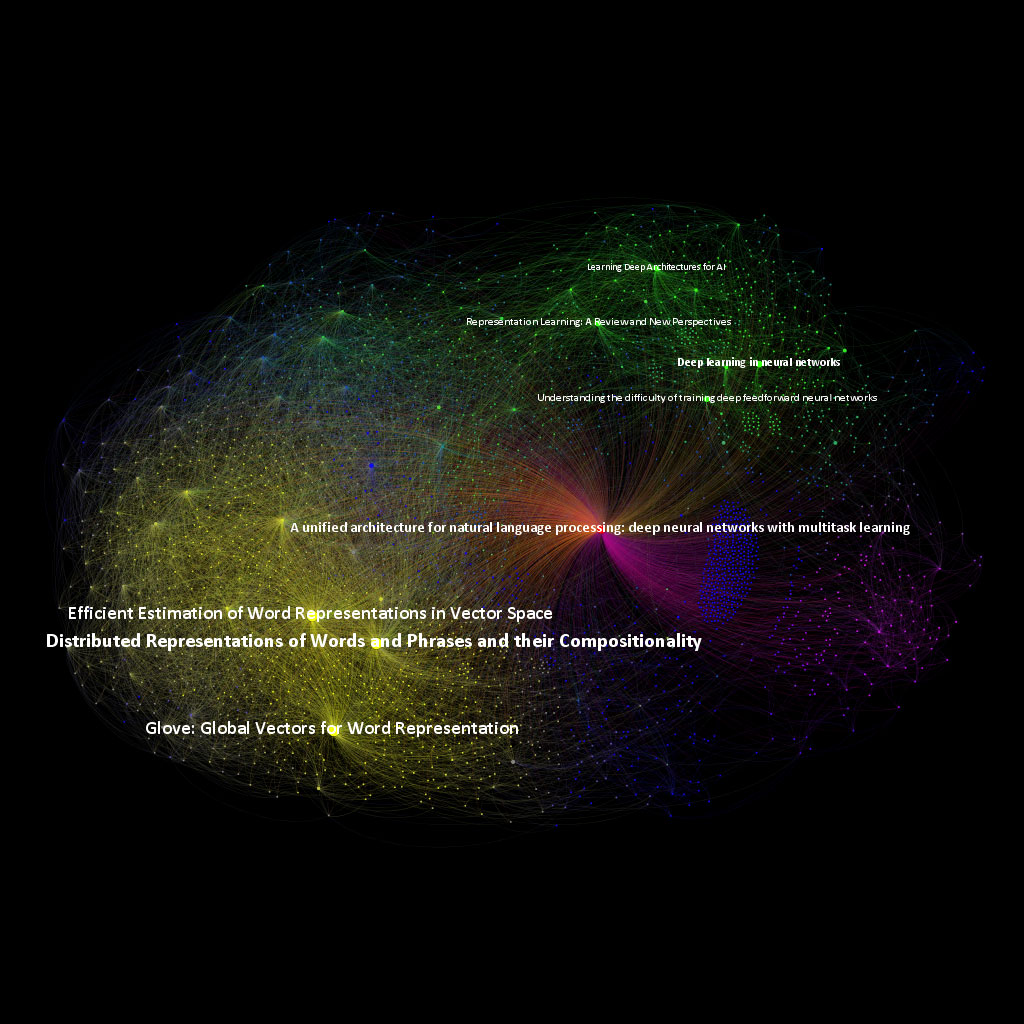}
    \end{minipage}
     \begin{minipage}[t]{0.5\textwidth}
    \centering
    \includegraphics[width = 0.9 \linewidth]{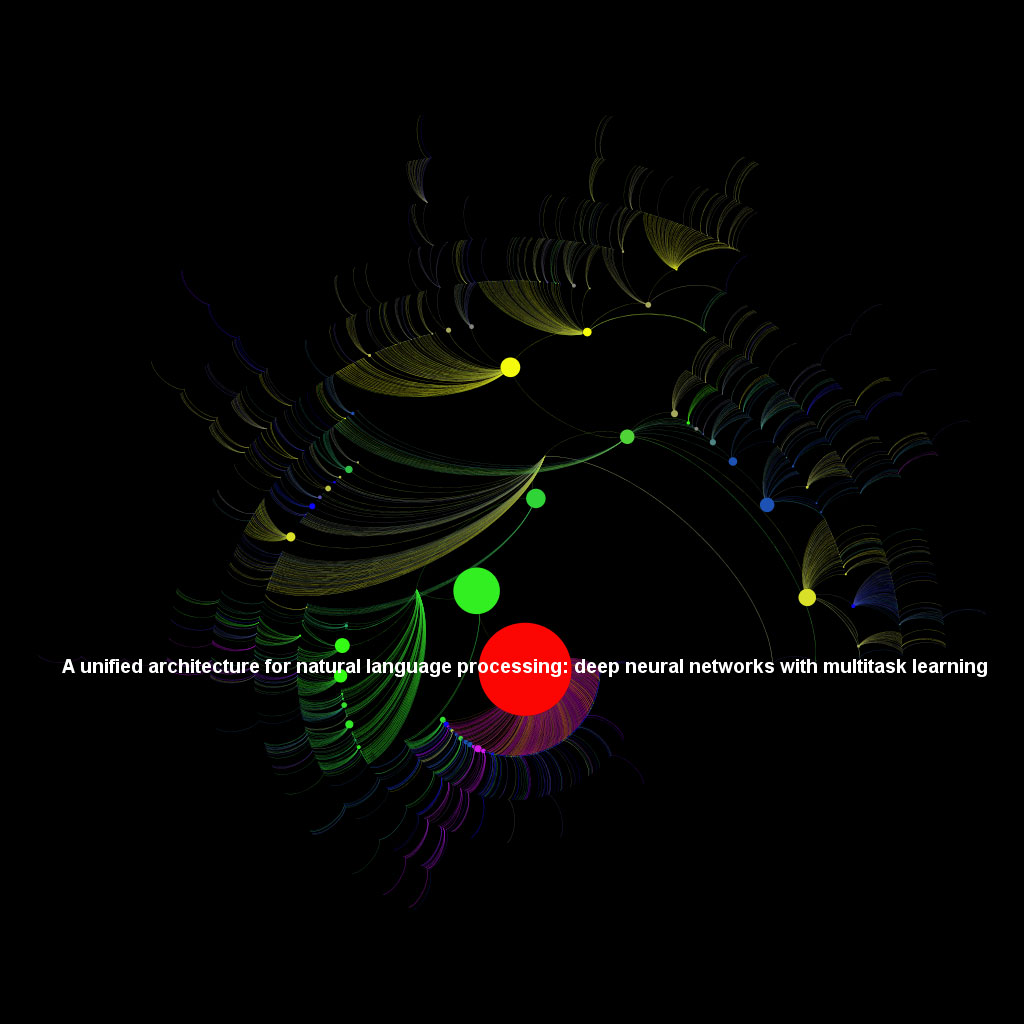}
    \end{minipage}

    \vspace{5mm}

    \end{subfigure}
    \begin{subfigure}{0.6\linewidth}
    \includegraphics[width = \linewidth]{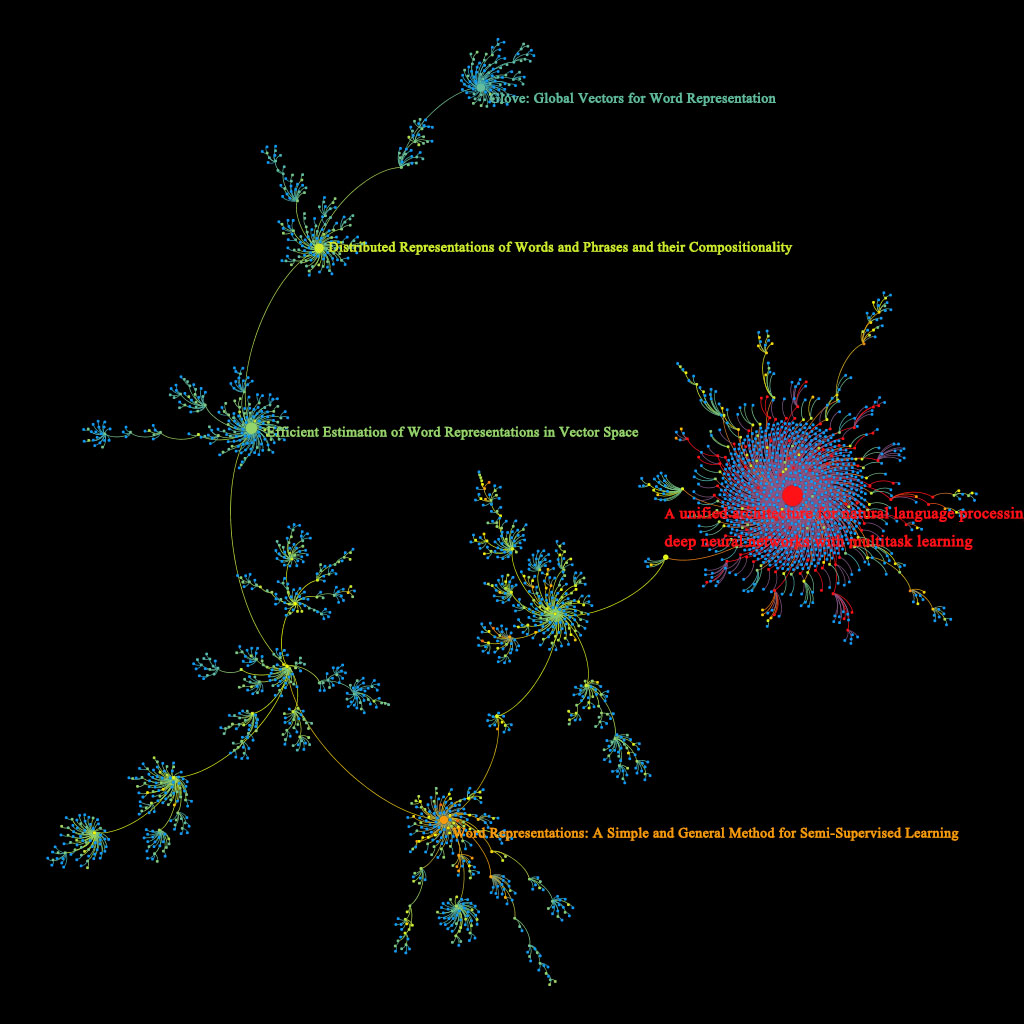}
    \end{subfigure}
    \begin{subfigure}{0.35\linewidth}
    \centering
    \includegraphics[width = 0.9 \linewidth]{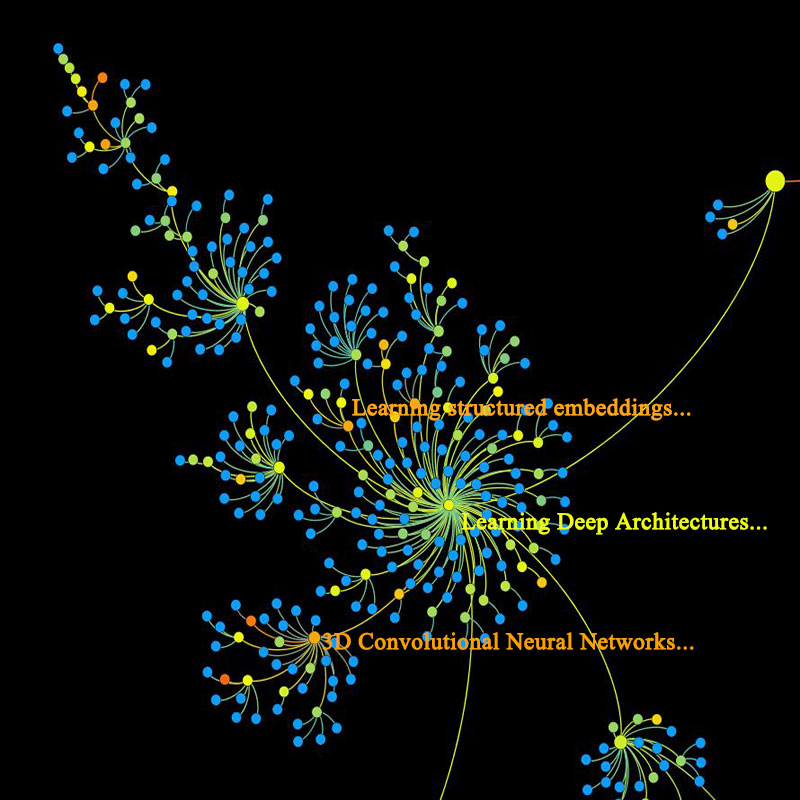}
    \end{subfigure}
    \caption{A unified architecture for NLP: Galaxy map, current skeleton tree and its regional zoom. Papers with more than 400 in-topic citations are labelled by title in the skeleton tree. Except the pioneering work, corresponding nodes' size is amplified by 3 times.}
    \label{fig:223688399-2020}
\end{figure}

\noindent The skeleton tree of this topic manifests a gradual structural advancement in line with a constantly small $T_{structure}^t$ (Fig. \ref{fig:223688399-tree_evo}). Its popular child papers have unanimously dedicated themselves to one single research sub-direction, which is portrayed by the steadily-growing big branch (Fig. \ref{fig:223688399-2020}). \\

\noindent Now we closely examine the internal heat distribution and its latest skeleton tree (Fig. \ref{fig:223688399-2020}). The pioneering work is the only heat source. Interestingly, half of the most popular child papers have a knowledge temperature below average. In fact, they all cited another popular child paper, WRSSL. In terms of idea inheritance, they are less close to the pioneering work than WRSSL. A bigger portion of original idea has caused their relatively low knowledge temperature. We see a clear node knowledge temperature decline from the root to leaves. This corresponds with the general rule "the older the hotter" (Fig. 5(i)). As the topic contains 2 articles published earlier than the pioneering work and they have few in-topic citations, the average node knowledge temperature for the oldest papers is not maximal. In addition, the blue nodes that surround the pioneering work and the most popular child papers are papers with few or without any in-topic citations. However, even if we set aside the oldest papers and the aforementioned coldest papers, the general rule is violated. Hit paper `Learning Deep Architectures for AI' (LDAAI) published in 2009 is colder than, for instance, its child papers `3D Convolutional Neural Networks for Human Action Recognition' published in 2013 and `Learning structured embeddings of knowledge bases' published in 2011. These 2 child papers are represented as orange nodes yet LDAAI is coloured yellow. This is mainly due to their relatively different research focus as their in-topic citations do not overlap with one another. Similarly, popular child paper EEWRVS is slightly colder than its descendant, DRWPC. These counter examples also illustrate that the general rule "the more influential the hotter" is very weak in this topic (Fig. \ref{fig:citation_T}(i)).\\

\noindent We observe in addition certain clustering effect in the skeleton tree (Table \ref{tab:223688399-clustering}). For example, all child papers of `Throughput-Optimized OpenCL-based FPGA Accelerator for Large-Scale Convolutional Neural Networks' have a research interest towards accelerator. This confirms the effectiveness of our skeleton tree extraction algorithm.\\

\begin{table}
    \centering
    \begin{tabular}{p{15cm} p{1cm}}
        \hline
        title & year\\
        \hline
        Throughput-Optimized OpenCL-based \textcolor{red}{FPGA Accelerator} for Large-Scale Convolutional Neural Networks & 2016 \\

        Automatic code generation of convolutional neural networks in \textcolor{orange}{FPGA} implementation & 2016\\

        Throughput-Optimized \textcolor{orange}{FPGA Accelerator} for Deep Convolutional Neural Networks & 2017\\

        Escher: A CNN \textcolor{orange}{Accelerator} with Flexible Buffering to Minimize Off-Chip Transfer & 2017\\

         Towards Efficient Hardware \textcolor{orange}{Acceleration} of Deep Neural Networks on FPGA & 2018\\

         UniCNN: A Pipelined \textcolor{orange}{Accelerator} Towards Uniformed Computing for CNNs & 2018\\
        \hline
    \end{tabular}
    \caption{A unified architecture for NLP: Clustering effect example. First line is the parent paper and the rest children.}
    \label{tab:223688399-clustering}
\end{table}

\subsubsection{Bose-Einstein condensation in a gas of sodium atoms}

\begin{figure}[htbp]
\centering
\begin{subfigure}[t]{0.7\linewidth}
\centering
\includegraphics[width=\linewidth]{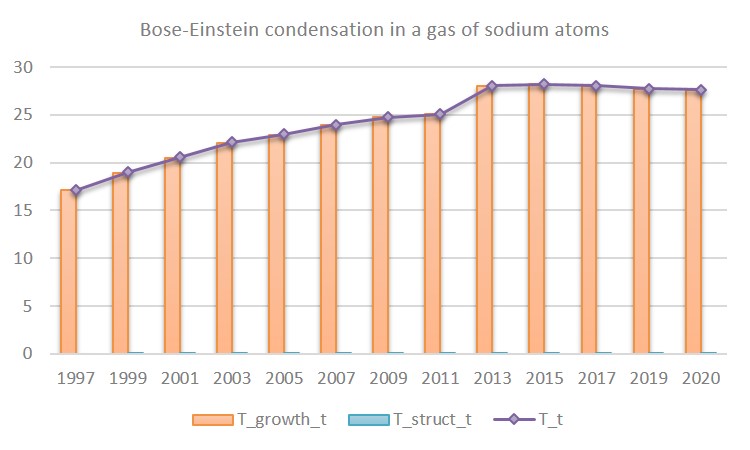}
\end{subfigure}

\begin{subfigure}[t]{0.7\linewidth}
\centering
\begin{tabular}{ccccccccc}
\hline
year & $|V^t|$ & $|E^t|$ & $n_t$ & $V_t$ & ${UsefulInfo}^t$ & $T_{growth}^t$ & $T_{struct}^t$ & $T^t$\\
\hline
1997 & 89 & 139 & 71.133 & 89 & 17.867 & 17.114 &   & 17.114 \\

1999 & 228 & 445 & 164.84 & 228 & 63.16 & 18.92 & 0.056 & 18.976 \\

2001 & 399 & 933 & 265.993 & 399 & 133.007 & 20.518 & 0.048 & 20.566 \\

2003 & 596 & 1600 & 369.024 & 596 & 226.976 & 22.092 & 0.036 & 22.127 \\

2005 & 784 & 2351 & 467.427 & 784 & 316.573 & 22.942 & 0.024 & 22.966 \\

2007 & 1014 & 3395 & 579.603 & 1014 & 434.397 & 23.93 & 0.024 & 23.954 \\

2009 & 1213 & 4326 & 670.455 & 1213 & 542.546 & 24.747 & 0.017 & 24.764 \\

2011 & 1454 & 5310 & 794.184 & 1454 & 659.816 & 25.043 & 0.016 & 25.059 \\

2013 & 1708 & 6450 & 833.549 & 1708 & 874.451 & 28.028 & 0.013 & 28.041 \\

2015 & 1905 & 7430 & 924.707 & 1905 & 980.293 & 28.179 & 0.012 & 28.191 \\

2017 & 2066 & 8141 & 1009.101 & 2066 & 1056.899 & 28.005 & 0.021 & 28.026 \\

2019 & 2296 & 9013 & 1132.976 & 2296 & 1163.024 & 27.72 & 0.021 & 27.741 \\

2020 & 2338 & 9171 & 1157.694 & 2338 & 1180.306 & 27.624 & 0.005 & 27.629 \\
\hline
\end{tabular}
\end{subfigure}
\caption{Bose-Einstein condensation: topic statistics and knowledge temperature evolution}
\label{fig:15804200_chart}
\end{figure}

Founded in 1995, this topic thrived for some 20 years before starting to stagnate since 2013 (Fig. \ref{fig:15804200_chart}). While most of the highest-cited child papers within the topic came between 1997 and 2003, several came after 2006, namely `Bose-Einstein condensation of exciton polaritons' (BECEP) published in 2006 in Nature, `Production of Cold Molecules via Magnetically Tunable Feshbach Resonances' published in 2006 in Reviews of Modern Physics, and 'Bose-Einstein condensation of photons in an optical microcavity' (BECPOM) published in 2010 in Nature. The relay among these popular child papers maintained the topic's flourishing for 20 years. In addition, the topic was most prolific between 2010 and 2012, with annual publication number all exceeding 5\% of current topic size. The increasing inflow of knowledge, together with the exposure brought by the aforementioned popular child papers, contributed to a slightly bigger climb in $T^t$ and $T_{growth}^t$ between 2011 and 2013. After that, the topic has not so far welcomed any superstars that have incited remarkable development. Yet it still has a rather stable knowledge accumulation judging from basic statistics. Hence overall $T_{growth}^t$ ceased to go up and so is $T^t$.\\

\begin{figure}[htbp]
\begin{subfigure}{\textwidth}
\begin{minipage}[t]{0.33\linewidth}
\includegraphics[width = \linewidth]{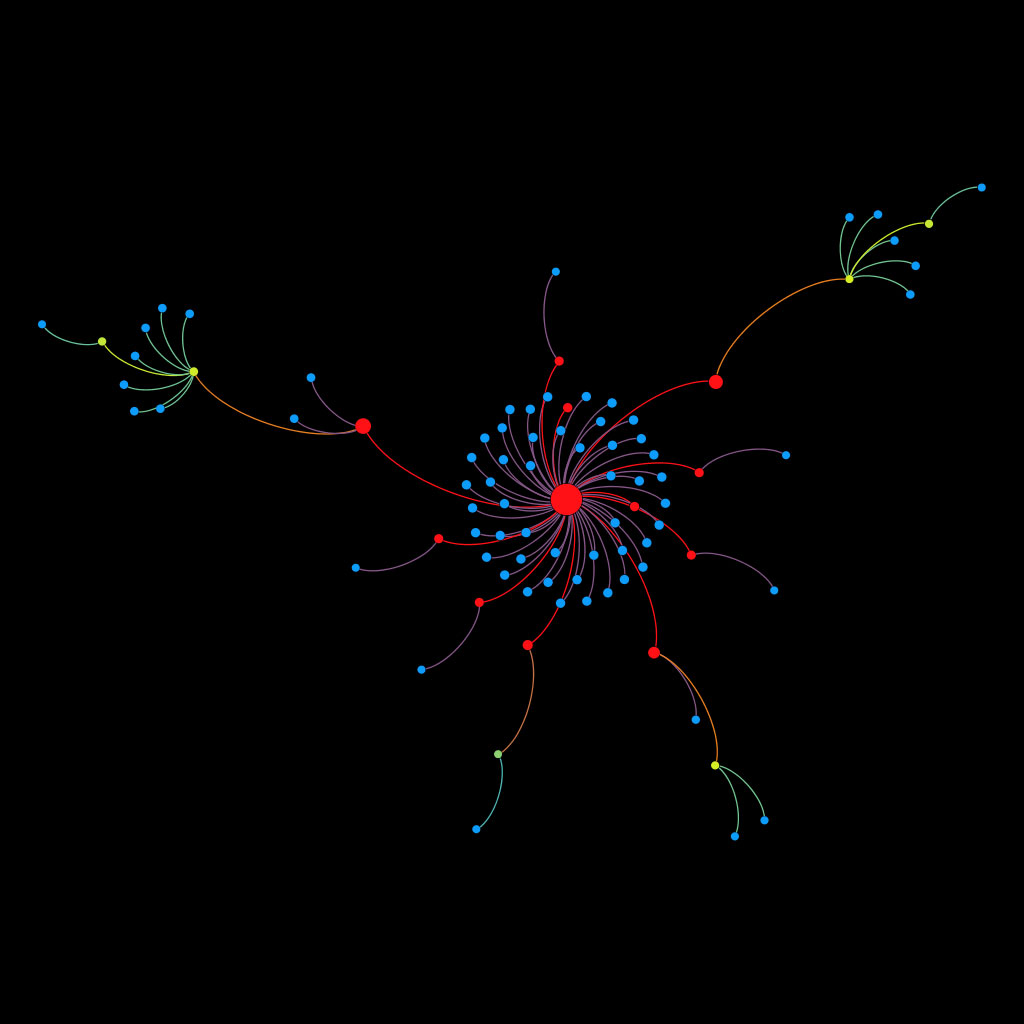}
\caption{Skeleton tree until 1997}
\end{minipage}
\begin{minipage}[t]{0.33\linewidth}
\includegraphics[width = \linewidth]{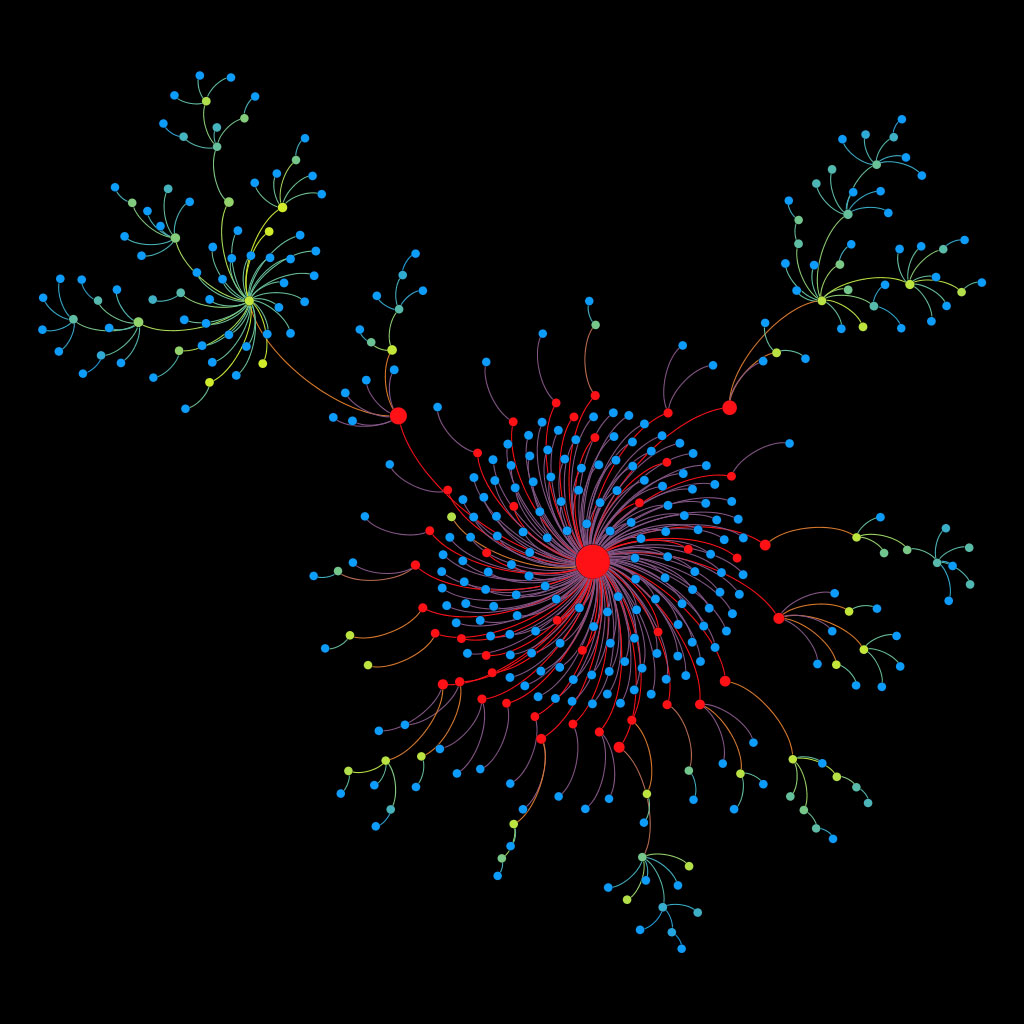}
\caption{Skeleton tree until 2001}
\end{minipage}
\begin{minipage}[t]{0.33\linewidth}
\includegraphics[width = \linewidth]{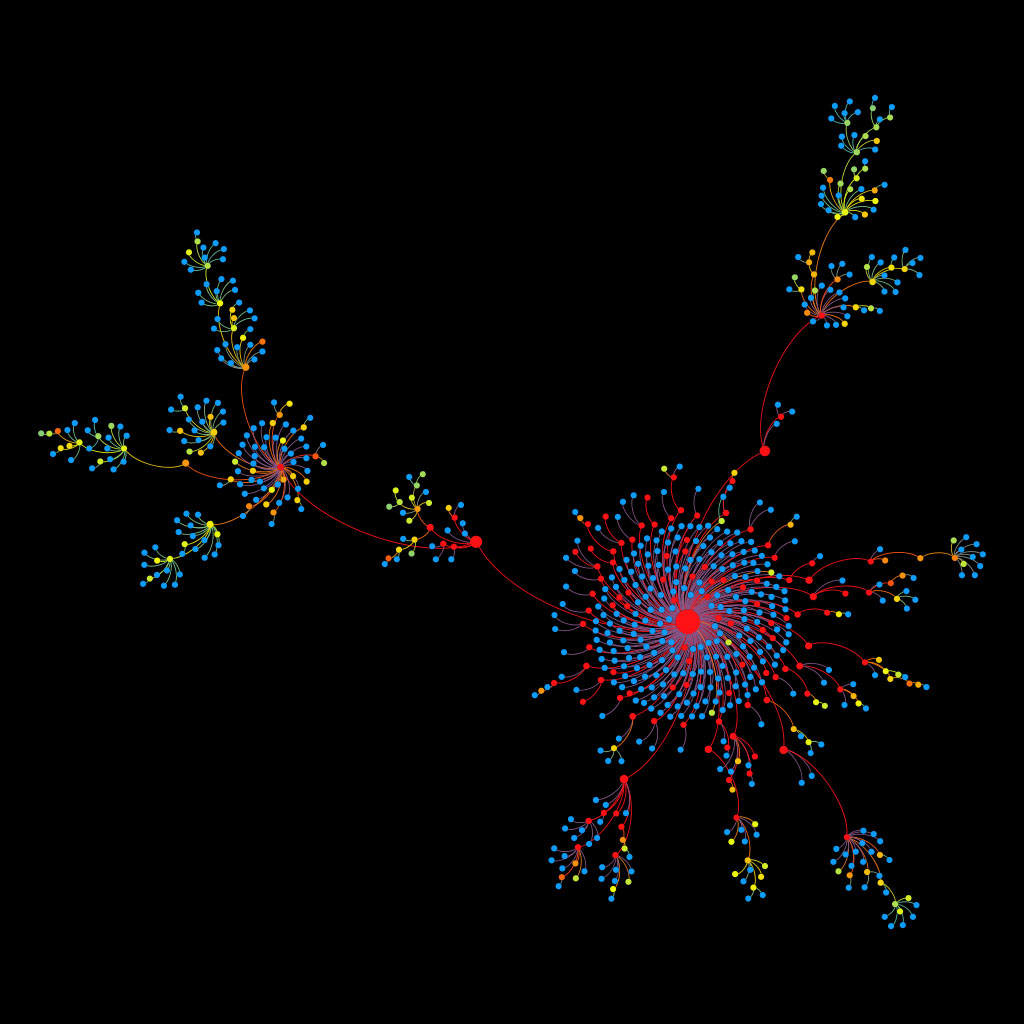}
\caption{Skeleton tree until 2005}
\end{minipage}
\end{subfigure}
\vspace{2mm}
\begin{subfigure}{\textwidth}
\begin{minipage}[t]{0.33\linewidth}
\includegraphics[width = \linewidth]{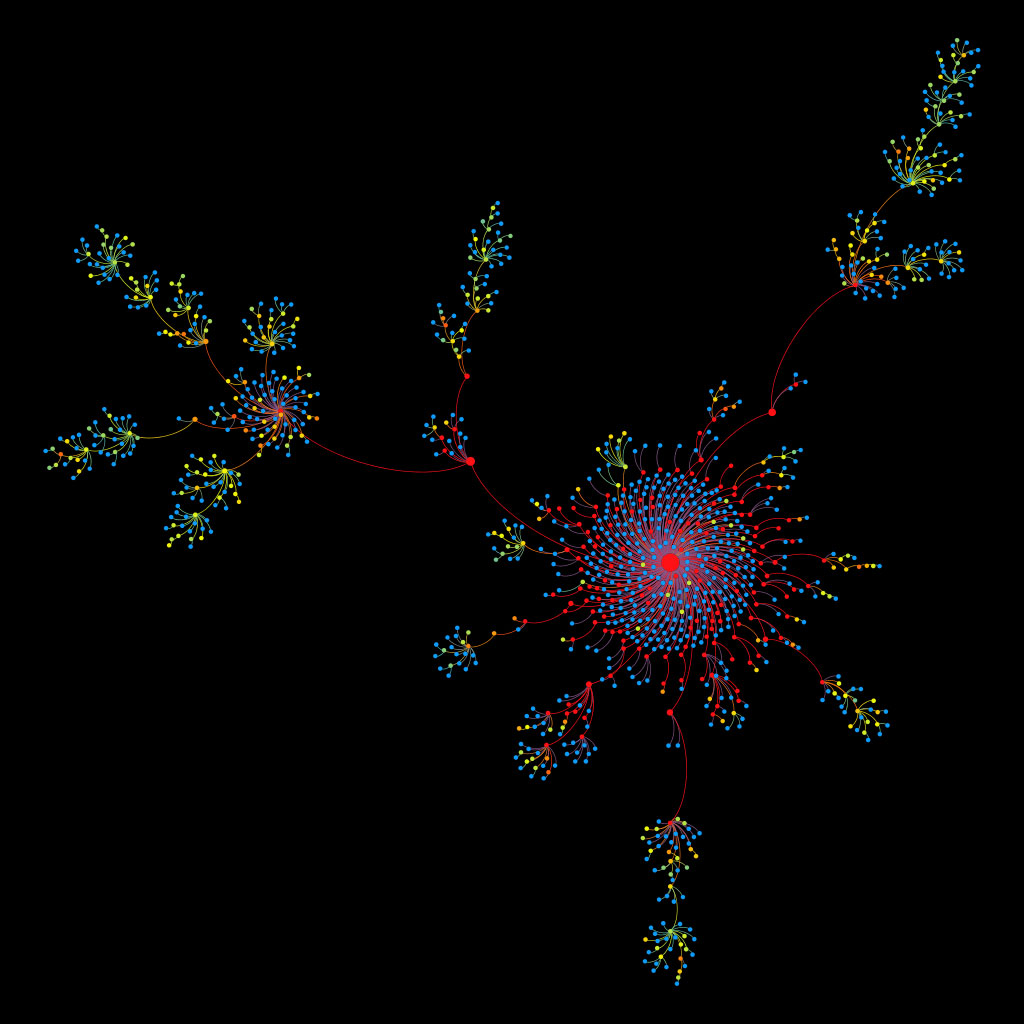}
\caption{Skeleton tree until 2009}
\end{minipage}
\begin{minipage}[t]{0.33\linewidth}
\includegraphics[width = \linewidth]{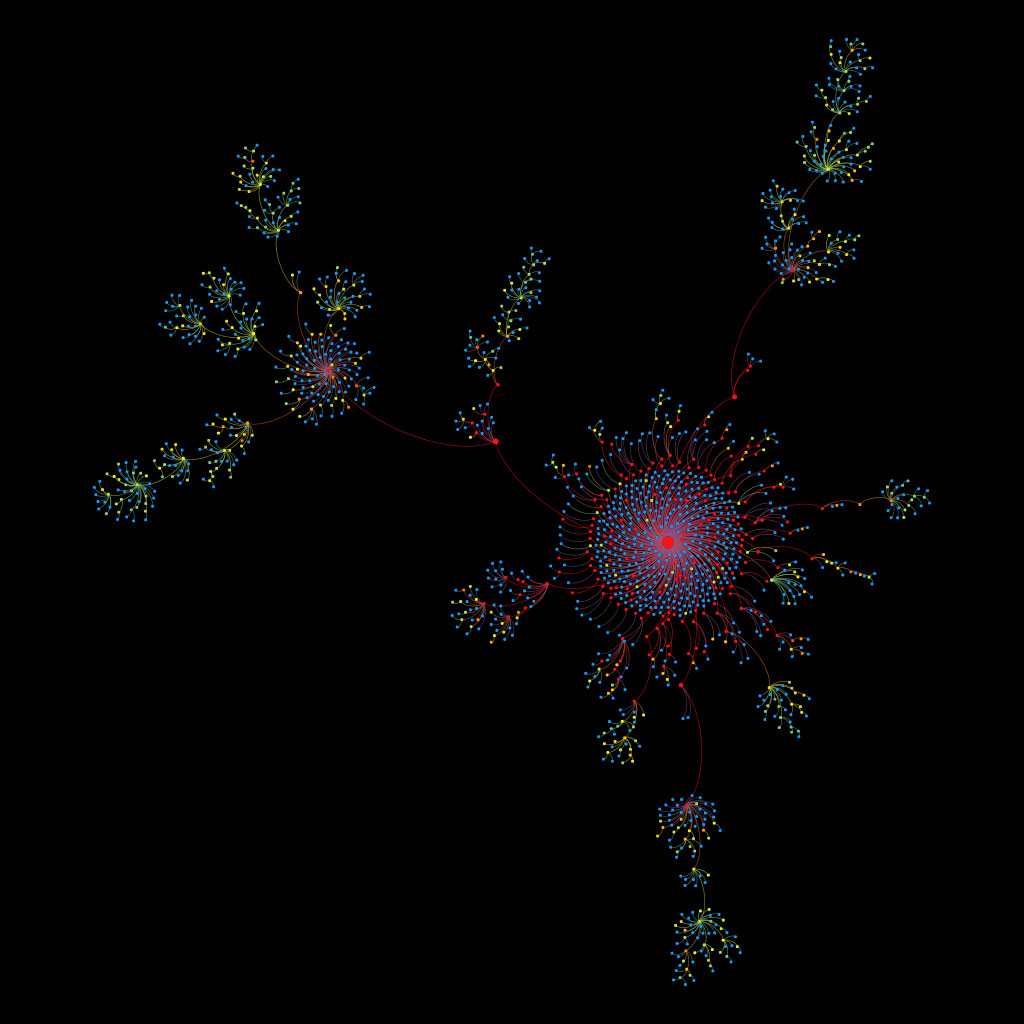}
\caption{Skeleton tree until 2013}
\end{minipage}
\begin{minipage}[t]{0.33\linewidth}
\includegraphics[width = \linewidth]{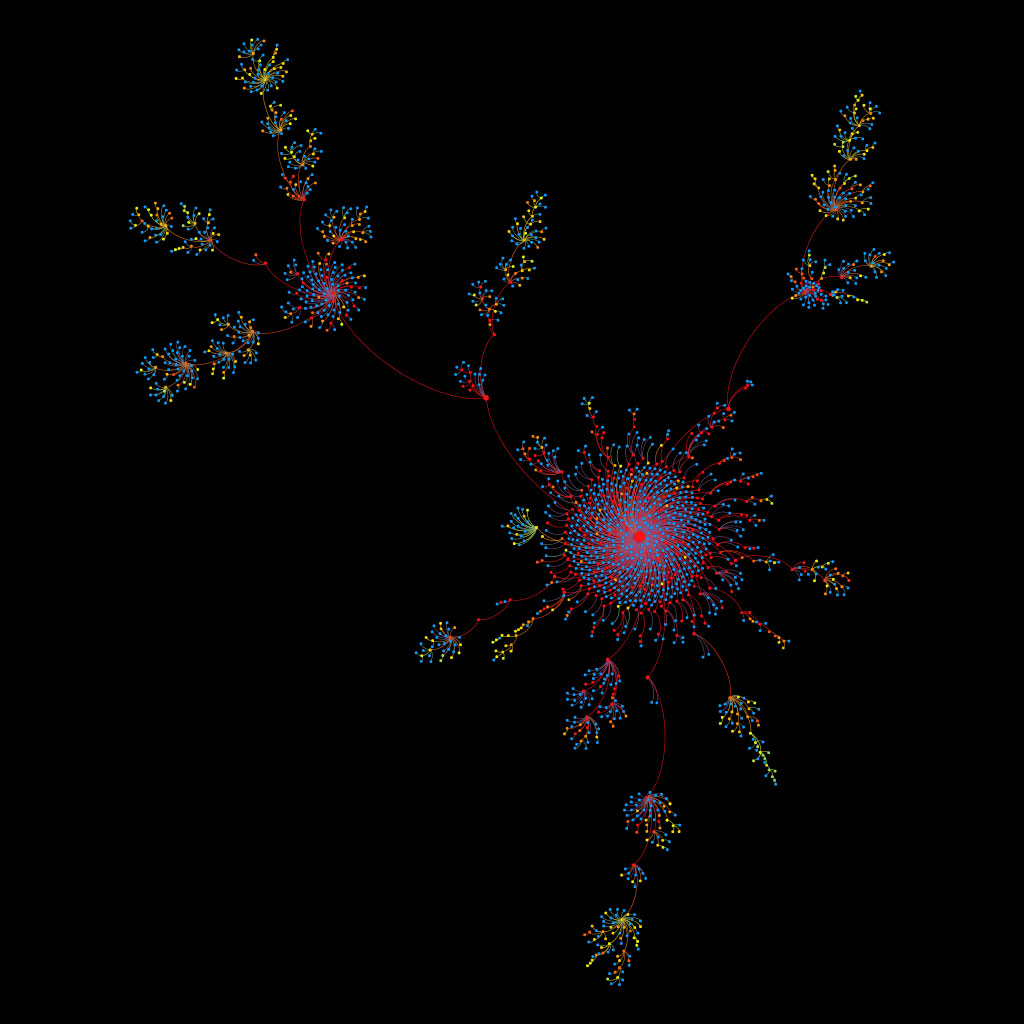}
\caption{Skeleton tree until 2017}
\end{minipage}
\end{subfigure}
\caption{Bose-Einstein condensation: Skeleton tree evolution}
\label{fig:15804200-tree_evo}
\end{figure}

\noindent $T_{structure}^t$ is higher in early days, which corresponds with a multi-dimensional growth in skeleton tree thanks to influential child papers published around 2000 (Fig. \ref{fig:15804200-tree_evo}). After 2013, skeleton tree has fixed its structure. We observe few visible changes in skeleton tree, namely some development in the research direction jointly led by popular child papers BECEP and BECPOM and a new small research branch deriving from the school of thought led by child papers `Second-Order Corrections to Mean Field Evolution of Weakly Interacting Bosons. I.' published in 2010 and its rather successful descendant `Derivation of the Cubic NLS and Gross-Pitaevskii Hierarchy from Manybody Dynamics in d = 3 Based on Spacetime Norms' published in 2014.\\

\begin{figure}[htbp]
    \centering
    \begin{subfigure}{\linewidth}
    \begin{minipage}[t]{0.5\textwidth}
    \centering
    \includegraphics[width = 0.9 \linewidth]{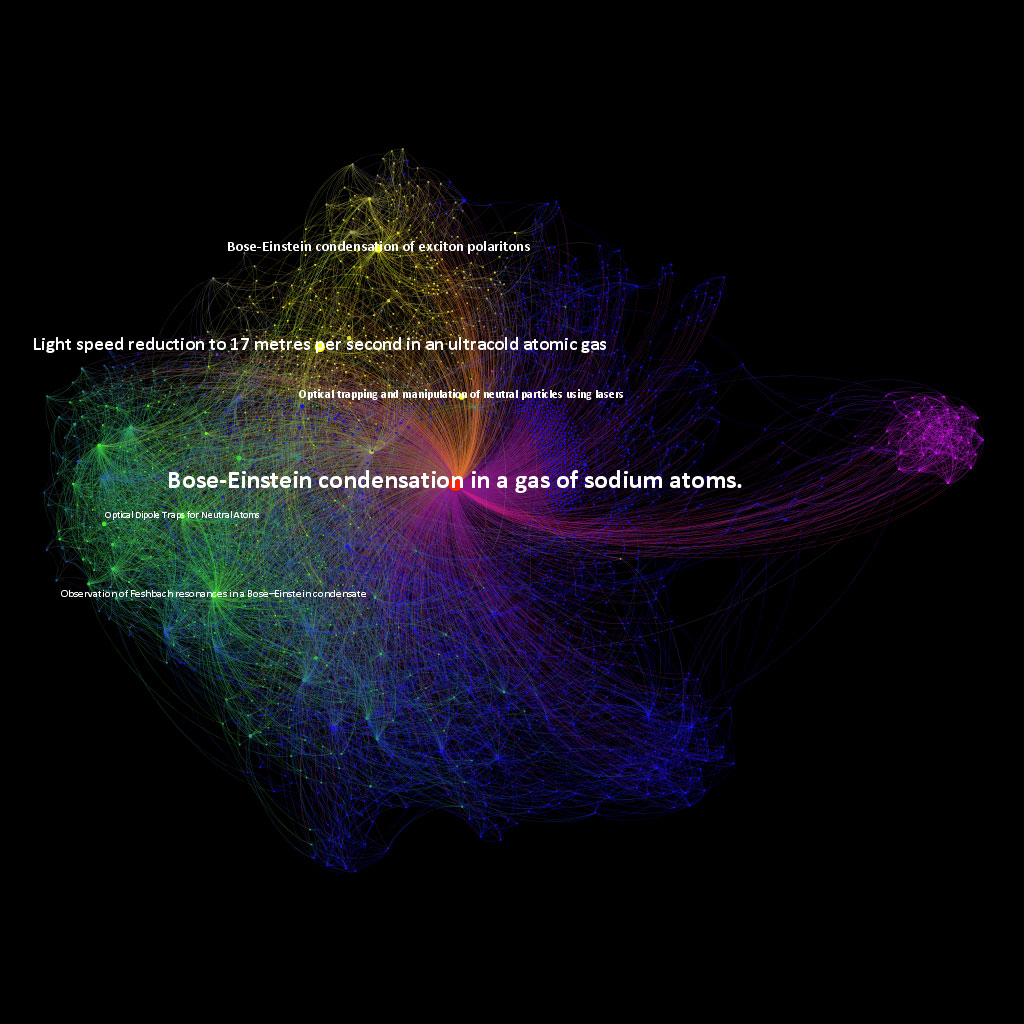}
     \end{minipage}
     \begin{minipage}[t]{0.5\textwidth}
    \centering
    \includegraphics[width = 0.9 \linewidth]{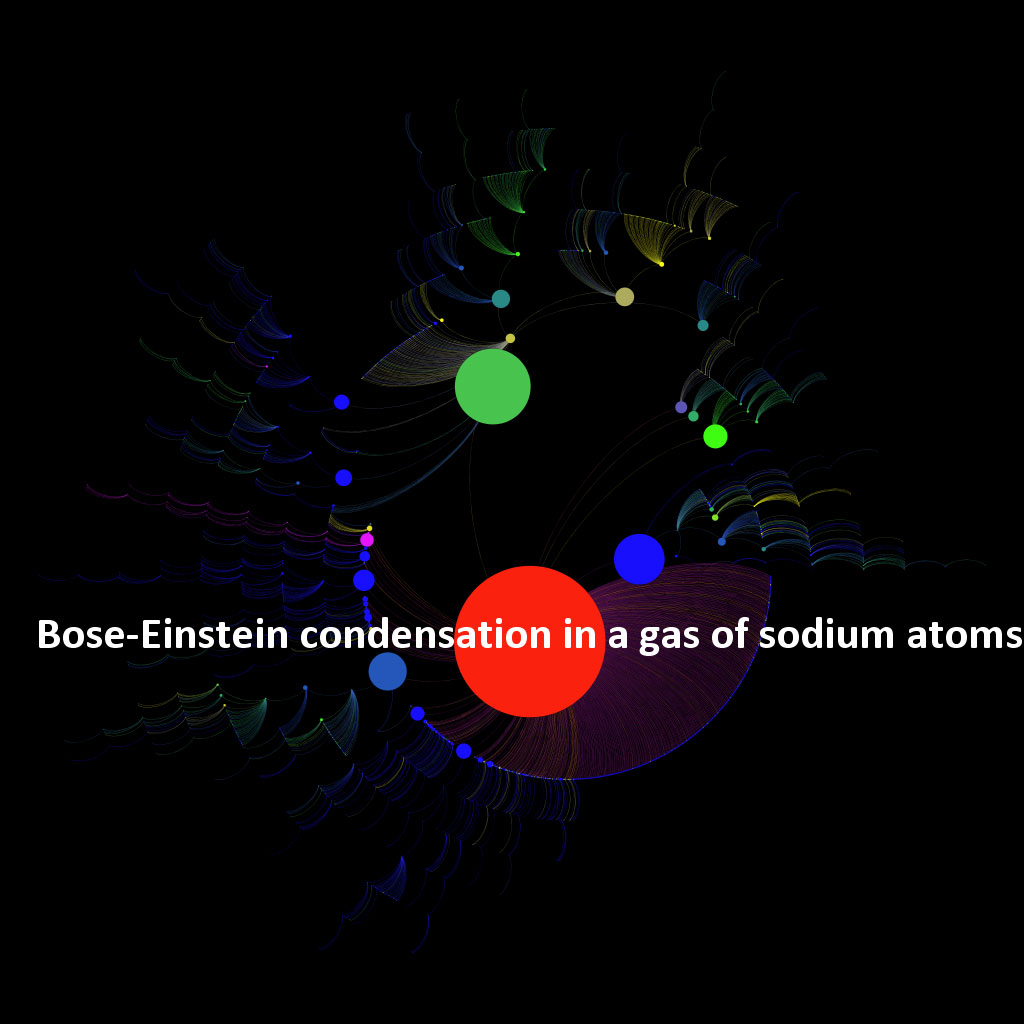}
     \end{minipage}
    \end{subfigure}

    \vspace{5mm}

    \begin{subfigure}{0.6\linewidth}
    \includegraphics[width = \linewidth]{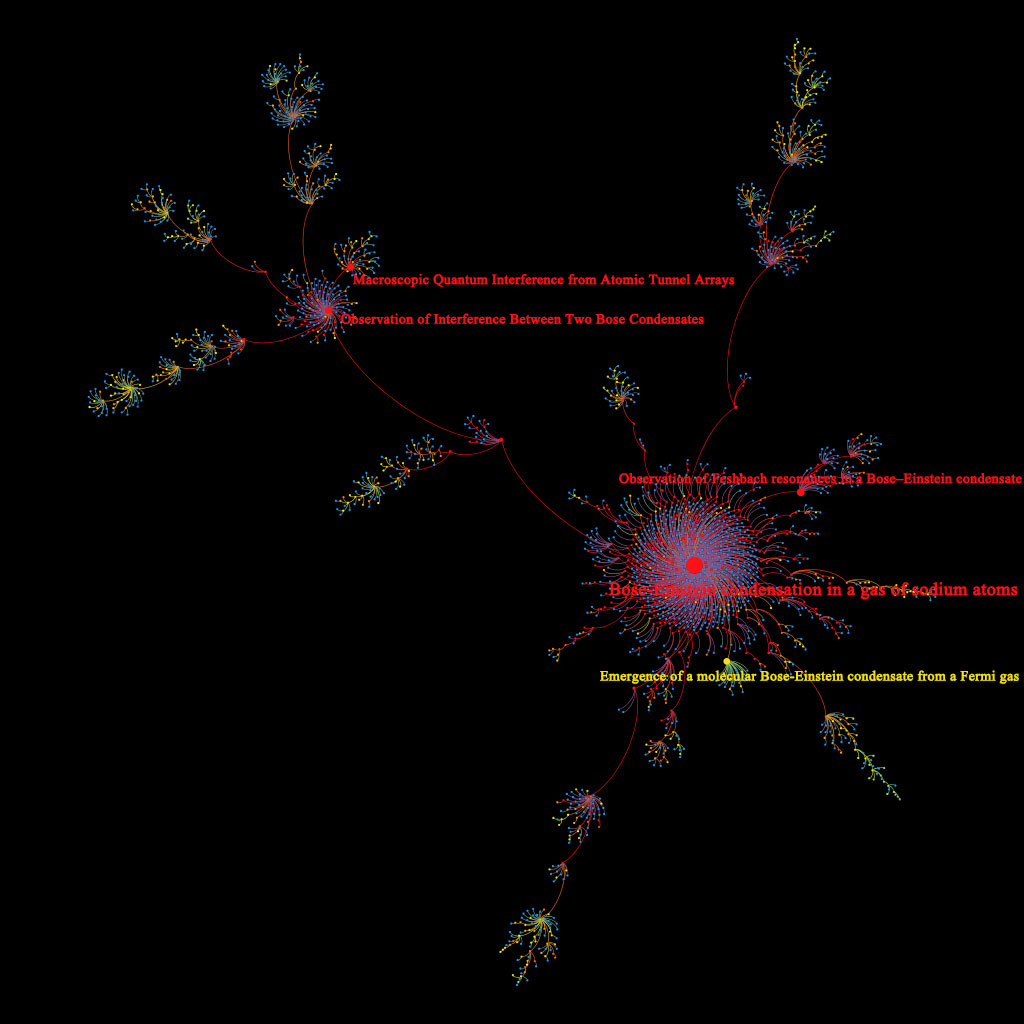}
    \end{subfigure}
    \begin{subfigure}{0.35\linewidth}
    \centering
    \includegraphics[width = 0.9\linewidth]{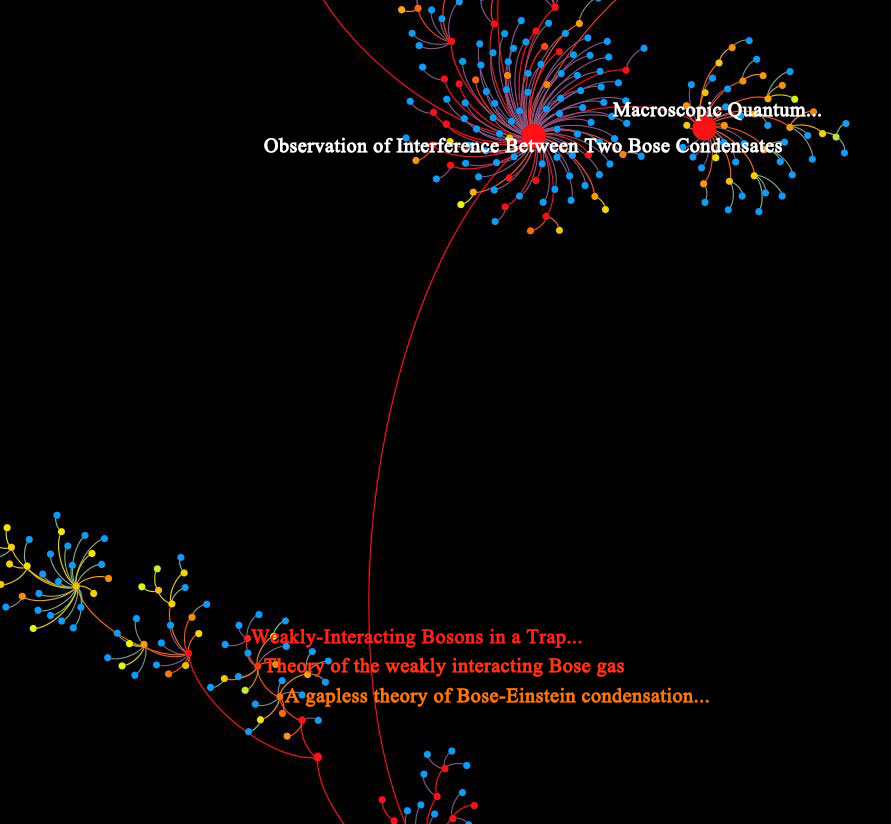}
    \end{subfigure}
    \caption{Bose-Einstein condensation: Galaxy map, current skeleton tree and its regional zoom. Papers with more than 150 in-topic citations are labelled by title in the skeleton tree. Except the pioneering work, corresponding nodes' size is amplified by 3 times.}
    \label{fig:15804200-2020}
\end{figure}

\noindent Now we closely examine its internal heat distribution together with its latest skeleton tree (Fig. \ref{fig:15804200-2020}). After more than 20 years of development, the heat has fully propagated to recent research directions led by popular child papers. Popular child papers are among the hottest articles and the child papers published during the flourishing period are relatively hot in general (Fig. 5(j)). The knowledge temperature decrease from cores to ends is clear. This corresponds with the general rule "the older the hotter". The blue nodes that surround the pioneering work and popular child papers in main clusters are papers with few or without any in-topic citations. However, there are exceptions. Paper `A gapless theory of Bose-Einstein condensation in dilute gases at finite temperature' published in 2000 is colder than its child paper `Theory of the weakly interacting Bose gas' (TWIBS) published in 2004. TWIBS is also slightly colder than its direct child paper in current skeleton tree `Weakly-Interacting Bosons in a Trap within Approximate Second Quantization Approach' (WIBTASQ) published in 2007. This is mainly due to their relatively different research focus as most of their in-topic citations do not overlap with one another. As WIBTASQ is the least developed among the three in terms of citations, this counter examples also illustrates that the general rule "the more influential the hotter" is weak (Fig. \ref{fig:citation_T}(j)).\\

\noindent We find the knowledge temperature evolution of child paper BECEP particularly interesting. Despite topic's stagnation starting from around 2013 and 2014, its knowledge temperature has been constantly on the rise since its publication, from 60.4 in 2006 to 83.5 in 2020. Its rising temperature demonstrates its above-average recent development compared to the entire topic. \\

\noindent We observe in addition certain clustering effect in the skeleton tree (Table \ref{tab:15804200-clustering}). For example, all child papers of `Comparative analysis of electric field influence on the quantum wells with different boundary conditions: II. Thermodynamic properties' have a research interest towards thermodynamics. This confirms the effectiveness of our skeleton tree extraction algorithm.\\

\begin{table}
    \centering
    \begin{tabular}{p{15cm} p{1cm}}
        \hline
        title & year\\
        \hline
        Comparative analysis of electric field influence on the quantum wells with different boundary conditions: II. \textcolor{red}{Thermodynamic properties} & 2015 \\

        Theory of the Robin quantum wall in a linear potential. II. \textcolor{orange}{Thermodynamic properties} & 2016\\

        Comparative analysis of electric field influence on the quantum wells with different boundary conditions.: I. \textcolor{orange}{Energy spectrum}, \textcolor{orange}{quantum information entropy} and polarization & 2015\\

        \textcolor{orange}{Thermodynamic Properties} of the 1D Robin Quantum Well & 2018\\
        \hline
    \end{tabular}
    \caption{Bose-Einstein condensation: Clustering effect example. First line is the parent paper and the rest children.}
    \label{tab:15804200-clustering}
\end{table}

\subsection{Awakened topics}

\subsubsection{Long short-term memory}

After a boom right after its birth, the topic hibernated for as long as 10 years before having an explosive growth. As is shown by the basic statistics, the topic's expansion in the first 15 years is much slower than recently. Apart from publication quantity difference, we also observe an obvious discrepancy in article's contribution to topic's flourishing. Few child papers turned out to be popular among topic members. Child paper `Learning to Forget: Continual Prediction with LSTM' (LFCP) published in 2000 is the only superstar the topic had for a long time. It successfully extended the pioneering work's idea and founded a new research focus, represented by the branch pointing to the bottom-left in skeleton tree (Fig. \ref{fig:56158074-tree_evo}(b,c,d)). Although the research branch seemed small by 2001, it already meant something compared to the then topic size. The evolution in knowledge structure led to a high $T_{structure}^t$. The remaining popular child papers, namely 2 published in 2003, `Kalman filters improve LSTM network performance in problems unsolvable by traditional recurrent nets' and `Learning precise timing with lstm recurrent networks', arriving later unanimously focused on LFCP's idea. Together they contributed to the maturation of this new sub-field and maintained partly the heat-level of the entire topic. The situation changed after 2010. The artificial intelligence frenzy pulled the topic under the spotlight. Thanks to the favorable background, the topic welcomed numerous popular child papers during 2013 and 2016, for instance, `Sequence to Sequence Learning with Neural Networks' (S2SNN) ,`Neural Machine Translation by Jointly Learning to Align and Translate' (NMTAT) and `Deep Residual Learning for Image Recognition' (DRLIR). While inheriting the essence of LFCP, they brought alone considerable amount of new knowledge, introduced new sub-topics and produced the renaissance of this old topic (Fig. \ref{fig:56158074-2020}, \ref{fig:56158074-tree_evo}(d,e,f)). Consequently, we see a slightly higher $T_{structure}^t$ around 2015 owing to the knowledge structure enrichment and a soar in $T^t$ starting from 2017. The long interval between the birth and the peak of impact and popularity makes us define this research field as an awakened topic.\\

\noindent There is a tiny cluster isolated from the majority of the skeleton tree (Fig. \ref{fig:56158074-2020} in the top-middle of current skeleton tree). This is because the topic contains several child papers published at the same time or evenly a bit earlier than the pioneering work. Comparatively speaking, their work is not very intimately related to that of the pioneering article. Therefore, altogether with some of their closest descendants, they were disconnected from the pioneering work during the skeleton tree construction.\\

\begin{figure}[htbp]
\centering
\begin{subfigure}[t]{0.7\linewidth}
\centering
\includegraphics[width=\linewidth]{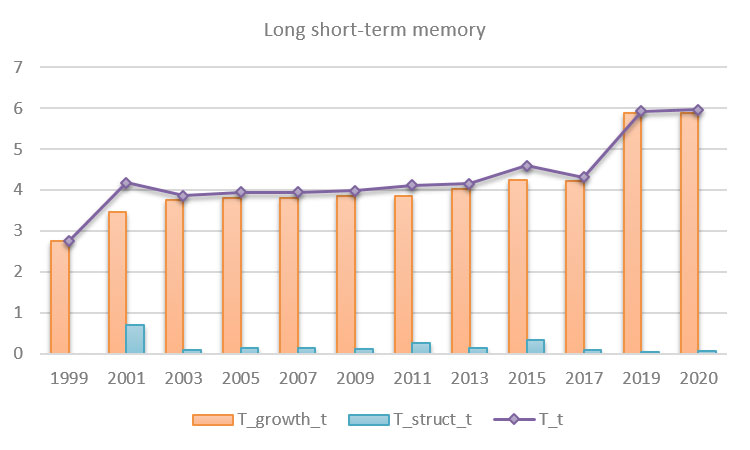}
\end{subfigure}

\begin{subfigure}[t]{\linewidth}
\centering
\begin{tabular}{ccccccccc}
\hline
year & $|V^t|$ & $|E^t|$ & $n_t$ & $V_t$ & ${UsefulInfo}^t$ & $T_{growth}^t$ & $T_{struct}^t$ & $T^t$\\
\hline
1999 & 17 & 21 & 14.333 & 17 & 2.667 & 2.747 &   & 2.747 \\

2001 & 56 & 109 & 37.44 & 56 & 18.56 & 3.465 & 0.714 & 4.179 \\

2003 & 102 & 237 & 62.773 & 102 & 39.227 & 3.764 & 0.1 & 3.864 \\

2005 & 156 & 407 & 95.061 & 156 & 60.939 & 3.801 & 0.143 & 3.944 \\

2007 & 230 & 682 & 140.092 & 230 & 89.908 & 3.803 & 0.139 & 3.942 \\

2009 & 331 & 1010 & 198.412 & 331 & 132.588 & 3.864 & 0.121 & 3.985 \\

2011 & 422 & 1414 & 253.302 & 422 & 168.698 & 3.859 & 0.261 & 4.12 \\

2013 & 568 & 2129 & 326.996 & 568 & 241.004 & 4.024 & 0.133 & 4.156 \\

2015 & 1323 & 7166 & 722.591 & 1323 & 600.409 & 4.241 & 0.348 & 4.589 \\

2017 & 5912 & 35684 & 3239.903 & 5912 & 2672.097 & 4.227 & 0.09 & 4.316 \\

2019 & 15279 & 90463 & 6023.461 & 15279 & 9255.539 & 5.876 & 0.046 & 5.921 \\

2020 & 16777 & 98553 & 6610.64 & 16777 & 10166.36 & 5.879 & 0.075 & 5.954 \\
\hline
\end{tabular}
\end{subfigure}
\caption{Long short-term memory: topic statistics and knowledge temperature evolution}
\end{figure}

\begin{figure}[htbp]
\begin{subfigure}{\textwidth}
\begin{minipage}[t]{0.33\linewidth}
\includegraphics[width = \linewidth]{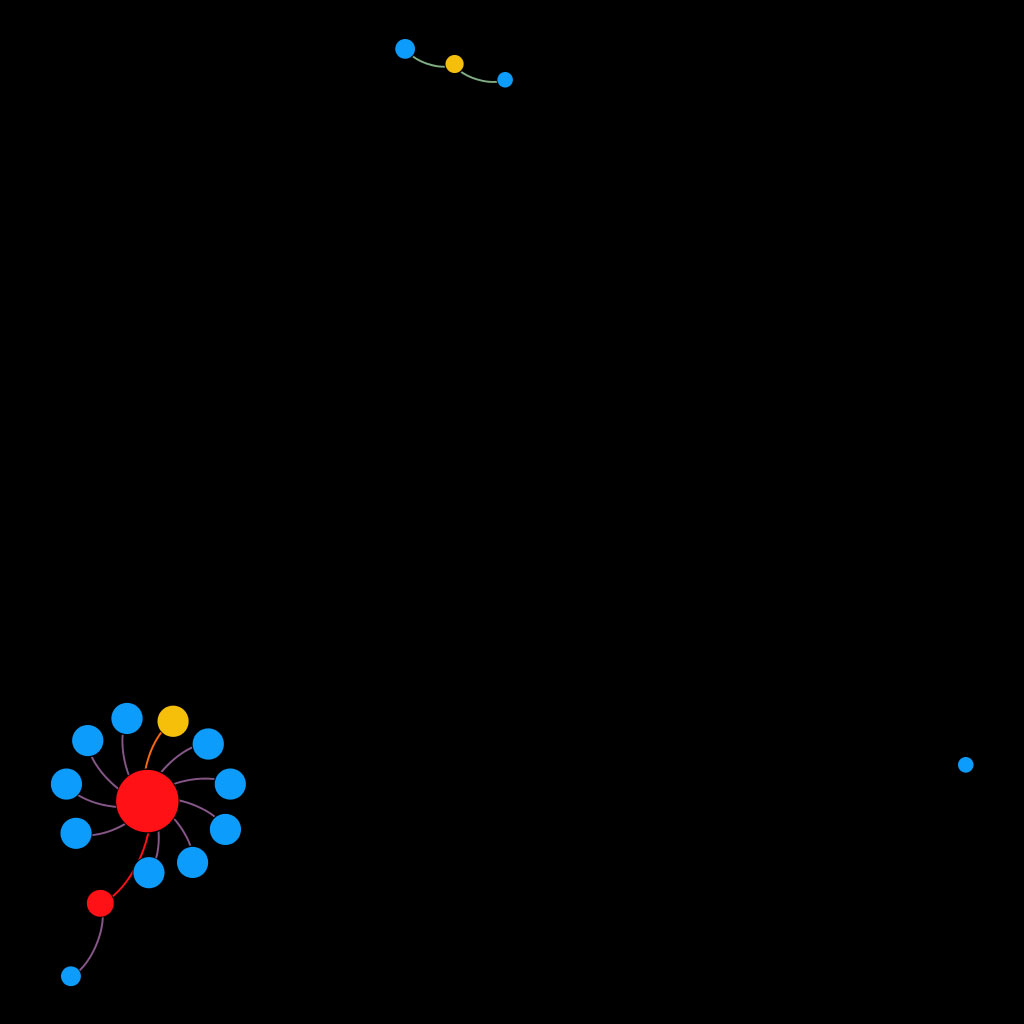}
\caption{Skeleton tree until 1999}
\end{minipage}
\begin{minipage}[t]{0.33\linewidth}
\includegraphics[width = \linewidth]{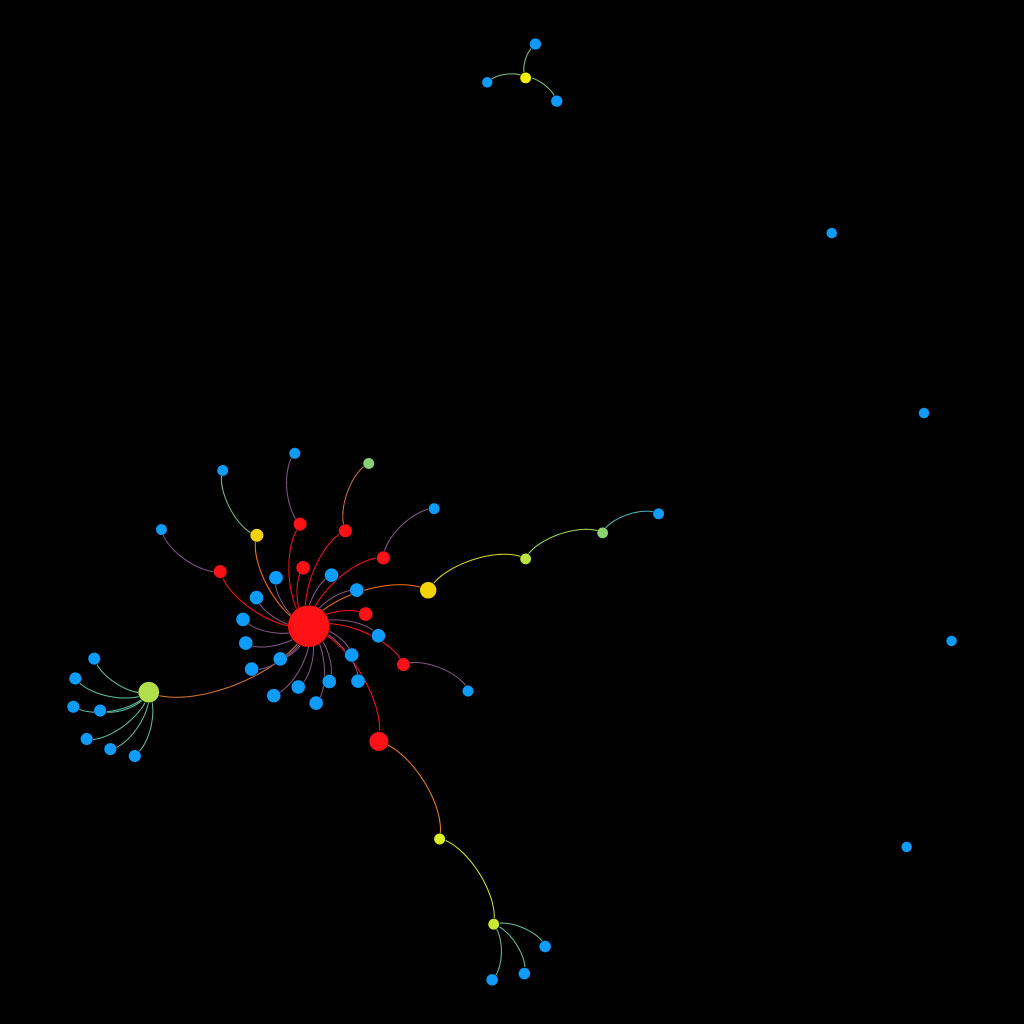}
\caption{Skeleton tree until 2001}
\end{minipage}
\begin{minipage}[t]{0.33\linewidth}
\includegraphics[width = \linewidth]{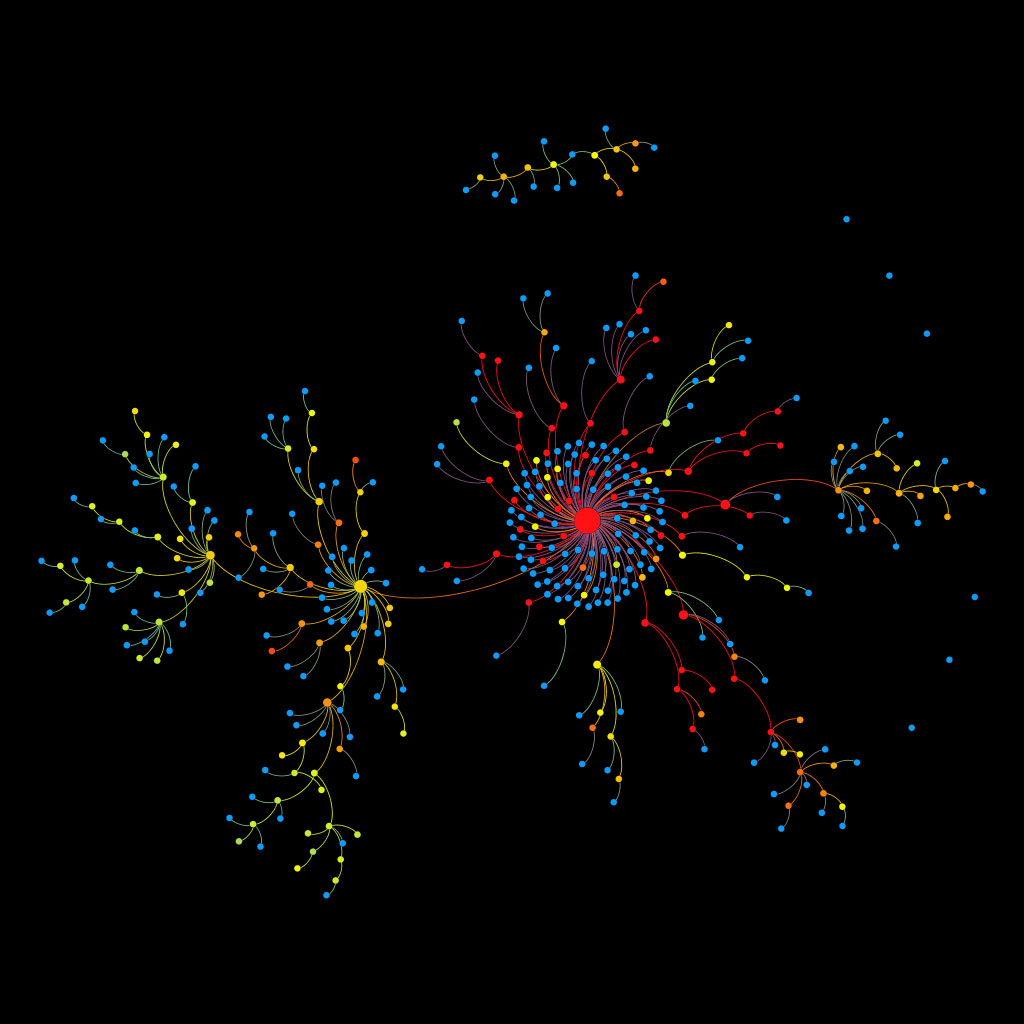}
\caption{Skeleton tree until 2011}
\end{minipage}
\end{subfigure}
\vspace{2mm}
\begin{subfigure}{\textwidth}
\begin{minipage}[t]{0.33\linewidth}
\includegraphics[width = \linewidth]{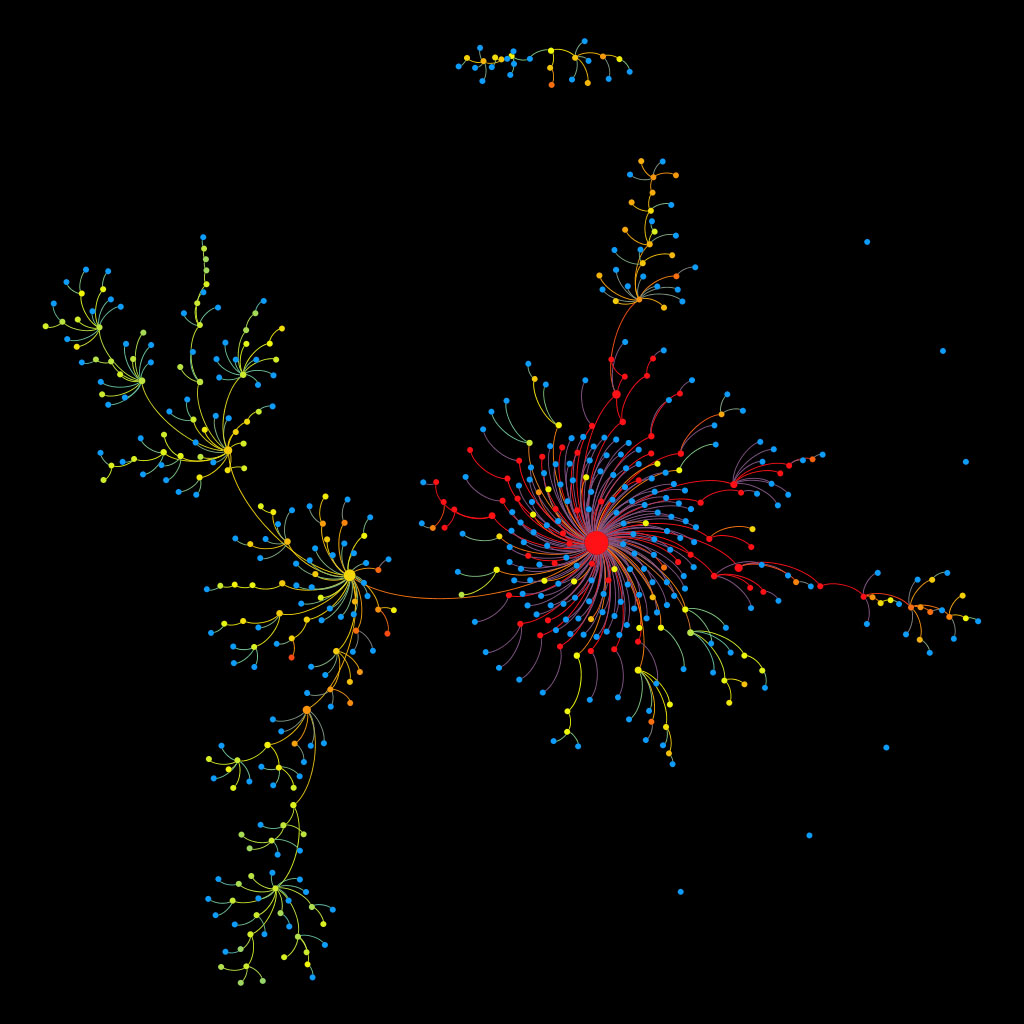}
\caption{Skeleton tree until 2013}
\end{minipage}
\begin{minipage}[t]{0.33\linewidth}
\includegraphics[width = \linewidth]{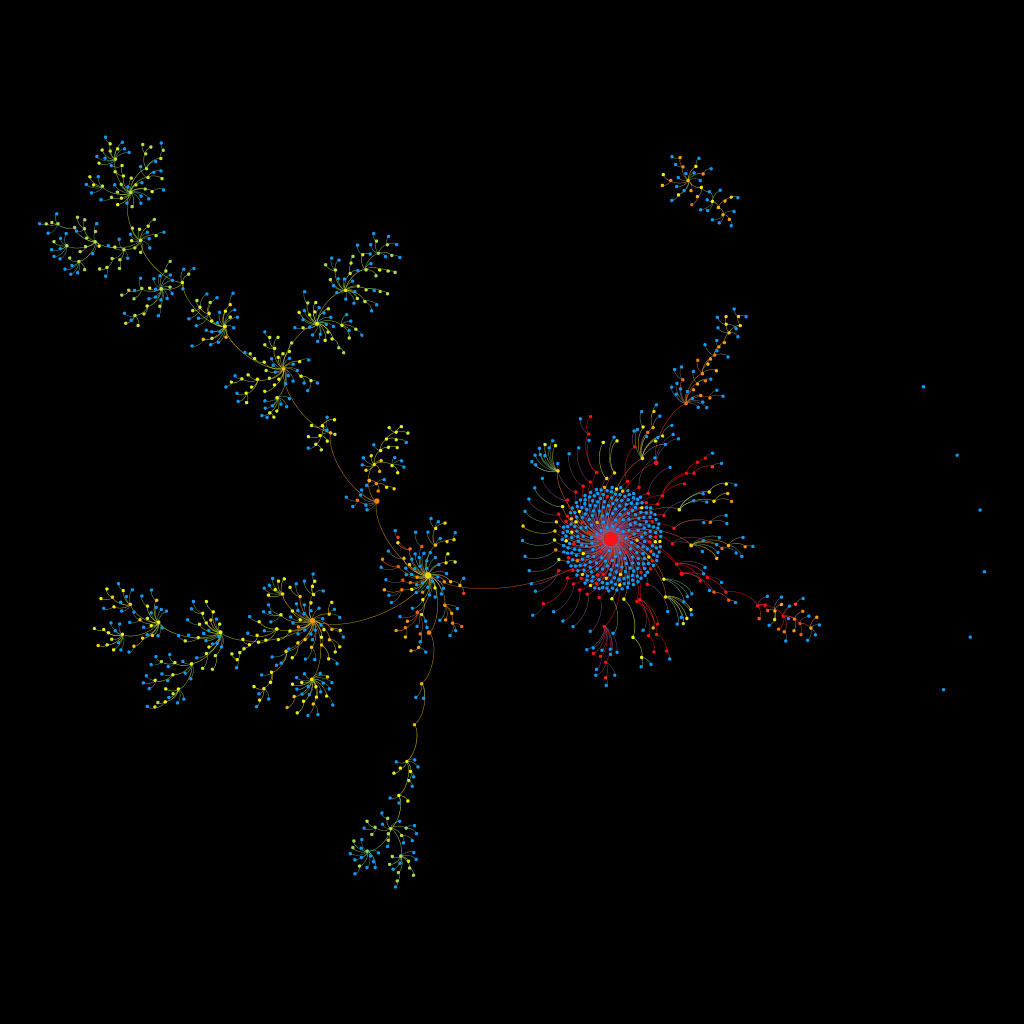}
\caption{Skeleton tree until 2015}
\end{minipage}
\begin{minipage}[t]{0.33\linewidth}
\includegraphics[width = \linewidth]{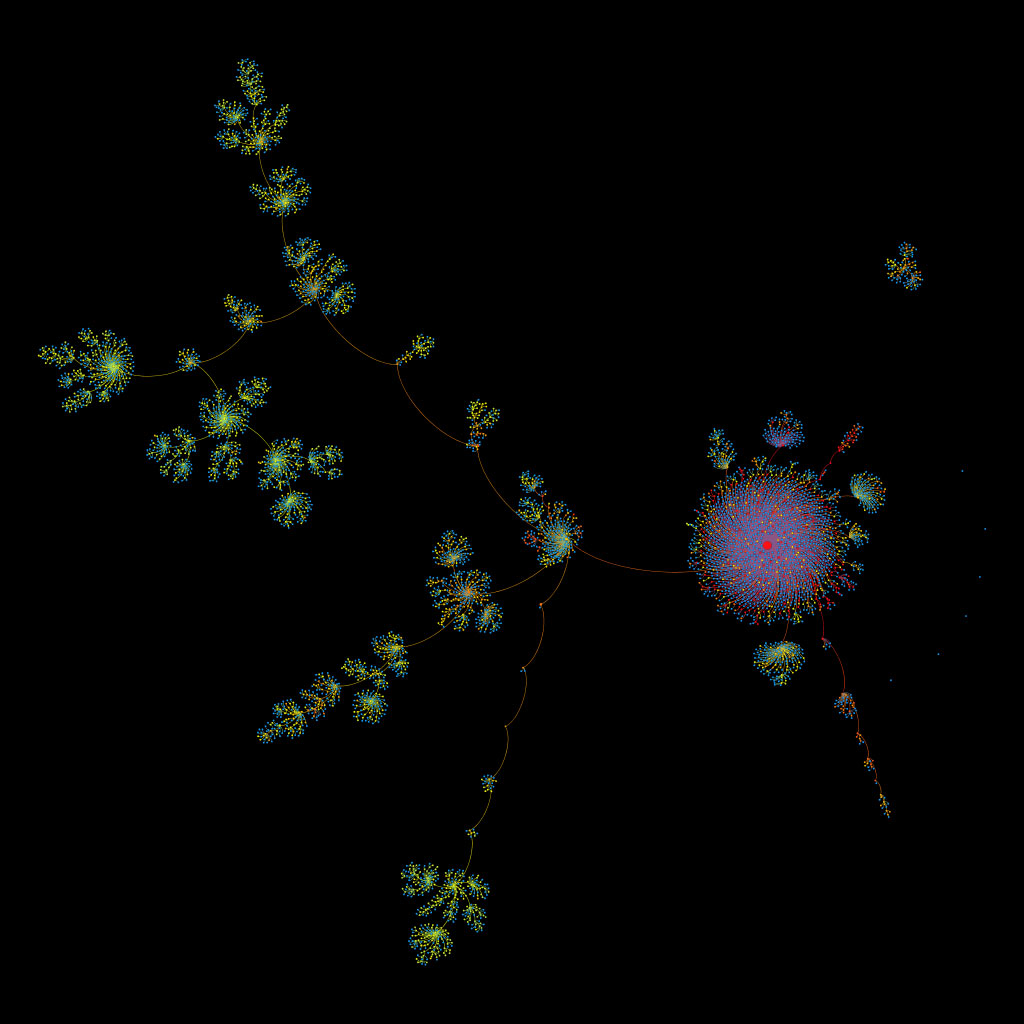}
\caption{Skeleton tree until 2017}
\end{minipage}
\end{subfigure}
\caption{Long short-term memory: Skeleton tree evolution}
\label{fig:56158074-tree_evo}
\end{figure}

\noindent Now we examine the heat distribution within the topic (Fig. \ref{fig:56158074-2020}). The pioneering work remains the only heat source so far. Although this topic has a long history, its flourishing took place a few years ago. It needs more time to have a thorough heat diffusion within the topic. That is why most popular child papers have a node knowledge temperature around or a bit above average. At present, most of the hottest articles are located around the pioneering work the central cluster. The knowledge temperature decline from the core to ends is obvious. This corresponds with the general rule "the older the hotter" (Fig. 5(k)). Note that the blue nodes surrounding the pioneering work and popular child papers in non-trivial clusters are papers with few or without any in-topic citations. The low average temperature for the oldest papers is due to their loose connection to the topic majority as they were published no later than the pioneering work and have had few child papers within the topic. However, even if we let alone these papers, age is not guarantee of a bigger impact and popularity. For instance, 2 popular children papers of LFCP are slightly hotter than itself. They are `Kalman filters improve LSTM network performance in problems unsolvable by traditional recurrent nets' published in 2003 and `Modeling systems with internal state using evolino' published in 2005. Both are coloured orange-red. Similarly, article `Generating Text with Recurrent Neural Networks' published in 2011 is also slightly colder than its child, `Understanding the exploding gradient problem', which was published in 2012. Their temperature difference is mainly owing to their research focus, as is reflected by their distinct citation patterns. These counter examples also illustrate that the general rule "the more influential the hotter" is weak (Fig. \ref{fig:citation_T}(k)). \\

\noindent We find the knowledge temperature evolution of LFCP particularly interesting. Its knowledge temperature dropped from 6.53 to 5.08 from 2001 to 2005. The decrease rate is greater than that of topic knowledge temperature. This is because its followers had little development, thus overall the bundle led by Learning to forget had a slower development than the entire topic. Its temperature has been on the rise since 2007. In particular, the increase has greatly accelerated from 2015. We attribute its surge to the arrival of several popular child papers published between 2014 and 2016: S2SNN (2014), `Empirical Evaluation of Gated Recurrent Neural Networks on Sequence Modeling' (2014), NMTAT (2015) and DRLIR (2016) (Fig. \ref{fig:56158074-2020}). Their instantaneous popularity has brought learning to forget back to scientists' attention. Recall that these papers also contributed a lot to the knowledge temperature leap of the entire topic starting from 2017.\\

\begin{figure}[htbp]
\centering
    \begin{subfigure}{\textwidth}
    \begin{minipage}[t]{0.55\textwidth}
    \centering
    \includegraphics[width = 0.9\linewidth]{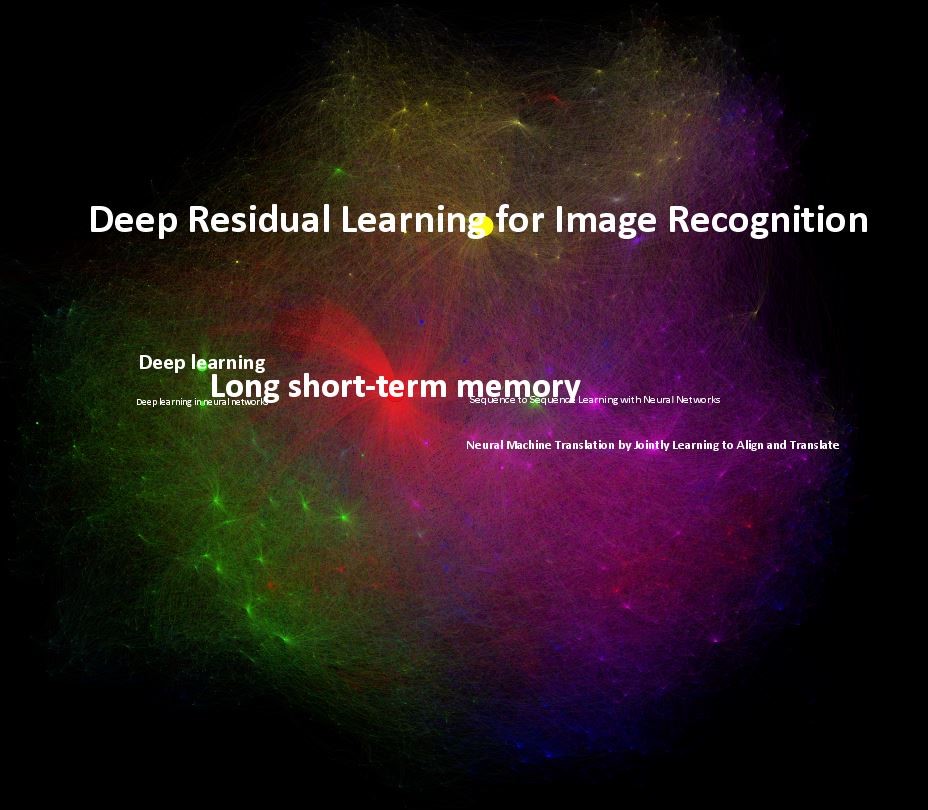}
    \end{minipage}
    \begin{minipage}[t]{0.45\textwidth}
    \centering
    \includegraphics[width = 0.95\linewidth]{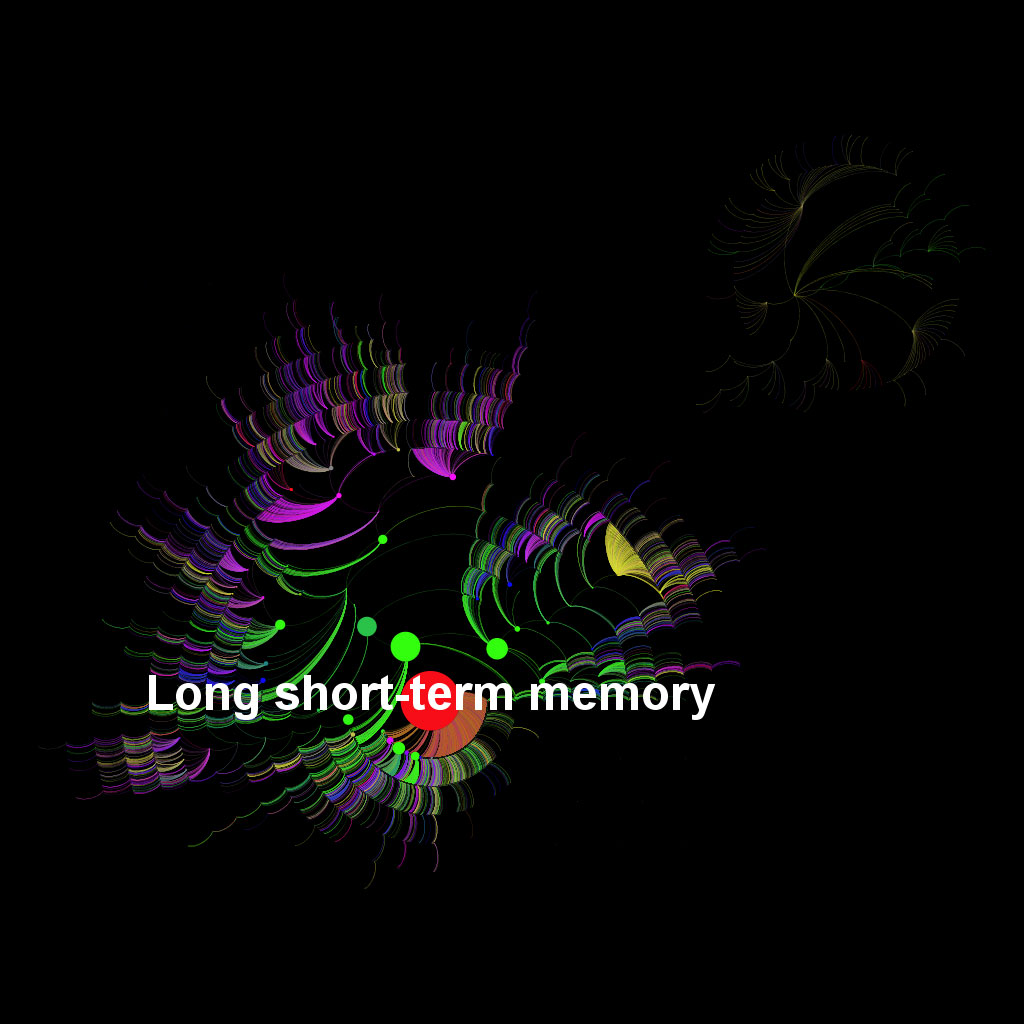}
    \end{minipage}
    \end{subfigure}

    \vspace{5mm}

    \begin{subfigure}{0.6\textwidth}
    \centering
    \includegraphics[width = \linewidth]{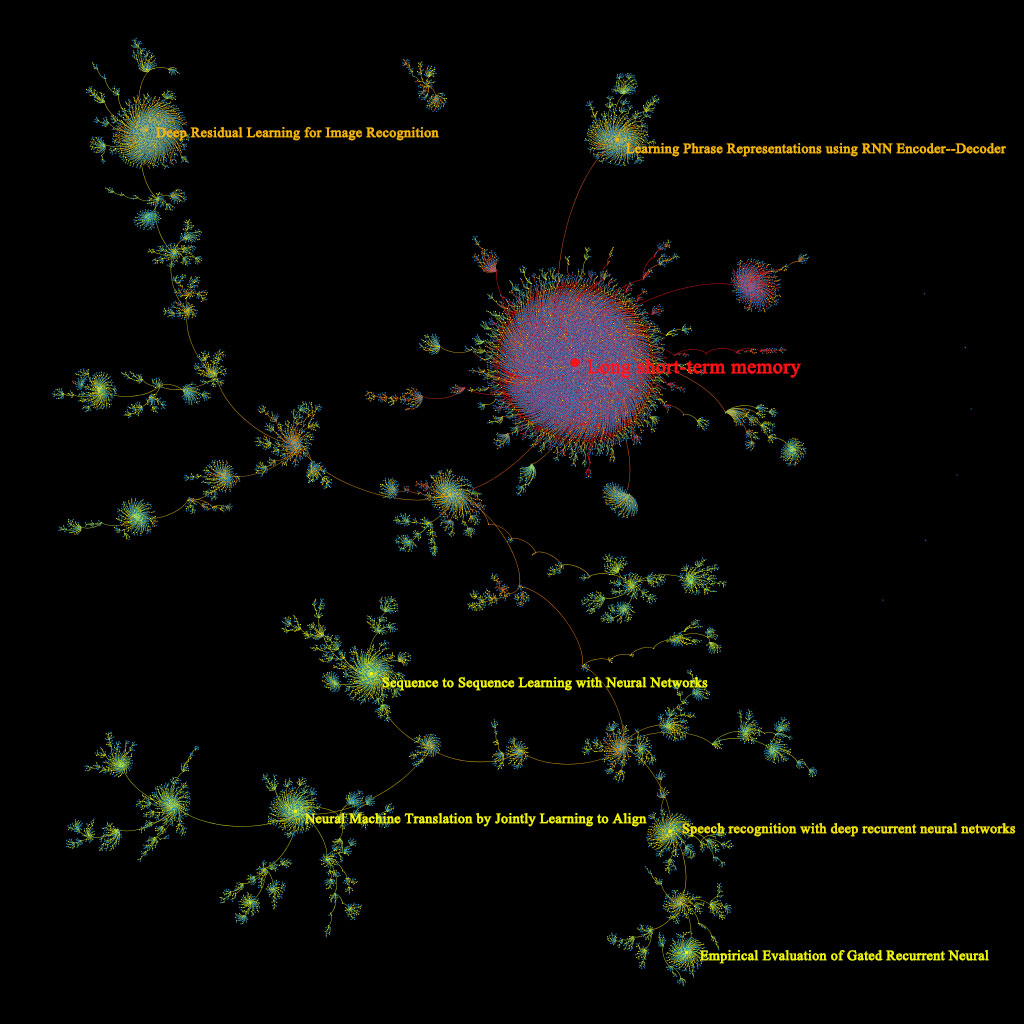}
    \end{subfigure}
    \begin{subfigure}{0.35\textwidth}
    \centering
    \begin{minipage}{\textwidth}
    \centering
    \includegraphics[width = 0.9\linewidth]{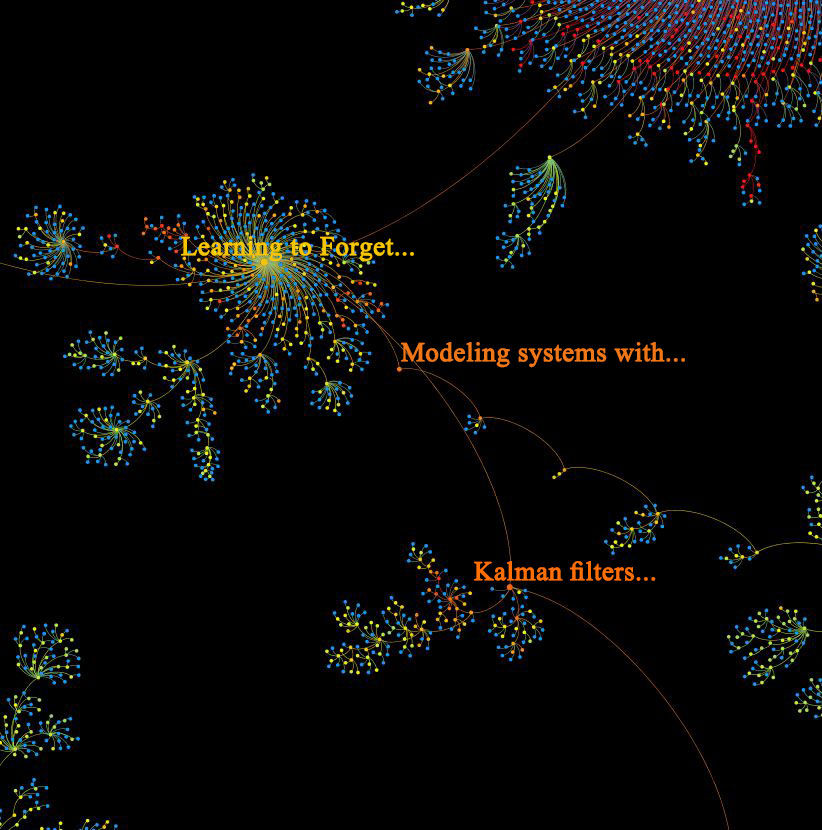}
    \end{minipage}
    \begin{minipage}{\textwidth}
    \centering
    \includegraphics[width = 0.9\linewidth]{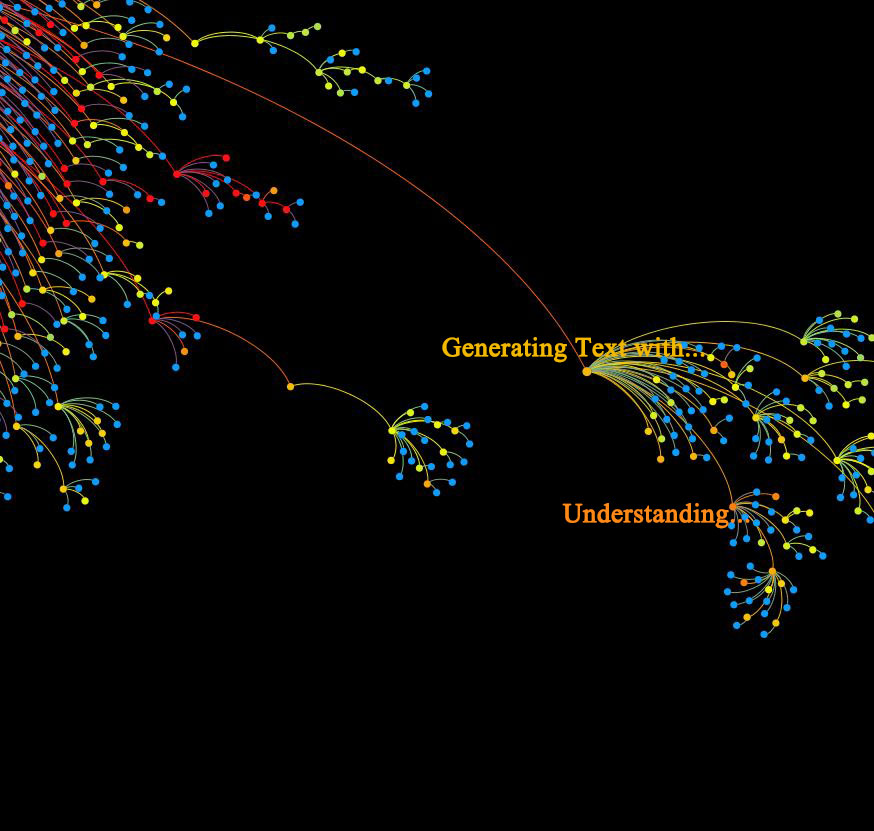}
    \end{minipage}
    \end{subfigure}
\caption{Long short-term memory: Galaxy map, current skeleton tree and its regional zoom. Papers with more than 1000 in-topic citations are labelled by title in the skeleton tree. Except the pioneering work, corresponding nodes' size is amplified by 3 times.}
\label{fig:56158074-2020}
\end{figure}

\noindent We observe in addition certain clustering effect in the skeleton tree. For example, almost all child papers of `Developing a Long Short-Term Memory (LSTM) based model for predicting water table depth in agricultural areas' deal with earth science and agriculture and `Visual Reasoning with a General Conditioning Layer' leads a handful of articles specialising in visual reasoning (Table \ref{tab:56158074-clustering}). We also identify some bundles dealing with energy forecast and financial trading. All these observations confirm the effectiveness of our skeleton tree extraction algorithm. Moreover, these aforementioned bundles were born no earlier than 2018, thus they are also good illustrations of some latest research hotspots in the topic.\\

\begin{table}
    \centering
    \begin{tabular}{p{15cm} p{1cm}}
        \hline
        title & year\\
        \hline
        Developing a Long Short-Term Memory (LSTM) based model for predicting \textcolor{red}{water table} depth in agricultural areas & 2018 \\

         \textcolor{orange}{Stream-Flow} Forecasting of Small Rivers Based on LSTM & 2020\\

        Developing a Long Short-Term Memory-based signal processing method for Coriolis \textcolor{orange}{mass flowmeter} & 2019\\

         Direct Multistep \textcolor{orange}{Wind Speed} Forecasting Using LSTM Neural Network Combining EEMD and Fuzzy Entropy & 2019\\

        Dynamic neural network modelling of \textcolor{orange}{soil moisture} content for predictive irrigation scheduling & 2018\\

        SMArtCast: Predicting \textcolor{orange}{soil moisture} interpolations into the future using Earth observation data in a deep learning framework & 2020\\

        Short-Term \textcolor{orange}{Streamflow} Forecasting for Para\'{i}ba do Sul River Using Deep Learning & 2019\\

        Synthetic well logs generation via Recurrent Neural Networks & 2018\\

        \textcolor{orange}{Reservoir} Facies Classification using Convolutional Neural Networks & 2019\\

        Comparative applications of data-driven models representing \textcolor{orange}{water table} fluctuations & 2019\\
        \hline
    \end{tabular}

    \vspace{2mm}

    \begin{tabular}{p{15cm} p{1cm}}
        \hline
        title & year\\
        \hline
        FiLM: \textcolor{red}{Visual Reasoning} with a General Conditioning Layer & 2018 \\

        LEARNING TO \textcolor{orange}{COLOR FROM LANGUAGE} & 2018\\

        Feature-wise transformations & 2018\\

        RAVEN: A Dataset for Relational and Analogical \textcolor{orange}{Visual rEasoNing} & 2019\\

        A Dataset and Architecture for \textcolor{orange}{Visual Reasoning} with a Working Memory & 2018\\

        Cycle-Consistency for Robust \textcolor{orange}{Visual Question Answering} & 2019\\

        On Self Modulation for Generative Adversarial Networks & 2019\\

        Interactive Sketch \& Fill: Multiclass \textcolor{orange}{Sketch-to-Image Translation} & 2019\\

        TapNet: Neural Network Augmented with Task-Adaptive Projection for Few-Shot Learning & 2019\\

        \textcolor{orange}{Predicting Taxi Demand} Based on \textcolor{orange}{3D} Convolutional Neural Network and Multi-task Learning & 2019\\
        \hline
    \end{tabular}

    \caption{Long short-term memory: Clustering effect example. First line is the parent paper and the rest children.}
    \label{tab:56158074-clustering}
\end{table}

\subsubsection{Particle swarm optimization}

\begin{figure}[htbp]
\centering
\begin{subfigure}[t]{0.7\linewidth}
\centering
\includegraphics[width=\linewidth]{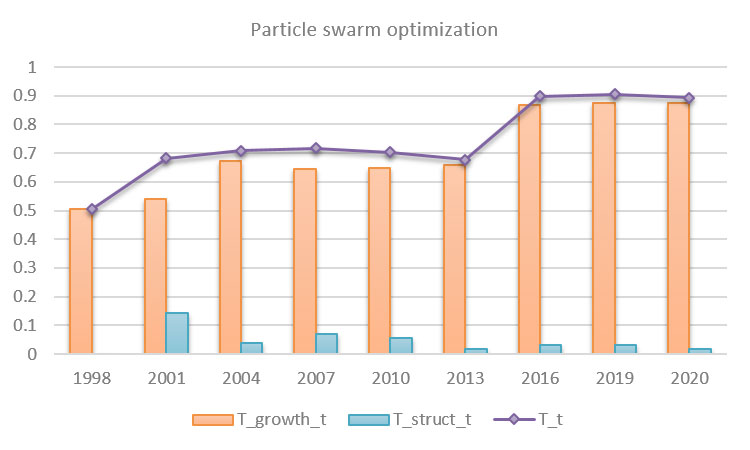}
\end{subfigure}
\begin{subfigure}[t]{\linewidth}
\centering
\begin{tabular}{ccccccccc}
\hline
year & $|V^t|$ & $|E^t|$ & $n_t$ & $V_t$ & ${UsefulInfo}^t$ & $T_{growth}^t$ & $T_{struct}^t$ & $T^t$\\
\hline
1998 & 16 & 32 & 10.9 & 16 & 5.1 & 0.504 &   & 0.504 \\

2001 & 69 & 215 & 43.86 & 69 & 25.14 & 0.54 & 0.142 & 0.683 \\

2004 & 494 & 2272 & 252.724 & 494 & 241.276 & 0.671 & 0.037 & 0.708 \\

2007 & 2818 & 13285 & 1500.051 & 2818 & 1317.949 & 0.645 & 0.071 & 0.717 \\

2010 & 9186 & 44357 & 4877.391 & 9186 & 4308.609 & 0.647 & 0.056 & 0.703 \\

2013 & 17705 & 90172 & 9243.861 & 17705 & 8461.139 & 0.658 & 0.019 & 0.676 \\

2016 & 26159 & 143862 & 10349.479 & 26159 & 15809.521 & 0.868 & 0.0305 & 0.899 \\

2019 & 31436 & 180700 & 12357.104 & 31436 & 19078.897 & 0.874 & 0.032 & 0.906 \\

2020 & 31800 & 183342 & 12502.285 & 31800 & 19297.715 & 0.874 & 0.019 & 0.893 \\
\hline
\end{tabular}
\end{subfigure}
\caption{Particle swarm optim: topic statistics and knowledge temperature evolution}
\label{fig:199411215_chart}
\end{figure}

\begin{figure}[htbp]
\centering
    \begin{subfigure}{\textwidth}
    \begin{minipage}[t]{0.5\textwidth}
    \centering
    \includegraphics[width = 0.9\linewidth]{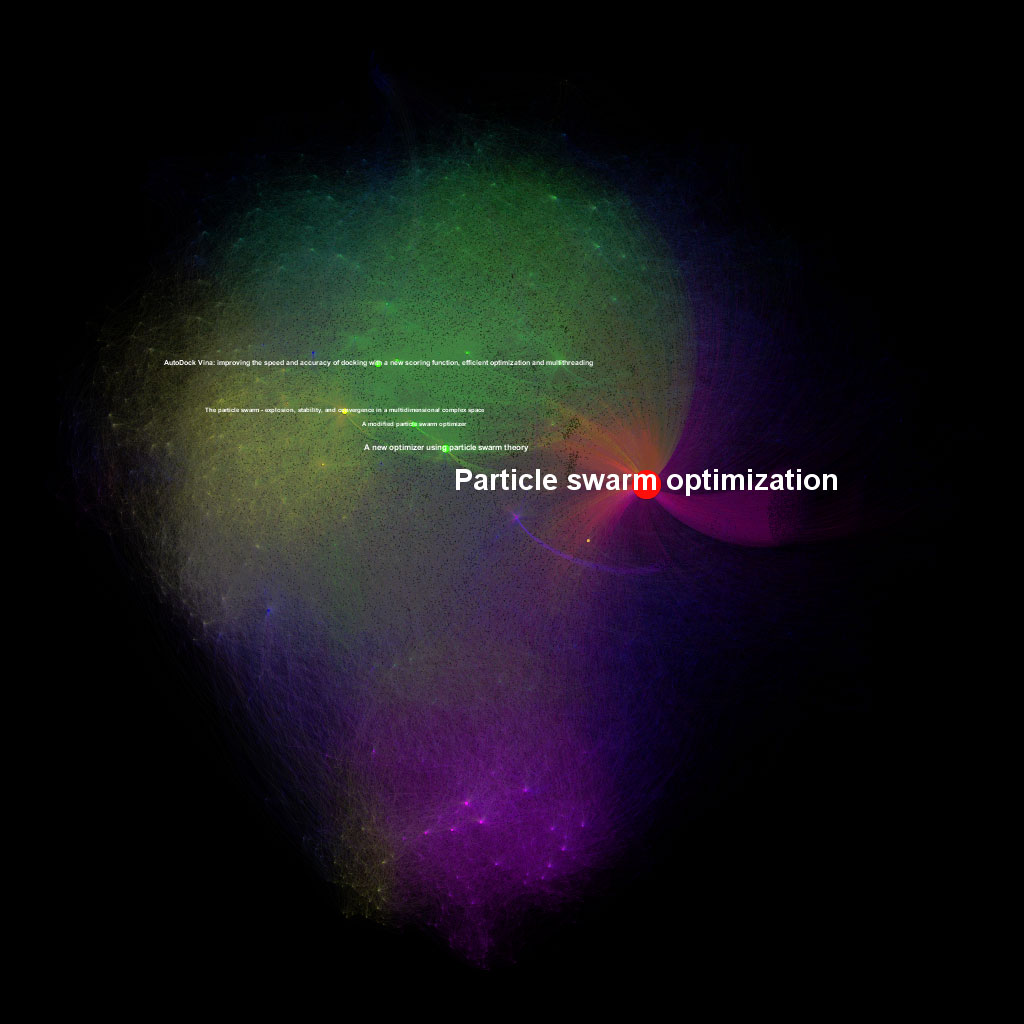}
    \end{minipage}
    \begin{minipage}[t]{0.5\textwidth}
    \centering
    \includegraphics[width = 0.9\linewidth]{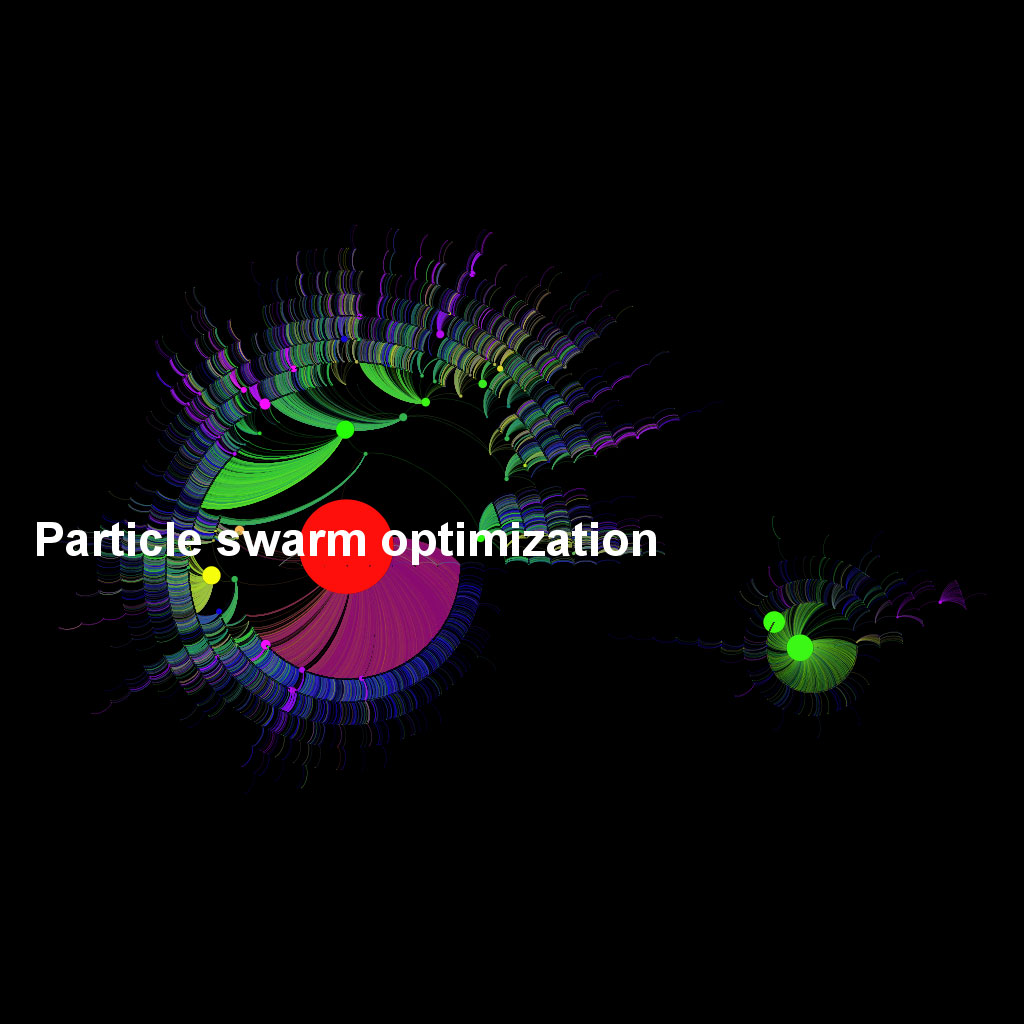}
    \end{minipage}
    \end{subfigure}

    \vspace{5mm}

    \begin{subfigure}{0.6\textwidth}
    \centering
    \includegraphics[width = \linewidth]{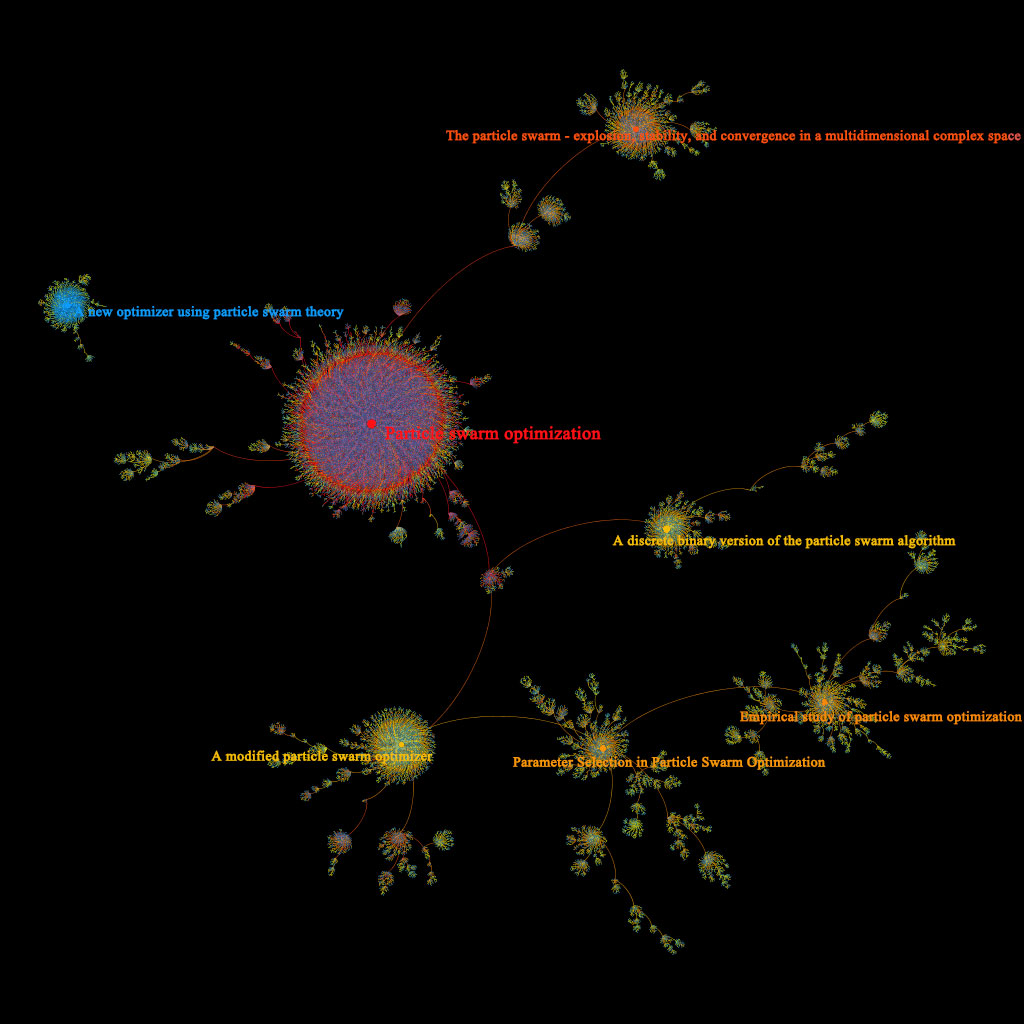}
    \end{subfigure}
    \begin{subfigure}{0.35\textwidth}
    \centering
    \begin{minipage}{\textwidth}
    \centering
    \includegraphics[width = 0.9\linewidth]{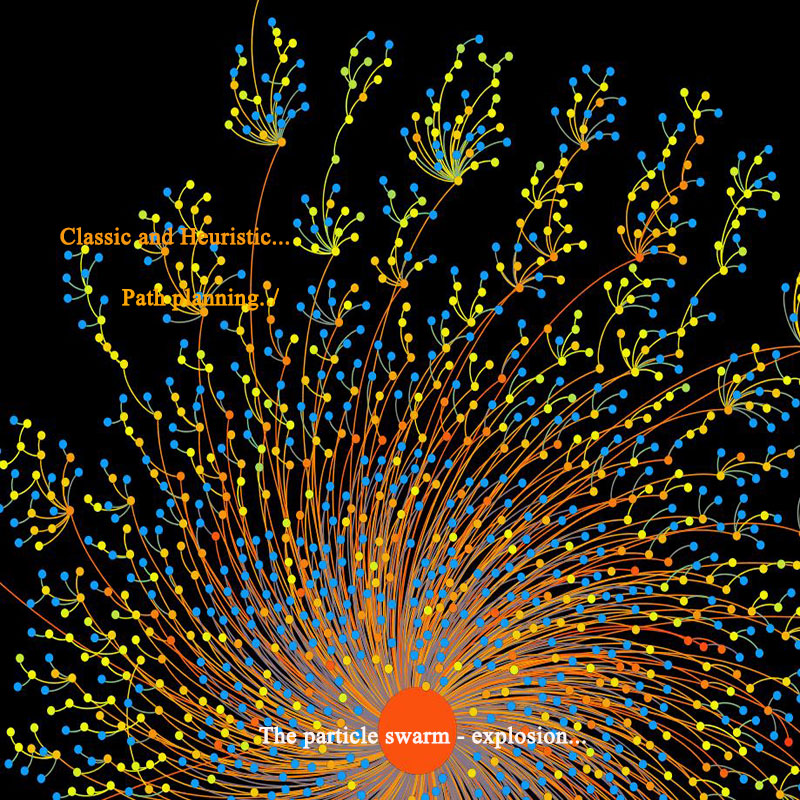}
    \end{minipage}
    \begin{minipage}{\textwidth}
    \centering
      \includegraphics[width = 0.9\linewidth]{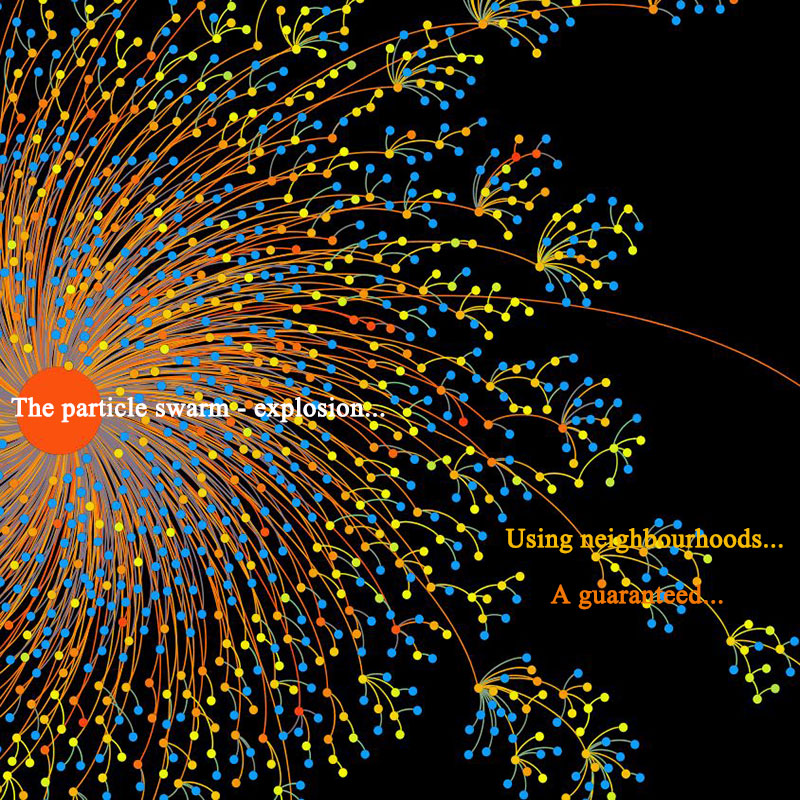}
    \end{minipage}
    \end{subfigure}
\caption{Particle swarm optim: Galaxy map and current skeleton tree. Papers with more than 1700 in-topic citations are labelled by title in the skeleton tree. Except the pioneering work, corresponding nodes' size is amplified by 5 times.}
\label{fig:199411215-2020}
\end{figure}

\begin{figure}[htbp]
\begin{subfigure}{\textwidth}
\begin{minipage}[t]{0.33\linewidth}
\includegraphics[width = \linewidth]{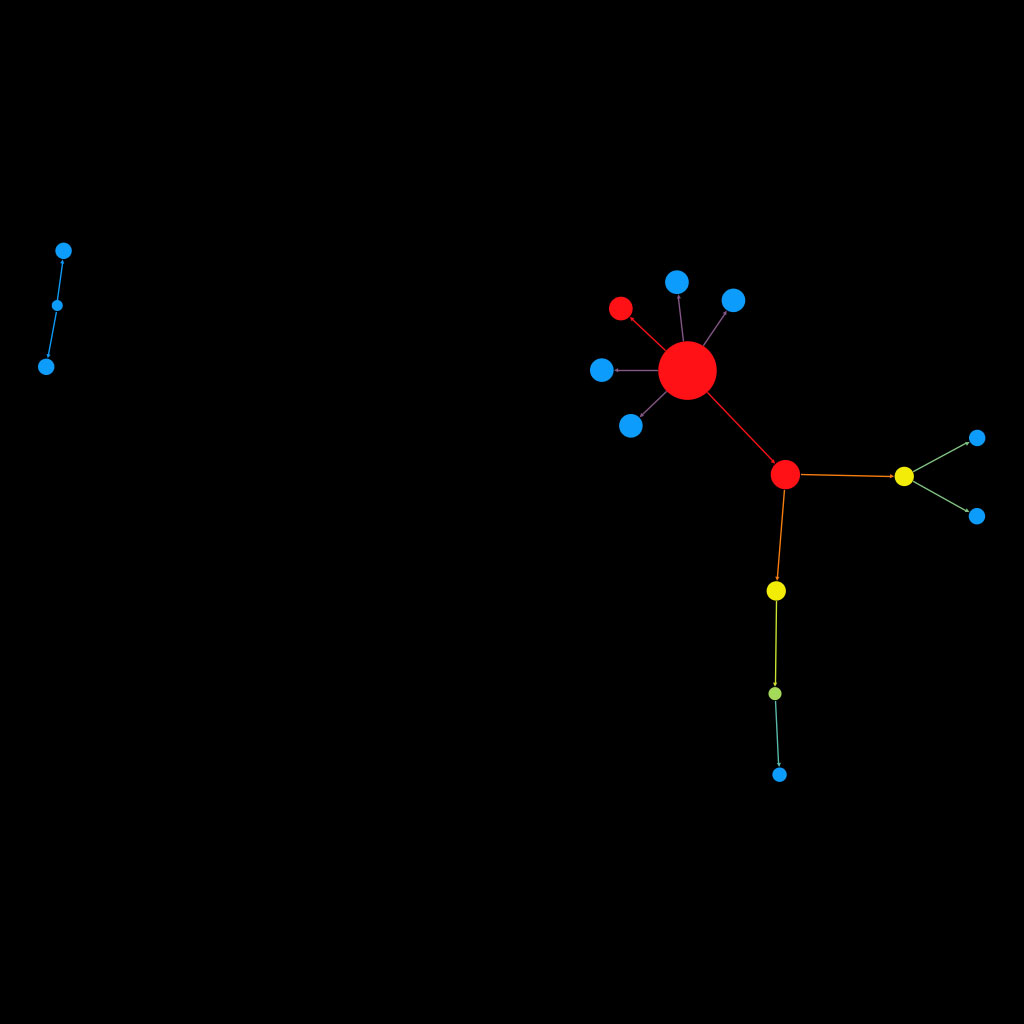}
\caption{Skeleton tree until 1998}
\end{minipage}
\begin{minipage}[t]{0.33\linewidth}
\includegraphics[width = \linewidth]{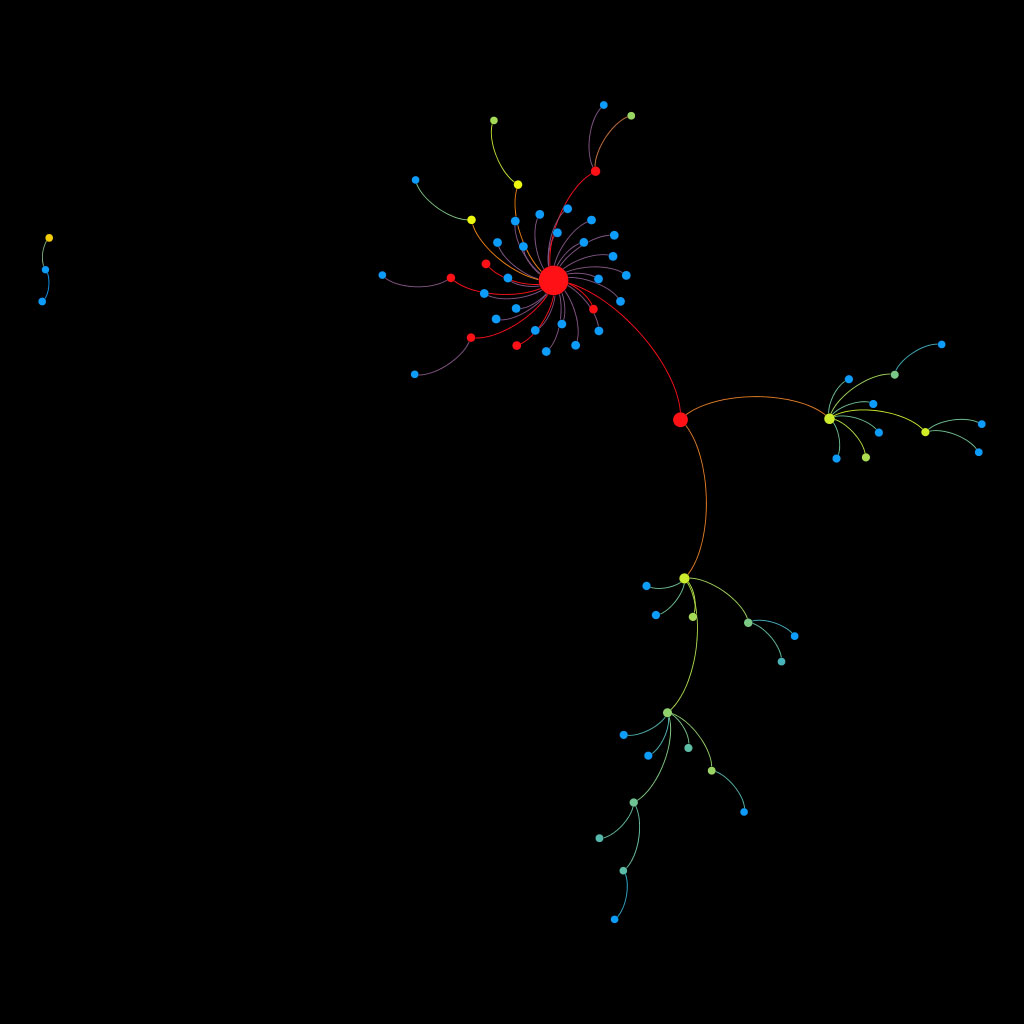}
\caption{Skeleton tree until 2001}
\end{minipage}
\begin{minipage}[t]{0.33\linewidth}
\includegraphics[width = \linewidth]{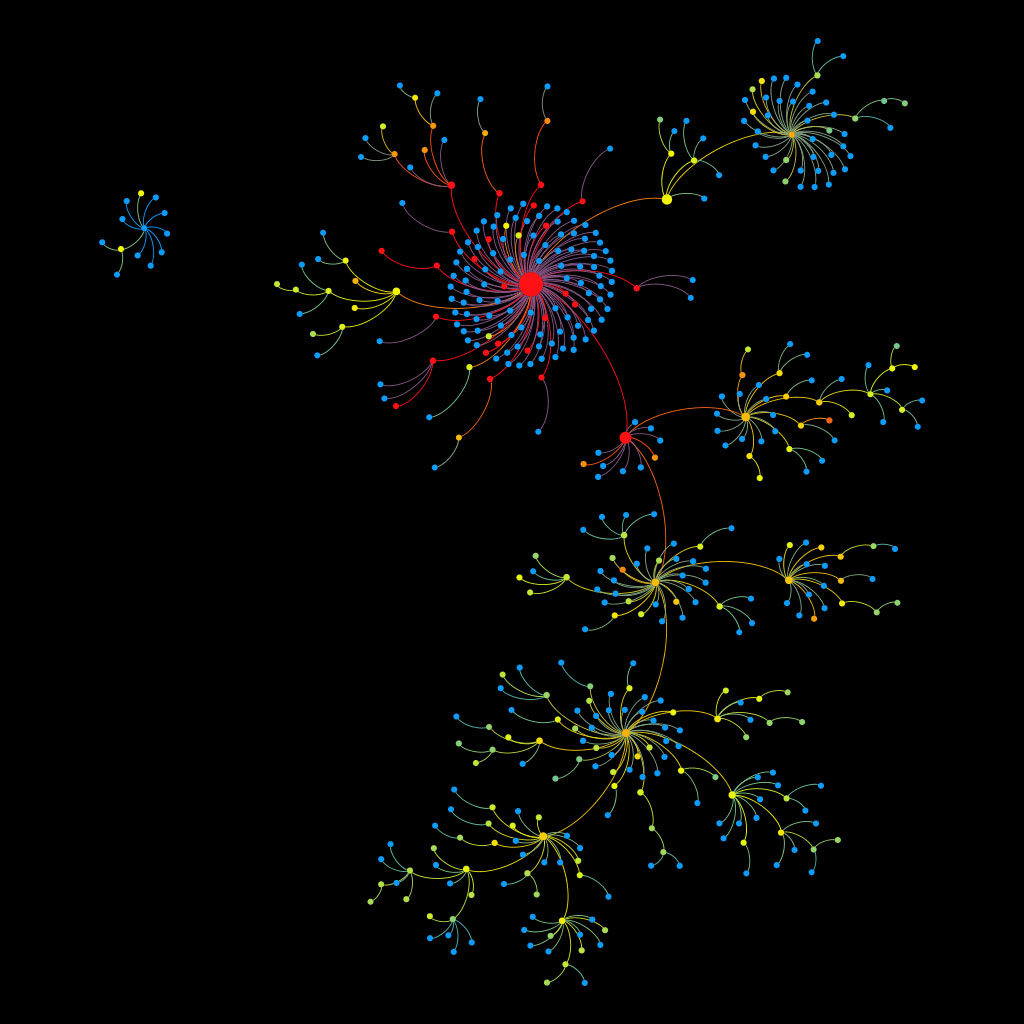}
\caption{Skeleton tree until 2004}
\end{minipage}
\end{subfigure}
\vspace{2mm}
\begin{subfigure}{\textwidth}
\begin{minipage}[t]{0.33\linewidth}
\includegraphics[width = \linewidth]{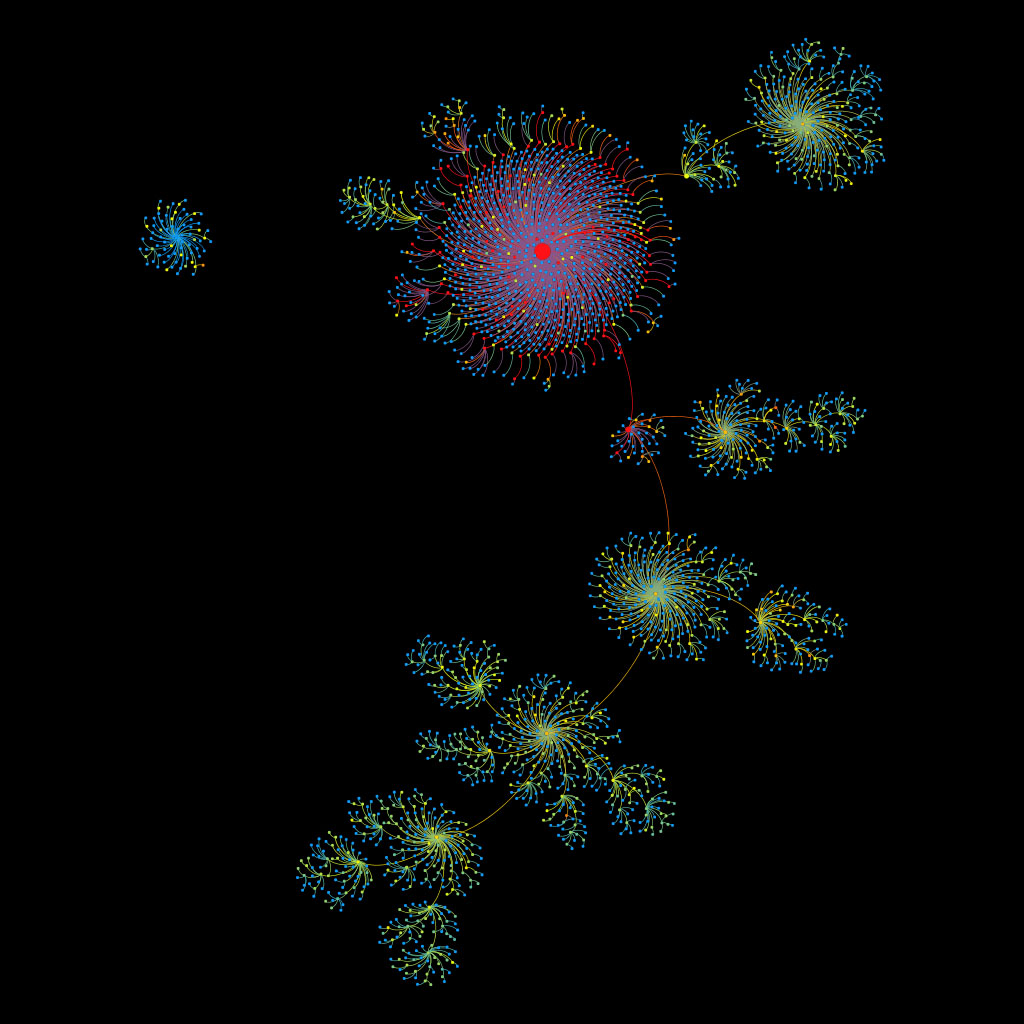}
\caption{Skeleton tree until 2007}
\end{minipage}
\begin{minipage}[t]{0.33\linewidth}
\includegraphics[width = \linewidth]{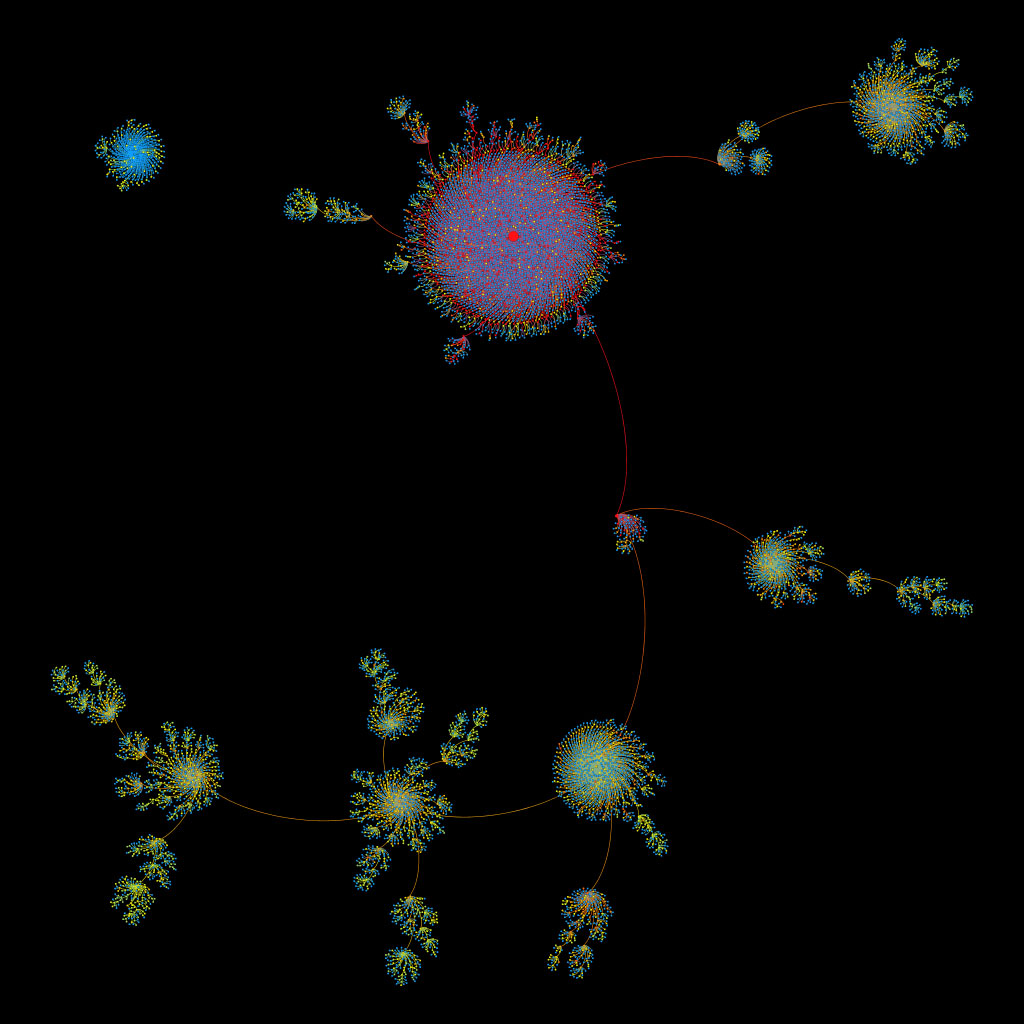}
\caption{Skeleton tree until 2010}
\end{minipage}
\begin{minipage}[t]{0.33\linewidth}
\includegraphics[width = \linewidth]{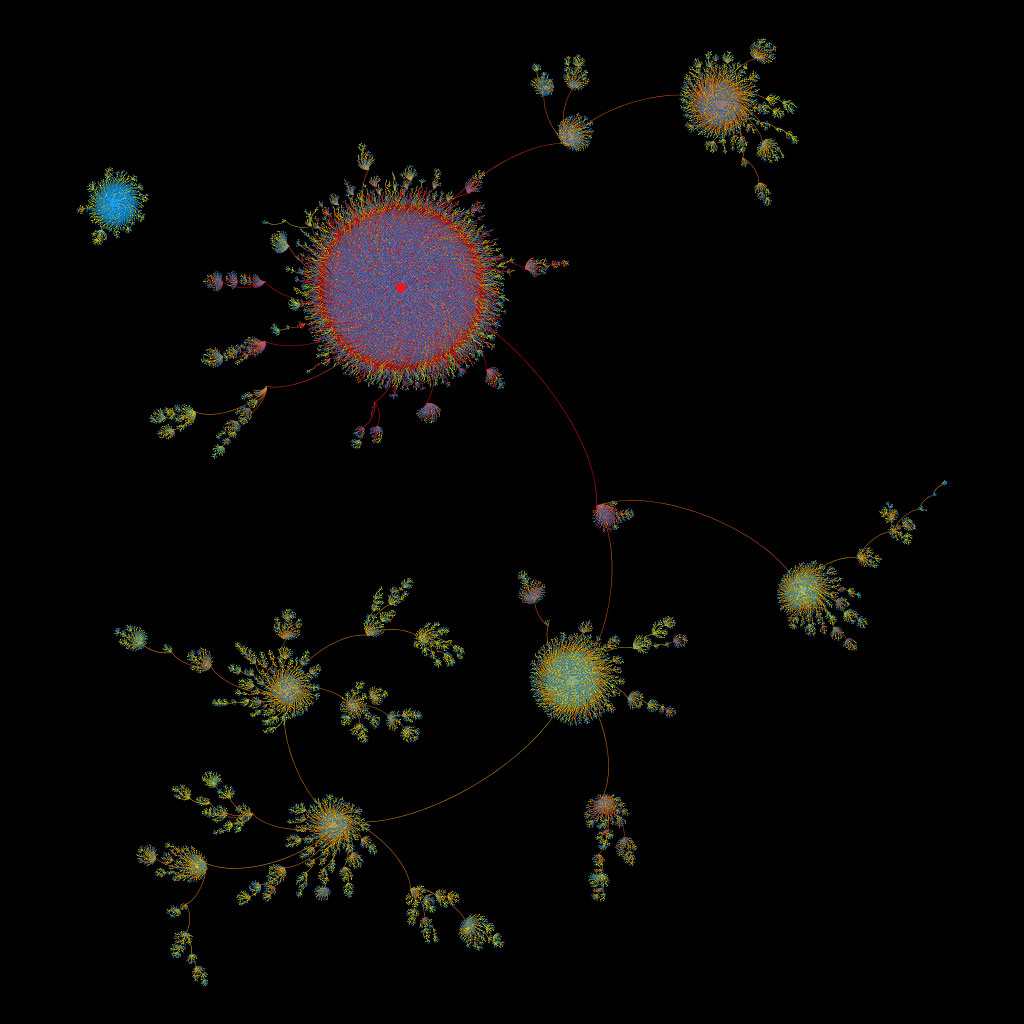}
\caption{Skeleton tree until 2016}
\end{minipage}
\end{subfigure}
\caption{Particle swarm optim: Skeleton tree evolution}
\label{fig:199411215-tree_evo}
\end{figure}

The topic gained popularity and expanded its impact steadily from its birth until around 2004 largely under the joint efforts of the pioneering work and several well-developed child papers published before 2000, namely `A modified particle swarm optimizer',`Empirical study of particle swarm optimization', and `Parameter Selection in Particle Swarm Optimization'. It is also these prominent child papers within the topic that lay the foundation of the skeleton tree (Fig. \ref{fig:199411215-2020}). Another 2 influential younger child papers, `Comparing inertia weights and constriction factors in particle swarm optimization' published in 2000 and `The particle swarm - explosion, stability, and convergence in a multidimensional complex space' published in 2002, opened up a smaller sub-topic, which is visualized as the smaller major arm that extend from the central cluster. Their arrival ensured topic's thriving in its first 10 years, which is reflected by a rising $T_{growth}^t$ and a relatively high $T_{structure}^t$ during that period. In comparison, nothing remarkable happened in the following 5 years. Papers published during this period simply extended the established sub-topics. As a result, $T^t$ and its components stagnated (Fig. \ref{fig:199411215_chart}). Next, the machine learning wave revitalized the topic. Starting from somewhere between 2010 and 2013, novel research focuses have been derived from the older sub-topics and some of them already had certain development (Fig. \ref{fig:199411215-tree_evo} (e,f)). This phenomenon is illustrated by the increasingly rich end structure of skeleton tree. In addition, annual publication number reached record high for the year 2014. This trend resulted in $T^t$'s surge shortly after. As the tendency is cooling down now, so is the topic. Overall, this is a topic waken up by the AI booming. \\

\noindent There is a small cold cluster detached from the topic majority (Fig.\ref{fig:199411215-tree_evo} in the top-right of (f)). This cluster is led by popular child paper `A new optimizer using particle swarm theory' published in the same year as the pioneering work. Thus the two papers probably have different focus even though they bear resemblance in their ideas. Their divergences cause their separation in the skeleton tree and their distinct knowledge temperatures. The separated skeleton tree also accords with topic's galaxy map representation where it seems to be split into 2 parties (Fig. \ref{fig:199411215-2020}). \\

\noindent Now we closely examine the internal heat distribution together with its latest skeleton tree (Fig. \ref{fig:199411215-2020}). After 25 years of development, the heat has already fulled diffused to the entire topic, as most popular child papers that founded recent research focuses have a knowledge temperature above average. They are the topic's heat sources. It is clear that node knowledge temperature decreases globally as the articles are located farther away from multiple research centers. This fits the general rule "the older the hotter" (Fig. 5(l)). Note that the colder average knowledge temperatures among the oldest articles is caused by the "cold" popular child paper mentioned in the previous paragraph and the relatively independent research branch it leads.  This child paper is also responsible for the drastic average temperature plunge in most-cited papers (Fig. \ref{fig:citation_T}(l)). Besides, the blue nodes that surround the pioneering work and  popular child papers in non-trivial clusters are papers with few or without any in-topic citations. However, the general rule is violated even if we do not consider this "cold" research branch. For example, `Path planning for mobile robot using the particle swarm optimization with mutation operator' is slightly colder than its child paper `Classic and Heuristic Approaches in Robot Motion Planning A Chronological Review'. The former is coloured yellow-orange and the latter orange. Their temperature difference is mainly due to their different research focus, which is reflected by their distinct citations. Similarly, paper `Using neighbourhoods with the guaranteed convergence PSO' is also colder than its child paper `A guaranteed convergence dynamic double particle swarm optimizer'. The former is coloured orange and the latter orange-red. These counter examples illustrate that the general rule "the older the hotter" is not robust. \\

\noindent We observe in addition certain clustering effect in the skeleton tree. For example, almost all child papers of `A self-generating fuzzy system with ant and particle swarm cooperative optimization' deal with fuzzy rule (Table \ref{tab:199411215-clustering}). This confirms the effectiveness of our skeleton tree extraction algorithm.\\

\begin{table}
    \centering
    \begin{tabular}{p{15cm} p{1cm}}
        \hline
        title & year\\
        \hline
       A self-generating \textcolor{red}{fuzzy system} with ant and particle swarm cooperative optimization & 2009 \\

         ANFIS modelling of a twin rotor system using particle swarm optimisation and RLS & 2010\\

        Improving \textcolor{orange}{fuzzy knowledge} integration with particle swarmoptimization & 2010\\

         Designing \textcolor{orange}{Fuzzy-Rule-Based Systems} Using Continuous Ant-Colony Optimization & 2010\\

        \textcolor{orange}{Fuzzy Neural Networks} Learning by Variable-Dimensional Quantum-behaved Particle Swarm Optimization Algorithm & 2013\\

        Modeling and OnLine Control of Nonlinear Systems using \textcolor{orange}{Neuro- Fuzzy} Learning tuned by Metaheuristic Algorithms & 2014\\
        \hline
    \end{tabular}

    \caption{Particle swarm optim: Clustering effect example. First line is the parent paper and the rest children.}
    \label{tab:199411215-clustering}
\end{table}

\subsection{Rise-fall-cycle topics}
\subsubsection{On random graphs, I}
\begin{figure}[htbp]
\centering
\begin{subfigure}[t]{0.7\linewidth}
\centering
\includegraphics[width=\linewidth]{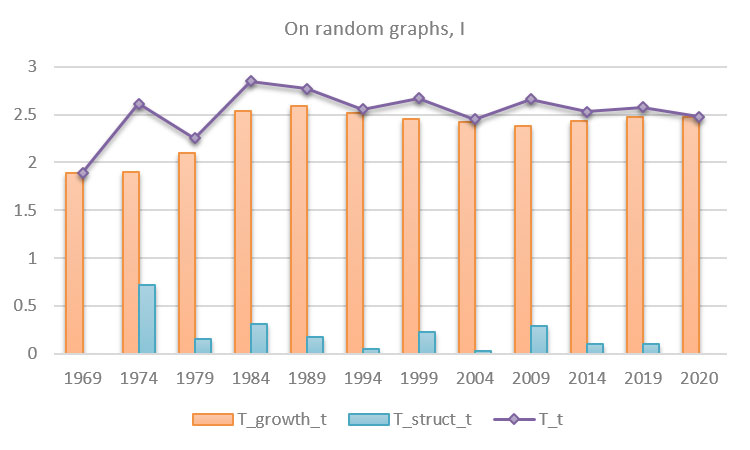}
\end{subfigure}

\begin{subfigure}[t]{\linewidth}
\centering
\begin{tabular}{ccccccccc}
\hline
year & $|V^t|$ & $|E^t|$ & $n_t$ & $V_t$ & ${UsefulInfo}^t$ & $T_{growth}^t$ & $T_{struct}^t$ & $T^t$\\
\hline
1969 & 5 & 6 & 4 & 5 & 1 & 1.892 &   & 1.892 \\

1974 & 14 & 20 & 11.167 & 14 & 2.833 & 1.898 & 0.714 & 2.612 \\

1979 & 30 & 57 & 21.606 & 30 & 8.394 & 2.102 & 0.151 & 2.253 \\

1984 & 59 & 157 & 35.23 & 59 & 23.77 & 2.535 & 0.314 & 2.849 \\

1989 & 87 & 240 & 50.815 & 87 & 36.185 & 2.592 & 0.176 & 2.768 \\

1994 & 111 & 288 & 66.871 & 111 & 44.129 & 2.513 & 0.045 & 2.558 \\

1999 & 135 & 334 & 83.416 & 135 & 51.584 & 2.45 & 0.222 & 2.672 \\

2004 & 311 & 832 & 194.093 & 311 & 116.907 & 2.426 & 0.024 & 2.45 \\

2009 & 1346 & 4172 & 856.782 & 1346 & 489.218 & 2.378 & 0.284 & 2.663 \\

2014 & 3312 & 10406 & 2063.311 & 3312 & 1248.689 & 2.43 & 0.099 & 2.529 \\

2019 & 5387 & 17095 & 3295.006 & 5387 & 2091.994 & 2.475 & 0.102 & 2.577 \\

2020 & 5389 & 17098 & 3294.798 & 5389 & 2094.202 & 2.476 & 0 & 2.476 \\
\hline
\end{tabular}
\end{subfigure}
\caption{On random graphs: topic statistics and knowledge temperature evolution}
\label{fig:176392498_chart}
\end{figure}

\begin{figure}[htbp]
\begin{subfigure}{\textwidth}
\begin{minipage}[t]{0.33\linewidth}
\includegraphics[width = \linewidth]{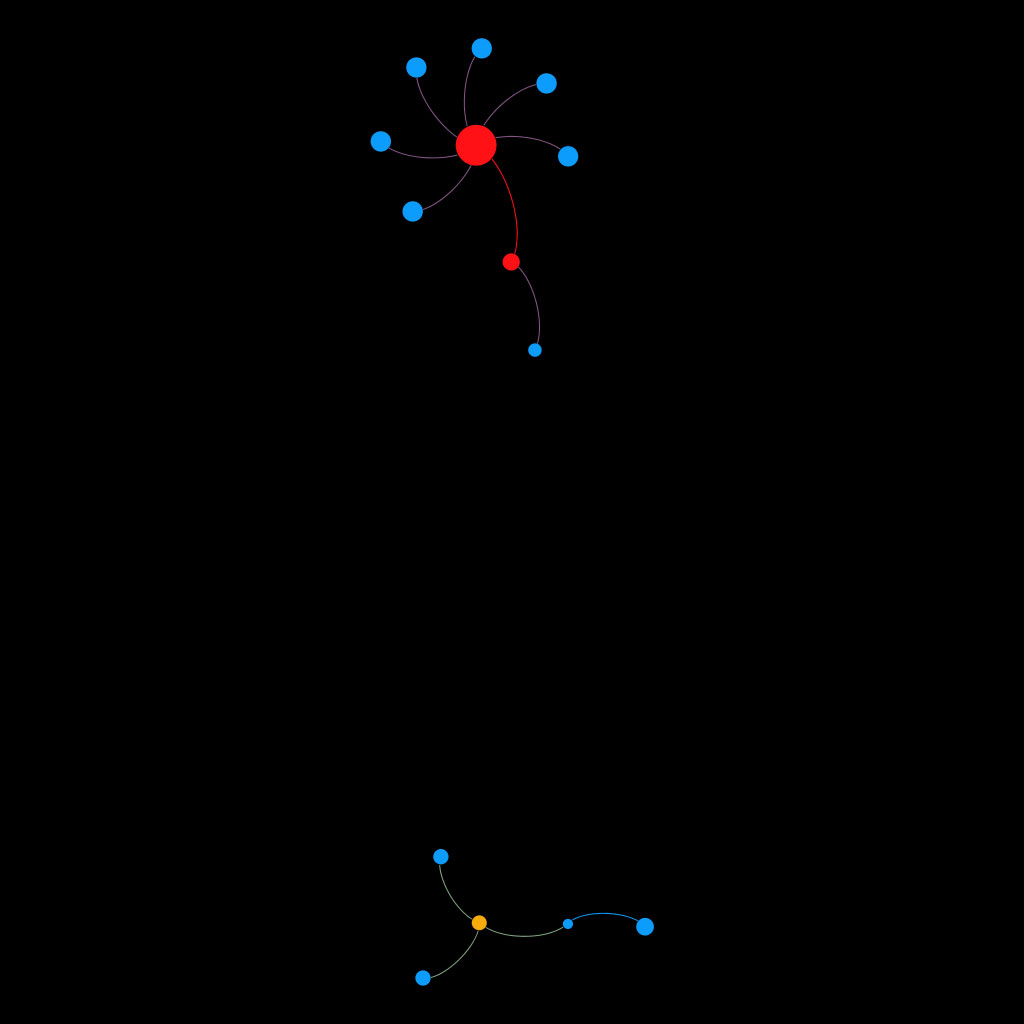}
\caption{Skeleton tree until 1974}
\end{minipage}
\begin{minipage}[t]{0.33\linewidth}
\includegraphics[width = \linewidth]{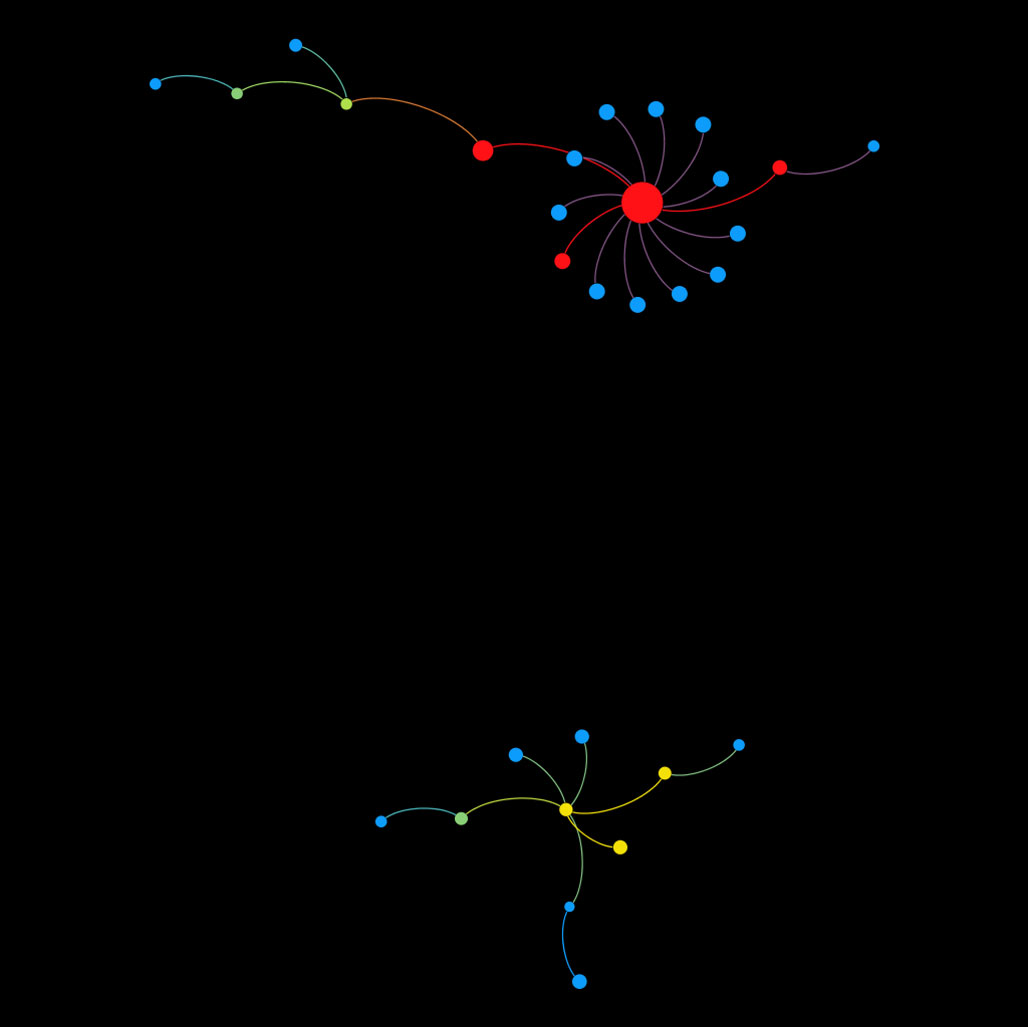}
\caption{Skeleton tree until 1979}
\end{minipage}
\begin{minipage}[t]{0.33\linewidth}
\includegraphics[width = \linewidth]{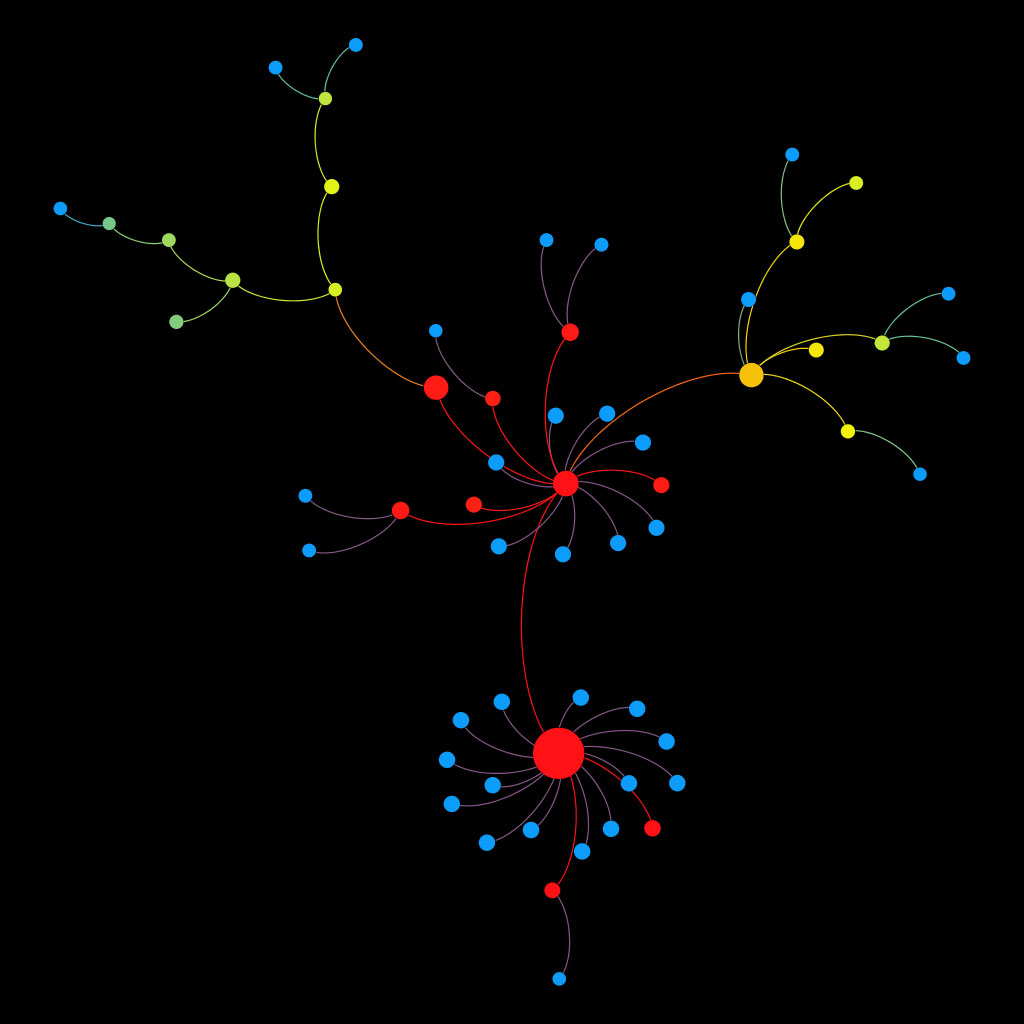}
\caption{Skeleton tree until 1984}
\end{minipage}
\end{subfigure}

\begin{subfigure}{\textwidth}
\begin{minipage}[t]{0.33\linewidth}
\includegraphics[width = \linewidth]{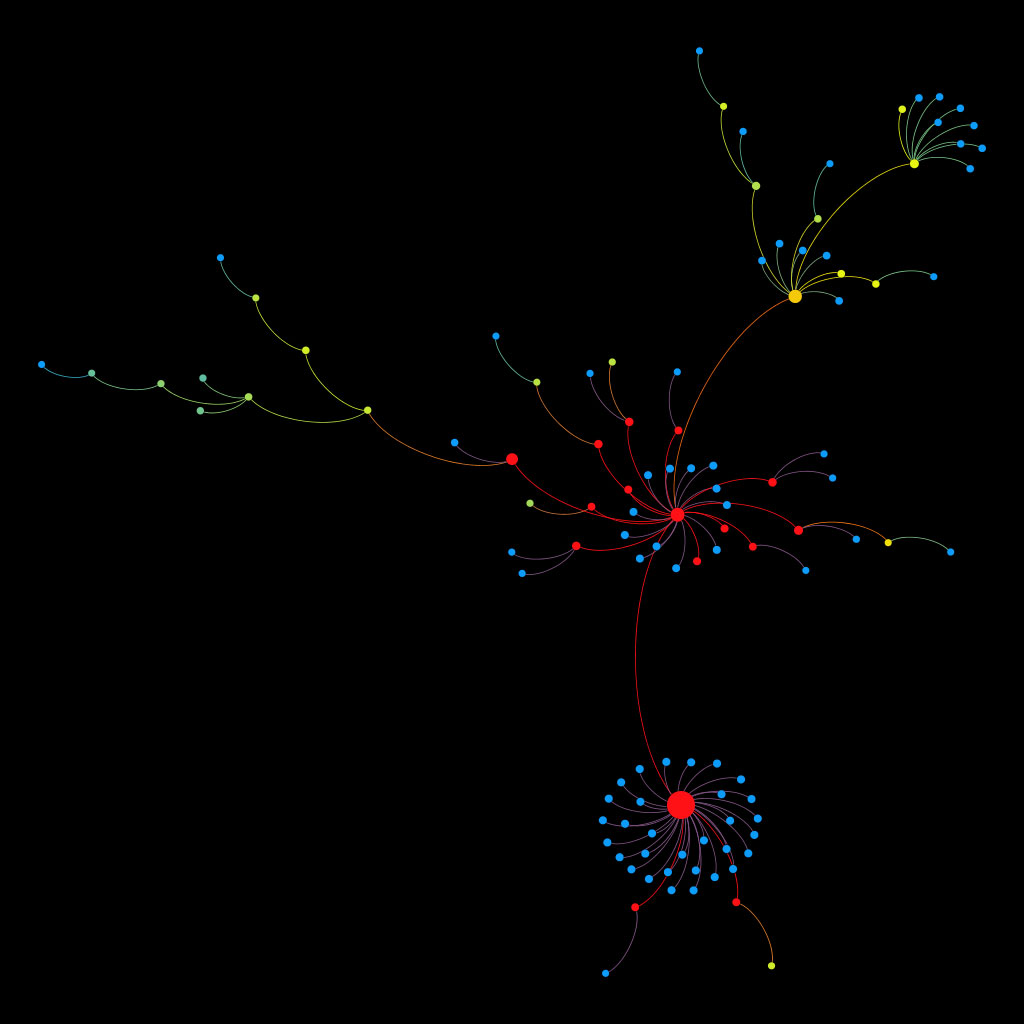}
\caption{Skeleton tree until 1994}
\end{minipage}
\begin{minipage}[t]{0.33\linewidth}
\includegraphics[width = \linewidth]{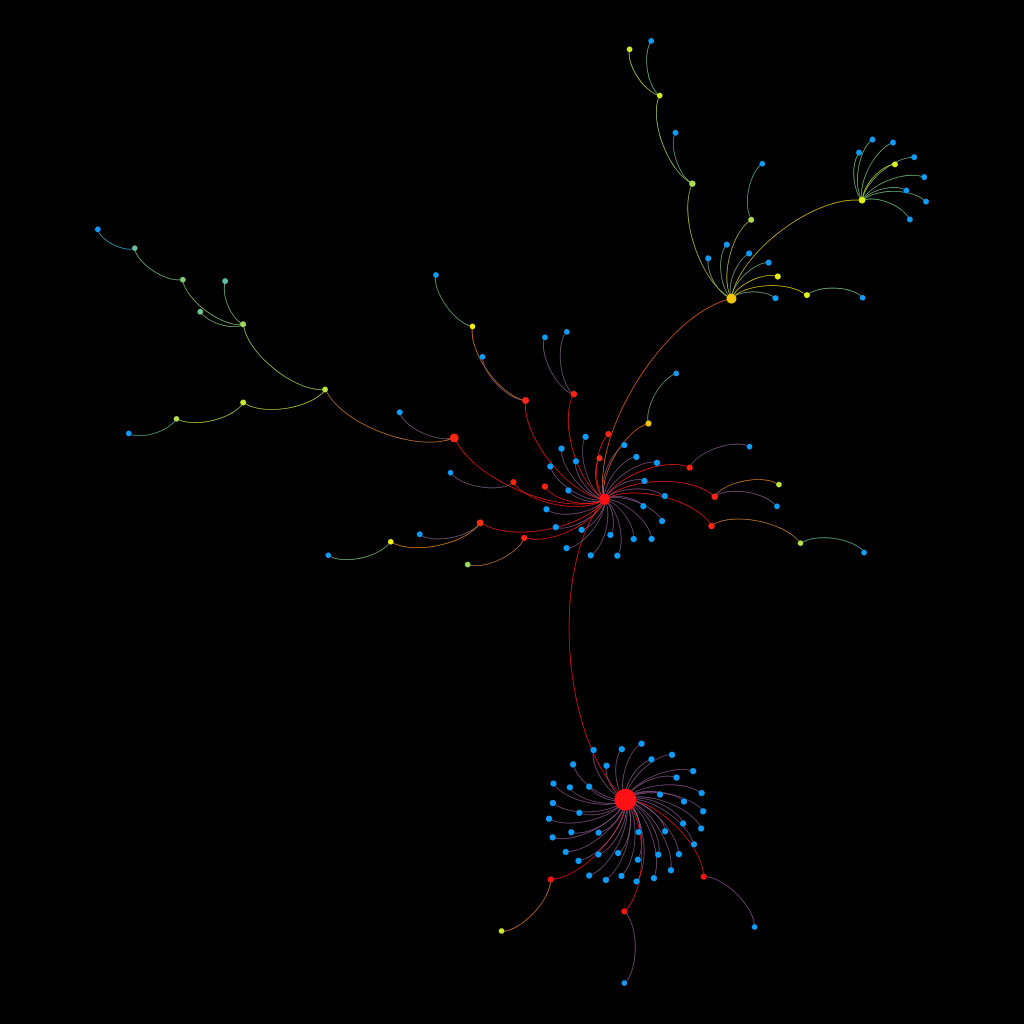}
\caption{Skeleton tree until 1999}
\end{minipage}
\begin{minipage}[t]{0.33\linewidth}
\includegraphics[width = \linewidth]{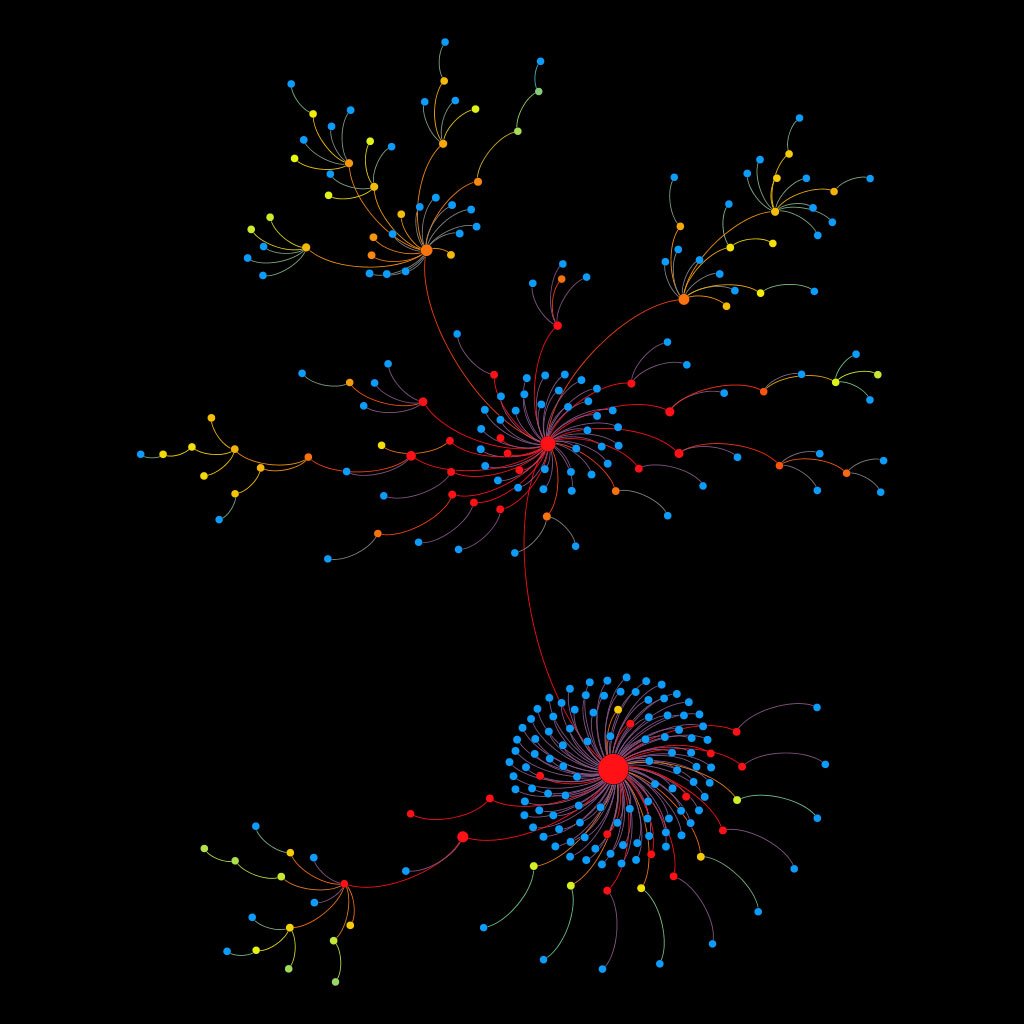}
\caption{Skeleton tree until 2004}
\end{minipage}
\end{subfigure}

\begin{subfigure}{\textwidth}
\begin{minipage}[t]{0.33\linewidth}
\includegraphics[width = \linewidth]{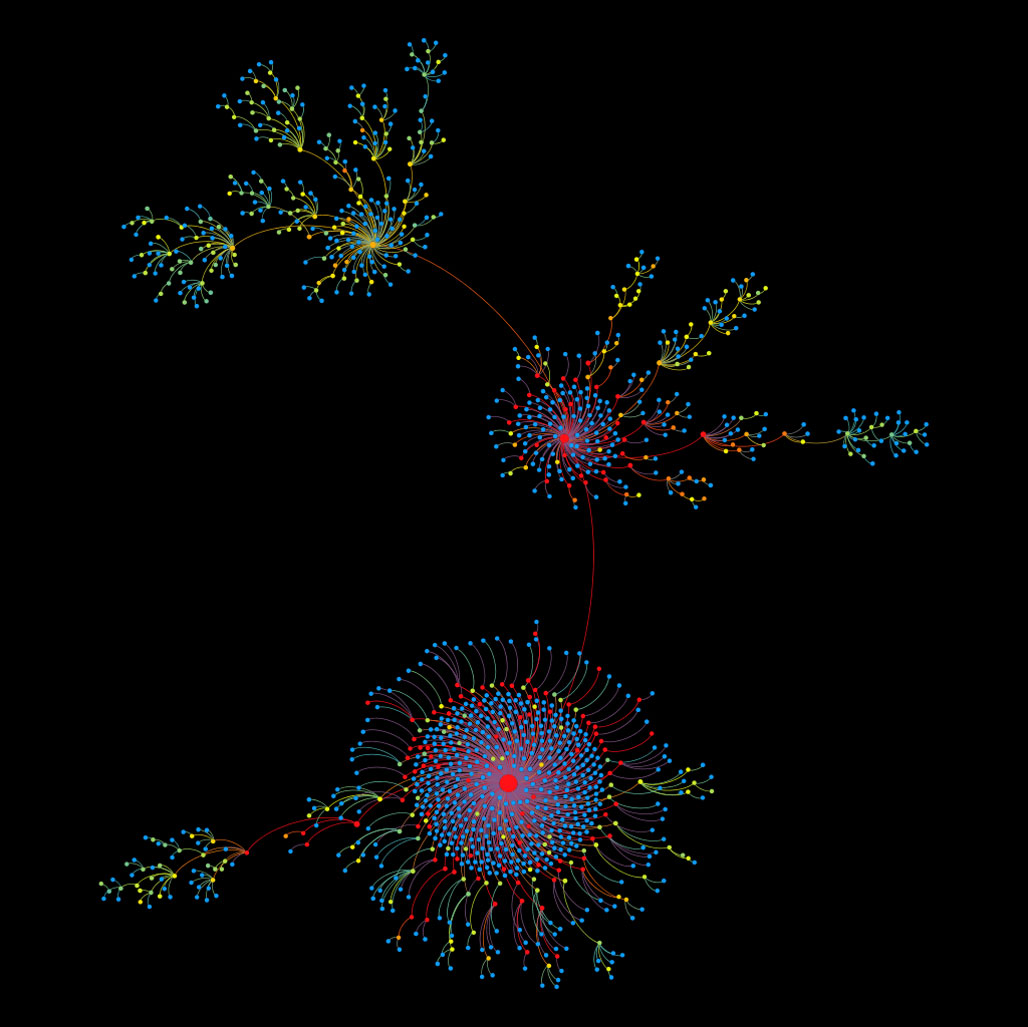}
\caption{Skeleton tree until 2009}
\end{minipage}
\begin{minipage}[t]{0.33\linewidth}
\includegraphics[width = \linewidth]{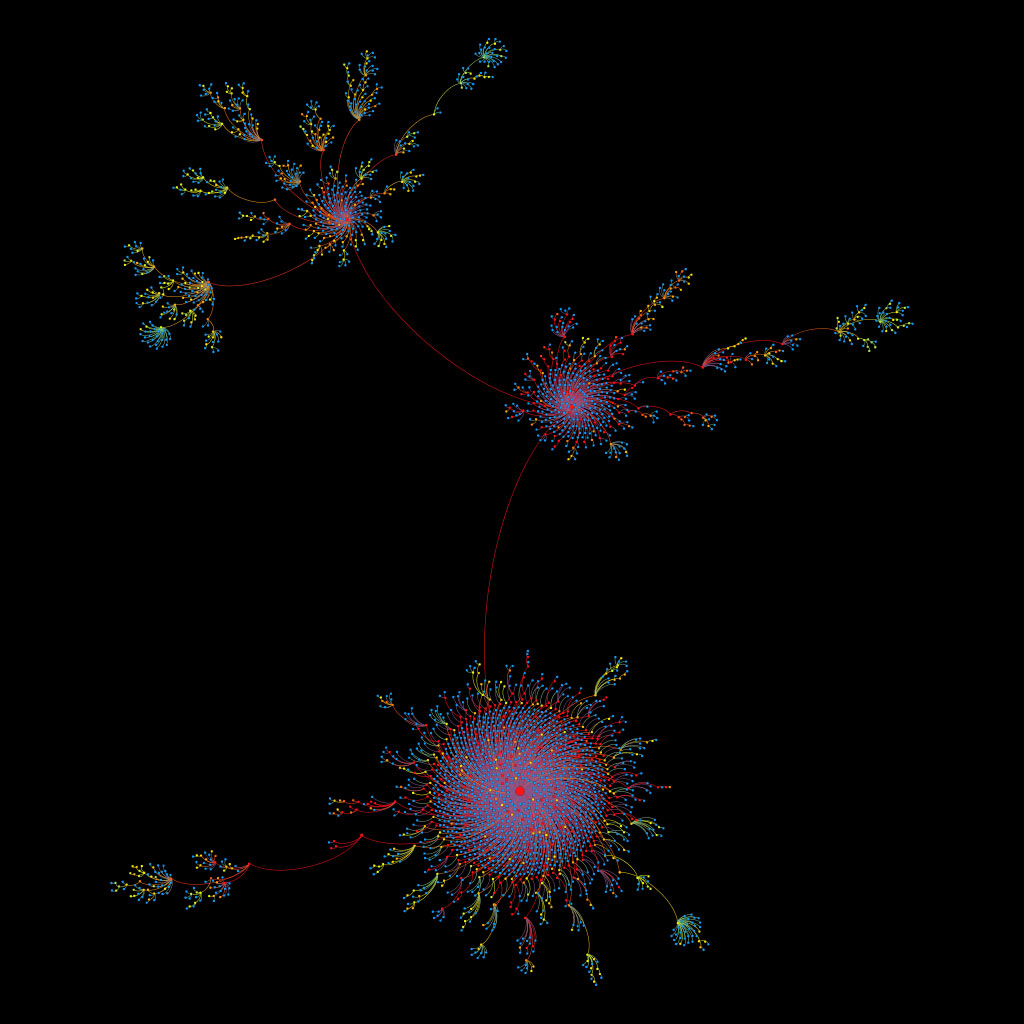}
\caption{Skeleton tree until 2014}
\end{minipage}
\begin{minipage}[t]{0.33\linewidth}
\includegraphics[width = \linewidth]{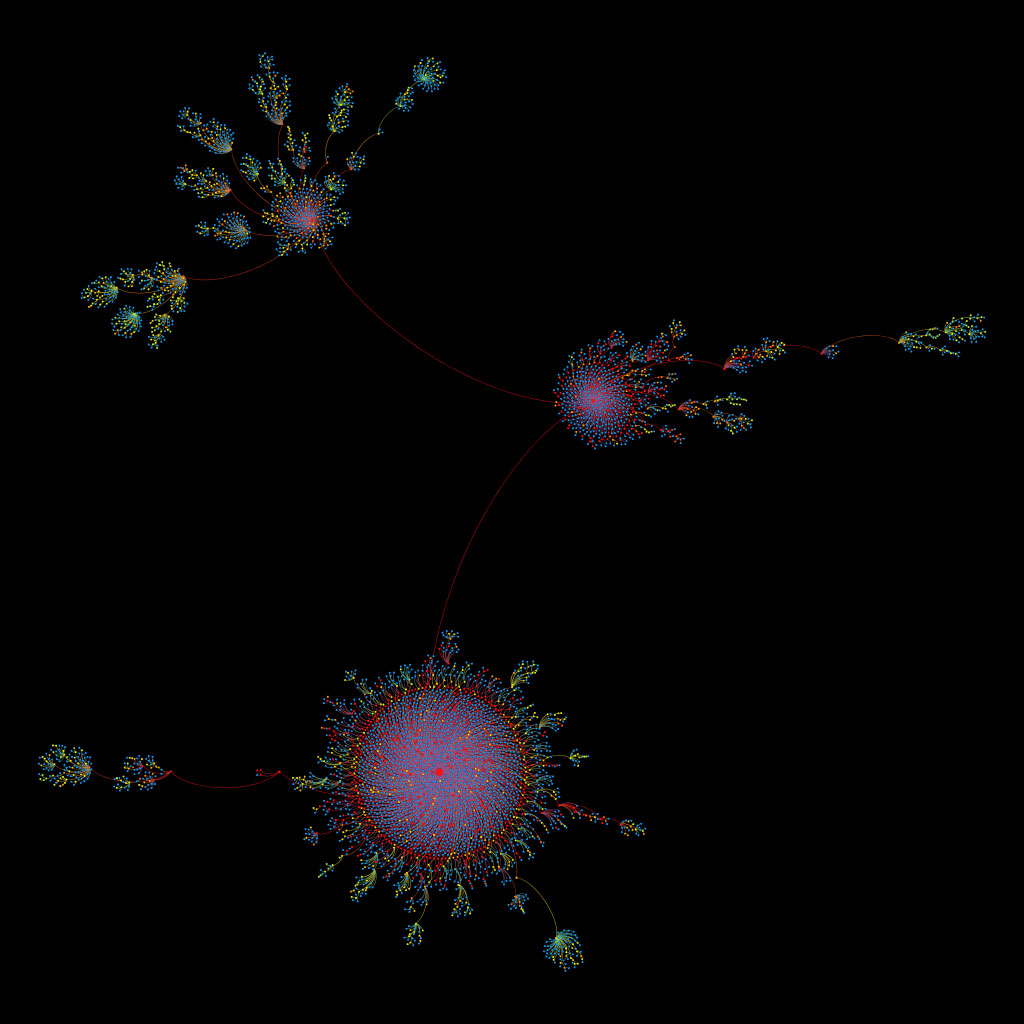}
\caption{Skeleton tree until 2019}
\end{minipage}
\end{subfigure}
\caption{On random graphs: Skeleton tree evolution}
\label{fig:176392498-tree_evo}
\end{figure}

As is shown by $T^t$ and $T_{growth}^t$, the impact and popularity evolution of this topic is a bit complicated (Fig. \ref{fig:176392498_chart}). The publication of popular child paper `On the evolution of random graphs' (OERG) in 1984 brought the first boom in the 1980s. This article combined its ancestors' ideas and successfully fused the previously separated parts in skeleton tree due to an atypical citation from an older article `On the existence of a factor of degree one of a connected random graph' (Fig. \ref{fig:176392498-tree_evo}(b,c)). This merge is the first significant evolution in knowledge structure and thus led to a spike in $T_{structure}^t$. Afterwards, the topic went relatively silent in the 1990s before a group of popular child papers came during 2001 and 2003. Among these articles, `Random graphs with arbitrary degree distributions and their applications' published in 2001 non-trivially furthered the study of OERG and introduced a new research focus into the topic, as is illustrated by the emergence of a third cluster in the skeleton tree (Fig. \ref{fig:176392498-tree_evo}(f,g)). Its followers and popular child papers, `Evolution of networks' published in 2002 and `The Structure and Function of Complex Networks' published in 2003 extended its idea and created several new research sub-fields. That is why we observe some splits derived from the young cluster  (Fig. \ref{fig:176392498-tree_evo}(g)). They successfully attracted a lot of attention in a short time and the topic has witnessed an accelerated expansion since around 2000. Together with their contribution to the topic knowledge pattern, this topic experienced another booming around 2010. Later, the topic kept its activity thanks to several young promising papers including `Measurement and analysis of online social networks' published in 2007, `Community detection in graphs' published in 2010 and `Catastrophic cascade of failures in interdependent networks' published in 2010. Although they opened up several new research orientations, there have not been a substantial subsequent development and the branches leading by them remain small in comparison to the principal clusters (Fig. \ref{fig:176392498-tree_evo}(h,f)). Consequently, they have mostly helped maintain the topic's visibility and its stable impact. \\

\begin{figure}[htbp]
\centering
    \begin{subfigure}{\textwidth}
    \begin{minipage}[t]{0.5\textwidth}
    \centering
    \includegraphics[width = 0.9\linewidth]{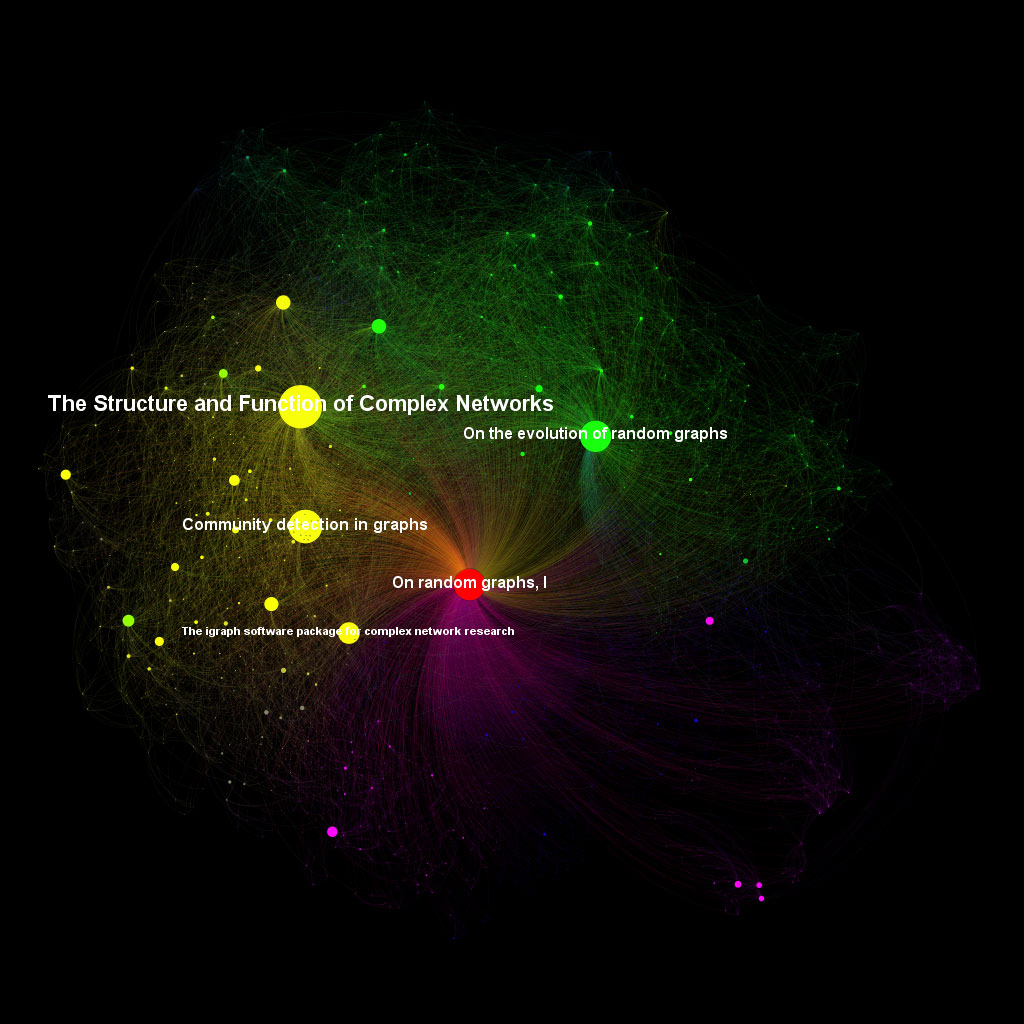}
    \end{minipage}
    \begin{minipage}[t]{0.5\textwidth}
    \centering
    \includegraphics[width = 0.9\linewidth]{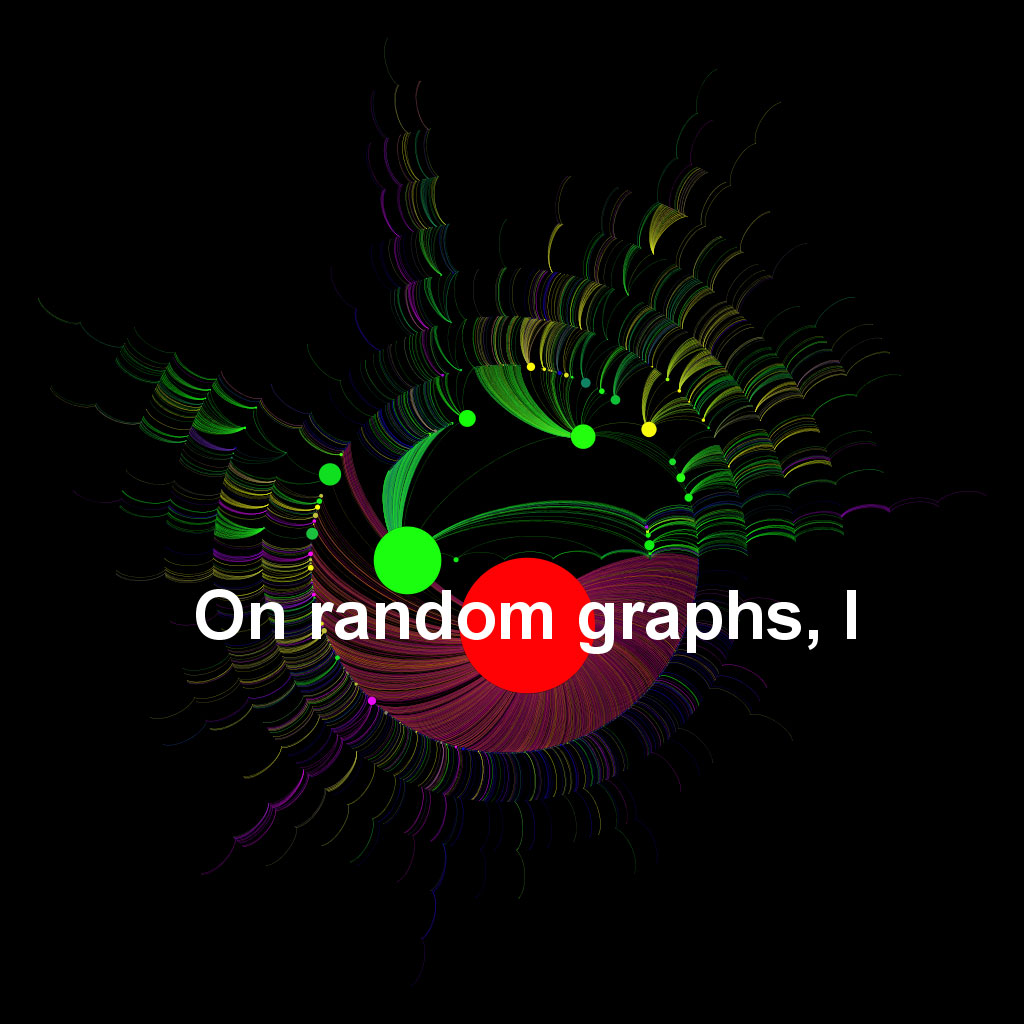}
    \end{minipage}
    \end{subfigure}

    \vspace{5mm}

    \begin{subfigure}{0.8\textwidth}
    \centering
    \includegraphics[width = 0.8\linewidth]{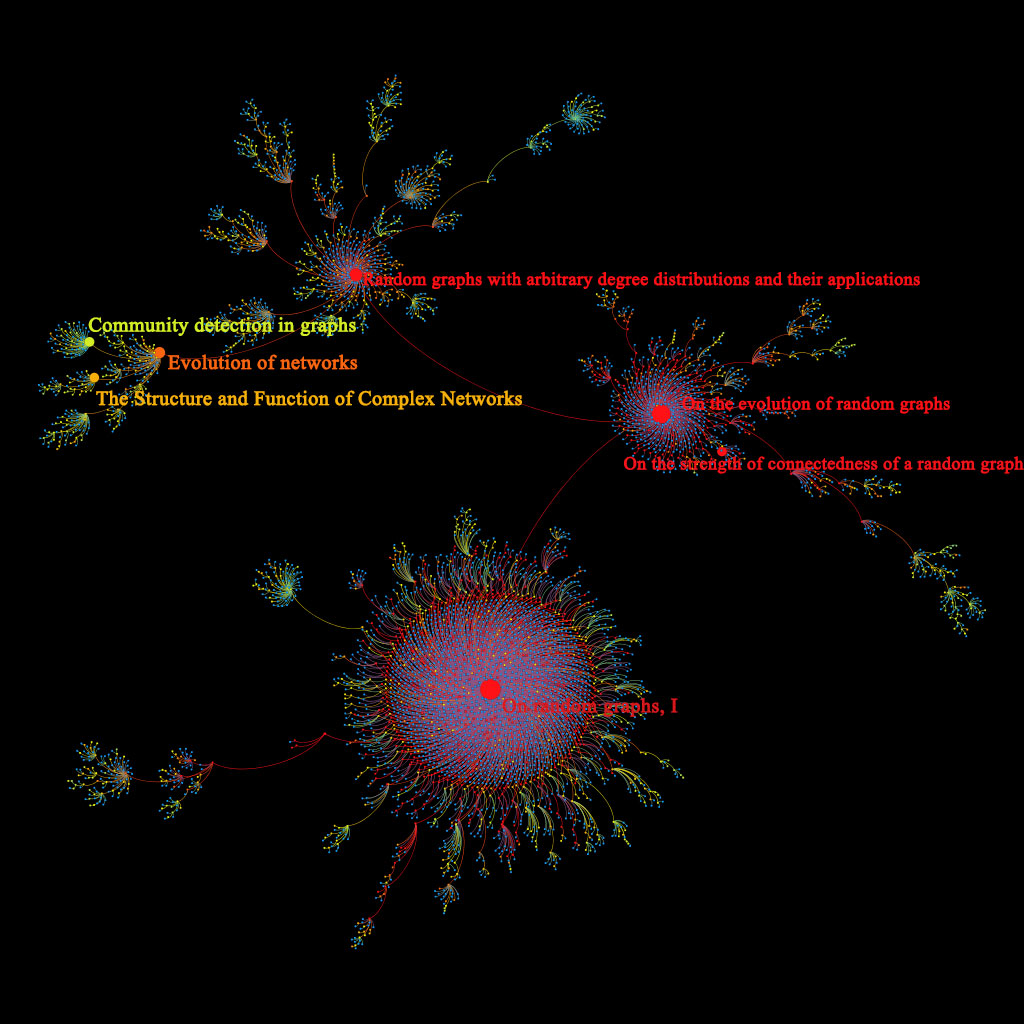}
    \end{subfigure}
\caption{On random graphs: Galaxy map and current skeleton tree. Papers with more than 200 in-topic citations are labelled by title in the skeleton tree. Except the pioneering work, corresponding nodes' size is amplified by 5 times.}
\label{fig:176392498-2020}
\end{figure}

\noindent Now we closely examine the internal heat distribution together with its latest skeleton tree (Fig. \ref{fig:176392498-2020}). The topic has a long development history. In each period, new research focuses emerged (Fig. \ref{fig:176392498-tree_evo} every line shows a period). Today, we see 3 major research focuses and their founders are all the heat sources. As the articles are located farther away from the pioneering paper or the sub-topic centers, their node knowledge temperature decreases globally. The blue nodes that surround the pioneering work and popular child papers in main clusters are papers with few or without any in-topic citations. Generally speaking, older papers are hotter than the younger (Fig. 5(m)). In comparison with other scientific topics, knowledge temperature fluctuates more among the "middle-aged" papers. This phenomenon is in line with the up and downs the topic experienced during their publication period. Besides, we also observe a general rule  "the more influential the hotter" in the topic (Fig. \ref{fig:citation_T}(m)) as the most-cited child papers are among the hottest articles. However, this rule is only robust for the most eminent child papers. \\

\noindent We observe in addition certain clustering effect in the skeleton tree (Table \ref{tab:176392498-clustering}). For example, all child papers of `False Beliefs in Unreliable Knowledge Networks' probe into knowledge network. This confirms the effectiveness of our skeleton tree extraction algorithm. Moreover, the small group was born in 2017, suggesting that their research focus, knowledge network, may be one of the latest hotspots within the topic.\\

\begin{table}
    \centering
    \begin{tabular}{p{15cm} p{1cm}}
        \hline
        title & year\\
        \hline
        False Beliefs in Unreliable \textcolor{red}{Knowledge Networks} & 2017 \\

         Communication Policies in \textcolor{orange}{Knowledge Networks} & 2018\\

        Experts in \textcolor{orange}{Knowledge Networks}: Central Positioning and Intelligent Selections & 2018\\

        How to facilitate \textcolor{orange}{knowledge} diffusion in complex networks: The roles of network structure, knowledge role distribution and selection rule & 2019\\
        \hline
    \end{tabular}

    \caption{On random graphs: Clustering effect example. First line is the parent paper and the rest children.}
    \label{tab:176392498-clustering}
\end{table}

\subsubsection{Collective dynamics of `small-world' networks}

\begin{figure}[htbp]
\centering
\begin{subfigure}[t]{0.7\linewidth}
\centering
\includegraphics[width=\linewidth]{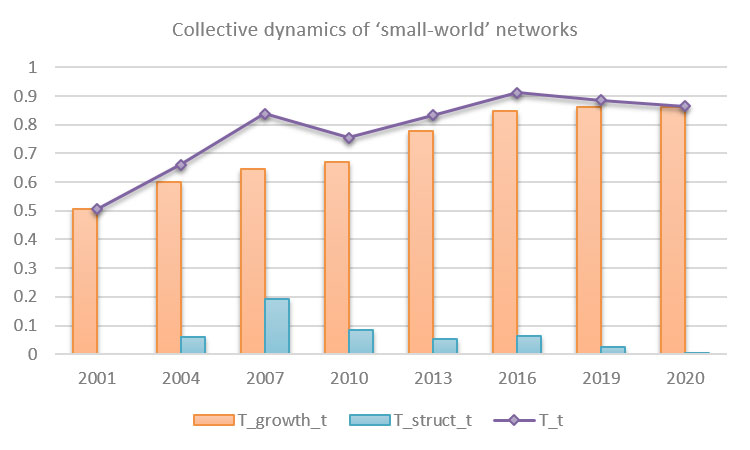}
\end{subfigure}

\begin{subfigure}[t]{\linewidth}
\centering
\begin{tabular}{ccccccccc}
\hline
year & $|V^t|$ & $|E^t|$ & $n_t$ & $V_t$ & ${UsefulInfo}^t$ & $T_{growth}^t$ & $T_{struct}^t$ & $T^t$\\
\hline
2001 & 246 & 754 & 169.095 & 246 & 76.905 & 0.505 &   & 0.505 \\

2004 & 1404 & 7192 & 812.506 & 1404 & 591.494 & 0.6 & 0.059 & 0.66 \\

2007 & 4209 & 26584 & 2260.505 & 4209 & 1948.495 & 0.647 & 0.192 & 0.839 \\

2010 & 8517 & 58620 & 4415.143 & 8517 & 4101.857 & 0.67 & 0.085 & 0.755 \\

2013 & 13998 & 104667 & 6245.327 & 13998 & 7752.673 & 0.779 & 0.054 & 0.833 \\

2016 & 20221 & 158518 & 8280.879 & 20221 & 11940.121 & 0.848 & 0.062 & 0.91 \\

2019 & 25313 & 204644 & 10197.242 & 25313 & 15115.759 & 0.863 & 0.023 & 0.886 \\

2020 & 25548 & 206643 & 10288.162 & 25548 & 15259.839 & 0.863 & 0.001 & 0.863 \\
\hline
\end{tabular}
\end{subfigure}
\caption{small-world: topic statistics and knowledge temperature evolution}
\label{fig:240908848_chart}
\end{figure}

\begin{figure}[htbp]
\begin{subfigure}{\textwidth}
\begin{minipage}[t]{0.33\linewidth}
\includegraphics[width = \linewidth]{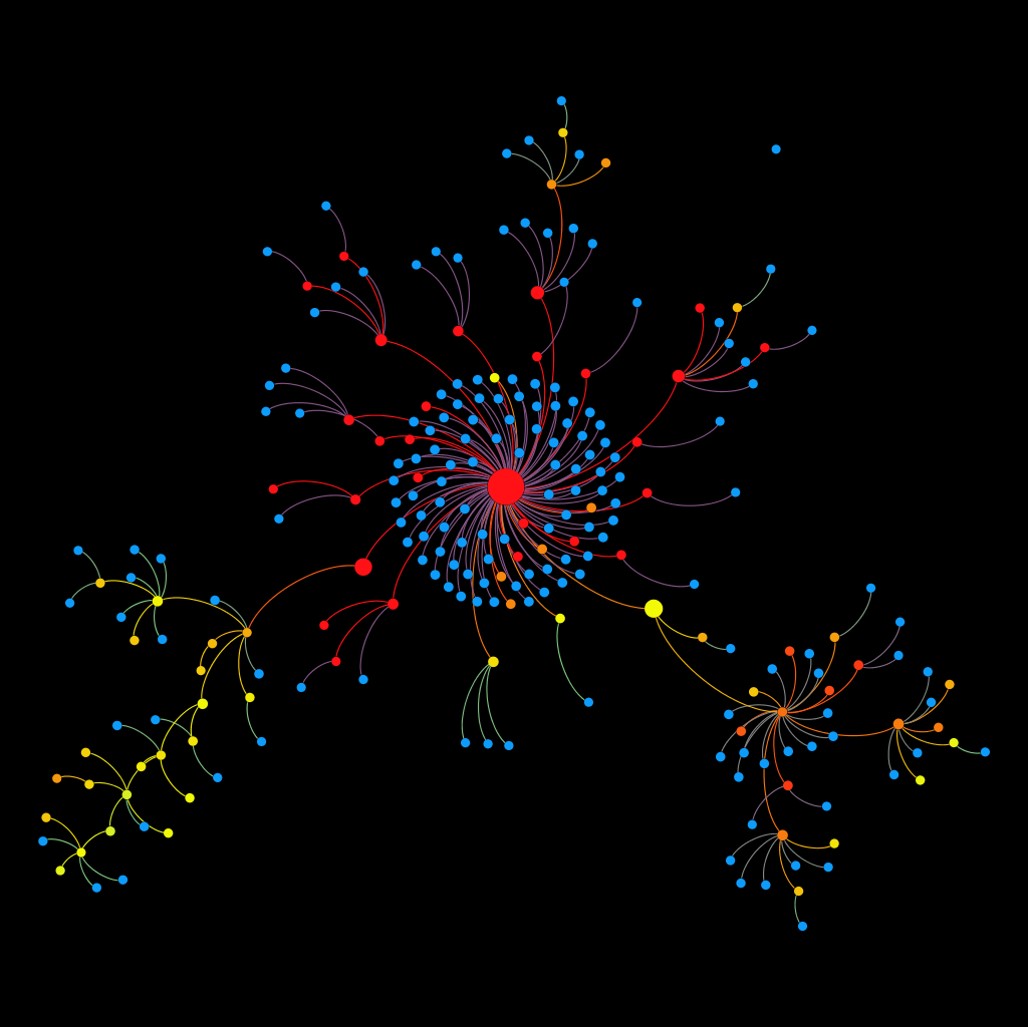}
\caption{Skeleton tree until 2001}
\end{minipage}
\begin{minipage}[t]{0.33\linewidth}
\includegraphics[width = \linewidth]{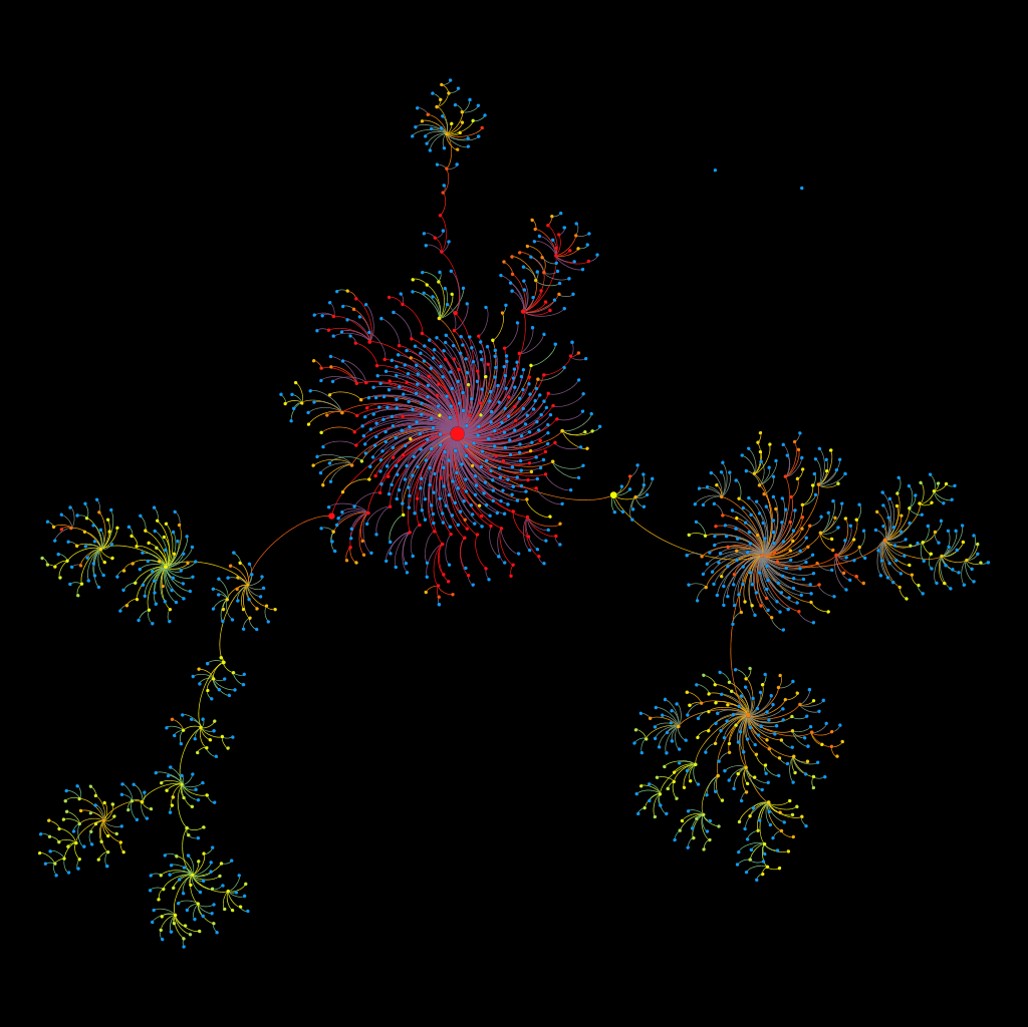}
\caption{Skeleton tree until 2004}
\end{minipage}
\begin{minipage}[t]{0.33\linewidth}
\includegraphics[width = \linewidth]{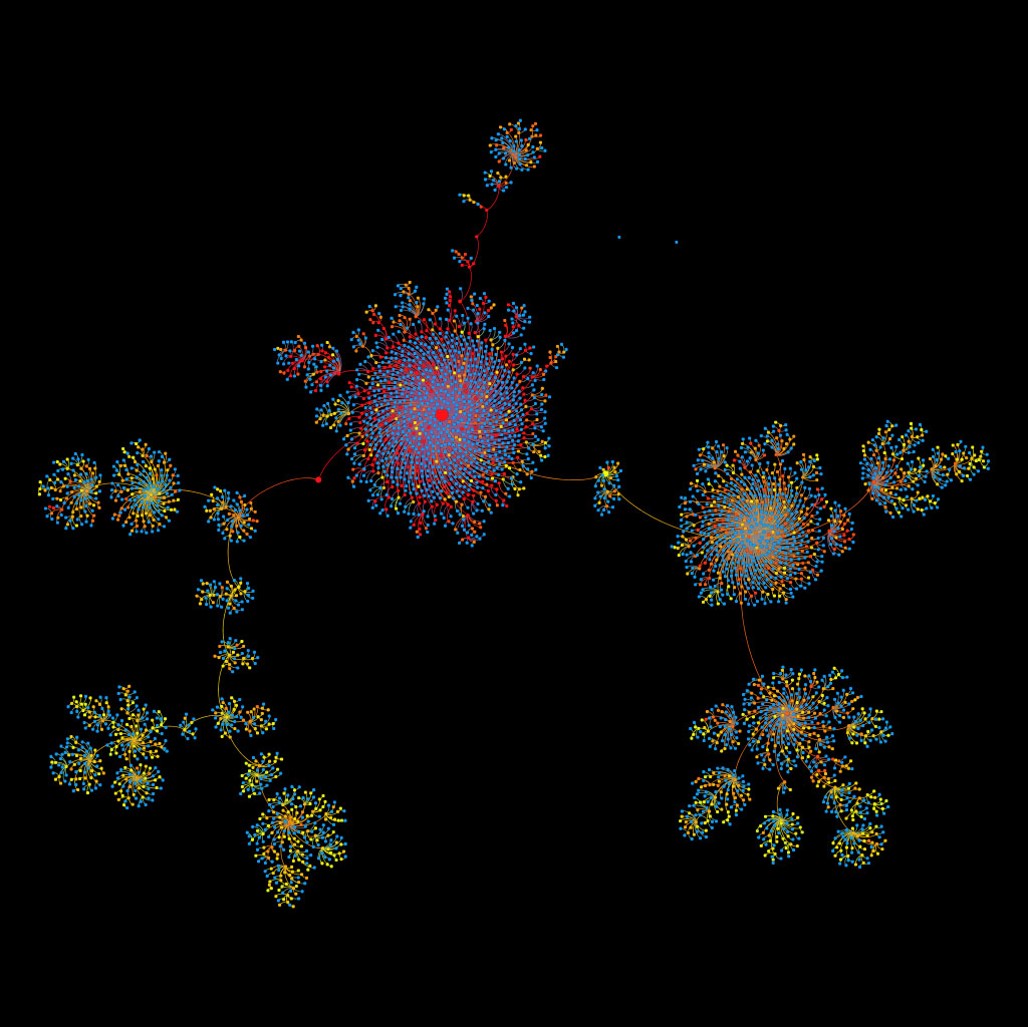}
\caption{Skeleton tree until 2007}
\end{minipage}
\end{subfigure}

\begin{subfigure}{\textwidth}
\begin{minipage}[t]{0.33\linewidth}
\includegraphics[width = \linewidth]{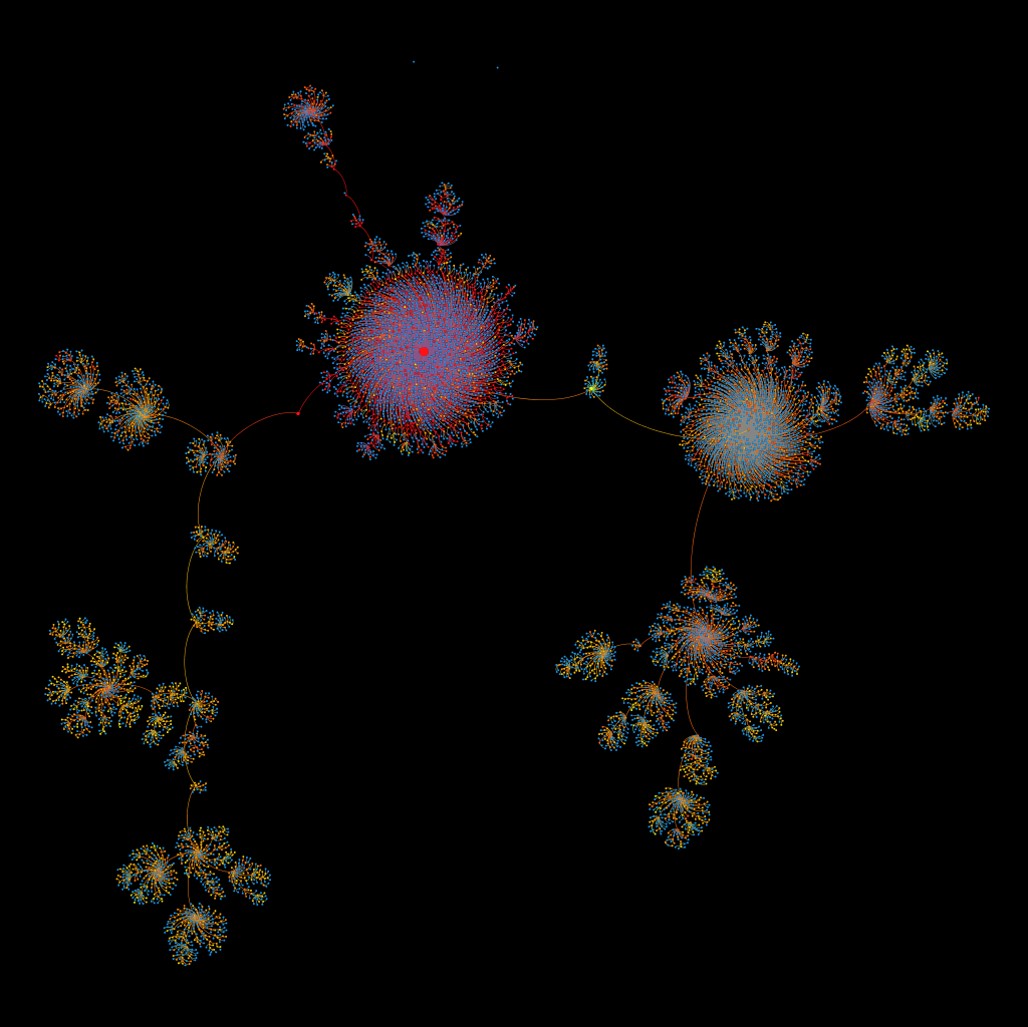}
\caption{Skeleton tree until 2010}
\end{minipage}
\begin{minipage}[t]{0.33\linewidth}
\includegraphics[width = \linewidth]{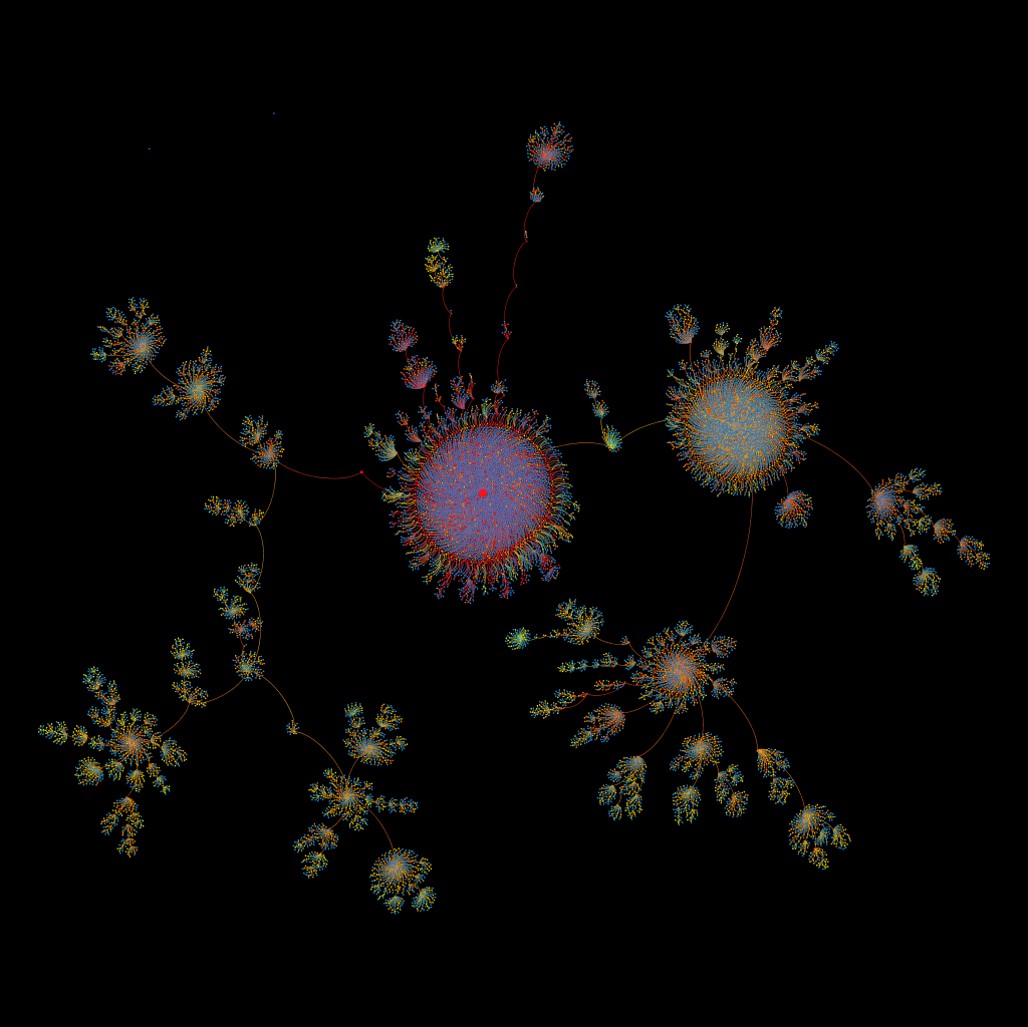}
\caption{Skeleton tree until 2013}
\end{minipage}
\begin{minipage}[t]{0.33\linewidth}
\includegraphics[width = \linewidth]{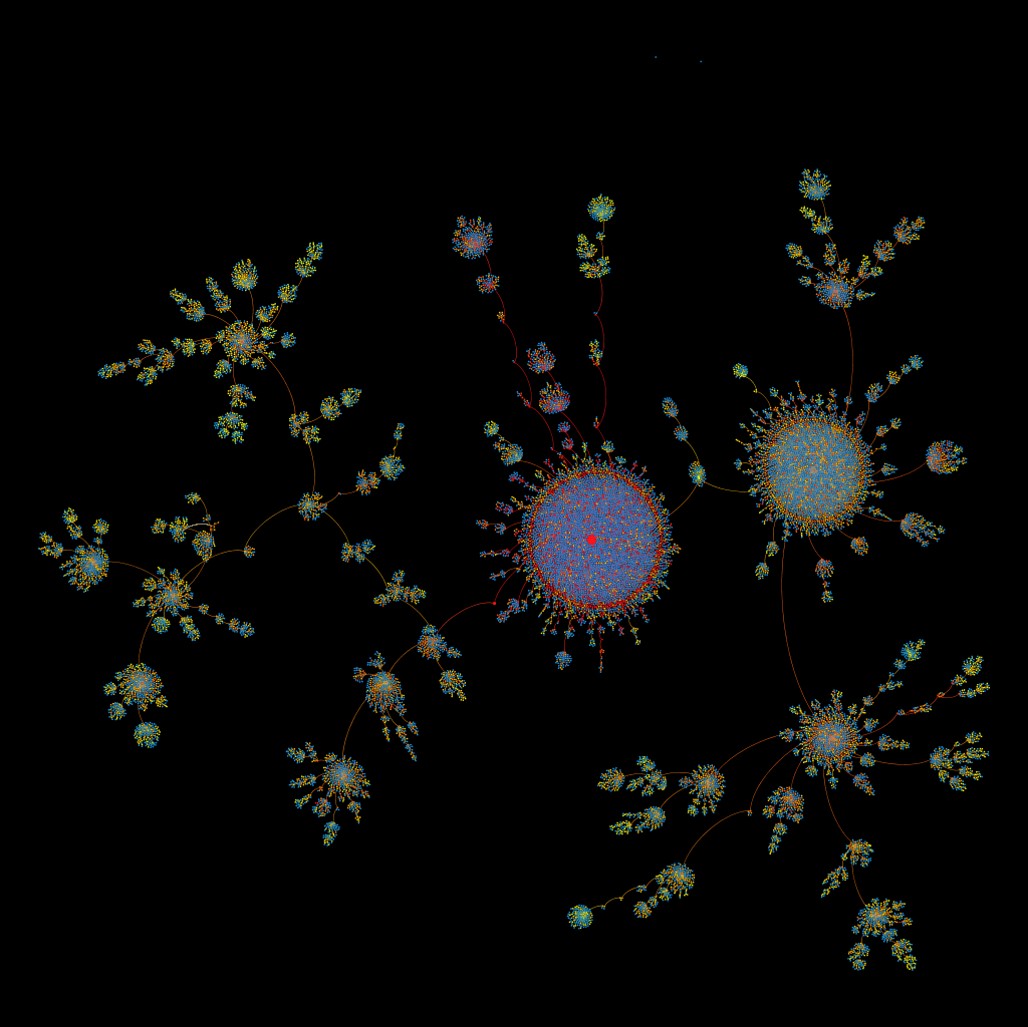}
\caption{Skeleton tree until 2016}
\end{minipage}
\end{subfigure}
\caption{small-world: Skeleton tree evolution}
\label{fig:240908848-tree_evo}
\end{figure}

\begin{figure}[htbp]
\centering
    \begin{subfigure}{\textwidth}
    \begin{minipage}[t]{0.5\textwidth}
    \centering
    \includegraphics[width = 0.9\linewidth]{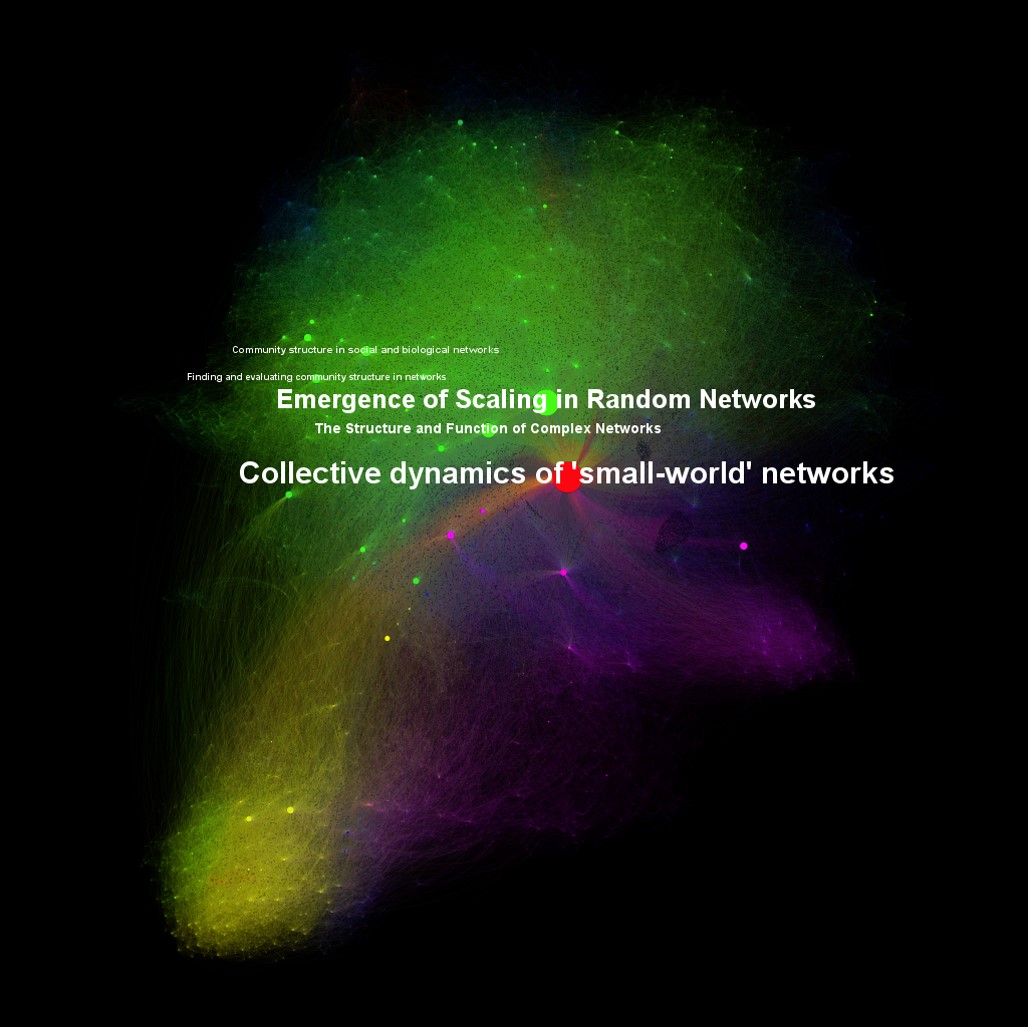}
    \end{minipage}
    \begin{minipage}[t]{0.5\textwidth}
    \centering
    \includegraphics[width = 0.9\linewidth]{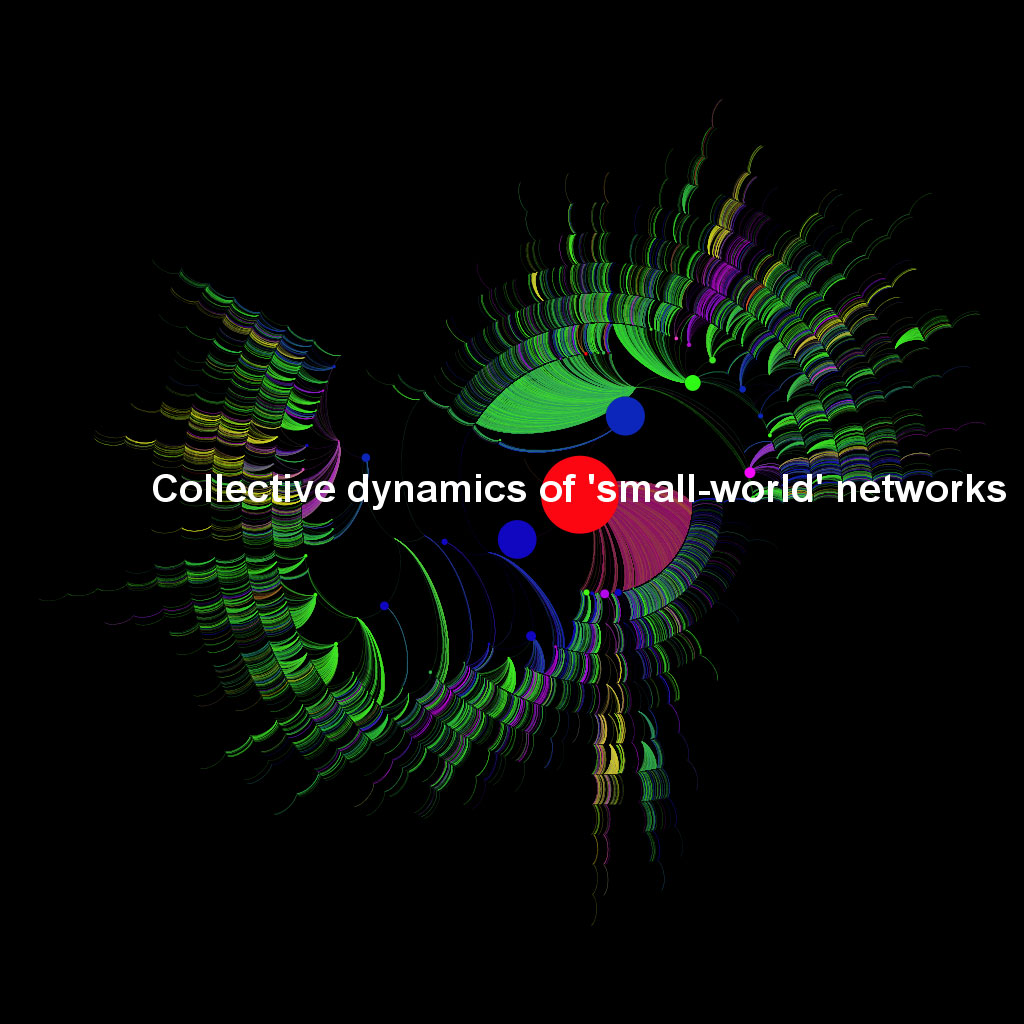}
    \end{minipage}
    \end{subfigure}

    \vspace{5mm}

    \begin{subfigure}{0.6\textwidth}
    \centering
    \includegraphics[width = \linewidth]{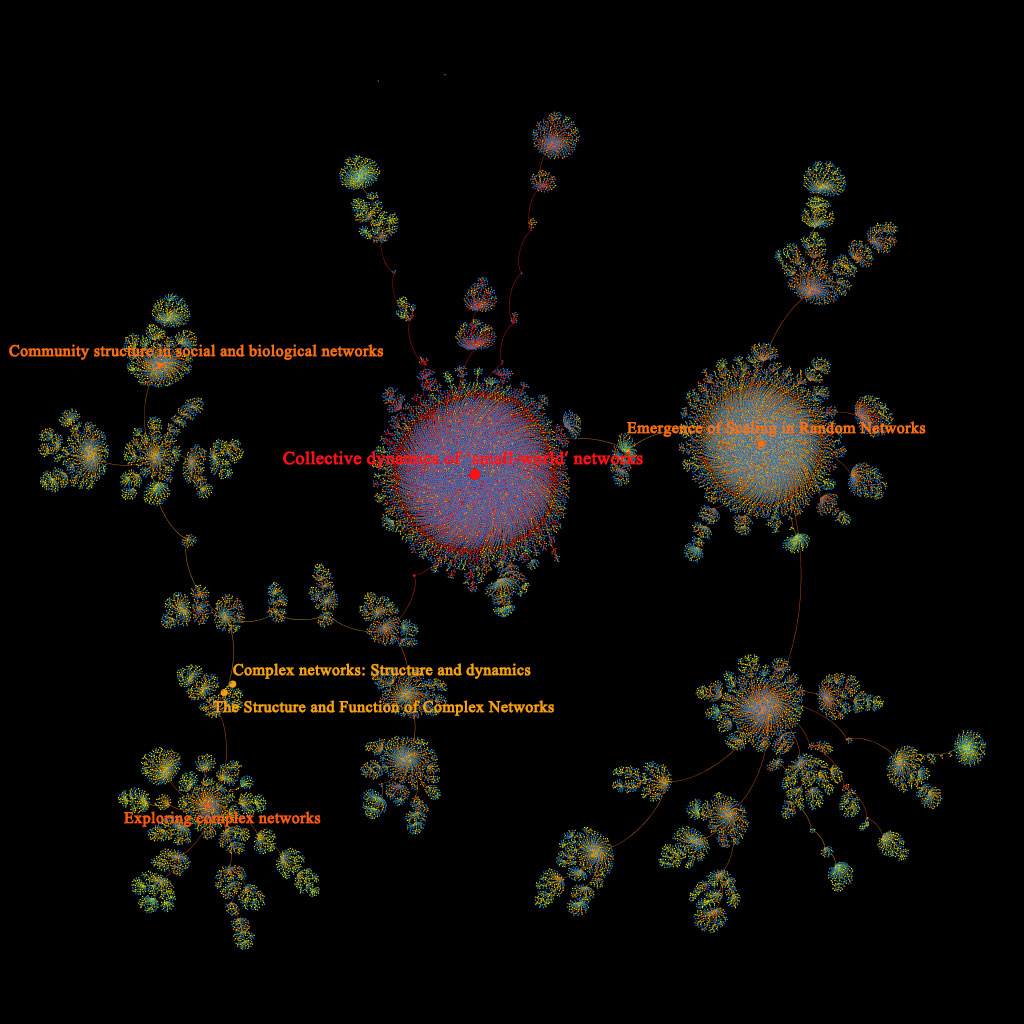}
    \end{subfigure}
    \begin{subfigure}{0.35\textwidth}
    \centering
    \includegraphics[width = 0.9\linewidth]{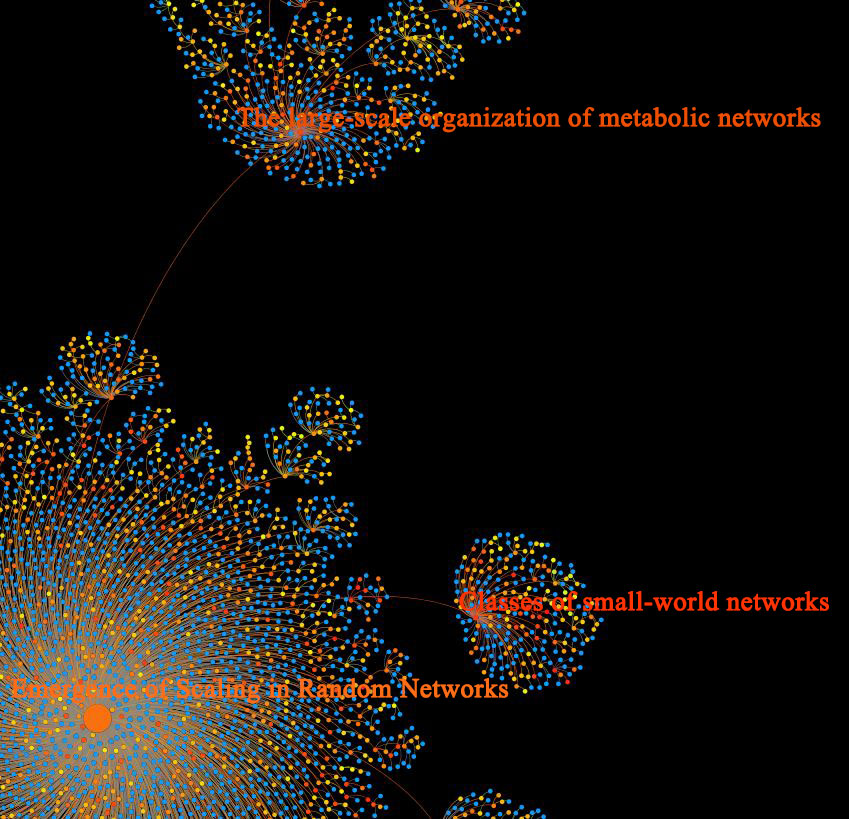}
    \end{subfigure}
\caption{small-world: Galaxy map, current skeleton tree and its regional zoom. Papers with more than 2000 in-topic citations are labelled by title in the skeleton tree. Except the pioneering work, corresponding nodes' size is amplified by 6 times.}
\label{fig:240908848-2020}
\end{figure}

As is shown by $T^t$, although the topic is heating up thanks to a robust knowledge accumulation, it has experienced multiple up and downs during the past 20 years due to short-term popularity fluctuations (Fig. \ref{fig:240908848_chart}). This topic has welcome 2 waves of popular child papers, the first coming between its birth and 2003 and the second batch being published around 2009 and 2010. The oldest popular articles, namely `Emergence of Scaling in Random Networks' (ESRN) published in 1999 in Science, `Exploring complex networks' published in 2001 in Nature and `Community structure in social and biological networks' published in 2002 shaped the fundamentals of topic knowledge structure together with the pioneering work by 2007 (Fig. \ref{fig:240908848-tree_evo}(c), \ref{fig:240908848-2020}). Their substantial contribution to the knowledge quantity and diversity led to a fast rise in both $T_{growth}^t$ and $T_{structure}^t$. As a result, the topic reached the first peak around 2007. For the following years, the short-term exposure increase brought by these eminent child papers gradually wore off and few child papers emerged as rising stars. The topic development during this period was primarily a fortification of its existing knowledge architecture. That is why the topic slightly cooled down during 2007 and 2010 despite a robust topic expansion and an on-going useful information accumulation. It was also during this down period when the younger popular child papers were published. Some of them, including `Complex brain networks: graph theoretical analysis of structural and functional systems' published in 2009 and `Complex network measures of brain connectivity: Uses and interpretations' published in 2010, introduced new research sub-fields closely related to the idea of the pioneering work. They both formed a non-trivial branch extending directly out of the central cluster (Fig. \ref{fig:240908848-tree_evo}(e,f)). Others continued to enrich the existing research fields created by former eminent child papers. For example, `Emergence of Scaling in Random Networks' demonstrated an exceptional capability to attract substantially more subsequent works even after 10 years of its publication thanks to the explosive growth of social networks. The new knowledge extension and the lasting refinement of the entire knowledge framework are portrayed by a flourishing topic skeleton tree with  multidimensional development and a steadily rising $T^t$ until 2016, a year when the topic hit the second peak. While the first golden age is essentially owing to a rapid internal growth, the second streak is largely propelled by favorable social trends, especially the prevalence of online social network and the popularization of brain or neuroscience. Recently, the short-term focus benefit has been dying out and no remarkable progress have been matured enough to cause a stir. Thus the topic is now seeing a small slip. \\

\noindent Now we closely examine the internal heat distribution together with its latest skeleton tree (Fig. \ref{fig:240908848-2020}). All popular child papers have a knowledge temperature above average. This shows that the heat diffusion within the topic is completed after over 20 years of development. Most research focuses derived from the original ideas of the pioneering work have had some substantial development. The ensemble makes up the majority of heat sources within the topic. Besides, we also spot few atypical heat sources. They are articles that connect non-trivial research directions in the skeleton tree. For example, paper `Combatting maelstroms in networks of communicating agents' published in 1999 connects the entire left research branch and the central cluster led by the pioneering work. It does not have any direct followers on skeleton tree, but it is the hottest node and its big structure entropy suggests that it is important to the entire knowledge framework. Its value lies exclusively in the enlightenment. As the articles are located farther away from these heat sources, their node knowledge temperature decreases. This accords with the general rule "the older the hotter" (Fig. 5(n)). Note that the average temperature for the oldest papers is not the highest. This is due to the presence of 3 "cold" articles published in the same year as the pioneering work. They either hardly inspired any subsequent works or failed to attract the attention of recent researches. Besides, the blue nodes that surround the pioneering work and the most popular child papers in principal clusters in the current skeleton tree are papers with little or no in-topic development. However, the general rule is violated even if we let alone the oldest articles.
For example, paper ESRN is slightly colder than its child papers, `The large-scale organization of metabolic networks.' published in 2000 in Nature and `Classes of small-world networks' published in 2000. Both are coloured red while ESRN is coloured orange-red. The temperature difference is mainly due to their different research focus, as is reflected by their distinct citations. The counter example also illustrates that the general rule "the more influential the hotter" is weak (Fig. \ref{fig:citation_T}(n)). Last but not the least, we find that most articles published in top journals such as Science and Nature have high knowledge temperatures and numerous citations. This accords with the prior study which points out the boosting effect of renowned journals on articles$^{30}$. \\

\noindent We observe in addition certain clustering effect in the skeleton tree (Table \ref{tab:240908848-clustering}). This confirms the effectiveness of our skeleton tree extraction algorithm. Moreover, these newly-formed small groups are very young, suggesting that their research focus may be among the latest hotspots within the topic.\\

\begin{table}
    \centering
    \begin{tabular}{p{15cm} p{1cm}}
        \hline
        title & year\\
        \hline
        \textcolor{red}{Robustness of Synchrony} in Complex Networks and Generalized \textcolor{red}{Kirchhoff} Indices & 2018 \\

        Impact of network topology on the \textcolor{orange}{stability of DC microgrids} & 2019\\

        The key player problem in complex oscillator networks and \textcolor{orange}{electric power grids}: Resistance centralities identify local vulnerabilities & 2019\\

        Quantifying transient spreading dynamics on networks & 2019\\

        Global \textcolor{orange}{robustness} versus local vulnerabilities in complex \textcolor{orange}{synchronous} networks & 2019\\
       \hline
    \end{tabular}

\vspace{2mm}

    \begin{tabular}{p{15cm} p{1cm}}
        \hline
        title & year\\
        \hline
        Multiplex \textcolor{red}{lexical} networks reveal patterns in early word acquisition in children & 2017 \\

        Multiplex model of mental \textcolor{orange}{lexicon} reveals explosive learning in humans & 2018\\

        How children develop their ability to combine \textcolor{orange}{words}: a network-based approach & 2019\\

        Multiplex model of mental \textcolor{orange}{lexicon} reveals explosive learning in humans & 2018\\

        Applying network theory to fables: complexity in Slovene \textcolor{orange}{belles-lettres} for different age groups & 2019\\

        Knowledge gaps in the early growth of \textcolor{orange}{semantic feature} networks & 2018\\

        The \textcolor{orange}{orthographic} similarity structure of English \textcolor{orange}{words}: Insights from network science & 2018\\

        Node Ordering for Rescalable Network Summarization (or, the Apparent Magic of \textcolor{orange}{Word} Frequency and Age of Acquisition in the Lexicon) & 2018\\

        spreadr: An R package to simulate spreading activation in a network & 2019\\
       \hline
    \end{tabular}

    \caption{small-world: Clustering effect example. First line is the parent paper and the rest children.}
    \label{tab:240908848-clustering}
\end{table}

\subsubsection{Latent dirichlet allocation}

\begin{figure}[htbp]
\centering
\begin{subfigure}[t]{0.7\linewidth}
\centering
\includegraphics[width=\linewidth]{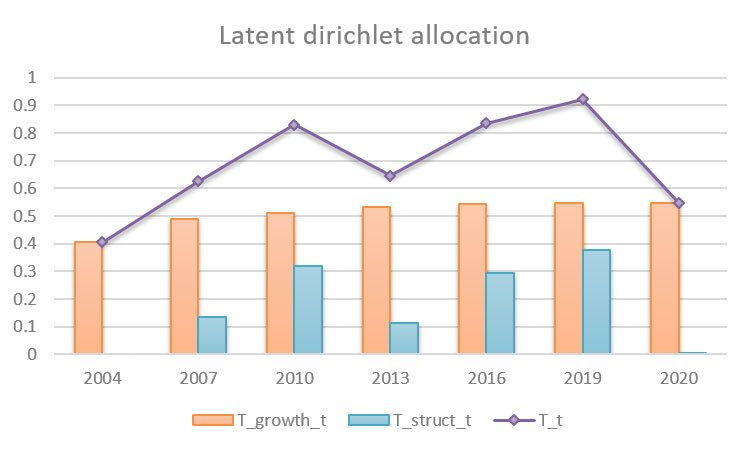}
\end{subfigure}

\begin{subfigure}[t]{\linewidth}
\centering
\begin{tabular}{ccccccccc}
\hline
year & $|V^t|$ & $|E^t|$ & $n_t$ & $V_t$ & ${UsefulInfo}^t$ & $T_{growth}^t$ & $T_{struct}^t$ & $T^t$\\
\hline
2004 & 95 & 190 & 63.402 & 95 & 31.598 & 0.406 &   & 0.406 \\

2007 & 554 & 1931 & 306.602 & 554 & 247.398 & 0.489 & 0.136 & 0.626 \\

2010 & 2287 & 10388 & 1211.237 & 2287 & 1075.763 & 0.511 & 0.319 & 0.83 \\

2013 & 6302 & 33738 & 3193.479 & 6302 & 3108.521 & 0.534 & 0.112 & 0.646 \\

2016 & 12945 & 75942 & 6459.428 & 12945 & 6485.572 & 0.542 & 0.293 & 0.835 \\

2019 & 18583 & 113483 & 9213.456 & 18583 & 9369.544 & 0.546 & 0.377 & 0.923 \\

2020 & 18813 & 114970 & 9330.17 & 18813 & 9482.83 & 0.546 & 0.002 & 0.548 \\
\hline
\end{tabular}
\end{subfigure}
\caption{LDA: topic statistics and knowledge temperature evolution}
\label{fig:162868488_chart}
\end{figure}

\begin{figure}[htbp]
\begin{subfigure}{\textwidth}
\begin{minipage}[t]{0.33\linewidth}
\includegraphics[width = \linewidth]{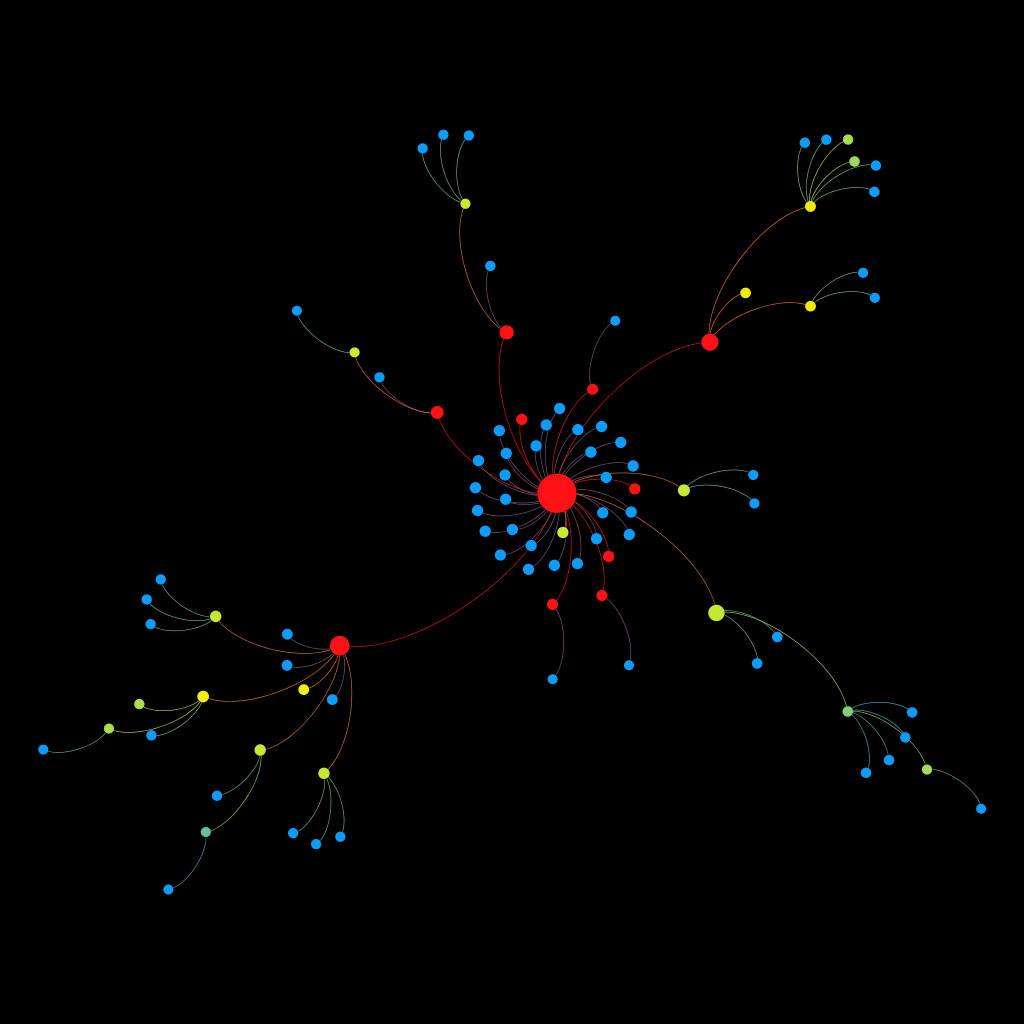}
\caption{Skeleton tree until 2004}
\end{minipage}
\begin{minipage}[t]{0.33\linewidth}
\includegraphics[width = \linewidth]{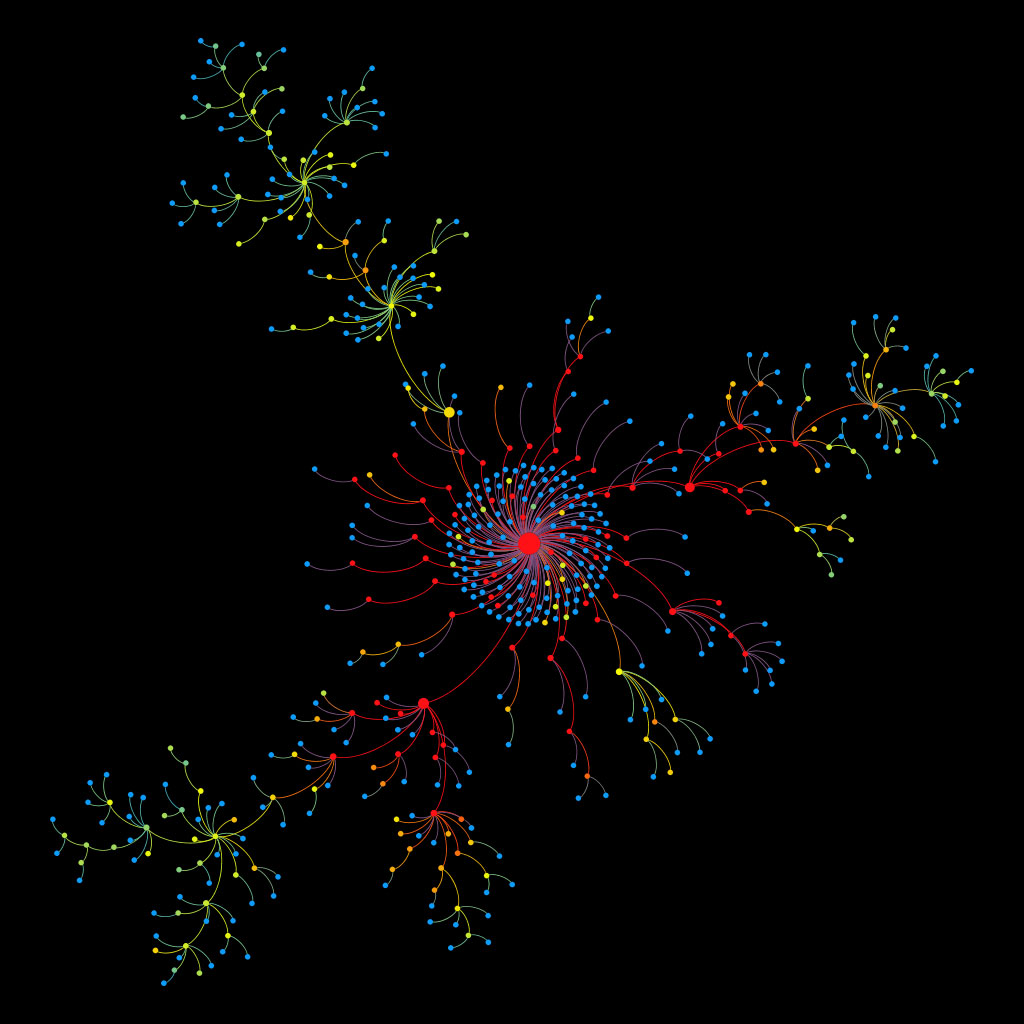}
\caption{Skeleton tree until 2007}
\end{minipage}
\begin{minipage}[t]{0.33\linewidth}
\includegraphics[width = \linewidth]{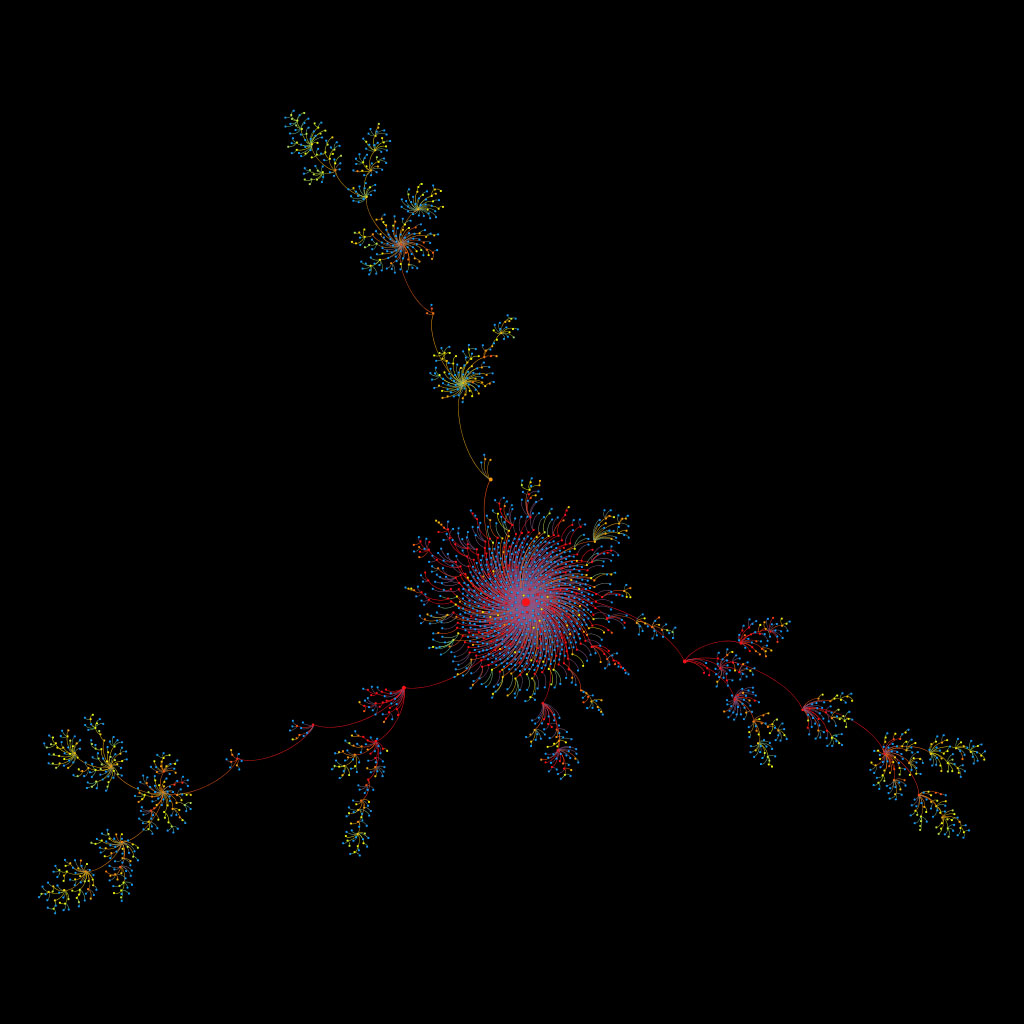}
\caption{Skeleton tree until 2010}
\end{minipage}
\end{subfigure}

\begin{subfigure}{\textwidth}
\begin{minipage}[t]{0.33\linewidth}
\includegraphics[width = \linewidth]{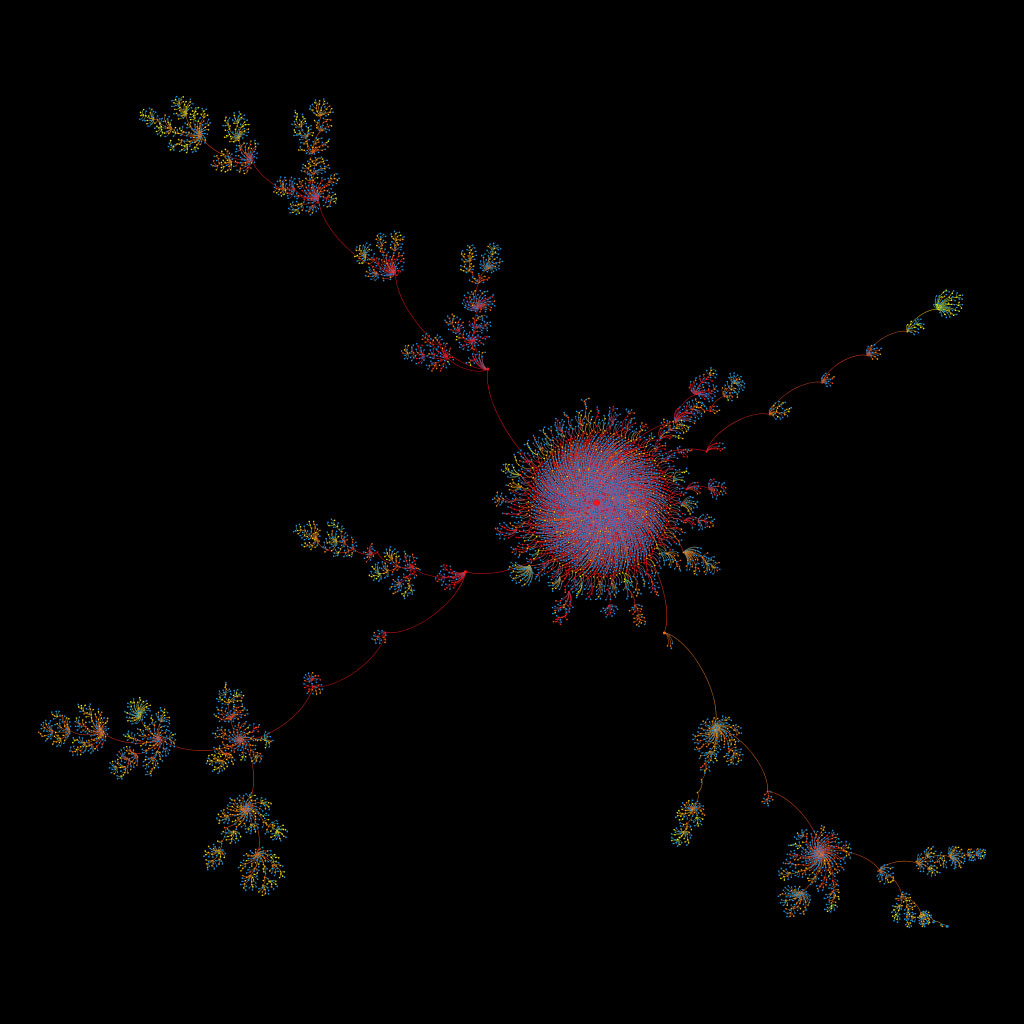}
\caption{Skeleton tree until 2013}
\end{minipage}
\begin{minipage}[t]{0.33\linewidth}
\includegraphics[width = \linewidth]{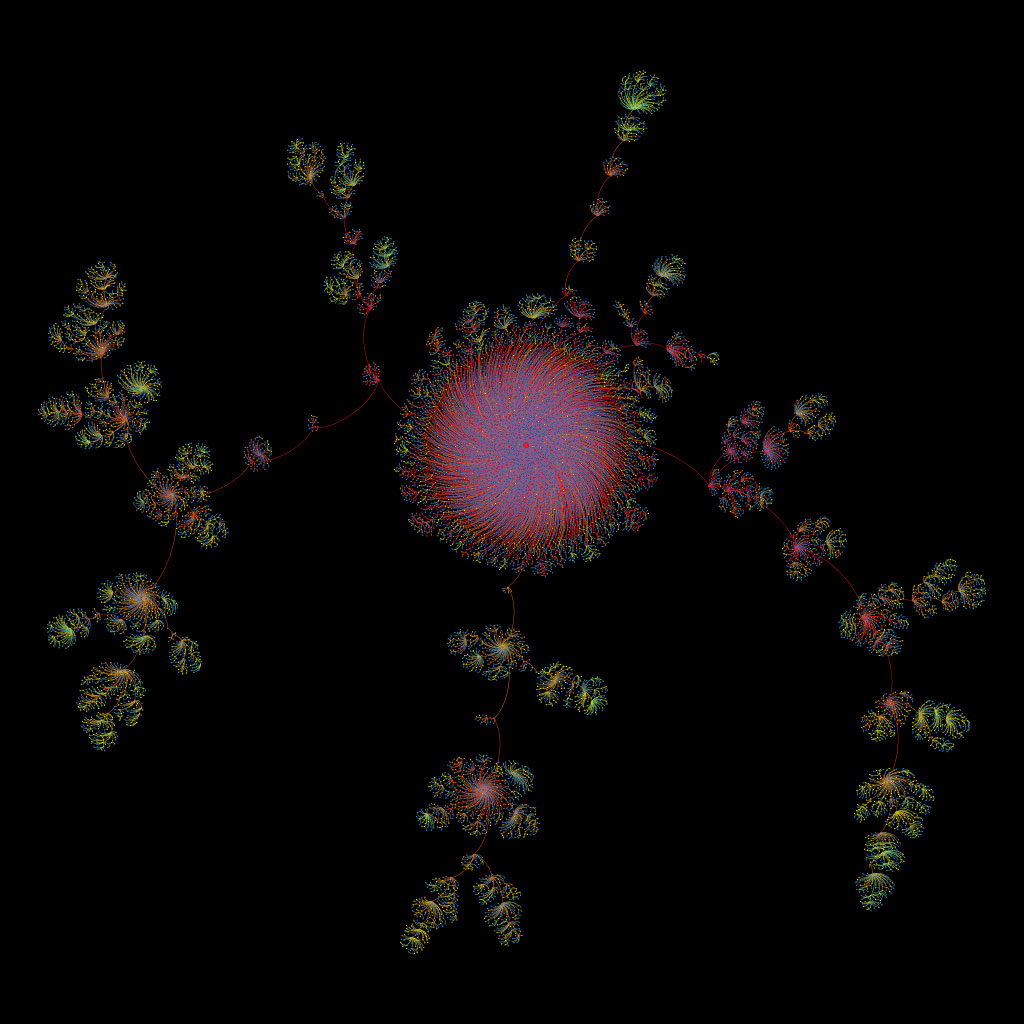}
\caption{Skeleton tree until 2016}
\end{minipage}
\begin{minipage}[t]{0.33\linewidth}
\includegraphics[width = \linewidth]{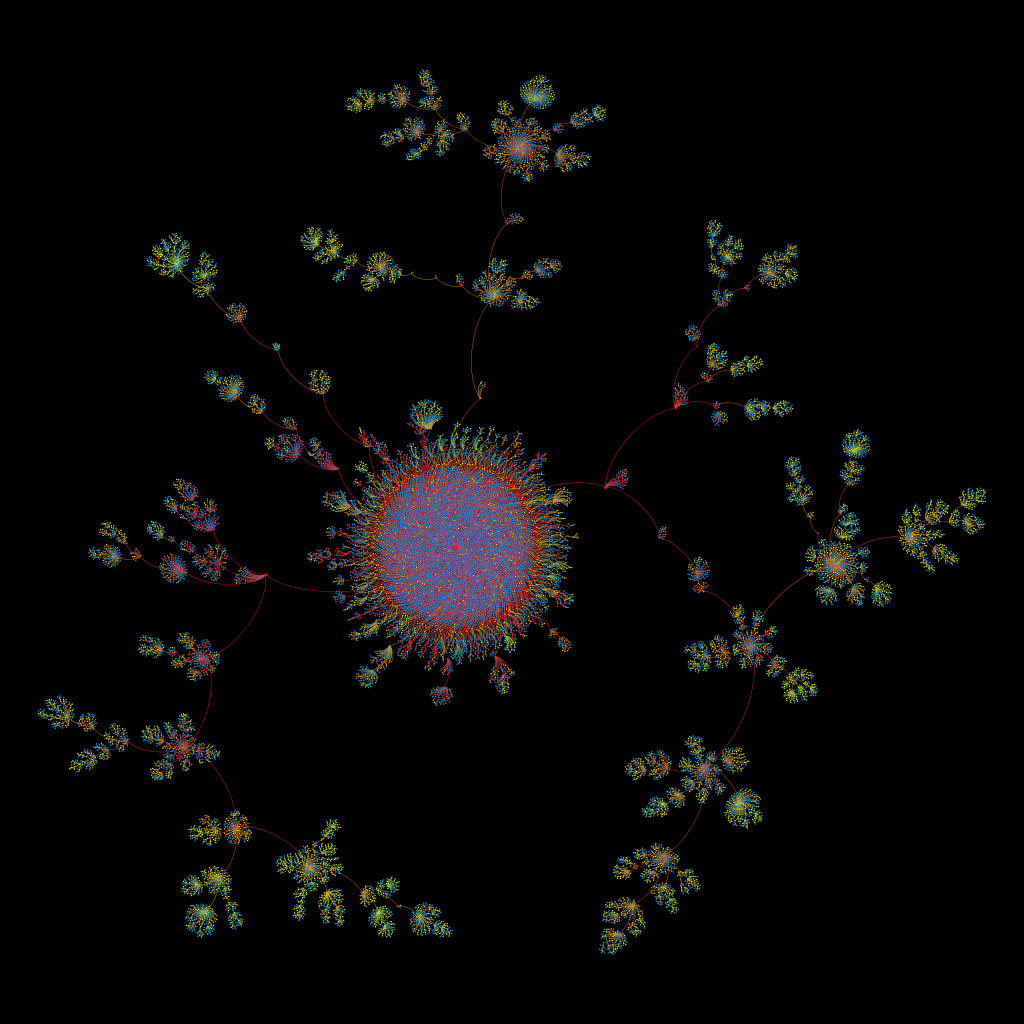}
\caption{Skeleton tree until 2019}
\end{minipage}
\end{subfigure}
\caption{LDA: Skeleton tree evolution}
\label{fig:162868488-tree_evo}
\end{figure}

\begin{figure}[htbp]
\centering
    \begin{subfigure}{\textwidth}
    \begin{minipage}[t]{0.5\textwidth}
    \centering
    \includegraphics[width = 0.9\linewidth]{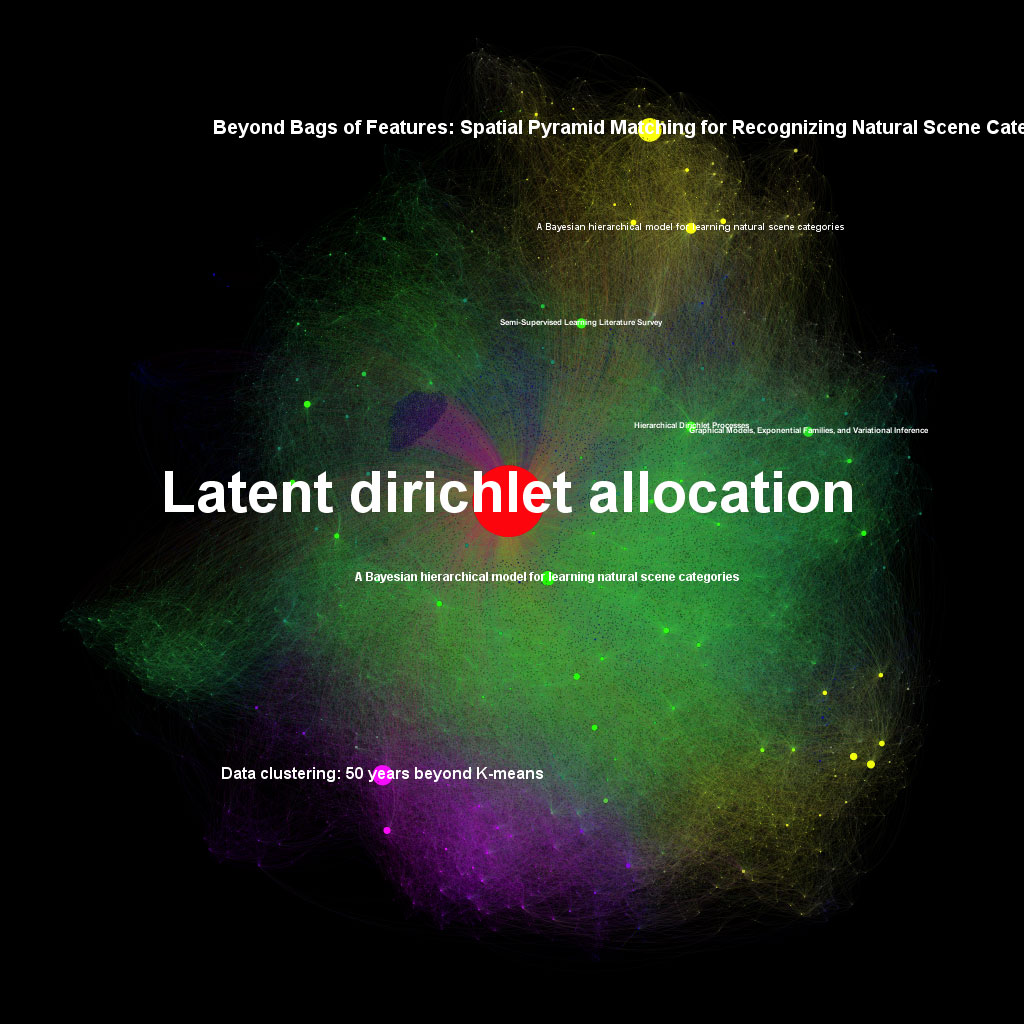}
    \end{minipage}
    \begin{minipage}[t]{0.5\textwidth}
    \centering
    \includegraphics[width = 0.9\linewidth]{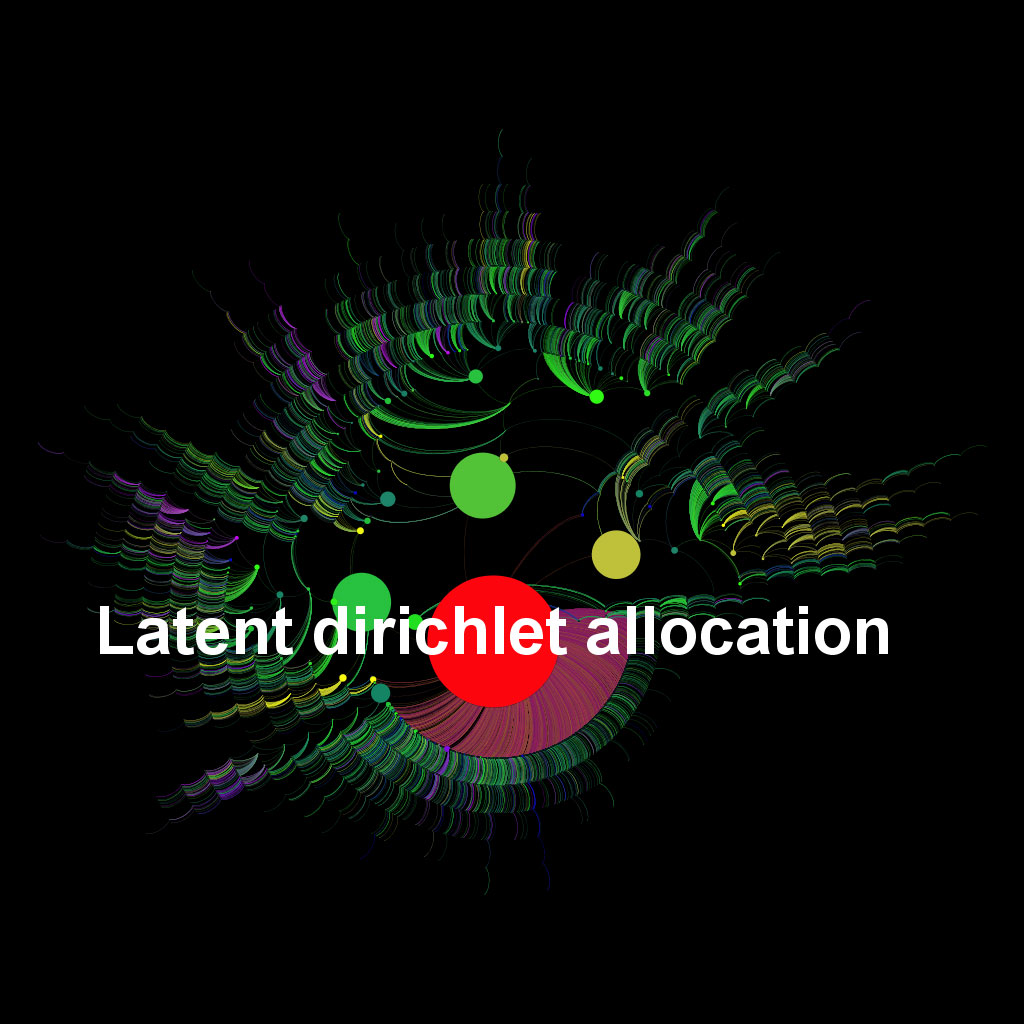}
    \end{minipage}
    \end{subfigure}

    \vspace{5mm}

    \begin{subfigure}{0.6\textwidth}
    \centering
    \includegraphics[width = \linewidth]{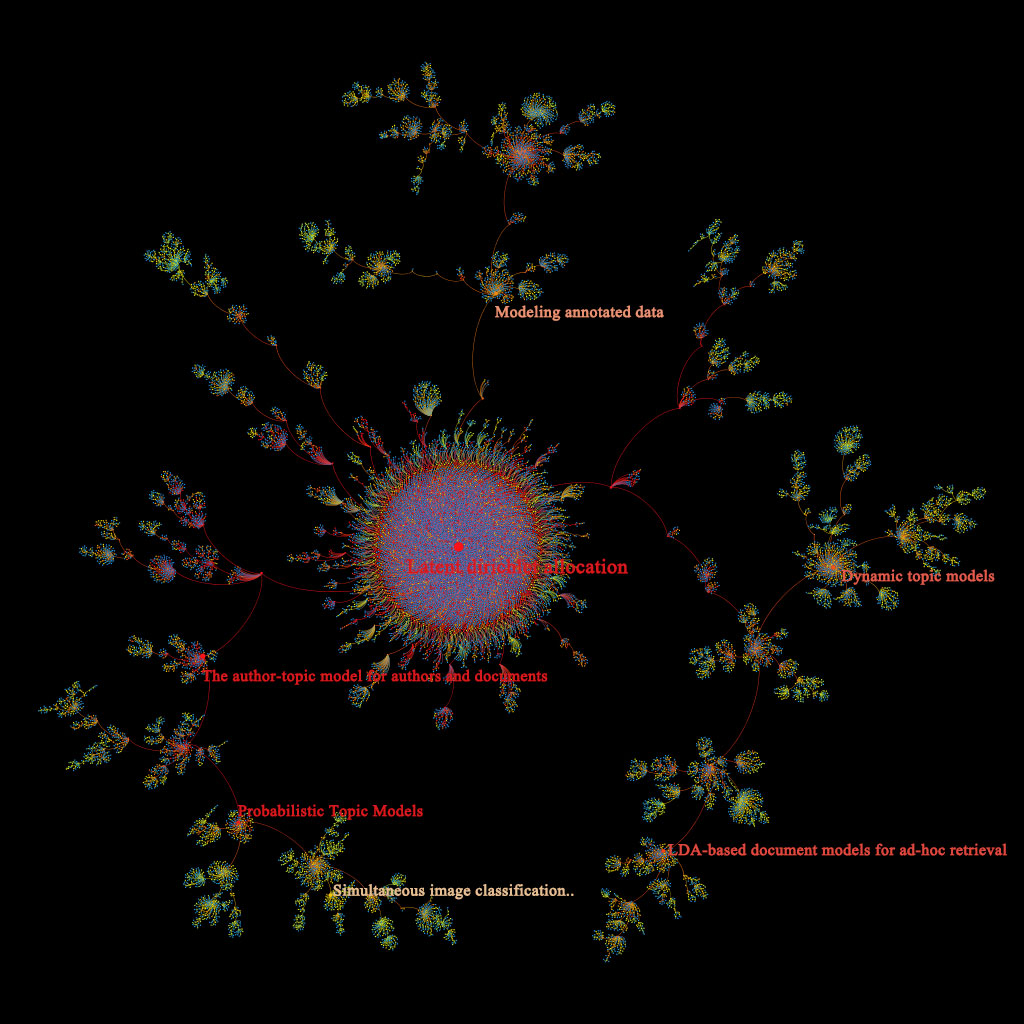}
    \end{subfigure}
    \begin{subfigure}{0.35\textwidth}
    \centering
    \includegraphics[width = 0.9\linewidth]{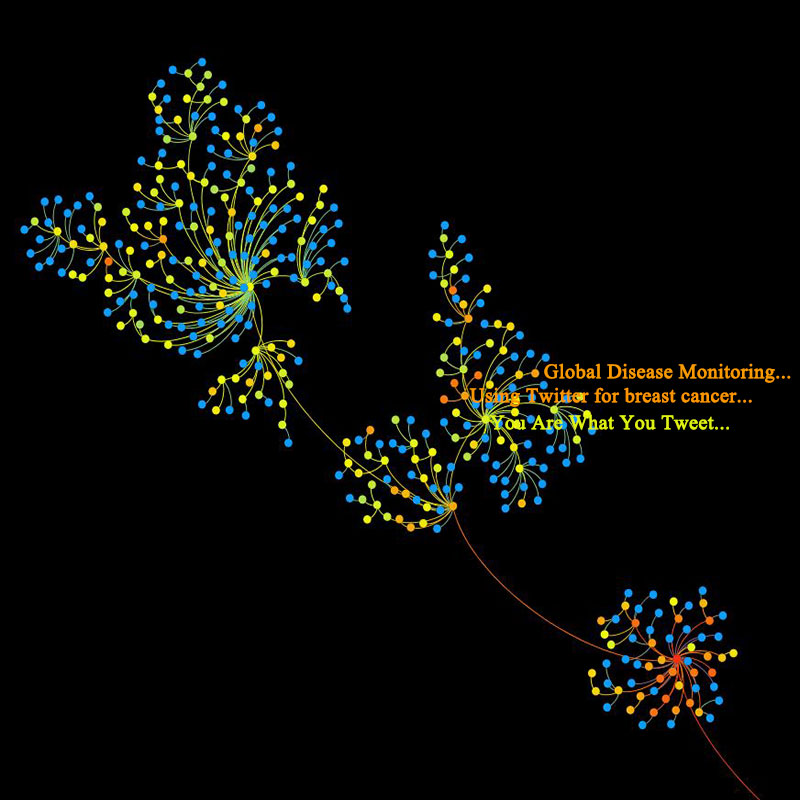}
    \end{subfigure}
\caption{LDA: Galaxy map, current skeleton tree and its regional zoom. Papers with more than 700 in-topic citations are labelled by title in the skeleton tree. Except the pioneering work, corresponding nodes' size is amplified by 5 times.}
\label{fig:162868488-2020}
\end{figure}

As is shown by $T^t$, the impact and popularity evolution of the topic fluctuates. After reaching the first peak around 2010, this field cooled down for a while before it became trendy again around 2019 (Fig. \ref{fig:162868488_chart}). In the long run, the topic has an increasing impact. The rise-and-fall pattern is largely due to the short-term popularity fluctuations, as is demonstrated by the variation of $T_{structure}^t$. In its first 10 years, the topic developed 3 principal research sub-fields, as is illustrated by the skeleton tree (Fig. \ref{fig:162868488-tree_evo} (a,b,c)). The advancement is largely owing to the the arrival of several influential child papers within the topic around 2005 and 2006: `A Bayesian hierarchical model for learning natural scene categories', `Hierarchical Dirichlet Processes' and 'Dynamic topic models' (Fig. \ref{fig:162868488-2020}). They increased the exposure of this topic, facilitated a rapid knowledge accumulation and enriched greatly the knowledge structure. Consequently, the topic had its first golden period. Afterwards, the sweeping trend of machine learning helped the topic gain more attention and fame. A new wave of popular papers joining between 2009 and 2012 gradually manifested their attractiveness, namely `Labeled LDA: A supervised topic model for credit attribution in multi-labeled corpora',`Reading Tea Leaves: How Humans Interpret Topic Models' and `Probabilistic topic models'. They extended the former research focuses and provided inspiration for novel, promising ideas. This is captured by the increasingly complex major branches in skeleton tree (Fig. \ref{fig:162868488-tree_evo} (e,f)). In particular, this wave brought a large amount of attention immediately to the topic and created a second glory. \\

\noindent Now we closely examine the internal heat distribution together with its latest skeleton tree (Fig. \ref{fig:162868488-2020}). After over 20 years of development, the original and recent research ideas have all had a rich development. The heat is therefore diffused to every corner of the skeleton tree with the help of popular child papers. Apart from multiple heat sources in the core of research branches, we also identify some hottest articles between principal clusters. For example, paper `Variational extensions to EM and multinomial PCA' published in 2002 connects the entire right branch and the central cluster. It does not have many direct followers within the topic, but it is the hottest node and it has a big structure entropy due to its knowledge bridging value. As the articles are located farther away from these "hit" papers, their node knowledge temperature decreases. This accords with the general rule "the older the hotter" (Fig. 5(o)). The blue nodes that surround the pioneering work and popular child papers in central parts are papers with few or without any in-topic followers. However, there are exceptions. Paper `You Are What You Tweet: Analyzing Twitter for Public Health' (YWTPH) published in 1998 is colder than its child papers, `Using Twitter for breast cancer prevention: an analysis of breast cancer awareness month' published in 2013 and `Global Disease Monitoring and Forecasting with Wikipedia' published in 2014. The latter two are coloured in orange-red while YWTPH is coloured in yellow-green. Their temperature difference lies primarily in their different research focus reflected by their distinct in-topic citations. This counter example also suggests that another general rule "the more influential the hotter" is not robust (Fig. \ref{fig:citation_T}(o)). \\

\noindent We observe in addition certain clustering effect in the skeleton tree (Table \ref{tab:162868488-clustering}). This confirms the effectiveness of our skeleton tree extraction algorithm. Moreover, these mini-groups are very young, suggesting that their research focus may be among the latest hotspots within the topic.\\

\begin{table}
    \centering
    \begin{tabular}{p{15cm} p{1cm}}
        \hline
        title & year\\
        \hline
        The spread of \textcolor{red}{true and false news} online & 2018 \\

        Assessing the Readiness of Academia in the Topic of \textcolor{orange}{False and Unverified Information} & 2019\\

        Ginger Cannot Cure Cancer: Battling \textcolor{orange}{Fake Health News} with a Comprehensive Data Repository & 2020\\

        Early Public Responses to the Zika-Virus on YouTube: Prevalence of and Differences Between \textcolor{orange}{Conspiracy Theory and Informational Videos} & 2018\\

        An opinion based cross-regional meteorological event detection model & 2019\\

        Investigating Italian \textcolor{orange}{disinformation spreading} on Twitter in the context of 2019 European elections & 2020\\
       \hline
    \end{tabular}

\vspace{2mm}

    \begin{tabular}{p{15cm} p{1cm}}
        \hline
        title & year\\
        \hline
        Automated \textcolor{red}{Text Analysis} for \textcolor{red}{Consumer Research} & 2018 \\

        Automated \textcolor{orange}{Text Analysis} & 2019\\

        Mining \textcolor{orange}{Product Relationships} for Recommendation Based on \textcolor{orange}{Cloud Service} Data & 2018\\

        \textcolor{orange}{Text mining analysis} roadmap (TMAR) for \textcolor{orange}{service} research & 2020\\

        Uniting the Tribes: Using \textcolor{orange}{Text} for \textcolor{orange}{Marketing} Insight: & 2019\\
       \hline
    \end{tabular}

    \caption{Clustering effect example. First line is the parent paper and the rest children.}
    \label{tab:162868488-clustering}
\end{table}

\subsubsection{A FUNDAMENTAL RELATION BETWEEN SUPERMASSIVE BLACK HOLES AND THEIR HOST GALAXIES}

\begin{figure}[htbp]
\centering
\begin{subfigure}[t]{0.7\linewidth}
\centering
\includegraphics[width=\linewidth]{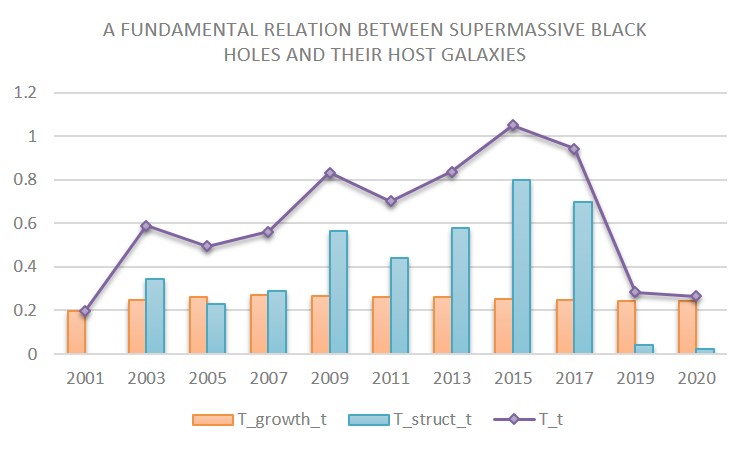}
\end{subfigure}

\begin{subfigure}[t]{0.7\linewidth}
\centering
\begin{tabular}{ccccccccc}
\hline
year & $|V^t|$ & $|E^t|$ & $n_t$ & $V_t$ & ${UsefulInfo}^t$ & $T_{growth}^t$ & $T_{struct}^t$ & $T^t$\\
\hline
2001 & 107 & 321 & 48.92 & 107 & 58.08 & 0.199 &   & 0.199 \\

2003 & 272 & 1278 & 100.136 & 272 & 171.864 & 0.247 & 0.342 & 0.589 \\

2005 & 481 & 3102 & 166.24 & 481 & 314.76 & 0.263 & 0.231 & 0.494 \\

2007 & 774 & 6584 & 259.743 & 774 & 514.257 & 0.271 & 0.291 & 0.562 \\

2009 & 1037 & 10438 & 353.838 & 1037 & 683.162 & 0.266 & 0.565 & 0.831 \\

2011 & 1296 & 13944 & 450.224 & 1296 & 845.776 & 0.261 & 0.44 & 0.702 \\

2013 & 1675 & 20757 & 586.139 & 1675 & 1088.861 & 0.26 & 0.576 & 0.836 \\

2015 & 1974 & 26387 & 714.358 & 1974 & 1259.642 & 0.251 & 0.799 & 1.05 \\

2017 & 2251 & 31260 & 831.841 & 2251 & 1419.159 & 0.246 & 0.697 & 0.943 \\

2019 & 2406 & 33494 & 902.186 & 2406 & 1503.814 & 0.242 & 0.041 & 0.283 \\

2020 & 2432 & 34120 & 911.152 & 2432 & 1520.848 & 0.242 & 0.022 & 0.264 \\
\hline
\end{tabular}
\end{subfigure}
\caption{BLACK HOLES: topic statistics and knowledge temperature evolution}
\label{fig:102900334_chart}
\end{figure}

The knowledge temperature evolution of this topic is quite unique. Not only $T^t$ manifests multiple local peaks every 6 years, but more importantly it is $T_{structure}^t$ that dominates the ups and downs of $T^t$ (Fig. \ref{fig:102900334_chart}). As for $T_{growth}^t$, its increase in the early days is due to the continual arrival of popular child papers within the topic until 2006. They brought a steady inflow of new knowledge that enriched the topic content. Almost all the popular papers published after 2008 have not so far achieved a comparable development. \\

\begin{figure}[htbp]
\begin{subfigure}{\textwidth}
\begin{minipage}[t]{0.33\linewidth}
\includegraphics[width = \linewidth]{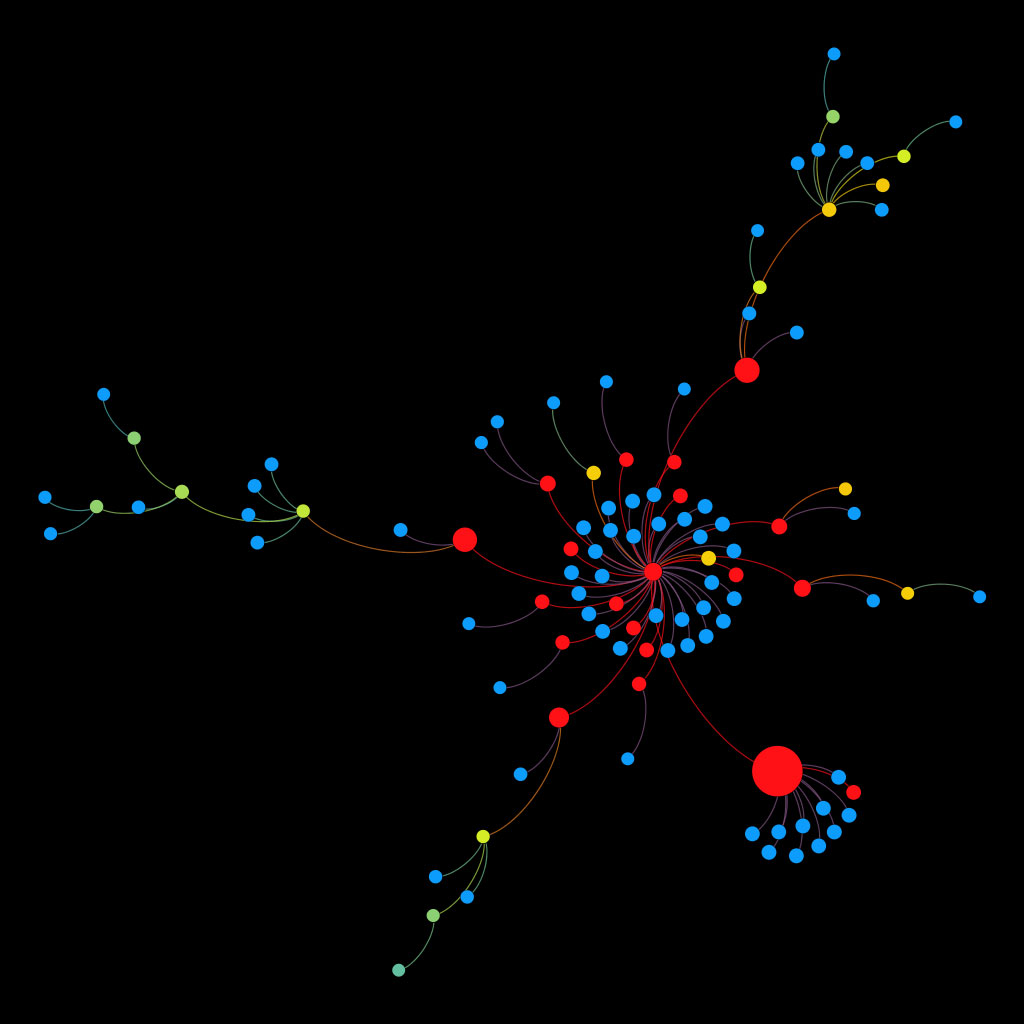}
\caption{Skeleton tree until 2001}
\end{minipage}
\begin{minipage}[t]{0.33\linewidth}
\includegraphics[width = \linewidth]{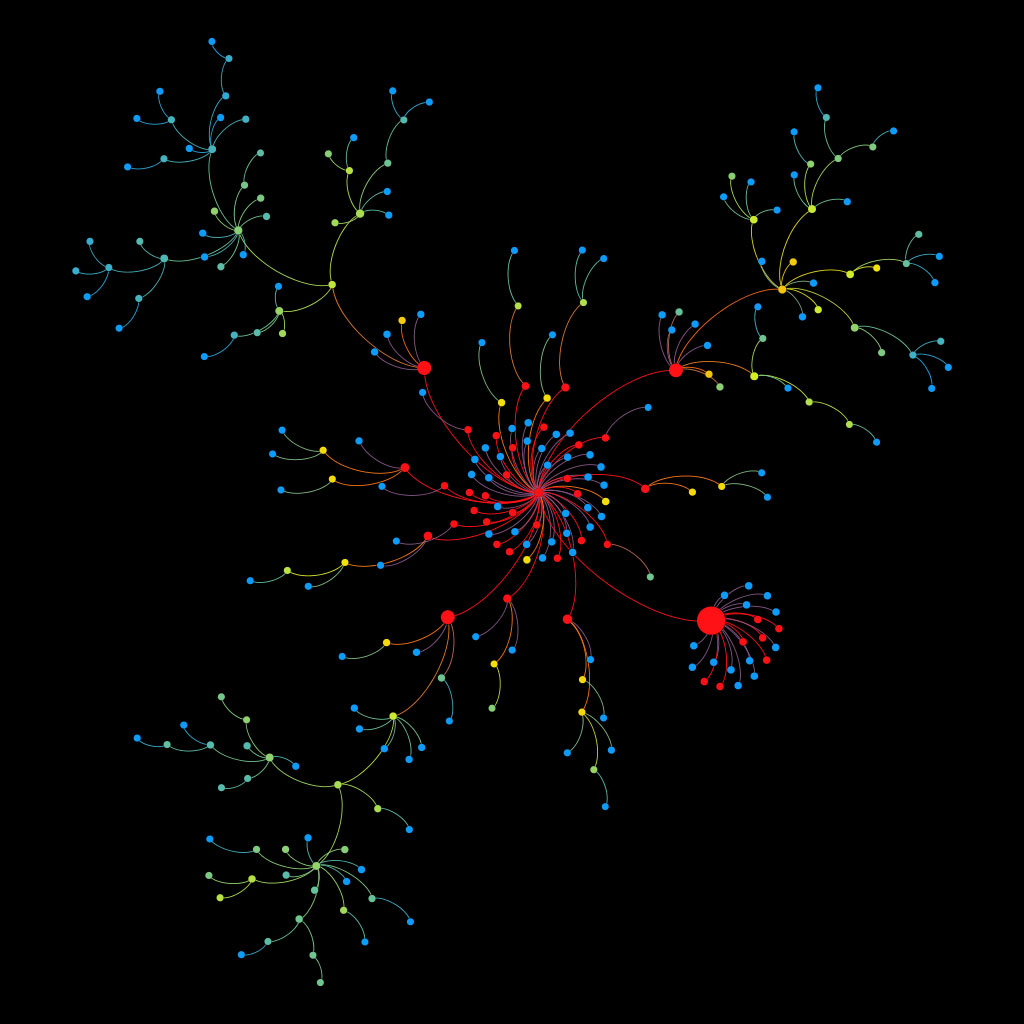}
\caption{Skeleton tree until 2003}
\end{minipage}
\begin{minipage}[t]{0.33\linewidth}
\includegraphics[width = \linewidth]{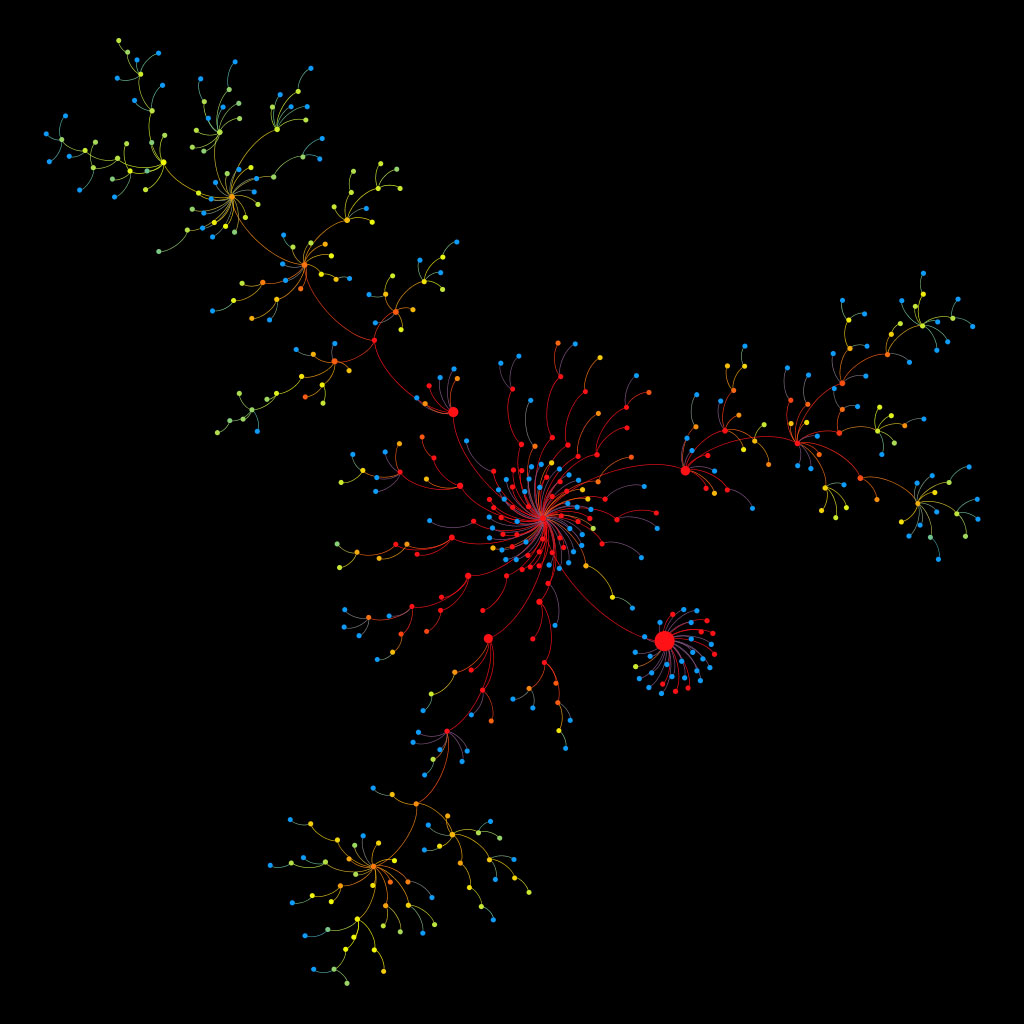}
\caption{Skeleton tree until 2005}
\end{minipage}
\end{subfigure}

\begin{subfigure}{\textwidth}
\begin{minipage}[t]{0.33\linewidth}
\includegraphics[width = \linewidth]{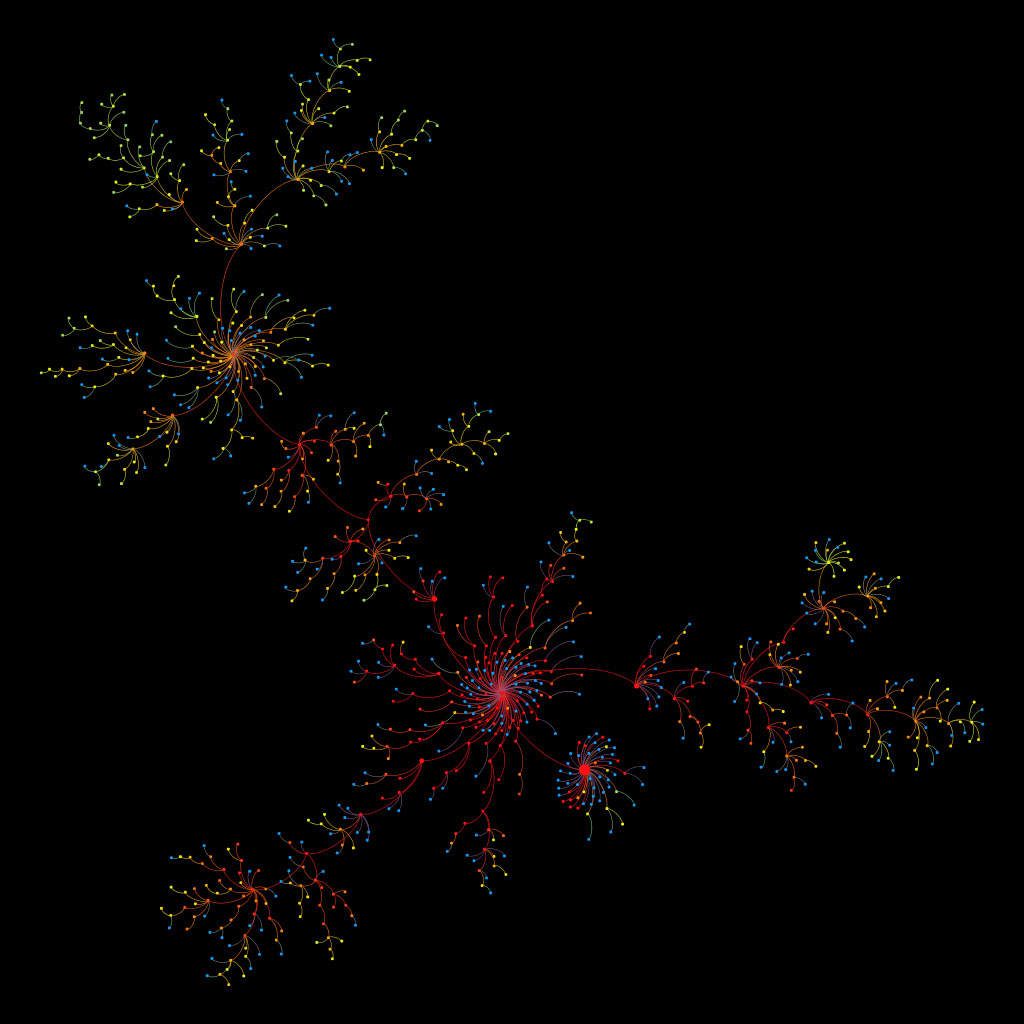}
\caption{Skeleton tree until 2009}
\end{minipage}
\begin{minipage}[t]{0.33\linewidth}
\includegraphics[width = \linewidth]{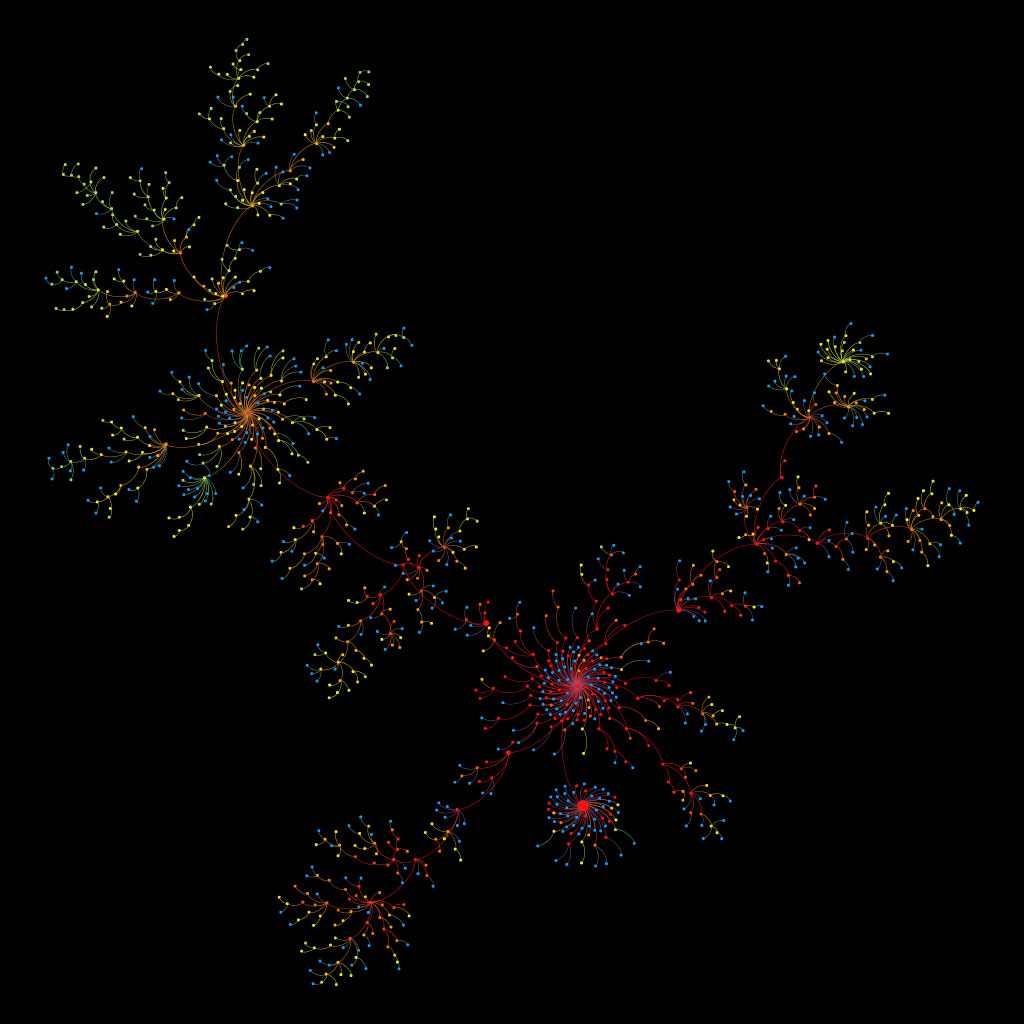}
\caption{Skeleton tree until 2011}
\end{minipage}
\begin{minipage}[t]{0.33\linewidth}
\includegraphics[width = \linewidth]{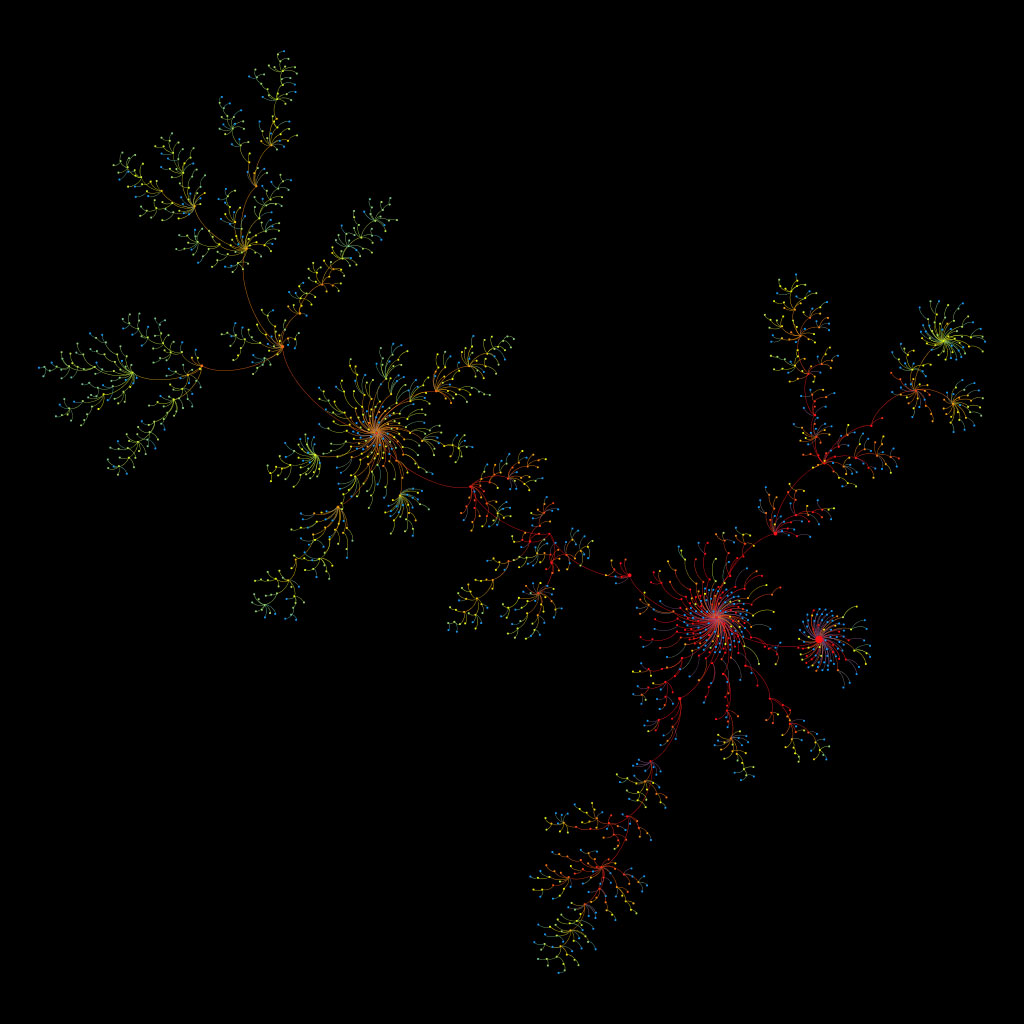}
\caption{Skeleton tree until 2015}
\end{minipage}
\end{subfigure}
\caption{BLACK HOLES: Skeleton tree evolution}
\label{fig:102900334-tree_evo}
\end{figure}

\begin{figure}[htbp]
\centering
    \begin{subfigure}{\textwidth}
    \begin{minipage}[t]{0.5\textwidth}
    \centering
    \includegraphics[width = 0.9\linewidth]{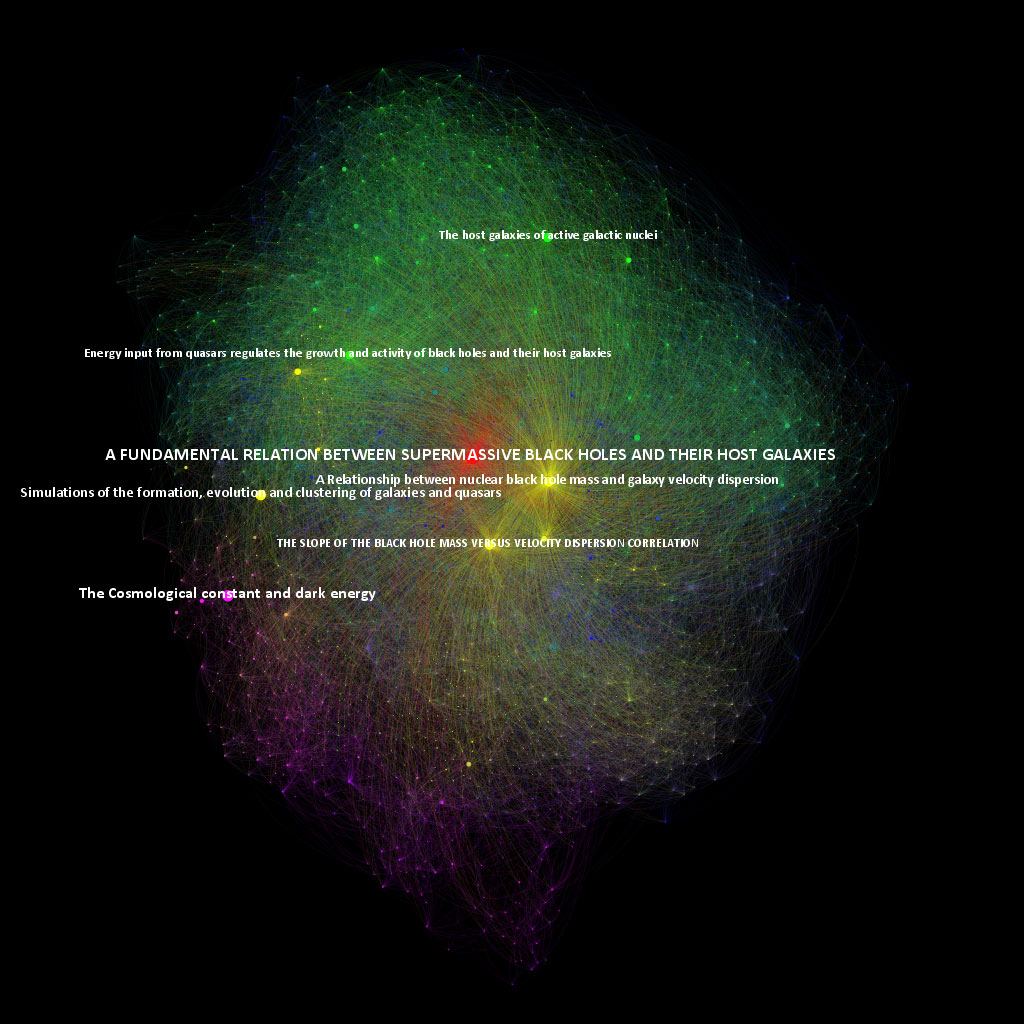}
    \end{minipage}
     \begin{minipage}[t]{0.5\textwidth}
     \centering
    \includegraphics[width = 0.9\linewidth]{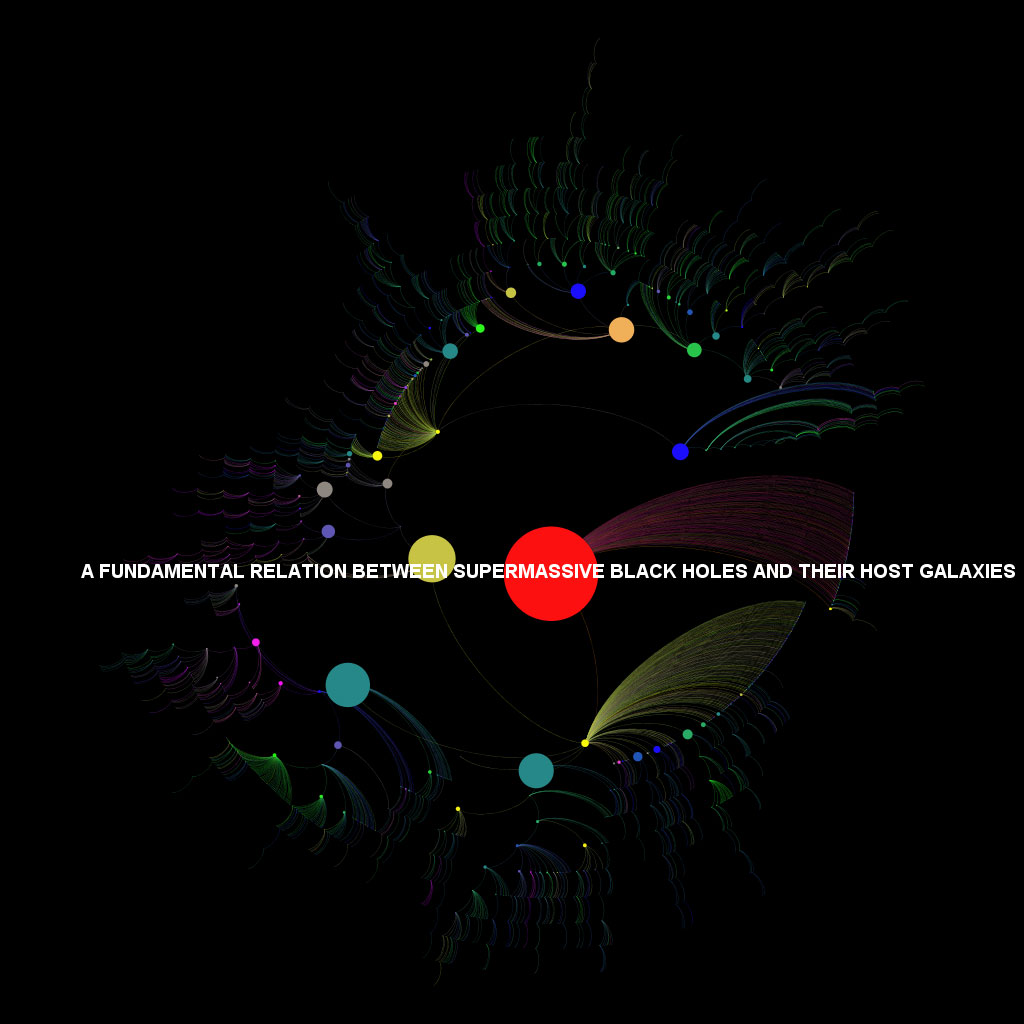}
    \end{minipage}
    \end{subfigure}

    \vspace{5mm}

    \begin{subfigure}{0.6\textwidth}
    \centering
    \includegraphics[width = \linewidth]{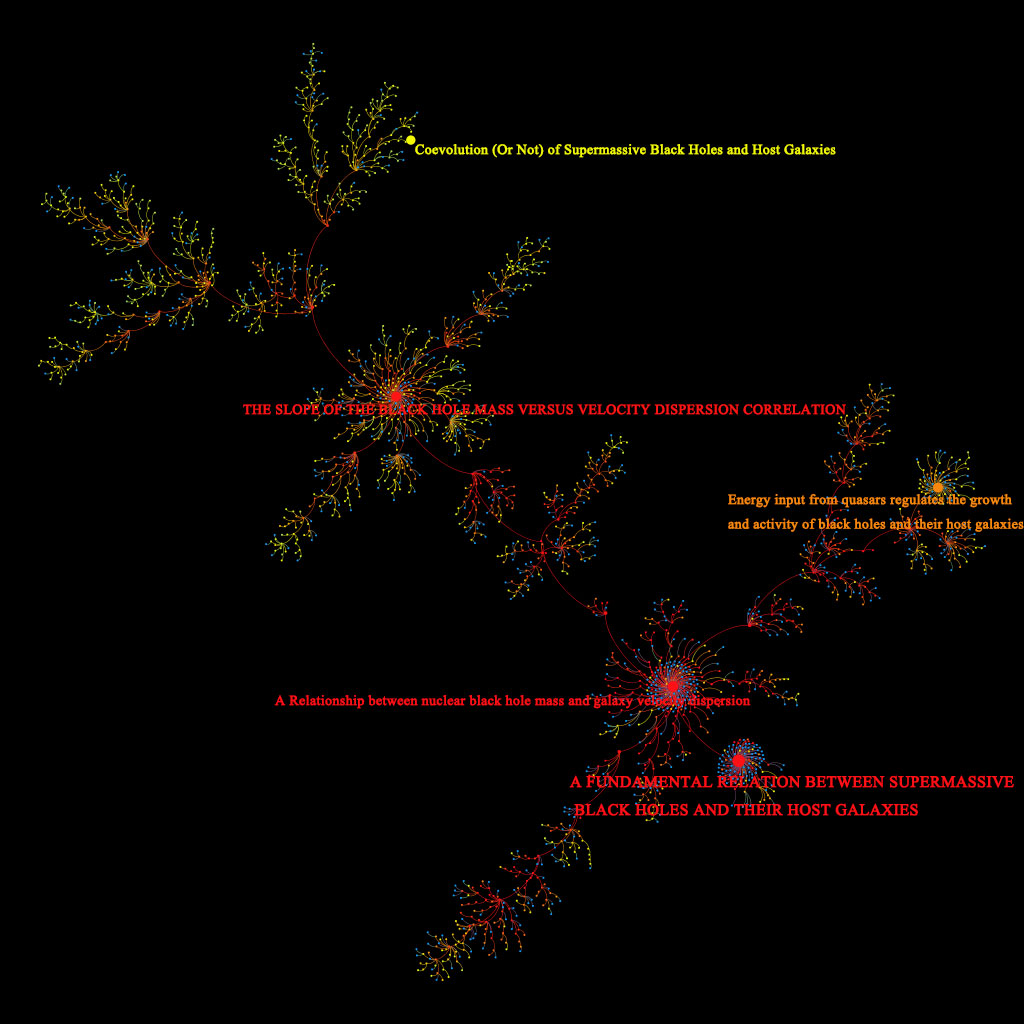}
    \end{subfigure}
    \begin{subfigure}{0.35\textwidth}
    \centering
    \includegraphics[width = 0.9\linewidth]{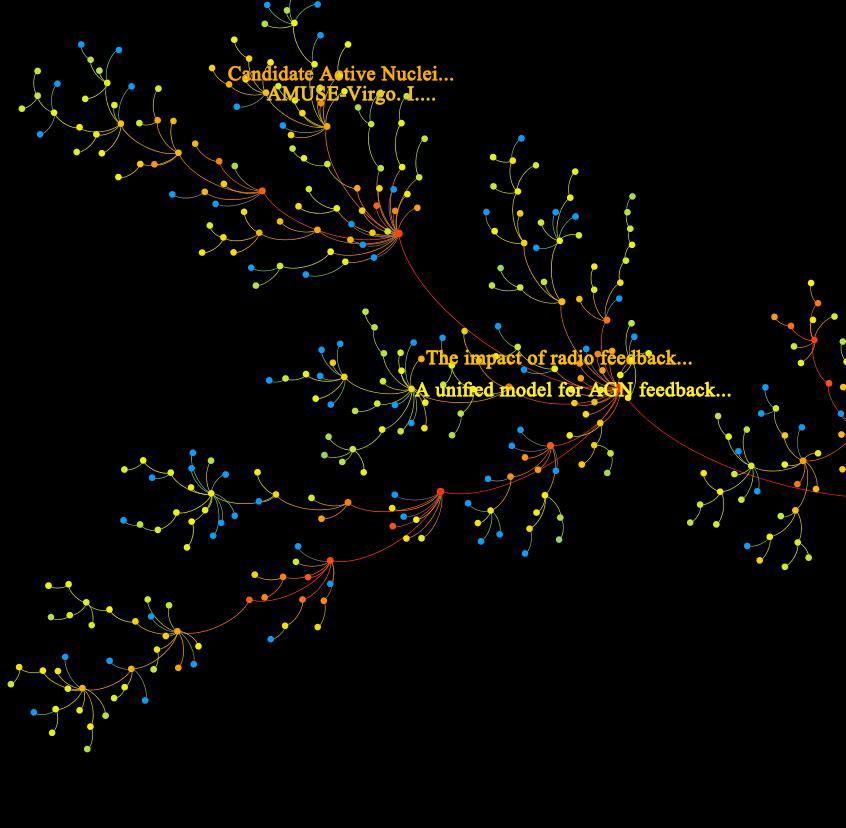}
    \end{subfigure}
\caption{BLACK HOLES: Galaxy map, current skeleton tree and its regional zoom. Papers with more than 340 in-topic citations are labelled by title in the skeleton tree. Except the pioneering work, corresponding nodes' size is amplified by 5 times.}
\label{fig:102900334-2020}
\end{figure}

\noindent The skeleton tree of this topic is also very special in that there are much fewer child papers surrounding the pioneering work, the biggest red node situated in bottom-right, than its prominent descendants, `A Relationship between nuclear black hole mass and galaxy velocity dispersion'(RNBHGVD) and `THE SLOPE OF THE BLACK HOLE MASS VERSUS VELOCITY DISPERSION CORRELATION' (Fig. \ref{fig:102900334-2020}). In fact, the pioneering work has never been the gravity center since the very beginning (Fig. \ref{fig:102900334-tree_evo}(a)). Great structural changes took place between 2001 and 2003. Firstly, we observe a significant development of 2 research directions. This is portrayed by the fast-growing left and right branches that derive from the cluster surrounded around the renowned child paper RNBHGVD. The root of these two primary branches, `On Black Hole Masses and Radio Loudness in Active Galactic Nuclei' and `Black Hole Mass Estimates from Reverberation Mapping and from Spatially Resolved Kinematics', established their indispensable role in knowledge pass-on. Secondly, the smaller branch pointing up-right in the middle of these 2 branches was initially led by paper `COOLING FLOWS AND QUASARS. II. DETAILED MODELS OF FEEDBACK-MODULATED ACCRETION FLOWS' (CFQMFMAF) in 2001. However, after 2 years this paper lost all of its followers in skeleton tree to paper `The correlation between black hole mass and bulge velocity dispersion in hierarchical galaxy formation models' published 1 year earlier  (Fig.\ref{fig:102900334-tree_evo}(b)). The latter only had 2 direct followers in 2001. The reason behind the structural transformation is probably because the articles inspired from paper `A Theoretical Model for the Mbh-$\sigma$ Relation for Supermassive Black Holes in Galaxies' (TMMRSBHG), the best-developped child paper of CFQMFMAF, during this period better characterise TMMRSBHG's research interests with their citation patterns. The additional citation information led to a distinct judgment about the most primordial inspiration source and thus caused the shift in the skeleton tree. Between 2003 and 2009, especially 2005 and 2009, the 3 principal research branches continued to grow. 2 out of the 3 ramified at their ends, suggesting the formation of new research sub-topics. The third $T_{structure}^t$ spike appeared around 2015. 2 out of the 3 principal branches manifested their lasting vigor by a non-trivial evolution at their ends especially during 2011 and 2015. Furthermore, till this end, one principal branch developed so well that it not only overshadowed the other 2 main branches but also claimed the core of the skeleton tree. Its rapid growth is partly thanks to the arrival of 2 popular child papers in 2013: `REVISITING THE SCALING RELATIONS OF BLACK HOLE MASSES AND HOST GALAXY PROPERTIES' and `Coevolution (Or Not) of Supermassive Black Holes and Host Galaxies' even though they themselves do not occupy strategic spots on the branch. Their direct contribution is rather implicit. But together with others they helped complete an obvious gravity shift in knowledge architecture, which is reflected by a surge in $T_{structure}^t$. \\

\begin{figure}[htbp]
\begin{subfigure}{\textwidth}
\centering
\begin{minipage}[t]{0.21\linewidth}
\includegraphics[width = \linewidth]{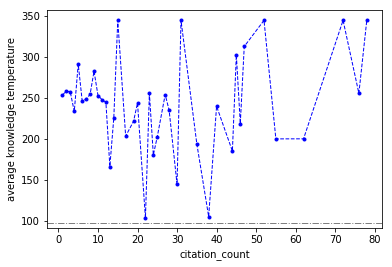}
\caption{}
\end{minipage}
\begin{minipage}[t]{0.21\linewidth}
\includegraphics[width = \linewidth]{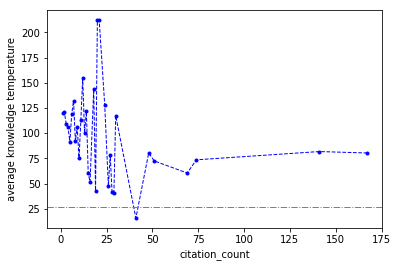}
\caption{}
\end{minipage}
\begin{minipage}[t]{0.21\linewidth}
\includegraphics[width = \linewidth]{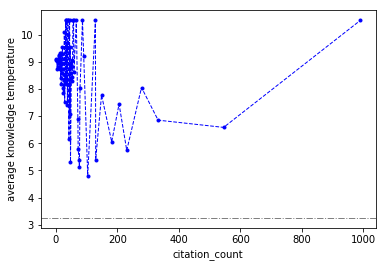}
\caption{}
\end{minipage}
\begin{minipage}[t]{0.21\linewidth}
\includegraphics[width = \linewidth]{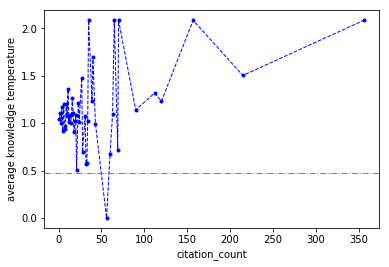}
\caption{}
\end{minipage}
\end{subfigure}

\begin{subfigure}{\textwidth}
\centering
\begin{minipage}[t]{0.21\linewidth}
\includegraphics[width = \linewidth]{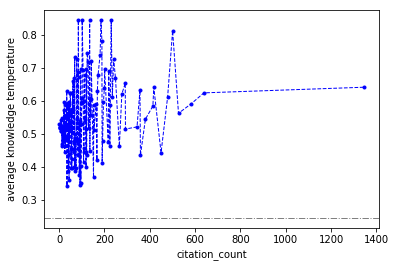}
\caption{}
\end{minipage}
\begin{minipage}[t]{0.21\linewidth}
\includegraphics[width = \linewidth]{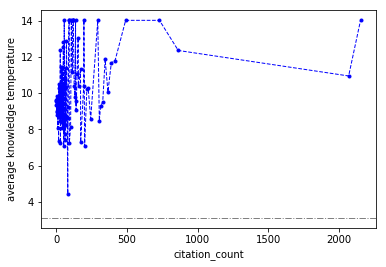}
\caption{}
\end{minipage}
\begin{minipage}[t]{0.21\linewidth}
\includegraphics[width = \linewidth]{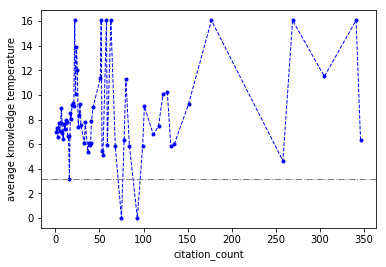}
\caption{}
\end{minipage}
\begin{minipage}[t]{0.21\linewidth}
\includegraphics[width = \linewidth]{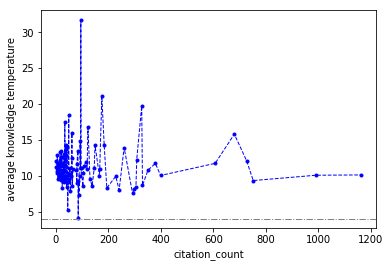}
\caption{}
\end{minipage}
\end{subfigure}

\begin{subfigure}{\textwidth}
\centering
\begin{minipage}[t]{0.21\linewidth}
\includegraphics[width = \linewidth]{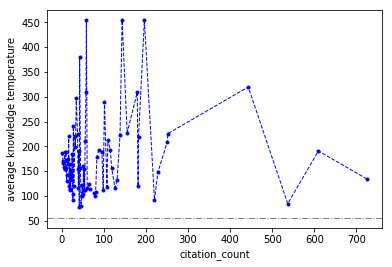}
\caption{}
\end{minipage}
\begin{minipage}[t]{0.21\linewidth}
\includegraphics[width = \linewidth]{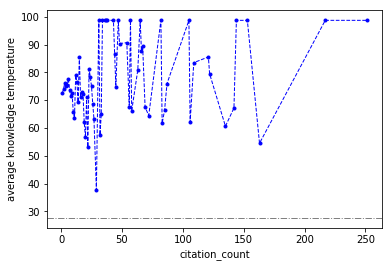}
\caption{}
\end{minipage}
\begin{minipage}[t]{0.21\linewidth}
\includegraphics[width = \linewidth]{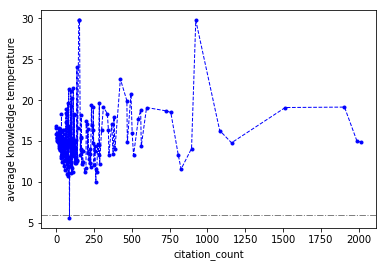}
\caption{}
\end{minipage}
\begin{minipage}[t]{0.21\linewidth}
\includegraphics[width = \linewidth]{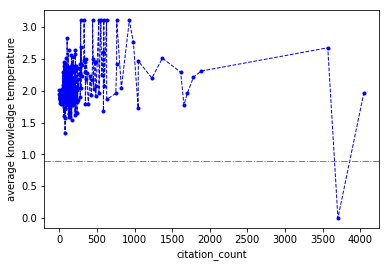}
\caption{}
\end{minipage}
\end{subfigure}

\begin{subfigure}{\textwidth}
\centering
\begin{minipage}[t]{0.21\linewidth}
\includegraphics[width = \linewidth]{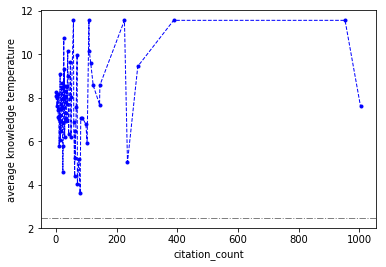}
\caption{}
\end{minipage}
\begin{minipage}[t]{0.21\linewidth}
\includegraphics[width = \linewidth]{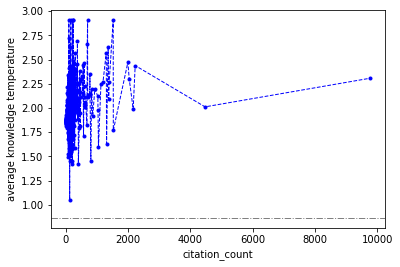}
\caption{}
\end{minipage}
\begin{minipage}[t]{0.21\linewidth}
\includegraphics[width = \linewidth]{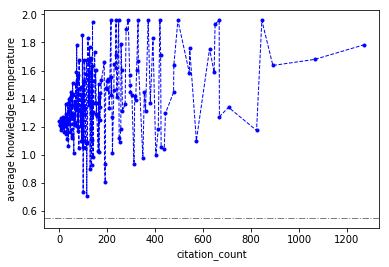}
\caption{}
\end{minipage}
\begin{minipage}[t]{0.21\linewidth}
\includegraphics[width = \linewidth]{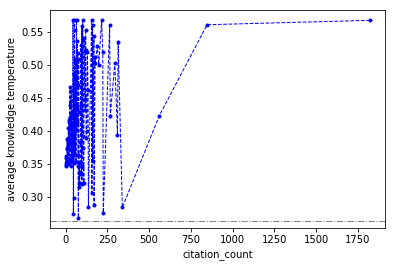}
\caption{}
\end{minipage}
\end{subfigure}
\caption{Relation between article in-topic citation and knowledge temperature. Grey dotted horizontal line marks the topic knowledge temperature in 2020. Articles with no citation and the pioneering work are excluded.}
\label{fig:citation_T}
\end{figure}

\noindent Now we closely examine the internal heat distribution and its latest skeleton tree (Fig. \ref{fig:102900334-2020}). The heat is already uniformly
diffused to major research sub-directions as most popular child papers have a knowledge temperature above average and some even become heat sources. It is clear that the periphery of skeleton tree is colder than the central parts. The blue nodes that surround the pioneering work and popular child papers in central parts are papers with few or without any in-topic citations. This observation accords with the general rule "the older the hotter" (Fig. 5(p)). The small drop in average knowledge temperatures among the oldest papers is due to the presence of several papers published in 2001 that had little inspiration to subsequent research. However, there are exceptions even if we ignore these old "cold" articles. For instance, paper `A unified model for AGN feedback in cosmological simulations of structure formation' published in 2007 is slightly colder than its child paper `The impact of radio feedback from active galactic nuclei in cosmological simulations : formation of disc galaxies' published in 2008. The former is coloured yellow-orange whereas the latter is coloured orange. Their difference in heat-level is mainly due to their slightly different research focus judging from their partially overlapped citations. Out of similar reason, paper `AMUSE-Virgo. I. Supermassive Black Holes in Low-Mass Spheroids' is also slightly colder than its child paper `Candidate Active Nuclei in Late-Type Spiral Galaxies'. These counter examples indicate that the other general rule "the more influential the hotter" is weak (Fig. \ref{fig:citation_T}(p)). \\

\noindent We observe in addition certain clustering effect in the skeleton tree (Table \ref{tab:102900334-clustering}). For example, all child papers of `Active galactic nuclei in the mid-IR: evolution and contribution to the cosmic infrared background' in current skeleton tree study Active galactic nuclei (AGN). This confirms the effectiveness of our skeleton tree extraction algorithm.\\

\begin{table}
    \centering
    \begin{tabular}{p{15cm} p{1cm}}
        \hline
        title & year\\
        \hline
        \textcolor{red}{Active galactic nuclei} in the mid-IR: evolution and contribution to the cosmic infrared background & 2006 \\

        The VVDS type-1 \textcolor{orange}{AGN} sample: the faint end of the luminosity function & 2007 \\

       The cosmological properties of \textcolor{orange}{AGN} in the XMM-Newton Hard Bright Survey & 2008\\

        VARIABILITY AND MULTIWAVELENGTH-DETECTED \textcolor{orange}{ACTIVE GALACTIC NUCLEI} IN THE GOODS FIELDS & 2011\\

        A multi-wavelength survey of \textcolor{orange}{AGN} in massive clusters: AGN distribution and host galaxy properties & 2014\\

        Using \textcolor{orange}{AGN} Variability Surveys to explore the AGN-Galaxy Connection & 2013\\
       \hline
    \end{tabular}

    \caption{Clustering effect example. First line is the parent paper and the rest children.}
    \label{tab:102900334-clustering}
\end{table}

\newpage

\newpage

\subsection{Topic Group}
A topic group is an ensemble of several closely-related topics. During a certain period, topics in a group can manifest distinct popularity and impact changes. Some may prosper while others stagnate or go downhill. When this is the case, our forest helping mechanism allows thriving topics to donate a small fraction of their vigor to their dying siblings. The heat exchange among topic group members somehow takes "background popularity and impact" into consideration. After forest helping, the knowledge temperatures of closely related topics have a more similar evolution and correspond better to idea inheritance and development.\\

\subsubsection{wireless network group}

\begin{figure}[htbp]
\centering
    \begin{subfigure}{\textwidth}
    \begin{minipage}[t]{0.5\textwidth}
    \centering
    \includegraphics[width = 0.9\linewidth]{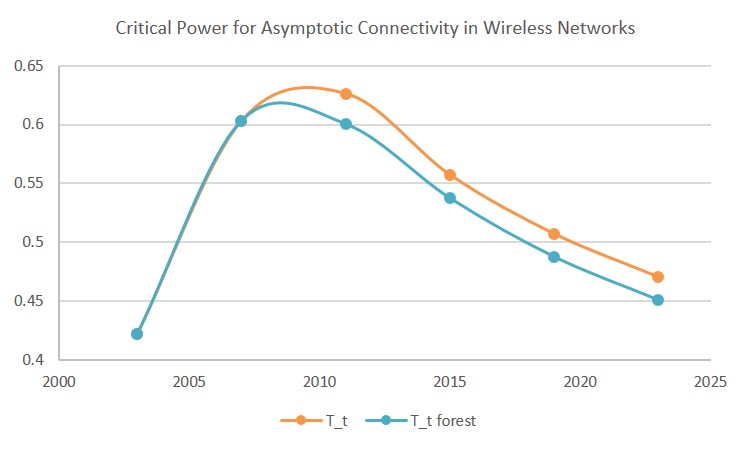}
    \end{minipage}
    \begin{minipage}[t]{0.5\textwidth}
    \centering
    \includegraphics[width = 0.9\linewidth]{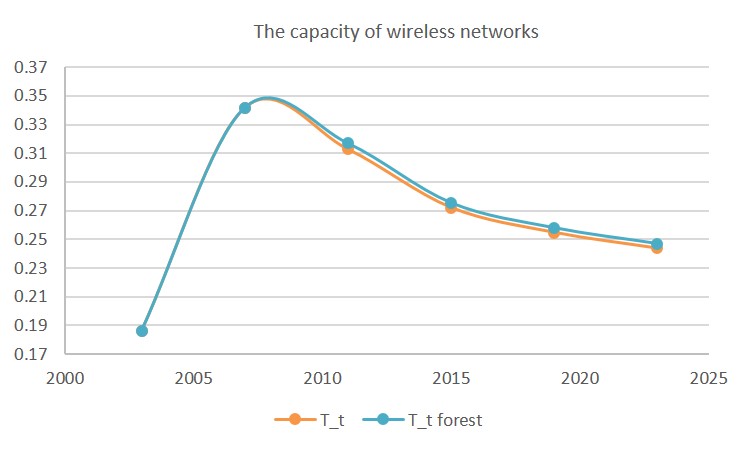}
    \end{minipage}
    \end{subfigure}
\caption{wireless network group: knowledge temperature evolution before and after forest helping}
\label{fig: wireless-group}
\end{figure}

The skeleton tree of topic led by `Critical Power for Asymptotic Connectivity in Wireless Networks' (CPACWN) reveals an indisputably intimate relation between the itself and the topic led by `The capacity of wireless networks' (CWN) (Fig. \ref{fig:62270017-2020}). Being the most prominent child paper of CPACWN, CWN substantially extended CPACWN's ideas and founded a new research focus. Its crucial role in topic's prosperity is also reflected by its high popularity and influence within the topic: it jointly inspired one third of the topic members, most of which were published during the flourishing period. Their similar knowledge temperature evolution also confirms their closeness. During forest helping, CPACWN's topic donated some of its heat to CWN's topic in early days. This behavior models the promotion effect brought by CPACWN's increasing impact and popularity. However, this did not help CWN's topic much because it had already a much bigger size. After the adjustment, their knowledge temperature evolution is more similar than before. Both topics were hottest in 2007 and 2008 (Fig. \ref{fig: wireless-group}). This corresponds better with their individual development and inherent connection. In fact, CMN achieved such a huge success that it took over its predecessor to be the new authority in their domain in just a few years. The dominating size of CWN's topic clearly makes it a better representative of background popularity and impact, which usually has a big influence on similar smaller topics. Therefore, the destiny of CPACWN's topic is to some extent determined by the development of CMN's topic. The rise-and-fall OF CWN's topic is thus an indicator of CPACWN's topic's flourishing.\\

\subsubsection{RNN gated unit group}

\begin{figure}[htbp]
\centering
    \begin{subfigure}{\textwidth}
    \begin{minipage}{0.5\textwidth}
      \centering
    \includegraphics[width = 0.85\linewidth]{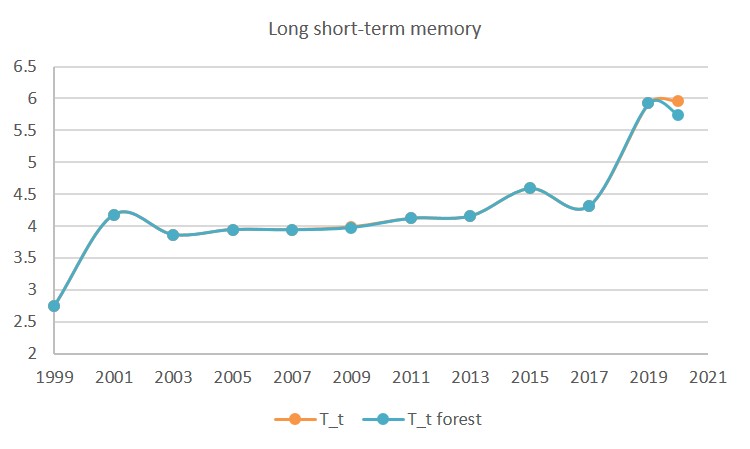}
    \end{minipage}
    \begin{minipage}{0.5\textwidth}
    \centering
    \includegraphics[width = 0.85\linewidth]{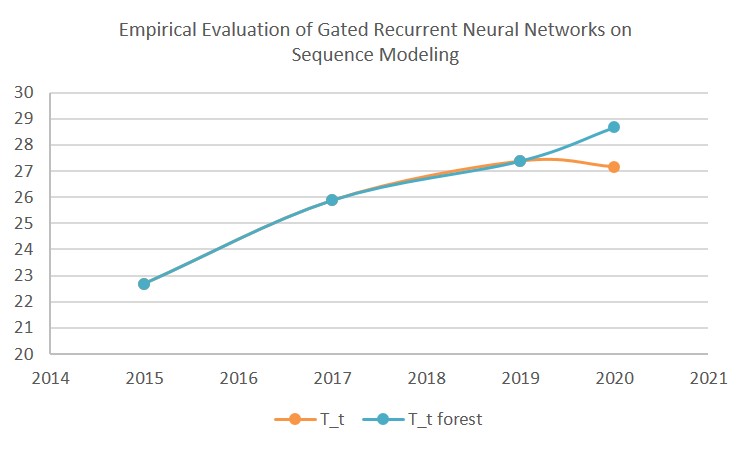}
    \end{minipage}
    \end{subfigure}
\caption{RNN gated unit group: knowledge temperature evolution before and after forest helping}
\label{fig: gated-unit-group}
\end{figure}

`Empirical Evaluation of Gated Recurrent Neural Networks on Sequence Modeling' (GRU) introduced a new research focus and made non-trivial contribution to the recent thriving of topic led by `Long short-term memory' (LSTM) (Fig. \ref{fig:56158074-2020}). In fact, nearly half of the papers that cite GRU also cite LSTM. Over the past 3 years, LSTM's topic has had a substantial development and a fast-growing impact and popularity thanks to a large number of new publications. In comparison, GRU's topic has shown signs of stagnation shortly after its initial glory. Today, the phenomenal size of LSTM's topic qualifies LSTM's authority claim in the domain. As a result, the prosperity of LSTM's topic is a nice representative of background popularity and impact, which usually has a big influence on similar smaller topics. While GRU helped with the flourishing of LSTM's topic in its early days, it is now LSTM's topic's turn to help maintain the heat-level of GRU's topic (Fig. \ref{fig: gated-unit-group}). A soaring background popularity and impact is favorable for GRU's topic future development, at least in a short term. For this topic group, the forest helping is just like the mechanism that we observe in the real nature: mother tree shares nutrients with its child trees so as to give them a better chance of survival.\\

\subsubsection{word embedding group}

\begin{figure}[htbp]
\centering
    \begin{subfigure}{\textwidth}
    \begin{minipage}{0.5\textwidth}
     \centering
    \includegraphics[width = 0.85\linewidth]{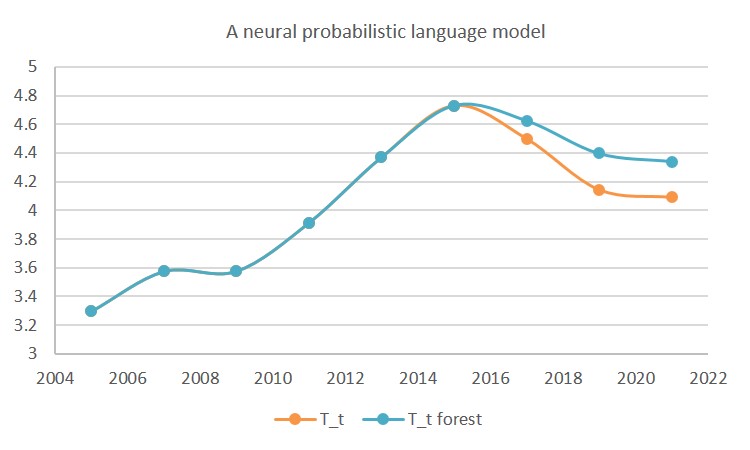}
    \end{minipage}
    \begin{minipage}{0.5\textwidth}
    \centering
    \includegraphics[width = 0.85\linewidth]{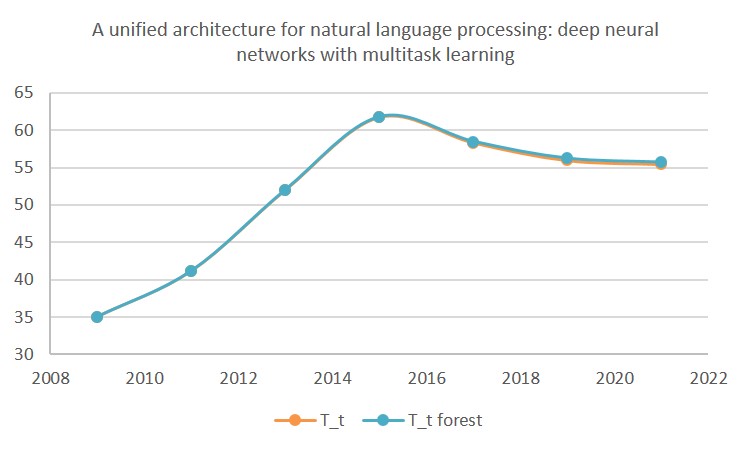}
    \end{minipage}
    \end{subfigure}

    \vspace{1mm}

     \begin{subfigure}{0.5\textwidth}
     \centering
    \includegraphics[width = 0.85\linewidth]{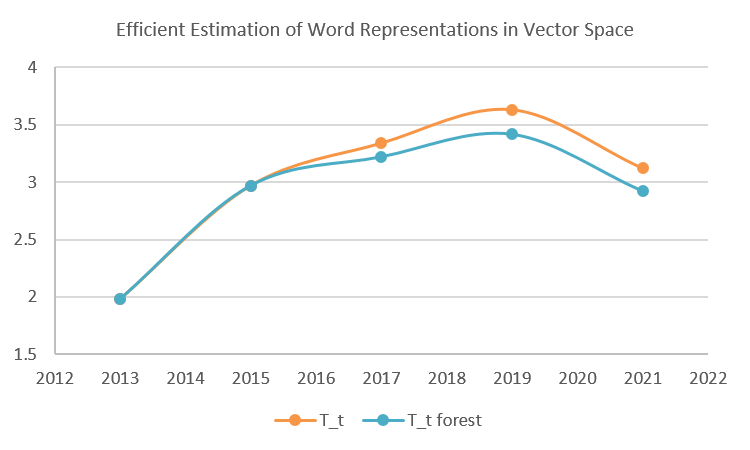}
    \end{subfigure}
\caption{word embedding group: knowledge temperature evolution before and after forest helping}
\label{fig: word-embedding-group}
\end{figure}

`Efficient Estimation of Word Representations in Vector Space' (EEWRVS) is the most influential child paper in both topics respectively led by `A neural probabilistic language model' (NPLM) and 'A unified architecture for natural language processing: deep neural networks with multitask learning' (UANLP). Furthermore, EEWRVS's topic is more than twice the size of NPLM's and UANLP's. EEWRVS has outperformed its parents and has established authority in this research field. The considerable size of EEWRVS's topic makes it a nice representation of background popularity and impact, which has an influence on smaller topics within the research field. Owing to its close relationship with NPLM's topic and UANLP's topic, the booming of EEWRVS's topic more or less increases their visibility and attracts research attention. Through forest helping, the "energy" from EEWRVS's topic slows down the perishing of NPLM's topic and UANLP's topic (Fig. \ref{fig: word-embedding-group}). The heat exchange models the boosting effect of the background, a bigger research field where the 3 belong to. \\

